%% file: BP_plus_MWPM.tex
\documentclass[accepted=2018-09-21,a4paper,onecolumn,superscriptaddress,12pt]{quantumarticle}
\pdfoutput=1
\usepackage{algpseudocode}
\usepackage[utf8]{inputenc}
\usepackage[english]{babel}
\usepackage[T1]{fontenc}
\usepackage{amsmath}
\usepackage{bbold}
\usepackage{hyperref}
\usepackage[framemethod=tikz]{mdframed}
\usepackage[numbers,sort&compress]{natbib}
\usepackage{nicefrac}
\usepackage{physics}
\usepackage{tikz}
\usepackage{pgfplots}
\usetikzlibrary{arrows.meta}

\usepackage[capitalise, noabbrev]{cleveref}

\providecommand{\id}{\hat{\mathbb{1}}}

\newlength\figureheight
\newlength\figurewidth
\pgfplotsset{compat=1.14, 
             every axis/.append style={
                    tick label style={/pgf/number format/fixed}
                    }}

\begin{document}
\title{Multi-path Summation for Decoding 2D\\Topological Codes}
\date{\today}
\author{Ben Criger}
\email{bcriger@gmail.com}
\affiliation{QuTech, TU Delft}
\affiliation{Institute for Globally Distributed Open Research and Education (IGDORE)}
\author{Imran Ashraf}
\affiliation{QuTech, TU Delft}

\begin{abstract}
Fault tolerance is a prerequisite for scalable quantum computing.
Architectures based on 2D topological codes are effective for near-term implementations of fault tolerance.
To obtain high performance with these architectures, we require a decoder which can adapt to the wide variety of error models present in experiments. 
The typical approach to the problem of decoding the surface code is to reduce it to minimum-weight perfect matching in a way that provides a suboptimal threshold error rate, and is specialized to correct a specific error model. 
Recently, optimal threshold error rates for a variety of error models have been obtained by methods which do not use minimum-weight perfect matching, showing that such thresholds can be achieved in polynomial time. 
It is an open question whether these results can also be achieved by minimum-weight perfect matching.
In this work, we use belief propagation and a novel algorithm for producing edge weights to increase the utility of minimum-weight perfect matching for decoding surface codes. 
This allows us to correct depolarizing errors using the rotated surface code, obtaining a threshold of $17.76 \pm 0.02 \%$.
This is larger than the threshold achieved by previous matching-based decoders ($14.88 \pm 0.02 \%$), though still below the known upper bound of $\sim 18.9\%$.
\end{abstract}

\section{Introduction}
Quantum information processing (QIP) provides a means of implementing certain algorithms with far lower memory and processing requirements than is possible in classical computing \cite{shor_factoring, ham_simulation}.
The analog nature of quantum operations suggests that devices which perform QIP will have to function correctly in the presence of small deviations between ideal operations and those that are actually implemented, a property called \emph{fault tolerance} \cite{gottesman, early_shor_steane}. 
In addition, many external sources of error exist in current QIP devices, making fault tolerance a practical necessity as well as a theoretical one.

In order to design fault-tolerant QIP protocols, we first require a set of quantum states for which the effect of small deviations can be reversed, called a \emph{quantum error-correcting code} \cite{early_qec}, analogous to \emph{classical} error-correcting codes \cite{macwilliams_sloane}.
One important difference between classical and quantum error correction is that quantum states are inherently continuous, whereas classical states are discrete. 
This suggests that noise processes on quantum computers will also be continuous, and therefore difficult to correct. 
However, the action of a projective measurement can effectively discretise the noise \cite[Chapter 2]{lidar_brun}, bringing the study of quantum error correction closer to classical coding theory. 
Unfortunately, the action of a projective measurement also ensures that the qubits used in a quantum code cannot be measured directly in order to determine where errors have occurred, since such a measurement would destroy any superposition of code states, discretising the logical state as well as the noise.
However, a large set of quantum error-correcting codes, called \emph{stabiliser codes} \cite{gottesman}, allow the measurement of multi-qubit operators which yield results depending only on which error has occurred, and not on the code state itself. 
These operators are the \emph{stabilisers} of the code, and the result from their measurement is called the \emph{syndrome}.
This syndrome indicates which stabilisers anticommute with the error. 

Typically, the stabilisers are taken to be \emph{Pauli operators} on $n$ qubits, tensor products of the operators $\id$, $X$, $Y$, and $Z$, where
\begin{equation}
\id = \begin{bmatrix}
1 & 0 \\ 0 & 1
\end{bmatrix}, \quad
X = \begin{bmatrix}
0 & 1 \\ 1 & 0
\end{bmatrix}, \quad
Y = \begin{bmatrix}
0 & -i \\ i & 0
\end{bmatrix}, \,\, \textrm{and} \,\,
Z = \begin{bmatrix}
1 & 0 \\ 0 & -1
\end{bmatrix}.
\end{equation}
Conveniently, noise discretisation implies that if a code can be used to correct the random application of these operators on a subset of the qubits, it can also correct arbitrary operators on the same subset, since the Pauli operators form a basis for the space of operators. 

When measuring the stabilisers, it is possible for the measurement apparatus to return an incorrect eigenvalue, or for the measurement procedure to produce errors on the qubits. 
In order to obtain fault tolerance in such a scenario, it is necessary to repeat the stabiliser measurements many times.
In this work, we restrict ourselves to a simpler model, in which Pauli errors occur at random on data qubits, and the measured eigenvalues are correct, and obtained without consequence.

In order to correct these Pauli errors, it is necessary to derive an error which reproduces the given syndrome, and is unlikely to alter the intended logical state if performed on the system. 
This process, called \emph{decoding}, is non-trivial, since a code with $n$ physical qubits will typically contain $\mathcal{O}(n)$ stabilisers, so the number of possible syndromes scales as $2^n$, prohibiting the use of a simple lookup table.
Also, many quantum codes are \emph{degenerate}, having multiple errors corresponding to a given syndrome, further complicating the process of decoding.

The purpose of this work is to improve the accuracy of a frequently-used method for decoding 2D \emph{homological codes} \cite{kitaev}.
The prototype for these codes is the 2D \emph{surface code}, which we review in the following section, along with the reduction of the decoding problem to minimum-weight perfect matching.
In \cref{sec:prior_work}, we review prior enhancements to the decoding algorithm of interest, as well as alternative algorithms. 
In \cref{sec:odds_calc,sec:bp}, we introduce the methods we propose to enhance the accuracy of surface code decoding and obtain increased threshold error probabilities. 
We discuss the effect of changing boundary conditions, as well as the additional complexity of decoding using these methods, and conclude in \cref{sec:d_and_c}.

 
\section{The Surface Code}

\begin{figure}[!ht]
\centering
\input{dist_9_surface_code.tex}
\caption{Surface code with distance $d = 9$.
Data qubits are placed at the vertices of a $d$-by-$d$ square tiling.
Logical operators can be placed along the sides of the square. 
Stabilisers of the form $ZZ$/$ZZZZ$ are supported on grey tiles, and stabilisers of the form $XX$/$XXXX$ are supported on white tiles. 
An $X$ or $Z$ error will anticommute with a stabiliser if the stabiliser is at an endpoint of a path of Pauli operators which is part of the error.
A $Y$ error is detected by both types of stabilisers.}
\label{fig:surface_code}
\end{figure}
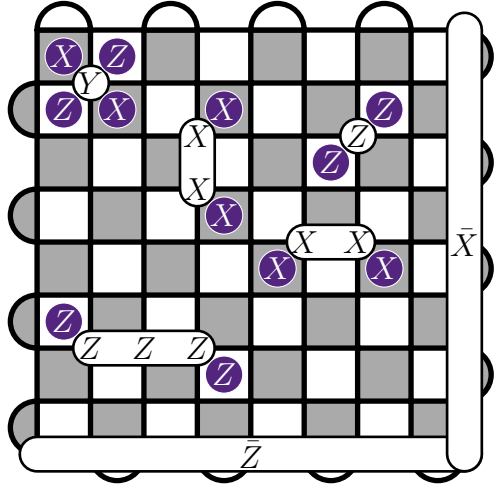

A surface code \cite{freedman_meyer, bravyi_kitaev, bombin_07} is supported on a square array of qubits, with stabilisers defined on individual square tiles, see \cref{fig:surface_code}. 
Specifically, we focus on the \emph{rotated} surface code, with the boundary conditions from \cref{fig:surface_code}. Earlier constructions involve tiling a surface with non-trivial topology, such as a torus \cite{kitaev}, or using an alternate open boundary condition with \emph{smooth} and \emph{rough} boundaries \cite{bravyi_kitaev}. 
We choose to study the rotated code, since it requires the fewest physical qubits per logical qubit. 

\begin{figure}[!ht]
\centering
\input{manhattan_distance.tex}
\caption{A pair of vertices separated by an edge whose minimum length is 5. 
A minimum-length path is composed of steps between neighbouring tiles, and any path which is consistent with the indicated direction will be minimum-length.}
\label{fig:manhattan_distance}
\end{figure}
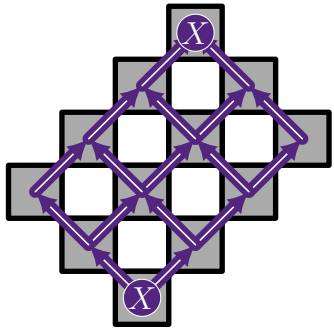

In any of these codes, pairs of stabilisers anticommute with chains of errors that connect the two tiles supporting the stabilisers, producing a set of locations where the syndrome is 1 (which we will call \emph{vertices} for the remainder of this work).
This mathematical structure reduces the decoding problem to that of finding a likely set of connecting errors which join the vertices in pairs (we will call these connecting errors \emph{paths} for the remainder of this work, and refer to abstract connections between vertices as \emph{edges}). 
To maximise the likelihood, we follow the derivation of \cite[Section IV D]{dklp}.
We begin with the assumption that only one type of error ($X$ or $Z$) can occur, for simplicity. The likelihood of an error $E$ is then a function of its weight $w(E)$ (the size of its support), the total number of qubits $n$, and the underlying probability of error $p$:
\begin{equation}
p(E) = p^{w(E)} (1 - p)^{n - w(E)} = (1 - p)^n \left( \frac{p}{1 - p} \right)^{w(E)}
\end{equation}
We assume that, if two vertices are connected by a path, the length of that path is minimal. 
This implies that the weight of the corresponding edge is given by the rotated Manhattan distance between the vertices it connects, see \cref{fig:manhattan_distance}.
Note that, due to the open boundary conditions, a single vertex can be connected to the boundary instead of another vertex. 
This is typically remedied by introducing \emph{virtual vertices} which are connected to each other with weight-0 edges, and to a designated vertex with a weight equal to the minimum path length between that vertex and the nearest boundary \cite{virtual_vertices}.

The total weight of an error is then the sum of the weights of the edges, so the likelihood can be expressed as a product over the edge set:
\begin{equation}
p(E) = (1 - p)^n \prod_{e \in \mathrm{edges}(E)} \left( \frac{p}{1 - p} \right)^{\abs{e}} \label{eq:original_likelihood}
\end{equation}
To maximise the likelihood, it suffices to maximise any function proportional to the likelihood, so we can use a simplified cost function:
\begin{flalign}
p(E) & \propto \prod_{e \in \mathrm{edges}(E)} \left( \frac{p}{1 - p} \right)^{\abs{e}} \\
 & \sim \sum_{e \in \mathrm{edges}(E)} \abs{e} \ln \left( \frac{p}{1 - p} \right) \\
 & \propto -\sum_{e \in \mathrm{edges}(E)} \abs{e} \label{eq:iidxz_obj_fun}
\end{flalign}
with the minus sign present since the \emph{odds} of a physical error $\nicefrac{p}{1-p} \ll 1$. 
This implies that, in order to maximise the likelihood of a prospective error, we should find a set of edges that connects the vertices in pairs (a \emph{perfect matching} in graph theory) with minimal weight. 
Finding minimal-weight perfect matchings is a well-known combinatorial problem, solved using the \emph{Blossom} algorithm \cite{edmonds, kolmogorov}. 

Decoding with this algorithm results in a threshold error rate of $\sim 10.3\%$ \cite{wang_harrington_preskill}, compared with a threshold rate of $\sim 10.9\%$ estimated using a correspondence between the toric code decoding threshold and the phase transition in a spin model \cite{PhysRevB.65.054425, PhysRevLett.87.047201}.
From this, we surmise that minimum-weight perfect matching (MWPM) decoding performs well, though not optimally, in the event that errors of a single type are distributed with identical probability on each qubit. 
In addition, there remain many scenarios in which the independent identically-distributed (IID) error model does not apply. 
In the following section, we summarise the current efforts to enhance MWPM decoding with both IID and more realistic models in mind.
\section{Prior Work}
\label{sec:prior_work}
\begin{figure}[!ht]
\centering
\input{yy_error.tex} \qquad
\input{iidxz_x_error.tex} \qquad
\input{iidxz_z_error.tex}
\caption{From left to right: a weight-2 $Y$ error acting on a distance-3 surface code, and the independent minimum-weight $X$ and $Z$ corrections.
The appropriate correction, $XX \cdot ZZ$, is not minimum-weight in the independent error model.}
\label{fig:yy_fail}
\end{figure}
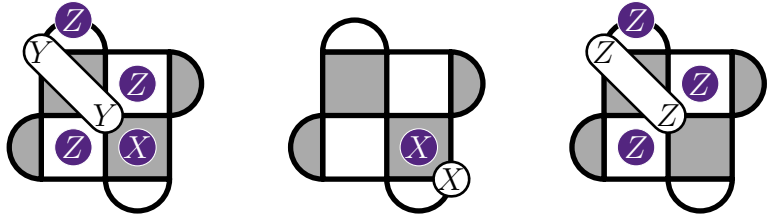

The derivation of the objective function in \cref{eq:iidxz_obj_fun} assumed that $X$ and $Z$ errors occur independently, with identical probability on each qubit. 
This implies that, if the probability of an $X$ or $Z$ error is $O(p)$, the probability of a $Y$ error is $O\left(p^2 \right)$, since $Y \sim XZ$. 
If the probabilities of $X$, $Y$, and $Z$ errors differ from this, the performance of the decoder will be decreased. 
Consider the frequently used example of \emph{depolarizing} noise, in which each type of error is applied with equal probability. 
A decoder which minimises the weight of $X$ and $Z$ errors separately may fail to minimise the relevant weight, resulting in a logical error, see \cref{fig:yy_fail}. 

To minimise the appropriate weight function, we need to avoid `double-counting' the weight of $Y$ errors. 
To accomplish this, Delfosse \cite{delfosse} and Fowler \cite{fowler_correlated_errors} reduce the weight of $X$ edges that cross $Z$ edges, since individual $Y$ errors result in crossed $X$ and $Z$ edges (see \cref{fig:surface_code}, upper left).
In order to determine where these crossings occur, the authors first calculate one of the matchings, $X$ or $Z$, then use the resulting edge set to determine edge costs for the other matching. 
In this way, Delfosse demonstrates that the threshold error rate against $Z$ errors can be raised for a homological code which has a naturally larger tolerance to $X$ errors. 

\begin{figure}[!ht]
\centering
\input{bbox_single.tex}
\qquad
\input{bbox_intersection.tex}
\caption{For a continuous chain of $Y$ errors, different minimum-weight $X$ and $Z$ paths can be found with different crossing numbers for the same syndrome.}
\label{fig:bbox_intersection}
\end{figure}
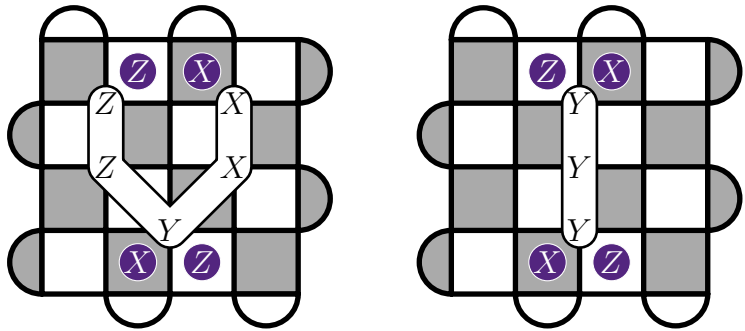

One obstacle for this approach is that $X$ and $Z$ paths can intersect more than once, see \cref{fig:bbox_intersection}.
It is not clear during decoding which error corresponding to a given edge should be selected in order to maximise the crossing number, thus minimising the total weight. 
Furthermore, there can be more than one minimum-weight perfect matching corresponding to a vertex set, further hampering any effort to obtain an accurate correction to the depolarizing channel using MWPM decoding.
Accounting for degeneracy caused by edge and matching multiplicity, it appears, will provide insight in this direction.

\begin{figure}[!ht]
\centering
\input{multiple_paths.tex}
\caption{Syndrome set with a single minimum-weight matching (weight 5) and 16 matchings with a single unit of excess weight. 
If the odds of an error $\nicefrac{p}{1-p} > \nicefrac{1}{16}$, then the higher-weight matching is more likely, see \cref{eq:StaceBarrettLikelihood}.}
\label{fig:multiple_paths}
\end{figure}
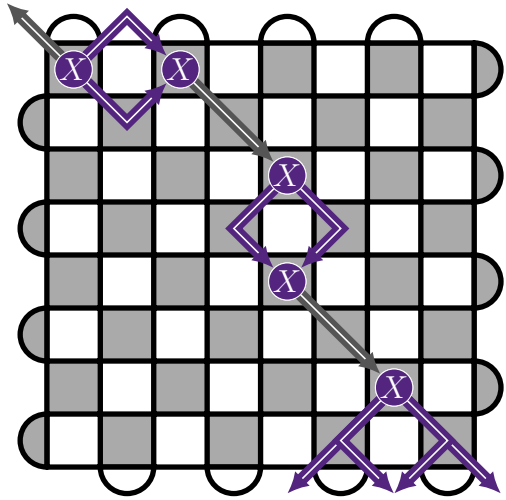

There have been multiple efforts to address these types of degeneracy. 
The first is due to Barrett, Stace and Doherty \cite{stace_barrett_09, stace_barrett, stace_barrett_pra}, and uses the IID error model.
They derive a modified objective function, accounting for the fact that there are multiple error configurations with the same weight corresponding to a given matching, see \cref{fig:multiple_paths}. 
If this number of minimum-weight configurations corresponding to the error $E$ is labeled $\Omega(E)$, then the derivation beginning with \cref{eq:original_likelihood} instead begins with the likelihood function
\begin{equation}
\Omega(E) (1 - p)^n \left( \frac{p}{1-p} \right)^{w(E)}.
\label{eq:StaceBarrettLikelihood}
\end{equation}
In order to use an MWPM-based decoder, it is necessary to express the likelihood function as a product over the set of edges. 
This requires us to express $\Omega(E)$ as a product over the set of edges. 
This is possible, since the number of paths joining a pair of vertices depends only on the positions of those vertices, and not on any property of the graph not related to the edge under consideration. 
See, for example, \cref{fig:multiple_paths}, in which the total number of weight-6 errors (16) is a product of the number of paths joining the paired vertices (2, 2, and 4). 
The probability of a minimum-weight error equivalent to $E$ can then be factored as before:
\begin{equation}
p(E) \propto \prod_{e \in \mathrm{edges}(E)} \Omega(e) \left( \frac{p}{1 - p} \right)^{\abs{e}}
\end{equation}
Taking the logarithm results in the final form of the objective function:
\begin{equation}
p(E) \propto - \sum_{e \in \mathrm{edges}(E)} \abs{e} - \frac{\ln \Omega(e)}{\ln \nicefrac{1 - p}{p}}.
\end{equation}
In order to calculate $\Omega(e)$ for an edge between two real vertices, Barrett/Stace/Doherty note that there are $\binom{\Delta_y + \Delta_x}{\Delta_x}$ paths between points on a rotated square tiling of the torus separated by $(\Delta_x, \Delta_y)$ (see also \cite{euler_problem_15}).

Using this modification to the edge weights, MWPM decoding results in a threshold near $10.3\%$ when correcting IID $X$/$Z$ noise.
This is similar to the increase in the threshold that can be obtained from solving an associated statistical physics model using matchings obtained from Metropolis sampling \cite{freund_grassberger}, which suggests a connection between the two methods. 
In addition, efforts have been made to improve the threshold of MWPM decoders by constraining the input graph to support errors from a specific coset, and by modifying the obtained solutions through Markov Chain Monte Carlo \cite{hutter_wootton_loss}.
However, the thresholds from modified MWPM decoding are still suboptimal, suggesting that further improvements are required. 

In a recent departure from MWPM-based decoding, Bravyi, et al. \cite{bravyi_decoders} have introduced polynomial-time maximum-likelihood decoders for surface codes which are based on efficient sub-theories of quantum mechanics. 
For IID $X$/$Z$ noise, they begin by matching each vertex to an arbitrary boundary, producing a correction which eliminates the syndrome, but results in a logical error with high likelihood (also known as a \emph{pure error} \cite{pure_errors}). 
They then provide a reduction from the problem of determining the probability that the pure error is an accurate correction to the calculation of the output of a \emph{matchgate circuit} \cite{valiant, bravyi_flo}.
For arbitrary stochastic Pauli noise, likelihood estimation reduces instead to the problem of contracting a tensor network state \cite{peps_mps}, which can also be solved in polynomial time, reminiscent of earlier decoders for topological codes which use the Renormalisation Group \cite{rg, hutter_loss_wootton}. 
Using these techniques, Bravyi et al. are able to obtain optimal thresholds against IID $X$/$Z$ and depolarizing noise, opening a gap between MWPM and maximum-likelihood decoding. 

This gap can be narrowed by introducing a generalised edge weight calculation, similar to the method of Barrett/Stace/Doherty, adapted to the case in which there are different probabilities of error on each qubit. 
We do this in the following section. 
\section{Odds Calculation}
\label{sec:odds_calc}
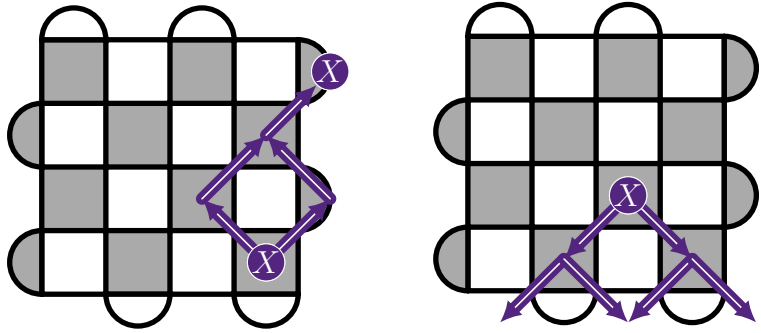
\begin{figure}[!ht]
\centering
\input{clipped_bbox.tex}
\qquad
\input{multiple_endpoints.tex}
\caption{Two cases in which the number of paths is not $\binom{\Delta_y + \Delta_x}{\Delta_x}$, due to open boundary conditions. 
Left: one otherwise-possible path between two syndromes cannot be realised since two of the qubits on the path are outside the lattice. 
Right (also see \cref{fig:multiple_paths}): Paths from a syndrome to the boundary can exit the lattice at multiple points, further preventing a simple calculation of $\Omega_e$.}
\label{fig:open_bc_cases}
\end{figure}

\begin{figure}[!ht]
\centering
\input{paper_iidxz_no_bp.tex}\\
\input{paper_iidxz_ext.tex}
\caption{Threshold of the rotated surface code correcting IID $X$/$Z$ errors. 
Above: Edge weights for MWPM are derived using the Manhattan distance, resulting threshold is $9.97 \pm 0.01 \%$, lower than the value of $10.3\%$ obtained using alternate boundary conditions in \cite{surface103}.
Below: Edge weights for MWPM are derived using path-counting, resulting in a threshold of $10.34 \pm 0.01 \%$, compensating for the disadvantage imposed by the rotated boundary conditions.
Error bars represent a $99\%$ confidence interval.
Solid lines are maximum-likelihood fits to data near the threshold.}
\label{fig:path_counting_threshold}
\end{figure}
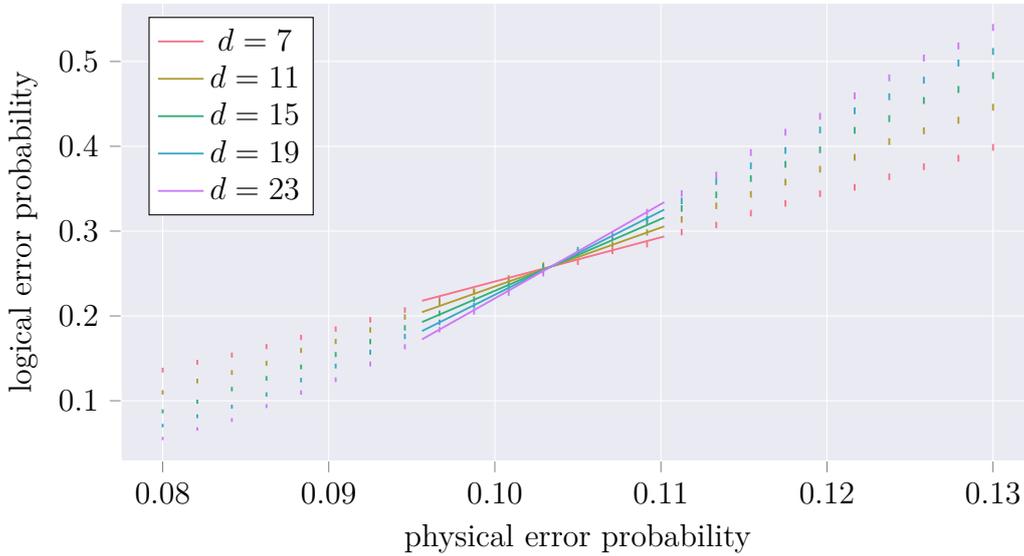

In order to more accurately approximate the odds of a path existing between two vertices, we first generalise the calculation of Barrett/Stace/Doherty to cases where a vertex is being matched to the boundary, or to where some paths cannot be realised, see \cref{fig:open_bc_cases}.
To count the number of paths on \emph{directed acyclic graphs} such as these, we can use the following function:
\begin{mdframed}[linecolor=quantumviolet, linewidth=2pt]
\begin{algorithmic}
\Function{num\_paths}{$g$, $v_i$, $F$} \Comment{$g$ directed acyclic graph, $v_i$ initial vertex, $F$ final vertex set};
\ForAll{$v \in$ \texttt{vertices}(g)}
\State{$\texttt{count}[ v ] \gets 0$}
\EndFor 
\State{$\texttt{count}[ v_i ] \gets 1$}
\While{\texttt{true}}
\If{$\texttt{all}(\texttt{count}[ f ] \neq 0 \,\, \texttt{for} \,\, f \in F)$}
\Return{$\texttt{sum}(\texttt{count}[ f ] \,\, \texttt{for} \,\, f \in F)$}
\EndIf
\ForAll{$v \in$ \texttt{vertices}(g)}
\If{$\texttt{all}(\texttt{count}[ i ] \neq 0 \,\, \texttt{for} \,\, i \in \texttt{incoming}(v))$}
\State{$\texttt{count}[v] \gets \texttt{sum}(\texttt{count}[ i ] \,\, \texttt{for} \,\, i \in \texttt{incoming}(v))$}
\EndIf
\EndFor
\EndWhile
\EndFunction
\end{algorithmic}
\end{mdframed}
If this function is applied to a closed bounding box (see \cref{fig:manhattan_distance}), the result will be $\binom{\Delta_y + \Delta_x}{\Delta_x}$, which simplifies the calculation for periodic boundary conditions, for which all bounding boxes are closed. 
For a generic directed acyclic graph, this function requires one operation for every edge of the input graph, and can be parallelized using \emph{message passing} \cite[Chapter 16]{mackay}, reducing the runtime to be proportional to the path length using one constant-size processor per vertex of the input graph. 
To be implemented serially, the input graph $g$ should first be \emph{topologically sorted} \cite[Section 22.4]{cormen2009introduction}, also requiring a number of operations at most linear in the graph's size. 
The key to the derivation of the path-counting algorithm is that the number of paths reaching a vertex $w$ is the sum of the number of paths reaching the vertices with edges incoming to $w$. 
Using this function to calculate $\Omega_e$, we can replicate the result of Barrett/Stace/Doherty using the rotated surface code, see \cref{fig:path_counting_threshold}.

To generalise to the case where $Y$ errors can occur with arbitrary probability, we first consider the role of detected $Z$ errors in determining where $X$ errors have occurred (see \cite{delfosse} for an in-depth discussion).
We assume that the physical error model has a fixed probability of $X$, $Y$, or $Z$ on each qubit, which is qubit-independent.
To see the role that detected $Z$ errors play, we imagine that we could determine with certainty the positions of errors that anticommute with the $Z$-detecting stabilisers. 
These errors must be either $Z$ or $Y$, and the qubits which have not been affected by one of these errors must have been affected by an $X$ error, if at all.
To use this information in determining the positions of $X$ errors, we use the \emph{conditional} probabilities of an $X$ error occurring, depending on the result of the $Z$ decoding:
\begin{equation}
p\left(X\textrm{ or }Y \middle \vert Y\textrm{ or }Z \right) = \dfrac{p_Y}{p_Y + p_Z}, \qquad
p\left(X\textrm{ or }Y \middle \vert \id\textrm{ or }X \right) = \dfrac{p_X}{p_X + p_{\id}}
\end{equation}
Thus, we see that attempting to account for $X$/$Z$ correlations in the error model will require us to consider error models in which there are different probabilities of error on each qubit, even when the underlying error model is qubit-independent.

The method we use to approximate conditional probabilities of error in the absence of perfect detection is explained in the following section.
For now, we generalise the derivation of the likelihood of an error beginning in \cref{eq:original_likelihood}, with a few approximations and definitions:
\begin{itemize}
\item We only consider minimum-length paths, which are restricted to the set of qubits within bounding boxes of the type seen in \cref{fig:manhattan_distance,fig:open_bc_cases}.
\item We consider two possibilities, either the vertices being considered are joined by a single path, or none of the qubits in the bounding box are in error.
\end{itemize}
The probability $p(E)$ (now referring to the probability of any error equivalent to $E$ under a change of minimum-length path) can then be expressed in terms of the odds of error on a given qubit, $o_q \equiv \frac{p_q}{1 - p_q}$:
\begin{flalign}
p(E) &= \prod_{q} (1 - p_q) \times \prod_{e \in \mathrm{edges}(E)} \sum_{p \in \mathrm{paths}(e)} \prod_{q \in p} o_q \\
&\sim \sum_{e \in \mathrm{edges}(E)} \log \left( \sum_{p \in \mathrm{paths}(e)} \prod_{q \in p} o_q \right)
\end{flalign}
We can divide the set $\mathrm{paths}(e)$ into disjoint subsets that pass through each of the vertices adjacent to the final vertex, assuming that there is only one (otherwise, we can divide the set $\mathrm{paths}(e)$ into disjoint subsets associated with individual final vertices, then proceed).
The path from each of these \emph{predecessor} vertices to the final vertex only traverses a single qubit $q_p$, so:
\begin{equation}
\sum_{p \in \mathrm{paths}(e)} \prod_{q \in p} o_q = \sum_{q_p} o_{q_p} \sum_{p' \in \mathrm{paths}(q_p)} \prod_{q \in p'} o_q,
\end{equation} 
where $\mathrm{paths}(q_p)$ is the set of paths leading from the initial vertex to the predecessor vertex $q_p$. 
This reduction is similar to the sum reduction used in path-counting, though now the sum is weighted by $o_{q_p}$.
The path-counting function can be easily modified to evaluate this quantity:
\begin{mdframed}[linecolor=quantumviolet, linewidth=2pt]
\begin{algorithmic}
\Function{path\_sum}{$g$, $v_i$, $F$, \texttt{o\_q}} \Comment{\texttt{o\_q} $o_q$ associated with edges $(v, v')$ of $g$};
\ForAll{$v \in$ \texttt{vertices}(g)}
\State{$\texttt{odds}[ v ] \gets 0$}
\EndFor 
\State{$\texttt{odds}[ v_i ] \gets 1$}
\While{\texttt{true}}
\If{$\texttt{all}(\texttt{odds}[ f ] \neq 0 \,\, \texttt{for} \,\, f \in F)$}
\Return{$\texttt{sum}(\texttt{odds}[ f ] \,\, \texttt{for} \,\, f \in F)$}
\EndIf
\ForAll{$v \in$ \texttt{vertices}(g)}
\If{$\texttt{all}(\texttt{odds}[ i ] \neq 0 \,\, \texttt{for} \,\, i \in \texttt{incoming}(v))$}
\State{$\texttt{odds}[v] \gets \texttt{sum}(\texttt{o\_q}[(i, v)] \ast \texttt{odds}[ i ] \,\, \texttt{for} \,\, i \in \texttt{incoming}(v))$}
\EndIf
\EndFor
\EndWhile
\EndFunction
\end{algorithmic}
\end{mdframed}
This produces an approximate cost function, since we have neglected the possibility that there is an error inside the bounding box that does not cause a syndrome (a randomly applied stabiliser), and we have failed to account for the large number of less likely matchings that are equivalent to the likeliest matchings up to a stabiliser. 
Nevertheless, we will see that surface code decoding can be improved, once we have an accurate set of estimates for the marginal probability of each type of error on each qubit. 
In the following section, we review belief propagation, a method which can provide these estimates.
\section{Belief Propagation}
\label{sec:bp}
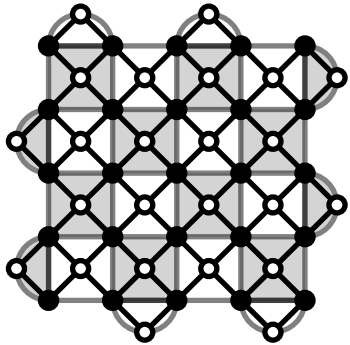
\begin{figure}[!ht]
\centering
\input{tanner_graph.tex}
\caption{Tanner graph for the $d=5$ surface code.
Qubit vertices are shown in black, stabiliser vertices are shown in white. 
Edges exist between stabiliser vertices and the qubit vertices in the support of the corresponding stabiliser. 
}
\label{fig:tanner_graph}
\end{figure}

Belief propagation (BP) is a message-passing algorithm for calculating marginal probabilities, which has been applied to decoding quantum codes correcting Pauli noise \cite{poulin_bp}.
To use BP for decoding, we first define the \emph{Tanner graph} corresponding to a surface code, which contains a \emph{qubit vertex} for every qubit in the code, and a \emph{check vertex} for every stabiliser check. 
An edge exists between each check vertex and each qubit vertex in its support, see \cref{fig:tanner_graph}.

The algorithm begins with a probability distribution $p(E_q)$ (also called a \emph{belief}), assigned to each qubit $q$, of the form $\left[ 1 - (p_Z + p_X + p_Y), \, p_Z, \, p_X, \, p_Y \right]$, equal to the \emph{prior distribution} of error probabilities on each qubit (called `the physical distribution' earlier).  
These beliefs are passed as messages to each of the neighbouring check vertices. 
Each check vertex $c$ calculates messages to be passed back to each of its neighbouring qubit vertices, one for each possible error $E_q$ on the qubit:
\begin{equation}
m_{c \rightarrow q}(E_q) \propto \sum_{E_{q'}, \, q'\in \mathrm{supp}(c) \backslash q } \left( \delta_{\mathrm{synd}_c, S_c \cdot E_c} \prod_{q' \in \mathrm{supp}(c) \backslash q} m_{q' \rightarrow c}(E_{q'}) \right).
\end{equation}
Here, proportionality indicates that the messages are to be normalized after they are calculated. 
Each qubit vertex then calculates a set of messages to be passed back:
\begin{equation}
m_{q  \rightarrow c}(E_q) \propto  p(E_q) \prod_{c' \in \mathrm{supp}(q) \backslash c} m_{c' \rightarrow q}(E_q),
\end{equation}
where proportionality again indicates that the messages are to be normalized. 

\begin{figure}[!ht]
\centering
\begin{minipage}[c][1.1\figureheight][c]{0.2\textwidth}
\input{x_upper_left.tex}\\
\input{x_top_middle.tex}
\vspace{1cm}
\end{minipage}
\begin{minipage}[c][1.1\figureheight][c]{0.79\textwidth}
\input{failed_beliefs.tex}
\end{minipage}
\caption{Left: Two possible errors which result in a split belief.
Right: Convergence of beliefs when the true error is a weight-one $X$ error on the upper-left qubit, the underlying error model has probabilities $\left[0.97,\,0.01,\,0.01,\,0.01 \right]$.
A nearly-identical set of beliefs is supported on the neighbouring qubit.}
\label{fig:bp_failure}
\end{figure}
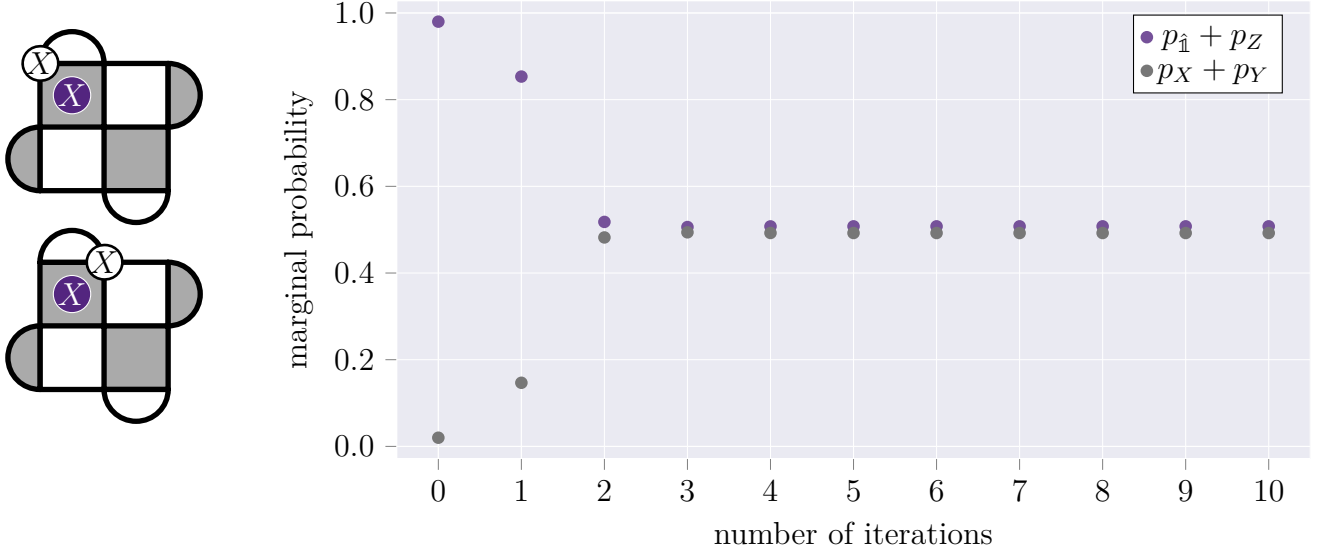

Once the algorithm has converged, we calculate the updated beliefs (the conditional probabilities):
\begin{equation}
b_q(E_q) = p(E_q) \prod_{c \in \mathrm{supp}(q)} m_{c \rightarrow q}(E_q).
\end{equation}
These beliefs can, in some instances, be used to provide an approximate maximum-likelihood error, but belief propagation is only guaranteed to provide an accurate estimate of likelihood on \emph{trees} (acyclic graphs). 
In decoding, where Tanner graphs have many short cycles, the product of the likeliest errors on each qubit may not conform with the syndrome. 
For an example of this, consider the scenario in \cref{fig:bp_failure} (also \cite{poulin_bp}). 
In this scenario, an $X$ error has occurred on the upper-left corner, causing a single vertex to appear in the neighbouring square. 
With equal likelihood, this vertex could have been caused by an $X$ error on the neighbouring qubit.
In this case, belief propagation outputs a marginal probability $p_X + p_Y$ slightly below $\nicefrac{1}{2}$ on each of these qubits.
We refer to this scenario as a \emph{split belief}. 
Taking the likeliest error on each qubit results in an error which does not conform with the syndrome, preventing the use of BP as a decoder.
However, using a split belief in the approximate cost calculation in \cref{sec:odds_calc} still provides a final \texttt{odds} near 2, resulting in a small negative edge weight, and a good chance that the vertex in question will be matched to the boundary, as it should be. 

\begin{figure}[!ht]
\centering
\input{paper_dep_no_bp.tex}\\
\input{paper_dep.tex}
\caption{Threshold for the rotated surface code correcting depolarizing errors.
Above: Using uniform edge weights, the threshold is $14.88 \pm 0.02 \%$, less than the value of $15.5 \pm 0.5 \%$ obtained using smooth/rough boundary conditions in \cite{virtual_vertices}.
Below: When using multi-path odds summation as discussed in \cref{sec:odds_calc}, the threshold improves to $17.76 \pm 0.02 \%$.
Error bars are $99\%$ confidence intervals.
Solid lines are maximum-likelihood fits to data points near the threshold.}
\label{fig:hi_threshold}
\end{figure}
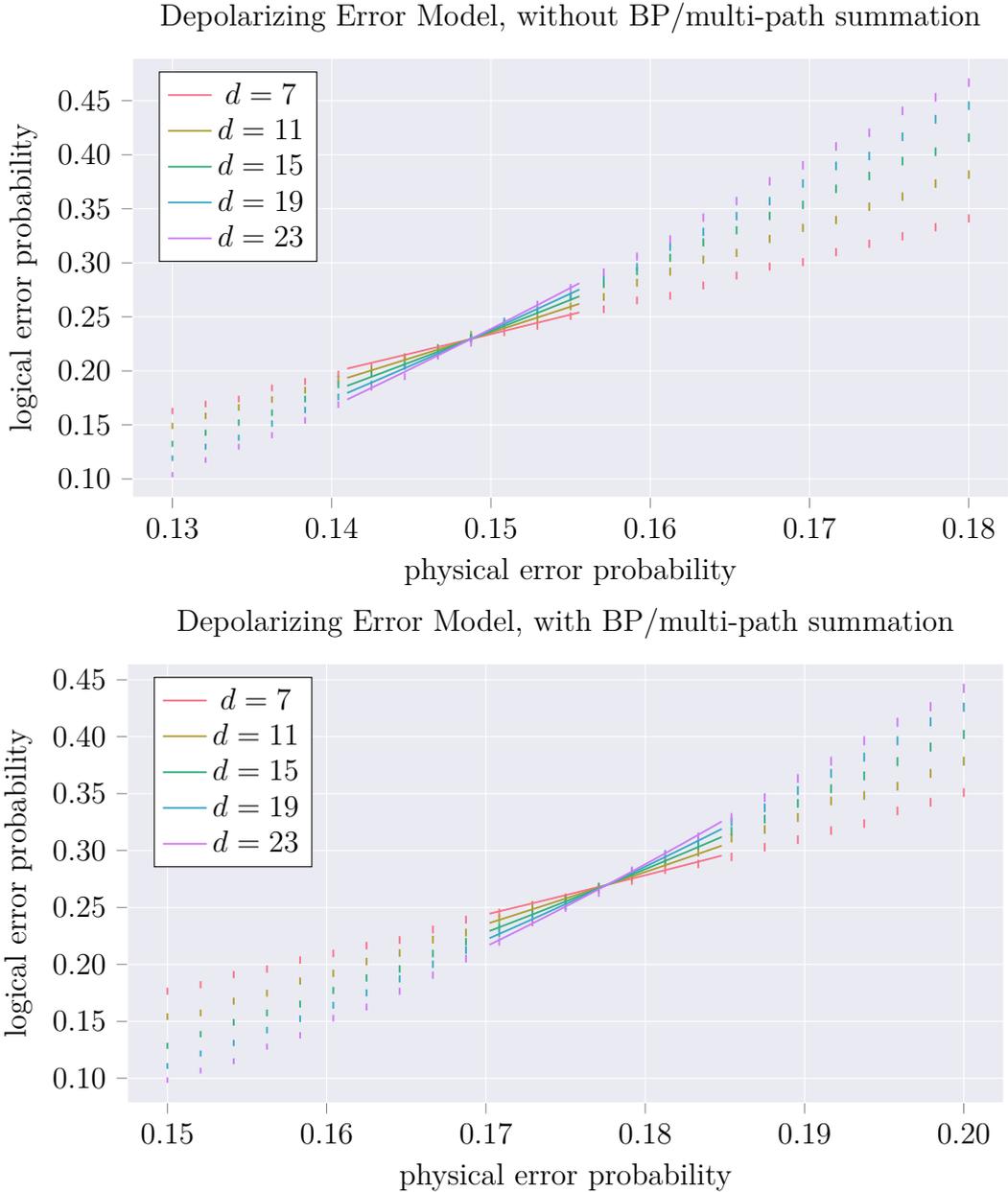

Using $d$ rounds of belief propagation and the cost calculation in \cref{sec:odds_calc}, we obtain a threshold of $17.76 \pm 0.02\%$ when correcting depolarizing noise, see \cref{fig:hi_threshold}. 
This is a significant improvement over normal MWPM decoding, though not optimal. 
In the following section, we discuss the complexity of the additional steps in the decoder, as well as potential enhancements and other applications. 

\section{Discussion/Conclusion}
\label{sec:d_and_c}
To our knowledge, little discussion exists in the literature regarding the effect of boundary conditions on the performance of surface code decoders. 
In addition, it is important to note the complexity of the additional steps used in the decoder studied in this work.
In this section, we address these topics, and conclude by pointing out opportunities for future research.
\subsection{Boundary Conditions}
Intuition drawn from the statistical physics of spin models which are related to surface codes motivates the belief that the decoding threshold of a surface code is independent of `finite-size' considerations such as boundary conditions and aspect ratio (if non-square lattices are used). 
Indeed, we see this belief to be approximately correct in the current work, though there are small deviations from decoding thresholds calculated elsewhere. 
To ensure that these deviations are related to boundary conditions, we re-evaluate the decoders we consider using smooth/rough boundary conditions in \cref{fig:planar_multipath,fig:planar_hi_threshold}, obtaining thresholds for comparison to \cref{fig:path_counting_threshold,fig:hi_threshold}.
\begin{figure}[!ht]
\centering
\input{paper_planar_iidxz_no_mp.tex}
\input{paper_planar_iidxz_mp.tex}
\caption{Replication of previous results using smooth/rough boundary conditions.
The threshold error rate of $10.30 \pm 0.01 \%$ is in accordance with earlier results.
This threshold increases to $10.59 \pm 0.01 \%$ using path-counting, comparable to the $\sim 10.6 \%$ observed by Barrett/Stace/Doherty (with periodic boundary conditions) \cite{stace_barrett_09,stace_barrett_pra}.
Error bars are $99 \%$ confidence intervals.
Solid lines are maximum-likelihood fits near the threshold.}
\label{fig:planar_multipath}
\end{figure}
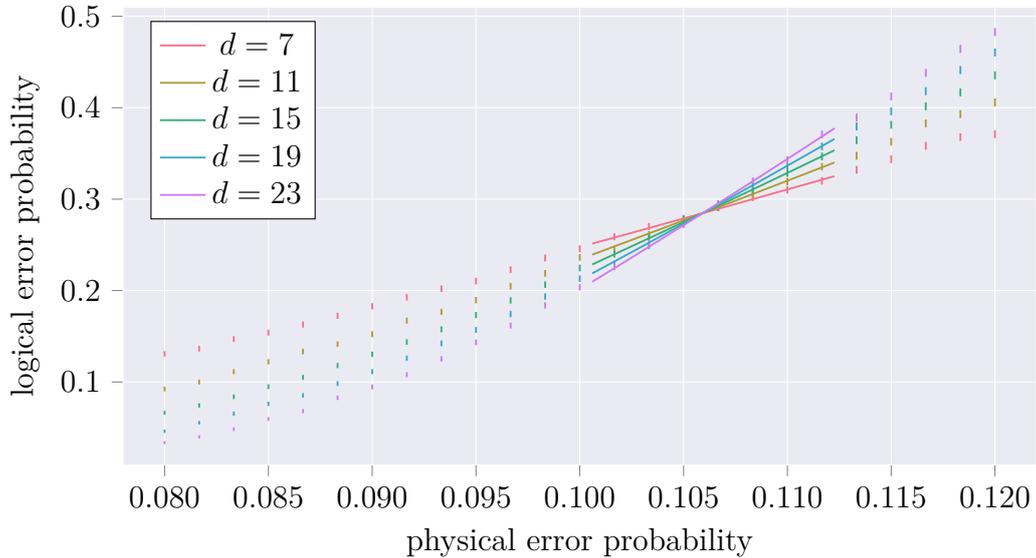
\begin{figure}[!ht]
\centering
\input{paper_planar_dep_no_mp.tex}
\input{paper_planar_dep_bp.tex}
\caption{Application of belief propagation/multi-path summation to the decoding of surface codes with smooth/rough boundaries which are subjected to depolarising noise.
A similar increase in threshold error rates is observed using these boundary conditions, with a threshold of $15.42 \pm 0.01 \%$ obtained using a standard MWPM decoder, compatible with the observation of $15.5 \pm 0.5 \%$ by \cite{virtual_vertices}.
This increases to $17.84 \pm 0.01 \%$ when belief propagation and multi-path summation are incorporated.
Error bars are $99 \%$ confidence intervals.
Solid lines are maximum-likelihood fits near the threshold.}
\label{fig:planar_hi_threshold}
\end{figure}
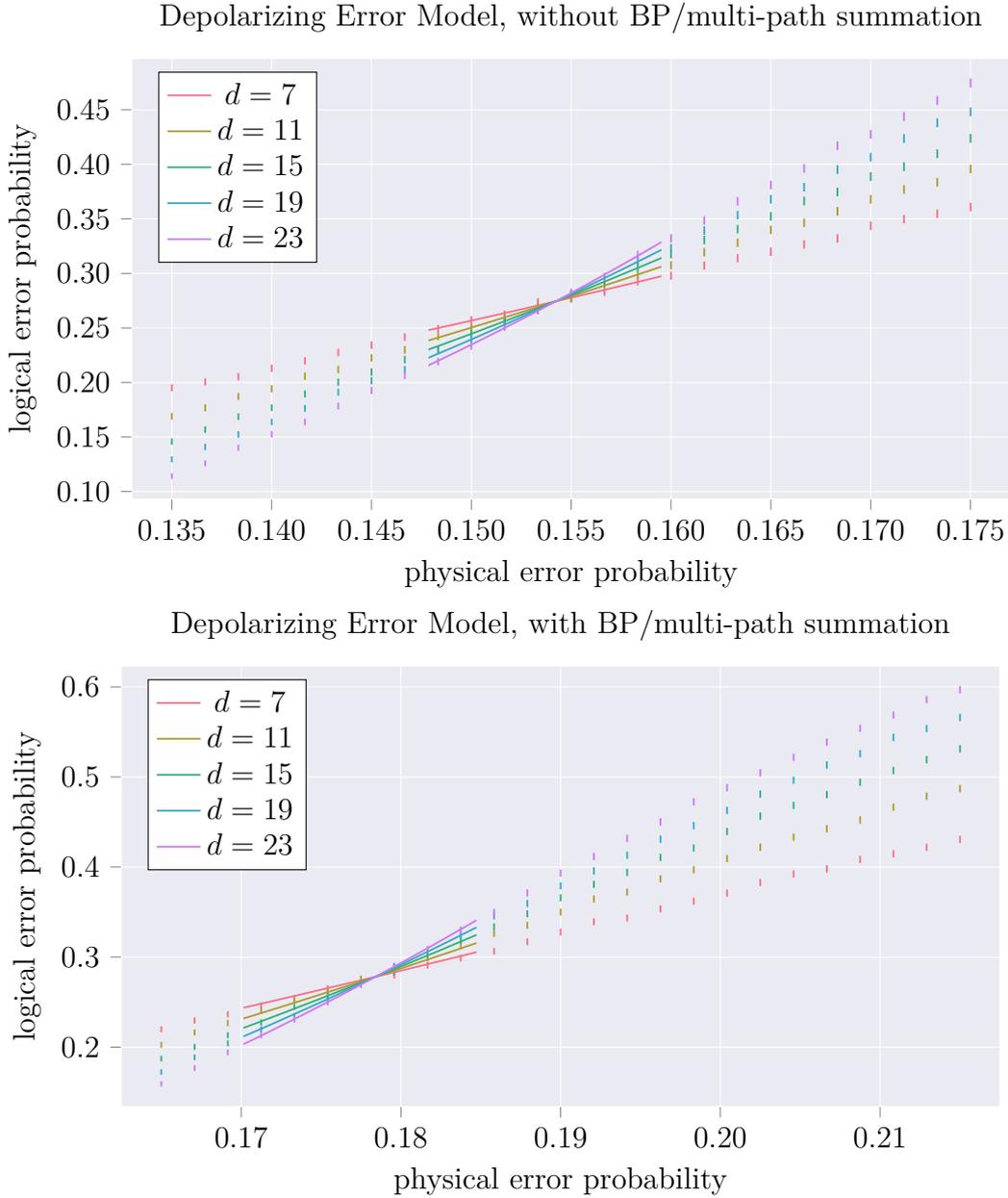

The fact that these thresholds exceed those for the rotated boundary conditions is confusing if the decoding threshold corresponds exactly with a thermodynamic property. 
However, this correspondence is only exact when using an \emph{optimal} decoder \cite{steve}. 
The suboptimality of the matching-based decoders studied here thus allows for boundary conditions to influence decoder performance.
Nonetheless, it is surprising that the change in threshold caused by the change in boundary condition should be discernible, and that it should be similar to the change in threshold caused by the change in decoding algorithm. 

\begin{figure}[ht!]
\centering
\hfill
\input{smooth_rough_traversal.tex}
\hfill
\input{rotated_traversal.tex}
\hfill
\hfill
\caption{Graphs for which paths between the two coloured vertices correspond to logical Paulis on distance-5 surface codes.
Left: Smooth/rough boundary conditions, right: rotated boundary conditions, with grey edges costing nothing to traverse.
With smooth/rough boundary conditions, horizontal edges cannot be used in a minimum-length path, resulting in a number of minimum-length paths equal to $d$. 
With rotated boundary conditions, every edge can be part of a minimum-length path, resulting in a number of minimum-length paths which can be calculated by path-counting as explained in \cref{sec:odds_calc} (in this instance, there are $52$ such paths), and lower-bounded by ${d \choose \left \lfloor \nicefrac{d}{2} \right \rfloor}$ (see text).}
\label{fig:traversal_path_sets}
\end{figure}
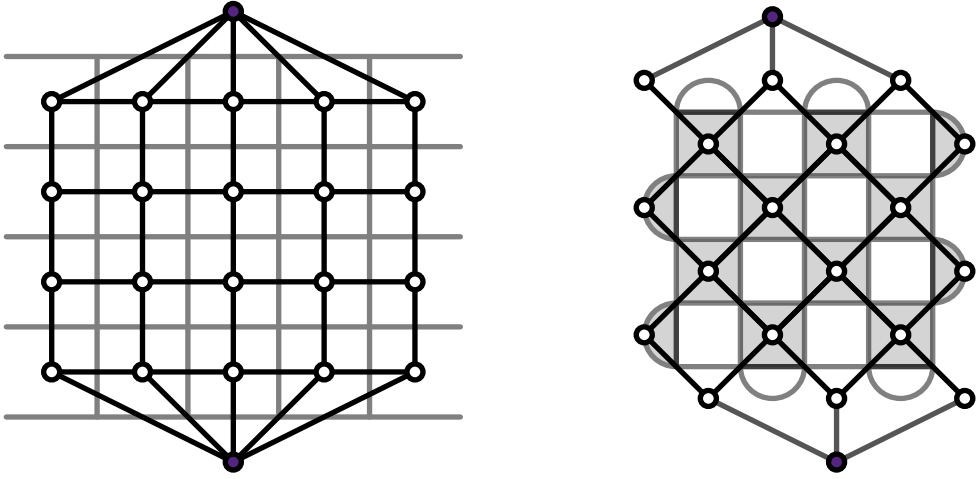

While we cannot explain this discrepancy completely, it can be explained partially by comparing the sets of minimum-length paths which traverse the two lattices, shown in \cref{fig:traversal_path_sets}.
When using smooth/rough boundary conditions, each vertex on the graph to be traversed only has one neighbouring edge which can be part of a minimum-length directed path from top to bottom, the edge immediately below. 
A minimum-length logical operator must, therefore, traverse the lattice directly from top to bottom.
Therefore, there are exactly $d$ minimum-length paths across the lattice. 
When using rotated boundary conditions, however, there are two edges neighbouring each vertex which can be part of a minimum-length path. 
The total number of minimum-length paths is therefore much larger, being bounded from below by ${d \choose \left \lfloor \nicefrac{d}{2} \right \rfloor}$, since the graph being traversed contains a $\left(\left \lfloor \nicefrac{d}{2} \right \rfloor \right)$-by-$\left(\left \lceil\nicefrac{d}{2} \right \rceil\right)$ bounding box as a subgraph.
The number of minimum-length paths on the rotated lattice can also be calculated exactly in linear time by path-counting, as explained in \cref{sec:odds_calc}. 

In order for a minimum-weight decoder to produce a logical error when a small number of (for example, $X$) errors occur on the lattice at random, it must be possible to produce a minimum-length path by traversing at least $\left(\left \lceil \nicefrac{d}{2} \right \rceil \right)$ edges corresponding to qubits on which errors have occurred. 
When the number of available minimum-length paths is much larger, intuition would say that should occur with higher probability. 
This qualitative difference sheds some light on the origin of the difference in threshold, though we cannot explain it fully. 

We note, in addition, that this difference in threshold has also been observed in predictions of logical error rates \cite{fowler_no_torus} and when using highly-correlated error models \cite{flammia_ultrahigh}. 
It is interesting to note that, for the two error models considered here, the difference in threshold between code families with different boundary conditions is smaller when the decoder is closer to optimality, though this is by no means conclusive.
Another confounding factor, which prevents us from attributing these deviations solely to boundary conditions, is the behaviour of the Blossom V MWPM implementation on graphs with non-unique minimum-weight perfect matchings, also noted by \cite{stace_barrett}, which may also play a role.

\subsection{Complexity}
The decoder studied in this work does not require any modification to the Blossom algorithm, but adds pre-calculation steps which determine the edge weights. 
The additional overhead from belief propagation is $O(d^3)$, since it acts for $d$ rounds and performs $O(d^2)$ operations in each round. 
The operations in each round of belief propagation can be carried out independently of one another, allowing parallelization to $O(d)$ time using $O(d^2)$ processors.
The path-summation function is more complex.
If all path-sums are calculated, $O(d^4)$ operations are required, since there are $O(d^4)$ ancilla pairs, each one requiring a constant number of operations for calculation, assuming that they are performed in the appropriate order.
A na\"ive parallel implementation of this algorithm would require one processor per pair of stabilisers ($O(d^4)$ processors) and $O(d)$ time, since the longest minimum-length paths are of $O(d)$ length. 
It is likely possible to reduce the number of processors, since each processor in the na\"ive parallelization only performs a calculation during one of the $O(d)$ timesteps of the algorithm.
It is also likely possible to reduce the overall complexity of decoding by combining the proposed edge weight evaluation techniques with methods for pruning the syndrome graph \cite{PhysRevLett.108.180501}, which is input to Blossom.
\subsection{Future Work}
To address the sub-optimality of the decoder, it is necessary to determine which of the approximations made in its definition are limiting accuracy. 
The first of these approximations is the set of marginal probabilities from belief propagation. 
The split belief in \cref{fig:bp_failure} produces an \texttt{odds} near 2, where it should be much higher in principle (in the absence of other nearby syndromes, or a strong prior). 
Efforts to decrease this inaccuracy by altering the Tanner graph or prior used in BP have failed. 
Immediate future work will likely focus on finding a modification to or substitute for the BP algorithm which remedies this inaccuracy.

In any event, it will be interesting to apply multi-path odds summation to the $(2+1)$-dimensional problem of decoding the surface code when incorporating multiple rounds of measurement to combat errors in the syndrome measurement circuit, as well as to the problem of decoding 2D colour codes, which can be reduced to multiple correlated instances of surface code decoding \cite{delfosse_colour_codes}.
 
\section*{Acknowledgements/Remarks}
The authors would like to thank Tom O'Brien and Brian Tarasinski for the motivation to study this problem, as well as Barbara Terhal, Kasper Duivenvoorden, Christophe Vuillot, Nikolas Breuckmann and Ben Brown for useful discussions.
Open-source software libraries used in this work include the Python libraries \texttt{numpy} \cite{numpy}, \texttt{networkx} \cite{networkx}, \texttt{scipy} \cite{scipy}, \texttt{matplotlib} \cite{matplotlib}, \texttt{seaborn} \cite{seaborn}, and \texttt{emcee} \cite{emcee}, as well as Blossom V \cite{kolmogorov} in C++.
Additional software produced for this work is available on Github \cite{code}.

\bibliographystyle{unsrtnat}
\bibliography{BP_plus_MWPM}
\end{document}

%% file: dist_9_surface_code.tex
\begin{tikzpicture}[x = 10pt, y = 10pt, gridline/.style = {black, line width = 2pt, line join = round, line cap = round}]
\begin{scope}[shift = {(0, 0)}]
\foreach \x/\y in {1/1, 1/5, 1/9, 1/13, 3/3, 3/7, 3/11, 3/15, 5/1, 5/5, 5/9, 5/13, 7/3, 7/7, 7/11, 7/15, 9/1, 9/5, 9/9, 9/13, 11/3, 11/7, 11/11, 11/15, 13/1, 13/5, 13/9, 13/13, 15/3, 15/7, 15/11, 15/15}{
    \fill[white] (\x, \y) rectangle +(2,2);
}
\foreach \x/\y in {1/3, 1/7, 1/11, 1/15, 3/1, 3/5, 3/9, 3/13, 5/3, 5/7, 5/11, 5/15, 7/1, 7/5, 7/9, 7/13, 9/3, 9/7, 9/11, 9/15, 11/1, 11/5, 11/9, 11/13, 13/3, 13/7, 13/11, 13/15, 15/1, 15/5, 15/9, 15/13}{
    \fill[quantumgray!50!white] (\x, \y) rectangle +(2,2);
}
\draw[gridline, step=2, shift = {(-1,-1)}] (2, 2) grid (18, 18);
\foreach \x/\y in {3/17, 7/17, 11/17, 15/17}{
    \filldraw[gridline, fill = white] (\x, \y) arc (0:180:1) -- cycle;
}
\foreach \x/\y in {17/3, 17/7, 17/11, 17/15}{
    \filldraw[gridline, fill = quantumgray!50!white] (\x, \y) arc (-90:90:1) -- cycle;
}
\foreach \x/\y in {1/3, 1/7, 1/11, 1/15}{
    \filldraw[gridline, fill = quantumgray!50!white] (\x, \y) arc (90:270:1) -- cycle;
}
\foreach \x/\y in {3/1, 7/1, 11/1, 15/1}{
    \filldraw[gridline, fill = white] (\x, \y) arc (180:360:1) -- cycle;
}
\end{scope}


\draw[black, line width=14pt, line cap=round, line join=round] (1, 1) -- ++(16, 0);
\draw[white, line width=12pt, line cap=round, line join=round] (1, 1) -- ++(8, 0) node[black]{$\bar{Z}$} -- ++(8, 0);

\draw[black, line width=14pt, line cap=round, line join=round] (17, 1) -- ++(0, 16);
\draw[white, line width=12pt, line cap=round, line join=round] (17, 1) -- ++(0, 8) node[black]{$\bar{X}$} -- ++(0, 8);

\draw[black, line width=14pt, line cap=round, line join=round] (3, 5) -- ++(2, 0) -- ++(2, 0)  ;
\draw[white, line width=12pt, line cap=round, line join=round] (3, 5) node[black]{$Z$} -- ++(2, 0) node[black]{$Z$} -- ++(2, 0) node[black]{$Z$};

\draw[black, line width=14pt, line cap=round, line join=round] (11,9) -- ++(2,0);
\draw[white, line width=12pt, line cap=round, line join=round] (11,9) node[black]{$X$} -- ++(2,0)node[black]{$X$};

\draw[black, line width=14pt, line cap=round, line join=round] (7, 11) -- ++(0, 2);
\draw[white, line width=12pt, line cap=round, line join=round] (7, 11) node[black]{$X$} -- ++(0, 2)node[black]{$X$};

\draw[black, line width=14pt, line cap=round, line join=round] (13, 13) -- ++(0, 0);
\draw[white, line width=12pt, line cap=round, line join=round] (13, 13) -- ++(0, 0)node[black]{$Z$};

\draw[black, line width=14pt, line cap=round, line join=round] (3, 15) -- ++(0, 0);
\draw[white, line width=12pt, line cap=round, line join=round] (3, 15) -- ++(0, 0)node[black]{$Y$};

\foreach \x/\y in {2/6, 8/4, 12/12, 14/14, 2/14, 4/16}{
    \filldraw[draw=white,fill=quantumviolet] (\x,\y) circle (7pt)node[white]{$Z$}; 
}
\foreach \x/\y in {8/10, 8/14, 10/8, 14/8, 2/16, 4/14}{
    \filldraw[draw=white,fill=quantumviolet] (\x,\y) circle (7pt)node[white]{$X$}; 
}
\end{tikzpicture}

%% file: manhattan_distance.tex
\begin{tikzpicture}[x = 10pt, y = 10pt, gridline/.style = {black, line width = 2pt, line join = round, line cap = round}]

\foreach \x/\y in {5/5, 7/3, 7/7, 9/5, 9/9, 11/7}{
    \filldraw[gridline, fill=white] (\x, \y) rectangle +(2,2);
}

\foreach \x/\y in {3/5, 5/3, 5/7, 7/1, 7/5, 7/9, 9/3, 9/7, 9/11, 11/5, 11/9, 13/7}{
    \filldraw[gridline, fill=quantumgray!50!white] (\x, \y) rectangle +(2,2);
}

\foreach \x/\y/\dx in {8/10/2, 12/10/-2, 6/8/2, 10/8/2, 10/8/-2, 14/8/-2, 4/6/2, 8/6/2, 8/6/-2, 12/6/2, 12/6/-2, 6/4/2, 6/4/-2, 10/4/2, 10/4/-2, 8/2/2, 8/2/-2}{
    \draw[quantumviolet, double, line width = 2pt, -{Latex[width=8pt, length=8pt]}, line cap = round] (\x, \y) -- ++(\dx, 2);
}

\foreach \x/\y in {8/2, 10/12}{
    \filldraw[draw=white, fill=quantumviolet] (\x, \y) circle (7pt) node[white]{$X$}; 
}

\end{tikzpicture}

%% file: yy_error.tex
\begin{tikzpicture}[x = 12pt, y = 12pt, gridline/.style = {black, line width = 2pt, line join = round, line cap = round}]
\begin{scope}[shift = {(0, 0)}]
\foreach \x/\y in {1/1, 3/3}{
    \fill[white] (\x, \y) rectangle +(2,2);
}
\foreach \x/\y in {1/3, 3/1}{
    \fill[quantumgray!50!white] (\x, \y) rectangle +(2,2);
}
\draw[gridline, step=2, shift = {(-1,-1)}] (2, 2) grid (6, 6);
\foreach \x/\y in {3/5}{
    \filldraw[gridline, fill = white] (\x, \y) arc (0:180:1) -- cycle;
}
\foreach \x/\y in {5/3}{
    \filldraw[gridline, fill = quantumgray!50!white] (\x, \y) arc (-90:90:1) -- cycle;
}
\foreach \x/\y in {1/3}{
    \filldraw[gridline, fill = quantumgray!50!white] (\x, \y) arc (90:270:1) -- cycle;
}
\foreach \x/\y in {3/1}{
    \filldraw[gridline, fill = white] (\x, \y) arc (180:360:1) -- cycle;
}
\end{scope}

\draw[black, line width=14pt, line cap=round, line join=round] (1, 5) -- ++(2, -2);
\draw[white, line width=12pt, line cap=round, line join=round] (1, 5) node[black]{$Y$} -- ++(2, -2)node[black]{$Y$};

\foreach \x/\y in {2/2, 4/4, 2/6}{
    \filldraw[draw=white, fill=quantumviolet] (\x, \y) circle (7pt) node[white]{$Z$}; 
}
\filldraw[draw=white, fill=quantumviolet] (4, 2) circle (7pt) node[white]{$X$};

\end{tikzpicture}

%% file: iidxz_x_error.tex
\begin{tikzpicture}[x = 12pt, y = 12pt, gridline/.style = {black, line width = 2pt, line join = round, line cap = round}]
\begin{scope}[shift = {(0, 0)}]
\foreach \x/\y in {1/1, 3/3}{
    \fill[white] (\x, \y) rectangle +(2,2);
}
\foreach \x/\y in {1/3, 3/1}{
    \fill[quantumgray!50!white] (\x, \y) rectangle +(2,2);
}
\draw[gridline, step=2, shift = {(-1,-1)}] (2, 2) grid (6, 6);
\foreach \x/\y in {3/5}{
    \filldraw[gridline, fill = white] (\x, \y) arc (0:180:1) -- cycle;
}
\foreach \x/\y in {5/3}{
    \filldraw[gridline, fill = quantumgray!50!white] (\x, \y) arc (-90:90:1) -- cycle;
}
\foreach \x/\y in {1/3}{
    \filldraw[gridline, fill = quantumgray!50!white] (\x, \y) arc (90:270:1) -- cycle;
}
\foreach \x/\y in {3/1}{
    \filldraw[gridline, fill = white] (\x, \y) arc (180:360:1) -- cycle;
}
\end{scope}

\draw[black, line width=14pt, line cap=round, line join=round] (5, 1) -- ++(0, 0)  ;
\draw[white, line width=12pt, line cap=round, line join=round] (5, 1) node[black]{$X$} -- ++(0, 0);

\filldraw[draw=white, fill=quantumviolet] (4, 2) circle (7pt) node[white]{$X$};

\end{tikzpicture}

%% file: iidxz_z_error.tex
\begin{tikzpicture}[x = 12pt, y = 12pt, gridline/.style = {black, line width = 2pt, line join = round, line cap = round}]
\begin{scope}[shift = {(0, 0)}]
\foreach \x/\y in {1/1, 3/3}{
    \fill[white] (\x, \y) rectangle +(2,2);
}
\foreach \x/\y in {1/3, 3/1}{
    \fill[quantumgray!50!white] (\x, \y) rectangle +(2,2);
}
\draw[gridline, step=2, shift = {(-1,-1)}] (2, 2) grid (6, 6);
\foreach \x/\y in {3/5}{
    \filldraw[gridline, fill = white] (\x, \y) arc (0:180:1) -- cycle;
}
\foreach \x/\y in {5/3}{
    \filldraw[gridline, fill = quantumgray!50!white] (\x, \y) arc (-90:90:1) -- cycle;
}
\foreach \x/\y in {1/3}{
    \filldraw[gridline, fill = quantumgray!50!white] (\x, \y) arc (90:270:1) -- cycle;
}
\foreach \x/\y in {3/1}{
    \filldraw[gridline, fill = white] (\x, \y) arc (180:360:1) -- cycle;
}
\end{scope}

\draw[black, line width=14pt, line cap=round, line join=round] (1, 5) -- ++(2, -2)  ;
\draw[white, line width=12pt, line cap=round, line join=round] (1, 5) node[black]{$Z$} -- ++(2, -2)node[black]{$Z$};

\foreach \x/\y in {2/2, 4/4, 2/6}{
    \filldraw[draw=white, fill=quantumviolet] (\x, \y) circle (7pt) node[white]{$Z$}; 
}

\end{tikzpicture}

%% file: bbox_single.tex
\begin{tikzpicture}[x = 12pt, y = 12pt, gridline/.style = {black, line width = 2pt, line join = round, line cap = round}]
\begin{scope}[shift = {(0, 0)}]
\foreach \x/\y in {1/1, 1/5, 3/3, 3/7, 5/1, 5/5, 7/3, 7/7}{
    \fill[white] (\x, \y) rectangle +(2,2);
}
\foreach \x/\y in {1/3, 1/7, 3/1, 3/5, 5/3, 5/7, 7/1, 7/5}{
    \fill[quantumgray!50!white] (\x, \y) rectangle +(2,2);
}
\draw[gridline, step=2, shift = {(-1,-1)}] (2, 2) grid (10, 10);
\foreach \x/\y in {3/9, 7/9}{
    \filldraw[gridline, fill = white] (\x, \y) arc (0:180:1) -- cycle;
}
\foreach \x/\y in {9/3, 9/7}{
    \filldraw[gridline, fill = quantumgray!50!white] (\x, \y) arc (-90:90:1) -- cycle;
}
\foreach \x/\y in {1/3, 1/7}{
    \filldraw[gridline, fill = quantumgray!50!white] (\x, \y) arc (90:270:1) -- cycle;
}
\foreach \x/\y in {3/1, 7/1}{
    \filldraw[gridline, fill = white] (\x, \y) arc (180:360:1) -- cycle;
}

\foreach \x/\y/\sym in {4/2/$X$, 6/8/$X$, 6/2/$Z$, 4/8/$Z$}{
    \filldraw[draw=white, fill=quantumviolet] (\x, \y) circle (7pt) node[white]{\sym}; 
}

\draw[black, line width=14pt, line cap=round, line join=round] (3, 7) -- ++(0, -2) -- ++(2, -2) -- ++(2, 2) -- ++(0, 2);
\draw[white, line width=12pt, line cap=round, line join=round] (3, 7) node[black]{$Z$} -- ++(0, -2) node[black]{$Z$} -- ++(2, -2) node[black]{$Y$} -- ++(2, 2) node[black]{$X$} -- ++(0, 2) node[black]{$X$} ;

\end{scope}
\end{tikzpicture}

%% file: bbox_intersection.tex
\begin{tikzpicture}[x = 12pt, y = 12pt, gridline/.style = {black, line width = 2pt, line join = round, line cap = round}]
\begin{scope}[shift = {(0, 0)}]
\foreach \x/\y in {1/1, 1/5, 3/3, 3/7, 5/1, 5/5, 7/3, 7/7}{
    \fill[white] (\x, \y) rectangle +(2,2);
}
\foreach \x/\y in {1/3, 1/7, 3/1, 3/5, 5/3, 5/7, 7/1, 7/5}{
    \fill[quantumgray!50!white] (\x, \y) rectangle +(2,2);
}
\draw[gridline, step=2, shift = {(-1,-1)}] (2, 2) grid (10, 10);
\foreach \x/\y in {3/9, 7/9}{
    \filldraw[gridline, fill = white] (\x, \y) arc (0:180:1) -- cycle;
}
\foreach \x/\y in {9/3, 9/7}{
    \filldraw[gridline, fill = quantumgray!50!white] (\x, \y) arc (-90:90:1) -- cycle;
}
\foreach \x/\y in {1/3, 1/7}{
    \filldraw[gridline, fill = quantumgray!50!white] (\x, \y) arc (90:270:1) -- cycle;
}
\foreach \x/\y in {3/1, 7/1}{
    \filldraw[gridline, fill = white] (\x, \y) arc (180:360:1) -- cycle;
}

\foreach \x/\y/\sym in {4/2/$X$, 6/8/$X$, 6/2/$Z$, 4/8/$Z$}{
    \filldraw[draw=white, fill=quantumviolet] (\x, \y) circle (7pt) node[white]{\sym}; 
}

\draw[black, line width=14pt, line cap=round, line join=round] (5, 3) -- ++(0, 2) -- ++(0, 2);
\draw[white, line width=12pt, line cap=round, line join=round] (5, 3) node[black]{$Y$} -- ++(0, 2) node[black]{$Y$} -- ++(0, 2) node[black]{$Y$};
\end{scope}
\end{tikzpicture}

%% file: multiple_paths.tex
\begin{tikzpicture}[x = 10pt, y = 10pt,
gridline/.style = {black, line width = 2pt, line join = round, line cap = round},
errow/.style={double, line width = 2pt, -{Latex[width=8pt, length=8pt]}, line cap = round}]
\begin{scope}[shift = {(0, 0)}]
\foreach \x/\y in {1/1, 1/5, 1/9, 1/13, 3/3, 3/7, 3/11, 3/15, 5/1, 5/5, 5/9, 5/13, 7/3, 7/7, 7/11, 7/15, 9/1, 9/5, 9/9, 9/13, 11/3, 11/7, 11/11, 11/15, 13/1, 13/5, 13/9, 13/13, 15/3, 15/7, 15/11, 15/15}{
    \fill[white] (\x, \y) rectangle +(2,2);
}
\foreach \x/\y in {1/3, 1/7, 1/11, 1/15, 3/1, 3/5, 3/9, 3/13, 5/3, 5/7, 5/11, 5/15, 7/1, 7/5, 7/9, 7/13, 9/3, 9/7, 9/11, 9/15, 11/1, 11/5, 11/9, 11/13, 13/3, 13/7, 13/11, 13/15, 15/1, 15/5, 15/9, 15/13}{
    \fill[quantumgray!50!white] (\x, \y) rectangle +(2,2);
}
\draw[gridline, step=2, shift = {(-1,-1)}] (2, 2) grid (18, 18);
\foreach \x/\y in {3/17, 7/17, 11/17, 15/17}{
    \filldraw[gridline, fill = white] (\x, \y) arc (0:180:1) -- cycle;
}
\foreach \x/\y in {17/3, 17/7, 17/11, 17/15}{
    \filldraw[gridline, fill = quantumgray!50!white] (\x, \y) arc (-90:90:1) -- cycle;
}
\foreach \x/\y in {1/3, 1/7, 1/11, 1/15}{
    \filldraw[gridline, fill = quantumgray!50!white] (\x, \y) arc (90:270:1) -- cycle;
}
\foreach \x/\y in {3/1, 7/1, 11/1, 15/1}{
    \filldraw[gridline, fill = white] (\x, \y) arc (180:360:1) -- cycle;
}
\end{scope}


\foreach \x/\y in {2/16, 6/16, 10/12, 10/8, 14/4}{
    \node (synd\x\y) [draw=white, fill=quantumviolet, circle, minimum size=14pt, inner sep=0pt] at (\x, \y)  {\textcolor{white}{$X$}}; 
}

\draw[errow, quantumgray] (synd216.north west) --++ (-2,2);
\draw[errow, quantumviolet] (synd216) --++ (2,2) --++ (synd616);
\draw[errow, quantumviolet] (synd216) --++ (2,-2) --++ (synd616);
\draw[errow, quantumgray] (synd616) --++ (synd1012);
\draw[errow, quantumviolet] (synd1012) --++ (2,-2) --++ (synd108);
\draw[errow, quantumviolet] (synd1012) --++ (-2,-2) --++ (synd108);
\draw[errow, quantumgray] (synd108) --++ (synd144);
\draw[errow, quantumviolet] (synd144) --++ (4,-4);
\draw[errow, quantumviolet] (synd144) --++ (-4,-4);
\draw[errow, quantumviolet] (synd144) --++ (2,-2) --++ (-2,-2);
\draw[errow, quantumviolet] (synd144) --++ (-2,-2) --++ (2,-2);

\foreach \x/\y in {2/16, 6/16, 10/12, 10/8, 14/4}{
    \node [draw=white, fill=quantumviolet, circle, minimum size=14pt, inner sep=0pt] at (\x, \y)  {\textcolor{white}{$X$}}; 
}

\end{tikzpicture}

%% file: clipped_bbox.tex
\begin{tikzpicture}[x = 12pt, y = 12pt, gridline/.style = {black, line width = 2pt, line join = round, line cap = round},
errow/.style={double, line width = 2pt, -{Latex[width=8pt, length=8pt]}, line cap = round}]
\begin{scope}[shift = {(0, 0)}]
\foreach \x/\y in {1/1, 1/5, 3/3, 3/7, 5/1, 5/5, 7/3, 7/7}{
    \fill[white] (\x, \y) rectangle +(2,2);
}
\foreach \x/\y in {1/3, 1/7, 3/1, 3/5, 5/3, 5/7, 7/1, 7/5}{
    \fill[quantumgray!50!white] (\x, \y) rectangle +(2,2);
}
\draw[gridline, step=2, shift = {(-1,-1)}] (2, 2) grid (10, 10);
\foreach \x/\y in {3/9, 7/9}{
    \filldraw[gridline, fill = white] (\x, \y) arc (0:180:1) -- cycle;
}
\foreach \x/\y in {9/3, 9/7}{
    \filldraw[gridline, fill = quantumgray!50!white] (\x, \y) arc (-90:90:1) -- cycle;
}
\foreach \x/\y in {1/3, 1/7}{
    \filldraw[gridline, fill = quantumgray!50!white] (\x, \y) arc (90:270:1) -- cycle;
}
\foreach \x/\y in {3/1, 7/1}{
    \filldraw[gridline, fill = white] (\x, \y) arc (180:360:1) -- cycle;
}
\end{scope}

\foreach \x/\y in {8/2, 10/8}{
    \node (synd\x\y) [draw=white, fill=quantumviolet, circle, minimum size=14pt, inner sep=0pt] at (\x, \y)  {\textcolor{white}{$X$}}; 
}

\draw[errow, quantumviolet] (synd82) --++ (2,2);
\draw[errow, quantumviolet] (synd82) --++ (-2,2);
\draw[errow, quantumviolet] (10,4) --++ (-2,2);
\draw[errow, quantumviolet] (6,4) --++ (2,2);
\draw[errow, quantumviolet] (8,6) --++ (synd108);

\foreach \x/\y in {8/2, 10/8}{
    \node (synd\x\y) [draw=white, fill=quantumviolet, circle, minimum size=14pt, inner sep=0pt] at (\x, \y)  {\textcolor{white}{$X$}}; 
}

\end{tikzpicture}

%% file: multiple_endpoints.tex
\begin{tikzpicture}[x = 12pt, y = 12pt, gridline/.style = {black, line width = 2pt, line join = round, line cap = round},
errow/.style={double, line width = 2pt, -{Latex[width=8pt, length=8pt]}, line cap = round}]
\begin{scope}[shift = {(0, 0)}]
\foreach \x/\y in {1/1, 1/5, 3/3, 3/7, 5/1, 5/5, 7/3, 7/7}{
    \fill[white] (\x, \y) rectangle +(2,2);
}
\foreach \x/\y in {1/3, 1/7, 3/1, 3/5, 5/3, 5/7, 7/1, 7/5}{
    \fill[quantumgray!50!white] (\x, \y) rectangle +(2,2);
}
\draw[gridline, step=2, shift = {(-1,-1)}] (2, 2) grid (10, 10);
\foreach \x/\y in {3/9, 7/9}{
    \filldraw[gridline, fill = white] (\x, \y) arc (0:180:1) -- cycle;
}
\foreach \x/\y in {9/3, 9/7}{
    \filldraw[gridline, fill = quantumgray!50!white] (\x, \y) arc (-90:90:1) -- cycle;
}
\foreach \x/\y in {1/3, 1/7}{
    \filldraw[gridline, fill = quantumgray!50!white] (\x, \y) arc (90:270:1) -- cycle;
}
\foreach \x/\y in {3/1, 7/1}{
    \filldraw[gridline, fill = white] (\x, \y) arc (180:360:1) -- cycle;
}
\end{scope}

\foreach \x/\y in {6/4}{
    \node (synd\x\y) [draw=white, fill=quantumviolet, circle, minimum size=14pt, inner sep=0pt] at (\x, \y)  {\textcolor{white}{$X$}}; 
}

\draw[errow, quantumviolet] (synd64) --++ (2,-2);
\draw[errow, quantumviolet] (synd64) --++ (-2,-2);
\draw[errow, quantumviolet] (8,2) --++ (-2,-2);
\draw[errow, quantumviolet] (8,2) --++ (2,-2);
\draw[errow, quantumviolet] (4,2) --++ (-2,-2);
\draw[errow, quantumviolet] (4,2) --++ (2,-2);

\foreach \x/\y in {6/4}{
    \node (synd\x\y) [draw=white, fill=quantumviolet, circle, minimum size=14pt, inner sep=0pt] at (\x, \y)  {\textcolor{white}{$X$}}; 
}

\end{tikzpicture}

%% file: paper_iidxz_no_bp.tex
\begin{tikzpicture}

\definecolor{color1}{rgb}{0.967797559291991,0.441274560091574,0.53581031550587}
\definecolor{color0}{rgb}{0.917647058823529,0.917647058823529,0.949019607843137}
\definecolor{color3}{rgb}{0.201253172212011,0.690792081537903,0.479667611892753}
\definecolor{color2}{rgb}{0.680418912779335,0.615149751467757,0.194054521114453}
\definecolor{color5}{rgb}{0.800493618642396,0.477033635337372,0.957954719600752}
\definecolor{color4}{rgb}{0.219799566082832,0.662515787685034,0.773209315931721}

\begin{axis}[
title={IID $X$/$Z$ Error Model, without multi-path summation},
xlabel={physical error probability},
ylabel={logical error probability},
xmin=0.0775, xmax=0.1325,
ymin=0.0356769674335804, ymax=0.596649509471933,
width=\figurewidth,
height=\figureheight,
xtick={0.07,0.08,0.09,0.1,0.11,0.12,0.13,0.14},
xticklabels={,0.08,0.09,0.10,0.11,0.12,0.13,},
ytick={0,0.1,0.2,0.3,0.4,0.5,0.6},
yticklabels={,0.1,0.2,0.3,0.4,0.5,},
tick align=outside,
tick pos=left,
xmajorgrids,
x grid style={white},
ymajorgrids,
y grid style={white},
axis line style={white},
axis background/.style={fill=color0},
legend pos=north west,
legend entries={$d = 7$, $d = 11$, $d = 15$, $d = 19$, $d = 23$}
]
\path [draw=color1, line width=0.7000000000000001pt] (axis cs:0.08,0.133574566071325)
--(axis cs:0.08,0.139070028444823);

\path [draw=color1, line width=0.7000000000000001pt] (axis cs:0.0820833333333333,0.143191625224946)
--(axis cs:0.0820833333333333,0.148851564465257);

\path [draw=color1, line width=0.7000000000000001pt] (axis cs:0.0841666666666667,0.15347626695563)
--(axis cs:0.0841666666666667,0.15929100795294);

\path [draw=color1, line width=0.7000000000000001pt] (axis cs:0.08625,0.162251591555564)
--(axis cs:0.08625,0.168201784090249);

\path [draw=color1, line width=0.7000000000000001pt] (axis cs:0.0883333333333333,0.173223166082935)
--(axis cs:0.0883333333333333,0.17932816037462);

\path [draw=color1, line width=0.7000000000000001pt] (axis cs:0.0904166666666667,0.185297703128931)
--(axis cs:0.0904166666666667,0.191557499177616);

\path [draw=color1, line width=0.7000000000000001pt] (axis cs:0.0925,0.19331836916349)
--(axis cs:0.0925,0.199684591420113);

\path [draw=color1, line width=0.7000000000000001pt] (axis cs:0.0945833333333333,0.205538032856673)
--(axis cs:0.0945833333333333,0.212049381760483);

\path [draw=color1, line width=0.7000000000000001pt] (axis cs:0.0966666666666667,0.215126066680857)
--(axis cs:0.0966666666666667,0.221753516902417);

\path [draw=color1, line width=0.7000000000000001pt] (axis cs:0.09875,0.226445945161478)
--(axis cs:0.09875,0.233189496700788);

\path [draw=color1, line width=0.7000000000000001pt] (axis cs:0.100833333333333,0.238414055999536)
--(axis cs:0.100833333333333,0.245273708856596);

\path [draw=color1, line width=0.7000000000000001pt] (axis cs:0.102916666666667,0.248650322181157)
--(axis cs:0.102916666666667,0.255606726136342);

\path [draw=color1, line width=0.7000000000000001pt] (axis cs:0.105,0.262950134484026)
--(axis cs:0.105,0.270041989976586);

\path [draw=color1, line width=0.7000000000000001pt] (axis cs:0.107083333333333,0.272586543857273)
--(axis cs:0.107083333333333,0.279755800228333);

\path [draw=color1, line width=0.7000000000000001pt] (axis cs:0.109166666666667,0.286973432148455)
--(axis cs:0.109166666666667,0.294249114727452);

\path [draw=color1, line width=0.7000000000000001pt] (axis cs:0.11125,0.298157859091701)
--(axis cs:0.11125,0.30552061765901);

\path [draw=color1, line width=0.7000000000000001pt] (axis cs:0.113333333333333,0.307097660558447)
--(axis cs:0.113333333333333,0.314518469784632);

\path [draw=color1, line width=0.7000000000000001pt] (axis cs:0.115416666666667,0.317517753826506)
--(axis cs:0.115416666666667,0.324996613711566);

\path [draw=color1, line width=0.7000000000000001pt] (axis cs:0.1175,0.330685578281314)
--(axis cs:0.1175,0.338241839044873);

\path [draw=color1, line width=0.7000000000000001pt] (axis cs:0.119583333333333,0.344404883995433)
--(axis cs:0.119583333333333,0.352038545637493);

\path [draw=color1, line width=0.7000000000000001pt] (axis cs:0.121666666666667,0.350771106252056)
--(axis cs:0.121666666666667,0.358433793223553);

\path [draw=color1, line width=0.7000000000000001pt] (axis cs:0.12375,0.364142108012926)
--(axis cs:0.12375,0.371872520753111);

\path [draw=color1, line width=0.7000000000000001pt] (axis cs:0.125833333333333,0.376980978734109)
--(axis cs:0.125833333333333,0.384759767023356);

\path [draw=color1, line width=0.7000000000000001pt] (axis cs:0.127916666666667,0.386762514754542)
--(axis cs:0.127916666666667,0.39458000348304);

\path [draw=color1, line width=0.7000000000000001pt] (axis cs:0.13,0.40062694711585)
--(axis cs:0.13,0.408492811393409);

\path [draw=color2, line width=0.7000000000000001pt] (axis cs:0.08,0.109657694614834)
--(axis cs:0.08,0.114717777046769);

\path [draw=color2, line width=0.7000000000000001pt] (axis cs:0.0820833333333333,0.121238801060392)
--(axis cs:0.0820833333333333,0.126521411018015);

\path [draw=color2, line width=0.7000000000000001pt] (axis cs:0.0841666666666667,0.130720408676638)
--(axis cs:0.0841666666666667,0.136167495501074);

\path [draw=color2, line width=0.7000000000000001pt] (axis cs:0.08625,0.145745854215445)
--(axis cs:0.08625,0.151444493895006);

\path [draw=color2, line width=0.7000000000000001pt] (axis cs:0.0883333333333333,0.156262698581629)
--(axis cs:0.0883333333333333,0.162125815128002);

\path [draw=color2, line width=0.7000000000000001pt] (axis cs:0.0904166666666667,0.172062152905435)
--(axis cs:0.0904166666666667,0.178147796977496);

\path [draw=color2, line width=0.7000000000000001pt] (axis cs:0.0925,0.18552990576443)
--(axis cs:0.0925,0.191799376922928);

\path [draw=color2, line width=0.7000000000000001pt] (axis cs:0.0945833333333333,0.200342498887362)
--(axis cs:0.0945833333333333,0.206795797132297);

\path [draw=color2, line width=0.7000000000000001pt] (axis cs:0.0966666666666667,0.213539348671607)
--(axis cs:0.0966666666666667,0.220147448673542);

\path [draw=color2, line width=0.7000000000000001pt] (axis cs:0.09875,0.228081038719789)
--(axis cs:0.09875,0.234843940478724);

\path [draw=color2, line width=0.7000000000000001pt] (axis cs:0.100833333333333,0.243270961125409)
--(axis cs:0.100833333333333,0.250178989531531);

\path [draw=color2, line width=0.7000000000000001pt] (axis cs:0.102916666666667,0.258344782213278)
--(axis cs:0.102916666666667,0.265397937266588);

\path [draw=color2, line width=0.7000000000000001pt] (axis cs:0.105,0.275556802569709)
--(axis cs:0.105,0.282745409160394);

\path [draw=color2, line width=0.7000000000000001pt] (axis cs:0.107083333333333,0.292207666557015)
--(axis cs:0.107083333333333,0.299522049575263);

\path [draw=color2, line width=0.7000000000000001pt] (axis cs:0.109166666666667,0.304669207995511)
--(axis cs:0.109166666666667,0.31207066700207);

\path [draw=color2, line width=0.7000000000000001pt] (axis cs:0.11125,0.320971768029567)
--(axis cs:0.11125,0.328469978134252);

\path [draw=color2, line width=0.7000000000000001pt] (axis cs:0.113333333333333,0.337622632016873)
--(axis cs:0.113333333333333,0.345217593219683);

\path [draw=color2, line width=0.7000000000000001pt] (axis cs:0.115416666666667,0.350384101859556)
--(axis cs:0.115416666666667,0.358046788831053);

\path [draw=color2, line width=0.7000000000000001pt] (axis cs:0.1175,0.367944426169237)
--(axis cs:0.1175,0.375684514019234);

\path [draw=color2, line width=0.7000000000000001pt] (axis cs:0.119583333333333,0.380135064532982)
--(axis cs:0.119583333333333,0.387933203041855);

\path [draw=color2, line width=0.7000000000000001pt] (axis cs:0.121666666666667,0.396679502312351)
--(axis cs:0.121666666666667,0.404535691480098);

\path [draw=color2, line width=0.7000000000000001pt] (axis cs:0.12375,0.41446235414772)
--(axis cs:0.12375,0.422366918864529);

\path [draw=color2, line width=0.7000000000000001pt] (axis cs:0.125833333333333,0.42543392867509)
--(axis cs:0.125833333333333,0.433367518721337);

\path [draw=color2, line width=0.7000000000000001pt] (axis cs:0.127916666666667,0.438195398517773)
--(axis cs:0.127916666666667,0.446148338783645);

\path [draw=color2, line width=0.7000000000000001pt] (axis cs:0.13,0.45445925811258)
--(axis cs:0.13,0.462450898817702);

\path [draw=color3, line width=0.7000000000000001pt] (axis cs:0.08,0.0898140443894038)
--(axis cs:0.08,0.0944484219895896);

\path [draw=color3, line width=0.7000000000000001pt] (axis cs:0.0820833333333333,0.103301147468024)
--(axis cs:0.0820833333333333,0.108225778362584);

\path [draw=color3, line width=0.7000000000000001pt] (axis cs:0.0841666666666667,0.115240232976644)
--(axis cs:0.0841666666666667,0.120406741616517);

\path [draw=color3, line width=0.7000000000000001pt] (axis cs:0.08625,0.129124015557577)
--(axis cs:0.08625,0.134542077052575);

\path [draw=color3, line width=0.7000000000000001pt] (axis cs:0.0883333333333333,0.144875094332321)
--(axis cs:0.0883333333333333,0.150554383792256);

\path [draw=color3, line width=0.7000000000000001pt] (axis cs:0.0904166666666667,0.161351806343002)
--(axis cs:0.0904166666666667,0.167282648658062);

\path [draw=color3, line width=0.7000000000000001pt] (axis cs:0.0925,0.176696530505621)
--(axis cs:0.0925,0.182849900346369);

\path [draw=color3, line width=0.7000000000000001pt] (axis cs:0.0945833333333333,0.19304746608874)
--(axis cs:0.0945833333333333,0.19940401323555);

\path [draw=color3, line width=0.7000000000000001pt] (axis cs:0.0966666666666667,0.211759128466108)
--(axis cs:0.0966666666666667,0.218347878248418);

\path [draw=color3, line width=0.7000000000000001pt] (axis cs:0.09875,0.22645562027129)
--(axis cs:0.09875,0.2331991718106);

\path [draw=color3, line width=0.7000000000000001pt] (axis cs:0.100833333333333,0.24668627488922)
--(axis cs:0.100833333333333,0.25362332862478);

\path [draw=color3, line width=0.7000000000000001pt] (axis cs:0.102916666666667,0.267700613401962)
--(axis cs:0.102916666666667,0.274831169333772);

\path [draw=color3, line width=0.7000000000000001pt] (axis cs:0.105,0.282039126144082)
--(axis cs:0.105,0.289276108283829);

\path [draw=color3, line width=0.7000000000000001pt] (axis cs:0.107083333333333,0.303353393061011)
--(axis cs:0.107083333333333,0.310745176957758);

\path [draw=color3, line width=0.7000000000000001pt] (axis cs:0.109166666666667,0.32028483523288)
--(axis cs:0.109166666666667,0.327783045337565);

\path [draw=color3, line width=0.7000000000000001pt] (axis cs:0.11125,0.339102923818185)
--(axis cs:0.11125,0.346707560130808);

\path [draw=color3, line width=0.7000000000000001pt] (axis cs:0.113333333333333,0.355705412256429)
--(axis cs:0.113333333333333,0.363397124557364);

\path [draw=color3, line width=0.7000000000000001pt] (axis cs:0.115416666666667,0.375703864238859)
--(axis cs:0.115416666666667,0.383482652528106);

\path [draw=color3, line width=0.7000000000000001pt] (axis cs:0.1175,0.393012635693415)
--(axis cs:0.1175,0.400849474641537);

\path [draw=color3, line width=0.7000000000000001pt] (axis cs:0.119583333333333,0.415855569960719)
--(axis cs:0.119583333333333,0.423769809787341);

\path [draw=color3, line width=0.7000000000000001pt] (axis cs:0.121666666666667,0.431471197198088)
--(axis cs:0.121666666666667,0.439414462354148);

\path [draw=color3, line width=0.7000000000000001pt] (axis cs:0.12375,0.449950656939956)
--(axis cs:0.12375,0.457922947425453);

\path [draw=color3, line width=0.7000000000000001pt] (axis cs:0.125833333333333,0.467298128833762)
--(axis cs:0.125833333333333,0.475289769538884);

\path [draw=color3, line width=0.7000000000000001pt] (axis cs:0.127916666666667,0.47957584318582)
--(axis cs:0.127916666666667,0.487586834110567);

\path [draw=color3, line width=0.7000000000000001pt] (axis cs:0.13,0.496671762224501)
--(axis cs:0.13,0.504682753149248);

\path [draw=color4, line width=0.7000000000000001pt] (axis cs:0.08,0.07340505814741)
--(axis cs:0.08,0.0776427562452834);

\path [draw=color4, line width=0.7000000000000001pt] (axis cs:0.0820833333333333,0.0861955533195302)
--(axis cs:0.0820833333333333,0.0907428549314035);

\path [draw=color4, line width=0.7000000000000001pt] (axis cs:0.0841666666666667,0.0993633777743377)
--(axis cs:0.0841666666666667,0.104210607790398);

\path [draw=color4, line width=0.7000000000000001pt] (axis cs:0.08625,0.116149693299019)
--(axis cs:0.08625,0.121335552158517);

\path [draw=color4, line width=0.7000000000000001pt] (axis cs:0.0883333333333333,0.13204589872095)
--(axis cs:0.0883333333333333,0.137512335765011);

\path [draw=color4, line width=0.7000000000000001pt] (axis cs:0.0904166666666667,0.149761024787631)
--(axis cs:0.0904166666666667,0.155517715126067);

\path [draw=color4, line width=0.7000000000000001pt] (axis cs:0.0925,0.16549275334275)
--(axis cs:0.0925,0.171491321426498);

\path [draw=color4, line width=0.7000000000000001pt] (axis cs:0.0945833333333333,0.183981888194431)
--(axis cs:0.0945833333333333,0.190232009133304);

\path [draw=color4, line width=0.7000000000000001pt] (axis cs:0.0966666666666667,0.205015576926798)
--(axis cs:0.0966666666666667,0.211526925830608);

\path [draw=color4, line width=0.7000000000000001pt] (axis cs:0.09875,0.224510923198978)
--(axis cs:0.09875,0.231235124518663);

\path [draw=color4, line width=0.7000000000000001pt] (axis cs:0.100833333333333,0.244548075620658)
--(axis cs:0.100833333333333,0.251465779136593);

\path [draw=color4, line width=0.7000000000000001pt] (axis cs:0.102916666666667,0.266549275334275)
--(axis cs:0.102916666666667,0.27366048104646);

\path [draw=color4, line width=0.7000000000000001pt] (axis cs:0.105,0.290156543276766)
--(axis cs:0.105,0.297461251185201);

\path [draw=color4, line width=0.7000000000000001pt] (axis cs:0.107083333333333,0.314034714294007)
--(axis cs:0.107083333333333,0.321494223959442);

\path [draw=color4, line width=0.7000000000000001pt] (axis cs:0.109166666666667,0.332204570521875)
--(axis cs:0.109166666666667,0.339770506395248);

\path [draw=color4, line width=0.7000000000000001pt] (axis cs:0.11125,0.352967356179493)
--(axis cs:0.11125,0.360639718260802);

\path [draw=color4, line width=0.7000000000000001pt] (axis cs:0.113333333333333,0.37919657888117)
--(axis cs:0.113333333333333,0.38698504228023);

\path [draw=color4, line width=0.7000000000000001pt] (axis cs:0.115416666666667,0.398856402020163)
--(axis cs:0.115416666666667,0.40671259118791);

\path [draw=color4, line width=0.7000000000000001pt] (axis cs:0.1175,0.41888387933203)
--(axis cs:0.1175,0.426798119158652);

\path [draw=color4, line width=0.7000000000000001pt] (axis cs:0.119583333333333,0.439249985487335)
--(axis cs:0.119583333333333,0.447202925753207);

\path [draw=color4, line width=0.7000000000000001pt] (axis cs:0.121666666666667,0.460341724878577)
--(axis cs:0.121666666666667,0.468333365583699);

\path [draw=color4, line width=0.7000000000000001pt] (axis cs:0.12375,0.47915013835407)
--(axis cs:0.12375,0.487161129278817);

\path [draw=color4, line width=0.7000000000000001pt] (axis cs:0.125833333333333,0.497261943923064)
--(axis cs:0.125833333333333,0.505272934847811);

\path [draw=color4, line width=0.7000000000000001pt] (axis cs:0.127916666666667,0.513148474235183)
--(axis cs:0.127916666666667,0.52115946515993);

\path [draw=color4, line width=0.7000000000000001pt] (axis cs:0.13,0.535091623289924)
--(axis cs:0.13,0.543083263995046);

\path [draw=color5, line width=0.7000000000000001pt] (axis cs:0.08,0.0611757193444146)
--(axis cs:0.08,0.0650747885988506);

\path [draw=color5, line width=0.7000000000000001pt] (axis cs:0.0820833333333333,0.0738791385282223)
--(axis cs:0.0820833333333333,0.0781265117359082);

\path [draw=color5, line width=0.7000000000000001pt] (axis cs:0.0841666666666667,0.0860988022214052)
--(axis cs:0.0841666666666667,0.0906461038332785);

\path [draw=color5, line width=0.7000000000000001pt] (axis cs:0.08625,0.101395150834962)
--(axis cs:0.08625,0.106281081290273);

\path [draw=color5, line width=0.7000000000000001pt] (axis cs:0.0883333333333333,0.118761972948393)
--(axis cs:0.0883333333333333,0.123996207356954);

\path [draw=color5, line width=0.7000000000000001pt] (axis cs:0.0904166666666667,0.136941504286074)
--(axis cs:0.0904166666666667,0.142495017318447);

\path [draw=color5, line width=0.7000000000000001pt] (axis cs:0.0925,0.156930281158691)
--(axis cs:0.0925,0.162803072814876);

\path [draw=color5, line width=0.7000000000000001pt] (axis cs:0.0945833333333333,0.177605990827996)
--(axis cs:0.0945833333333333,0.183769035778556);

\path [draw=color5, line width=0.7000000000000001pt] (axis cs:0.0966666666666667,0.198262350277676)
--(axis cs:0.0966666666666667,0.204696298302986);

\path [draw=color5, line width=0.7000000000000001pt] (axis cs:0.09875,0.222537200797229)
--(axis cs:0.09875,0.229242051897289);

\path [draw=color5, line width=0.7000000000000001pt] (axis cs:0.100833333333333,0.244857679134658)
--(axis cs:0.100833333333333,0.251775382650593);

\path [draw=color5, line width=0.7000000000000001pt] (axis cs:0.102916666666667,0.269713036242961)
--(axis cs:0.102916666666667,0.276853267284584);

\path [draw=color5, line width=0.7000000000000001pt] (axis cs:0.105,0.295593954991389)
--(axis cs:0.105,0.302937363339074);

\path [draw=color5, line width=0.7000000000000001pt] (axis cs:0.107083333333333,0.321155595116005)
--(axis cs:0.107083333333333,0.328663480330502);

\path [draw=color5, line width=0.7000000000000001pt] (axis cs:0.109166666666667,0.344733837729058)
--(axis cs:0.109166666666667,0.352367499371118);

\path [draw=color5, line width=0.7000000000000001pt] (axis cs:0.11125,0.368186303914549)
--(axis cs:0.11125,0.375936066874359);

\path [draw=color5, line width=0.7000000000000001pt] (axis cs:0.113333333333333,0.394899282106852)
--(axis cs:0.113333333333333,0.402745796164786);

\path [draw=color5, line width=0.7000000000000001pt] (axis cs:0.115416666666667,0.417442287969968)
--(axis cs:0.115416666666667,0.425356527796591);

\path [draw=color5, line width=0.7000000000000001pt] (axis cs:0.1175,0.441068906132085)
--(axis cs:0.1175,0.449031521507769);

\path [draw=color5, line width=0.7000000000000001pt] (axis cs:0.119583333333333,0.465450182859575)
--(axis cs:0.119583333333333,0.473441823564697);

\path [draw=color5, line width=0.7000000000000001pt] (axis cs:0.121666666666667,0.486222643627005)
--(axis cs:0.121666666666667,0.494233634551752);

\path [draw=color5, line width=0.7000000000000001pt] (axis cs:0.12375,0.504702103368873)
--(axis cs:0.12375,0.51271309429362);

\path [draw=color5, line width=0.7000000000000001pt] (axis cs:0.125833333333333,0.527390235879177)
--(axis cs:0.125833333333333,0.535381876584299);

\path [draw=color5, line width=0.7000000000000001pt] (axis cs:0.127916666666667,0.545637492985545)
--(axis cs:0.127916666666667,0.553609783471042);

\path [draw=color5, line width=0.7000000000000001pt] (axis cs:0.13,0.563217167514851)
--(axis cs:0.13,0.571150757561098);

\addplot [line width=0.7000000000000001pt, color1]
table {%
0.0914629643963 0.193687969116566
0.0916102707936101 0.194440770453131
0.0917575771909202 0.195194584936501
0.0919048835882303 0.195949412566677
0.0920521899855404 0.196705253343657
0.0921994963828505 0.197462107267442
0.0923468027801606 0.198219974338033
0.0924941091774707 0.198978854555428
0.0926414155747808 0.199738747919629
0.0927887219720909 0.200499654430634
0.092936028369401 0.201261574088445
0.0930833347667111 0.20202450689306
0.0932306411640212 0.202788452844481
0.0933779475613313 0.203553411942706
0.0935252539586414 0.204319384187737
0.0936725603559515 0.205086369579572
0.0938198667532616 0.205854368118213
0.0939671731505717 0.206623379803658
0.0941144795478818 0.207393404635909
0.0942617859451919 0.208164442614965
0.094409092342502 0.208936493740825
0.0945563987398121 0.209709558013491
0.0947037051371222 0.210483635432962
0.0948510115344323 0.211258725999238
0.0949983179317424 0.212034829712318
0.0951456243290525 0.212811946572204
0.0952929307263626 0.213590076578895
0.0954402371236727 0.21436921973239
0.0955875435209828 0.215149376032691
0.0957348499182929 0.215930545479797
0.095882156315603 0.216712728073708
0.0960294627129131 0.217495923814424
0.0961767691102232 0.218280132701945
0.0963240755075333 0.219065354736271
0.0964713819048434 0.219851589917402
0.0966186883021535 0.220638838245338
0.0967659946994636 0.221427099720079
0.0969133010967737 0.222216374341625
0.0970606074940838 0.223006662109976
0.0972079138913939 0.223797963025132
0.097355220288704 0.224590277087093
0.0975025266860141 0.225383604295859
0.0976498330833242 0.22617794465143
0.0977971394806343 0.226973298153806
0.0979444458779444 0.227769664802987
0.0980917522752545 0.228567044598973
0.0982390586725646 0.229365437541764
0.0983863650698747 0.23016484363136
0.0985336714671848 0.230965262867762
0.098680977864495 0.231766695250968
0.0988282842618051 0.232569140780979
0.0989755906591152 0.233372599457795
0.0991228970564253 0.234177071281417
0.0992702034537354 0.234982556251843
0.0994175098510455 0.235789054369074
0.0995648162483556 0.236596565633111
0.0997121226456657 0.237405090043952
0.0998594290429758 0.238214627601598
0.100006735440286 0.23902517830605
0.100154041837596 0.239836742157306
0.100301348234906 0.240649319155368
0.100448654632216 0.241462909300234
0.100595961029526 0.242277512591905
0.100743267426836 0.243093129030382
0.100890573824146 0.243909758615663
0.101037880221457 0.24472740134775
0.101185186618767 0.245546057226642
0.101332493016077 0.246365726252338
0.101479799413387 0.24718640842484
0.101627105810697 0.248008103744146
0.101774412208007 0.248830812210258
0.101921718605317 0.249654533823175
0.102069025002627 0.250479268582896
0.102216331399937 0.251305016489423
0.102363637797247 0.252131777542755
0.102510944194558 0.252959551742891
0.102658250591868 0.253788339089833
0.102805556989178 0.25461813958358
0.102952863386488 0.255448953224132
0.103100169783798 0.256280780011488
0.103247476181108 0.25711361994565
0.103394782578418 0.257947473026617
0.103542088975728 0.258782339254389
0.103689395373038 0.259618218628966
0.103836701770348 0.260455111150348
0.103984008167659 0.261293016818535
0.104131314564969 0.262131935633527
0.104278620962279 0.262971867595323
0.104425927359589 0.263812812703925
0.104573233756899 0.264654770959332
0.104720540154209 0.265497742361544
0.104867846551519 0.266341726910562
0.105015152948829 0.267186724606384
0.105162459346139 0.268032735449011
0.105309765743449 0.268879759438443
0.10545707214076 0.26972779657468
0.10560437853807 0.270576846857722
0.10575168493538 0.271426910287569
0.10589899133269 0.272277986864221
0.10604629773 0.273130076587678
};
\addplot [line width=0.7000000000000001pt, color2]
table {%
0.0914629643963 0.180512017311588
0.0916102707936101 0.181480715249284
0.0917575771909202 0.182451173015658
0.0919048835882303 0.18342339061071
0.0920521899855404 0.18439736803444
0.0921994963828505 0.185373105286848
0.0923468027801606 0.186350602367934
0.0924941091774707 0.187329859277697
0.0926414155747808 0.188310876016139
0.0927887219720909 0.189293652583258
0.092936028369401 0.190278188979056
0.0930833347667111 0.191264485203531
0.0932306411640212 0.192252541256684
0.0933779475613313 0.193242357138515
0.0935252539586414 0.194233932849024
0.0936725603559515 0.195227268388211
0.0938198667532616 0.196222363756076
0.0939671731505717 0.197219218952619
0.0941144795478818 0.198217833977839
0.0942617859451919 0.199218208831738
0.094409092342502 0.200220343514314
0.0945563987398121 0.201224238025568
0.0947037051371222 0.2022298923655
0.0948510115344323 0.20323730653411
0.0949983179317424 0.204246480531398
0.0951456243290525 0.205257414357364
0.0952929307263626 0.206270108012008
0.0954402371236727 0.207284561495329
0.0955875435209828 0.208300774807329
0.0957348499182929 0.209318747948006
0.095882156315603 0.210338480917362
0.0960294627129131 0.211359973715395
0.0961767691102232 0.212383226342106
0.0963240755075333 0.213408238797495
0.0964713819048434 0.214435011081562
0.0966186883021535 0.215463543194307
0.0967659946994636 0.21649383513573
0.0969133010967737 0.21752588690583
0.0970606074940838 0.218559698504609
0.0972079138913939 0.219595269932065
0.097355220288704 0.2206326011882
0.0975025266860141 0.221671692273012
0.0976498330833242 0.222712543186502
0.0977971394806343 0.22375515392867
0.0979444458779444 0.224799524499516
0.0980917522752545 0.22584565489904
0.0982390586725646 0.226893545127242
0.0983863650698747 0.227943195184121
0.0985336714671848 0.228994605069679
0.098680977864495 0.230047774783914
0.0988282842618051 0.231102704326828
0.0989755906591152 0.232159393698419
0.0991228970564253 0.233217842898688
0.0992702034537354 0.234278051927635
0.0994175098510455 0.23534002078526
0.0995648162483556 0.236403749471563
0.0997121226456657 0.237469237986543
0.0998594290429758 0.238536486330202
0.100006735440286 0.239605494502538
0.100154041837596 0.240676262503553
0.100301348234906 0.241748790333245
0.100448654632216 0.242823077991615
0.100595961029526 0.243899125478664
0.100743267426836 0.24497693279439
0.100890573824146 0.246056499938793
0.101037880221457 0.247137826911875
0.101185186618767 0.248220913713635
0.101332493016077 0.249305760344073
0.101479799413387 0.250392366803188
0.101627105810697 0.251480733090981
0.101774412208007 0.252570859207453
0.101921718605317 0.253662745152602
0.102069025002627 0.254756390926429
0.102216331399937 0.255851796528934
0.102363637797247 0.256948961960117
0.102510944194558 0.258047887219978
0.102658250591868 0.259148572308516
0.102805556989178 0.260251017225733
0.102952863386488 0.261355221971628
0.103100169783798 0.2624611865462
0.103247476181108 0.26356891094945
0.103394782578418 0.264678395181378
0.103542088975728 0.265789639241985
0.103689395373038 0.266902643131268
0.103836701770348 0.26801740684923
0.103984008167659 0.26913393039587
0.104131314564969 0.270252213771188
0.104278620962279 0.271372256975183
0.104425927359589 0.272494060007857
0.104573233756899 0.273617622869208
0.104720540154209 0.274742945559238
0.104867846551519 0.275870028077945
0.105015152948829 0.27699887042533
0.105162459346139 0.278129472601393
0.105309765743449 0.279261834606134
0.10545707214076 0.280395956439553
0.10560437853807 0.281531838101649
0.10575168493538 0.282669479592424
0.10589899133269 0.283808880911876
0.10604629773 0.284950042060007
};
\addplot [line width=0.7000000000000001pt, color3]
table {%
0.0914629643963 0.169376757199867
0.0916102707936101 0.170522995500218
0.0917575771909202 0.171671804322052
0.0919048835882303 0.172823183665368
0.0920521899855404 0.173977133530168
0.0921994963828505 0.175133653916451
0.0923468027801606 0.176292744824217
0.0924941091774707 0.177454406253466
0.0926414155747808 0.178618638204198
0.0927887219720909 0.179785440676413
0.092936028369401 0.180954813670111
0.0930833347667111 0.182126757185293
0.0932306411640212 0.183301271221957
0.0933779475613313 0.184478355780104
0.0935252539586414 0.185658010859735
0.0936725603559515 0.186840236460848
0.0938198667532616 0.188025032583445
0.0939671731505717 0.189212399227524
0.0941144795478818 0.190402336393087
0.0942617859451919 0.191594844080133
0.094409092342502 0.192789922288661
0.0945563987398121 0.193987571018673
0.0947037051371222 0.195187790270168
0.0948510115344323 0.196390580043146
0.0949983179317424 0.197595940337607
0.0951456243290525 0.198803871153551
0.0952929307263626 0.200014372490979
0.0954402371236727 0.201227444349889
0.0955875435209828 0.202443086730282
0.0957348499182929 0.203661299632158
0.095882156315603 0.204882083055518
0.0960294627129131 0.20610543700036
0.0961767691102232 0.207331361466686
0.0963240755075333 0.208559856454494
0.0964713819048434 0.209790921963786
0.0966186883021535 0.211024557994561
0.0967659946994636 0.212260764546819
0.0969133010967737 0.213499541620559
0.0970606074940838 0.214740889215783
0.0972079138913939 0.21598480733249
0.097355220288704 0.21723129597068
0.0975025266860141 0.218480355130354
0.0976498330833242 0.21973198481151
0.0977971394806343 0.220986185014149
0.0979444458779444 0.222242955738271
0.0980917522752545 0.223502296983877
0.0982390586725646 0.224764208750965
0.0983863650698747 0.226028691039537
0.0985336714671848 0.227295743849591
0.098680977864495 0.228565367181129
0.0988282842618051 0.229837561034149
0.0989755906591152 0.231112325408653
0.0991228970564253 0.23238966030464
0.0992702034537354 0.23366956572211
0.0994175098510455 0.234952041661063
0.0995648162483556 0.236237088121499
0.0997121226456657 0.237524705103418
0.0998594290429758 0.23881489260682
0.100006735440286 0.240107650631705
0.100154041837596 0.241402979178073
0.100301348234906 0.242700878245925
0.100448654632216 0.244001347835259
0.100595961029526 0.245304387946076
0.100743267426836 0.246609998578377
0.100890573824146 0.24791817973216
0.101037880221457 0.249228931407427
0.101185186618767 0.250542253604177
0.101332493016077 0.251858146322409
0.101479799413387 0.253176609562125
0.101627105810697 0.254497643323324
0.101774412208007 0.255821247606006
0.101921718605317 0.257147422410171
0.102069025002627 0.258476167735819
0.102216331399937 0.25980748358295
0.102363637797247 0.261141369951564
0.102510944194558 0.262477826841661
0.102658250591868 0.263816854253242
0.102805556989178 0.265158452186305
0.102952863386488 0.266502620640851
0.103100169783798 0.267849359616881
0.103247476181108 0.269198669114394
0.103394782578418 0.270550549133389
0.103542088975728 0.271904999673868
0.103689395373038 0.27326202073583
0.103836701770348 0.274621612319274
0.103984008167659 0.275983774424202
0.104131314564969 0.277348507050613
0.104278620962279 0.278715810198507
0.104425927359589 0.280085683867884
0.104573233756899 0.281458128058744
0.104720540154209 0.282833142771087
0.104867846551519 0.284210728004914
0.105015152948829 0.285590883760223
0.105162459346139 0.286973610037015
0.105309765743449 0.288358906835291
0.10545707214076 0.289746774155049
0.10560437853807 0.291137211996291
0.10575168493538 0.292530220359015
0.10589899133269 0.293925799243223
0.10604629773 0.295323948648914
};
\addplot [line width=0.7000000000000001pt, color4]
table {%
0.0914629643963 0.159557925224593
0.0916102707936101 0.160856724913665
0.0917575771909202 0.162158955722631
0.0919048835882303 0.163464617651492
0.0920521899855404 0.164773710700248
0.0921994963828505 0.166086234868898
0.0923468027801606 0.167402190157443
0.0924941091774707 0.168721576565883
0.0926414155747808 0.170044394094218
0.0927887219720909 0.171370642742447
0.092936028369401 0.172700322510571
0.0930833347667111 0.17403343339859
0.0932306411640212 0.175369975406504
0.0933779475613313 0.176709948534312
0.0935252539586414 0.178053352782016
0.0936725603559515 0.179400188149614
0.0938198667532616 0.180750454637106
0.0939671731505717 0.182104152244494
0.0941144795478818 0.183461280971776
0.0942617859451919 0.184821840818953
0.094409092342502 0.186185831786024
0.0945563987398121 0.187553253872991
0.0947037051371222 0.188924107079852
0.0948510115344323 0.190298391406608
0.0949983179317424 0.191676106853259
0.0951456243290525 0.193057253419804
0.0952929307263626 0.194441831106244
0.0954402371236727 0.195829839912579
0.0955875435209828 0.197221279838809
0.0957348499182929 0.198616150884933
0.095882156315603 0.200014453050952
0.0960294627129131 0.201416186336866
0.0961767691102232 0.202821350742675
0.0963240755075333 0.204229946268378
0.0964713819048434 0.205641972913977
0.0966186883021535 0.207057430679469
0.0967659946994636 0.208476319564857
0.0969133010967737 0.20989863957014
0.0970606074940838 0.211324390695317
0.0972079138913939 0.212753572940389
0.097355220288704 0.214186186305355
0.0975025266860141 0.215622230790217
0.0976498330833242 0.217061706394973
0.0977971394806343 0.218504613119624
0.0979444458779444 0.21995095096417
0.0980917522752545 0.22140071992861
0.0982390586725646 0.222853920012945
0.0983863650698747 0.224310551217175
0.0985336714671848 0.2257706135413
0.098680977864495 0.227234106985319
0.0988282842618051 0.228701031549233
0.0989755906591152 0.230171387233042
0.0991228970564253 0.231645174036746
0.0992702034537354 0.233122391960344
0.0994175098510455 0.234603041003837
0.0995648162483556 0.236087121167225
0.0997121226456657 0.237574632450508
0.0998594290429758 0.239065574853686
0.100006735440286 0.240559948376758
0.100154041837596 0.242057753019725
0.100301348234906 0.243558988782586
0.100448654632216 0.245063655665343
0.100595961029526 0.246571753667994
0.100743267426836 0.24808328279054
0.100890573824146 0.24959824303298
0.101037880221457 0.251116634395316
0.101185186618767 0.252638456877546
0.101332493016077 0.254163710479671
0.101479799413387 0.255692395201691
0.101627105810697 0.257224511043605
0.101774412208007 0.258760058005414
0.101921718605317 0.260299036087118
0.102069025002627 0.261841445288717
0.102216331399937 0.26338728561021
0.102363637797247 0.264936557051598
0.102510944194558 0.266489259612881
0.102658250591868 0.268045393294059
0.102805556989178 0.269604958095131
0.102952863386488 0.271167954016098
0.103100169783798 0.27273438105696
0.103247476181108 0.274304239217717
0.103394782578418 0.275877528498368
0.103542088975728 0.277454248898915
0.103689395373038 0.279034400419355
0.103836701770348 0.280617983059691
0.103984008167659 0.282204996819922
0.104131314564969 0.283795441700047
0.104278620962279 0.285389317700067
0.104425927359589 0.286986624819981
0.104573233756899 0.288587363059791
0.104720540154209 0.290191532419495
0.104867846551519 0.291799132899094
0.105015152948829 0.293410164498587
0.105162459346139 0.295024627217976
0.105309765743449 0.296642521057259
0.10545707214076 0.298263846016437
0.10560437853807 0.29988860209551
0.10575168493538 0.301516789294477
0.10589899133269 0.303148407613339
0.10604629773 0.304783457052096
};
\addplot [line width=0.7000000000000001pt, color5]
table {%
0.0914629643963 0.150688629772594
0.0916102707936101 0.152121826053074
0.0917575771909202 0.15355935543753
0.0919048835882303 0.155001217925961
0.0920521899855404 0.156447413518367
0.0921994963828505 0.157897942214747
0.0923468027801606 0.159352804015103
0.0924941091774707 0.160811998919434
0.0926414155747808 0.16227552692774
0.0927887219720909 0.163743388040021
0.092936028369401 0.165215582256277
0.0930833347667111 0.166692109576508
0.0932306411640212 0.168172970000714
0.0933779475613313 0.169658163528895
0.0935252539586414 0.171147690161051
0.0936725603559515 0.172641549897183
0.0938198667532616 0.174139742737289
0.0939671731505717 0.17564226868137
0.0941144795478818 0.177149127729427
0.0942617859451919 0.178660319881458
0.094409092342502 0.180175845137464
0.0945563987398121 0.181695703497446
0.0947037051371222 0.183219894961402
0.0948510115344323 0.184748419529334
0.0949983179317424 0.186281277201241
0.0951456243290525 0.187818467977122
0.0952929307263626 0.189359991856979
0.0954402371236727 0.19090584884081
0.0955875435209828 0.192456038928617
0.0957348499182929 0.194010562120399
0.095882156315603 0.195569418416156
0.0960294627129131 0.197132607815888
0.0961767691102232 0.198700130319595
0.0963240755075333 0.200271985927276
0.0964713819048434 0.201848174638933
0.0966186883021535 0.203428696454565
0.0967659946994636 0.205013551374173
0.0969133010967737 0.206602739397755
0.0970606074940838 0.208196260525312
0.0972079138913939 0.209794114756844
0.097355220288704 0.211396302092351
0.0975025266860141 0.213002822531834
0.0976498330833242 0.214613676075291
0.0977971394806343 0.216228862722723
0.0979444458779444 0.217848382474131
0.0980917522752545 0.219472235329513
0.0982390586725646 0.221100421288871
0.0983863650698747 0.222732940352203
0.0985336714671848 0.224369792519511
0.098680977864495 0.226010977790793
0.0988282842618051 0.227656496166051
0.0989755906591152 0.229306347645284
0.0991228970564253 0.230960532228491
0.0992702034537354 0.232619049915674
0.0994175098510455 0.234281900706832
0.0995648162483556 0.235949084601965
0.0997121226456657 0.237620601601073
0.0998594290429758 0.239296451704156
0.100006735440286 0.240976634911213
0.100154041837596 0.242661151222247
0.100301348234906 0.244350000637255
0.100448654632216 0.246043183156238
0.100595961029526 0.247740698779196
0.100743267426836 0.249442547506129
0.100890573824146 0.251148729337037
0.101037880221457 0.25285924427192
0.101185186618767 0.254574092310779
0.101332493016077 0.256293273453612
0.101479799413387 0.25801678770042
0.101627105810697 0.259744635051204
0.101774412208007 0.261476815505962
0.101921718605317 0.263213329064696
0.102069025002627 0.264954175727404
0.102216331399937 0.266699355494088
0.102363637797247 0.268448868364746
0.102510944194558 0.27020271433938
0.102658250591868 0.271960893417989
0.102805556989178 0.273723405600573
0.102952863386488 0.275490250887131
0.103100169783798 0.277261429277665
0.103247476181108 0.279036940772174
0.103394782578418 0.280816785370658
0.103542088975728 0.282600963073117
0.103689395373038 0.284389473879551
0.103836701770348 0.28618231778996
0.103984008167659 0.287979494804344
0.104131314564969 0.289781004922703
0.104278620962279 0.291586848145037
0.104425927359589 0.293397024471346
0.104573233756899 0.295211533901631
0.104720540154209 0.29703037643589
0.104867846551519 0.298853552074124
0.105015152948829 0.300681060816334
0.105162459346139 0.302512902662518
0.105309765743449 0.304349077612677
0.10545707214076 0.306189585666812
0.10560437853807 0.308034426824921
0.10575168493538 0.309883601087006
0.10589899133269 0.311737108453066
0.10604629773 0.3135949489231
};
\path [draw=white, fill opacity=0] (axis cs:0,0.0356769674335804)
--(axis cs:0,0.596649509471933);

\path [draw=white, fill opacity=0] (axis cs:1,0.0356769674335804)
--(axis cs:1,0.596649509471933);

\path [draw=white, fill opacity=0] (axis cs:0.0775,0)
--(axis cs:0.1325,0);

\path [draw=white, fill opacity=0] (axis cs:0.0775,1)
--(axis cs:0.1325,1);

\end{axis}

\end{tikzpicture}

%% file: paper_iidxz_ext.tex
\begin{tikzpicture}

\definecolor{color1}{rgb}{0.967797559291991,0.441274560091574,0.53581031550587}
\definecolor{color0}{rgb}{0.917647058823529,0.917647058823529,0.949019607843137}
\definecolor{color3}{rgb}{0.201253172212011,0.690792081537903,0.479667611892753}
\definecolor{color2}{rgb}{0.680418912779335,0.615149751467757,0.194054521114453}
\definecolor{color5}{rgb}{0.800493618642396,0.477033635337372,0.957954719600752}
\definecolor{color4}{rgb}{0.219799566082832,0.662515787685034,0.773209315931721}

\begin{axis}[
title={IID $X$/$Z$ Error Model, with multi-path summation},
xlabel={physical error probability},
ylabel={logical error probability},
xmin=0.0775, xmax=0.1325,
ymin=0.0291122088779112, ymax=0.568711812881871,
width=\figurewidth,
height=\figureheight,
xtick={0.07,0.08,0.09,0.1,0.11,0.12,0.13,0.14},
xticklabels={,0.08,0.09,0.10,0.11,0.12,0.13,},
ytick={0,0.1,0.2,0.3,0.4,0.5,0.6},
yticklabels={,0.1,0.2,0.3,0.4,0.5,},
tick align=outside,
tick pos=left,
xmajorgrids,
x grid style={white},
ymajorgrids,
y grid style={white},
axis line style={white},
axis background/.style={fill=color0},
legend pos=north west,
legend entries={$d = 7$, $d = 11$, $d = 15$, $d = 19$, $d = 23$}
]
\path [draw=color1, line width=0.7000000000000001pt] (axis cs:0.08,0.133368666313337)
--(axis cs:0.08,0.138958610413896);

\path [draw=color1, line width=0.7000000000000001pt] (axis cs:0.0820833333333333,0.142518574814252)
--(axis cs:0.0820833333333333,0.148258517414826);

\path [draw=color1, line width=0.7000000000000001pt] (axis cs:0.0841666666666667,0.15089849101509)
--(axis cs:0.0841666666666667,0.156778432215678);

\path [draw=color1, line width=0.7000000000000001pt] (axis cs:0.08625,0.160878391216088)
--(axis cs:0.08625,0.166908330916691);

\path [draw=color1, line width=0.7000000000000001pt] (axis cs:0.0883333333333333,0.171478285217148)
--(axis cs:0.0883333333333333,0.177668223317767);

\path [draw=color1, line width=0.7000000000000001pt] (axis cs:0.0904166666666667,0.181388186118139)
--(axis cs:0.0904166666666667,0.187708122918771);

\path [draw=color1, line width=0.7000000000000001pt] (axis cs:0.0925,0.192148078519215)
--(axis cs:0.0925,0.198608013919861);

\path [draw=color1, line width=0.7000000000000001pt] (axis cs:0.0945833333333333,0.203717962820372)
--(axis cs:0.0945833333333333,0.210317896821032);

\path [draw=color1, line width=0.7000000000000001pt] (axis cs:0.0966666666666667,0.216137838621614)
--(axis cs:0.0966666666666667,0.222887771122289);

\path [draw=color1, line width=0.7000000000000001pt] (axis cs:0.09875,0.225057749422506)
--(axis cs:0.09875,0.231887681123189);

\path [draw=color1, line width=0.7000000000000001pt] (axis cs:0.100833333333333,0.238357616423836)
--(axis cs:0.100833333333333,0.245327546724533);

\path [draw=color1, line width=0.7000000000000001pt] (axis cs:0.102916666666667,0.250787492125079)
--(axis cs:0.102916666666667,0.257877421225788);

\path [draw=color1, line width=0.7000000000000001pt] (axis cs:0.105,0.260057399426006)
--(axis cs:0.105,0.267227327726723);

\path [draw=color1, line width=0.7000000000000001pt] (axis cs:0.107083333333333,0.272847271527285)
--(axis cs:0.107083333333333,0.280137198628014);

\path [draw=color1, line width=0.7000000000000001pt] (axis cs:0.109166666666667,0.280807191928081)
--(axis cs:0.109166666666667,0.288157118428816);

\path [draw=color1, line width=0.7000000000000001pt] (axis cs:0.11125,0.295017049829502)
--(axis cs:0.11125,0.302466975330247);

\path [draw=color1, line width=0.7000000000000001pt] (axis cs:0.113333333333333,0.30339696603034)
--(axis cs:0.113333333333333,0.310906890931091);

\path [draw=color1, line width=0.7000000000000001pt] (axis cs:0.115416666666667,0.317376826231738)
--(axis cs:0.115416666666667,0.324986750132499);

\path [draw=color1, line width=0.7000000000000001pt] (axis cs:0.1175,0.328836711632884)
--(axis cs:0.1175,0.336506634933651);

\path [draw=color1, line width=0.7000000000000001pt] (axis cs:0.119583333333333,0.340246597534025)
--(axis cs:0.119583333333333,0.347986520134799);

\path [draw=color1, line width=0.7000000000000001pt] (axis cs:0.121666666666667,0.347736522634774)
--(axis cs:0.121666666666667,0.355516444835552);

\path [draw=color1, line width=0.7000000000000001pt] (axis cs:0.12375,0.360116398836012)
--(axis cs:0.12375,0.367956320436796);

\path [draw=color1, line width=0.7000000000000001pt] (axis cs:0.125833333333333,0.371956280437196)
--(axis cs:0.125833333333333,0.379846201537985);

\path [draw=color1, line width=0.7000000000000001pt] (axis cs:0.127916666666667,0.381936180638194)
--(axis cs:0.127916666666667,0.389866101338987);

\path [draw=color1, line width=0.7000000000000001pt] (axis cs:0.13,0.394746052539475)
--(axis cs:0.13,0.402725972740273);

\path [draw=color2, line width=0.7000000000000001pt] (axis cs:0.08,0.107338926610734)
--(axis cs:0.08,0.112428875711243);

\path [draw=color2, line width=0.7000000000000001pt] (axis cs:0.0820833333333333,0.120578794212058)
--(axis cs:0.0820833333333333,0.125928740712593);

\path [draw=color2, line width=0.7000000000000001pt] (axis cs:0.0841666666666667,0.13049869501305)
--(axis cs:0.0841666666666667,0.136038639613604);

\path [draw=color2, line width=0.7000000000000001pt] (axis cs:0.08625,0.141318586814132)
--(axis cs:0.08625,0.147038529614704);

\path [draw=color2, line width=0.7000000000000001pt] (axis cs:0.0883333333333333,0.156368436315637)
--(axis cs:0.0883333333333333,0.162328376716233);

\path [draw=color2, line width=0.7000000000000001pt] (axis cs:0.0904166666666667,0.166858331416686)
--(axis cs:0.0904166666666667,0.172978270217298);

\path [draw=color2, line width=0.7000000000000001pt] (axis cs:0.0925,0.180358196418036)
--(axis cs:0.0925,0.186668133318667);

\path [draw=color2, line width=0.7000000000000001pt] (axis cs:0.0945833333333333,0.19549804501955)
--(axis cs:0.0945833333333333,0.2019979800202);

\path [draw=color2, line width=0.7000000000000001pt] (axis cs:0.0966666666666667,0.212777872221278)
--(axis cs:0.0966666666666667,0.219487805121949);

\path [draw=color2, line width=0.7000000000000001pt] (axis cs:0.09875,0.225677743222568)
--(axis cs:0.09875,0.232527674723253);

\path [draw=color2, line width=0.7000000000000001pt] (axis cs:0.100833333333333,0.2409975900241)
--(axis cs:0.100833333333333,0.248007519924801);

\path [draw=color2, line width=0.7000000000000001pt] (axis cs:0.102916666666667,0.256117438825612)
--(axis cs:0.102916666666667,0.263267367326327);

\path [draw=color2, line width=0.7000000000000001pt] (axis cs:0.105,0.269377306226938)
--(axis cs:0.105,0.276627233727663);

\path [draw=color2, line width=0.7000000000000001pt] (axis cs:0.107083333333333,0.28029719702803)
--(axis cs:0.107083333333333,0.287647123528765);

\path [draw=color2, line width=0.7000000000000001pt] (axis cs:0.109166666666667,0.294647053529465)
--(axis cs:0.109166666666667,0.30209697903021);

\path [draw=color2, line width=0.7000000000000001pt] (axis cs:0.11125,0.309986900130999)
--(axis cs:0.11125,0.317546824531755);

\path [draw=color2, line width=0.7000000000000001pt] (axis cs:0.113333333333333,0.325966740332597)
--(axis cs:0.113333333333333,0.333626663733363);

\path [draw=color2, line width=0.7000000000000001pt] (axis cs:0.115416666666667,0.339486605133949)
--(axis cs:0.115416666666667,0.347216527834722);

\path [draw=color2, line width=0.7000000000000001pt] (axis cs:0.1175,0.353876461235388)
--(axis cs:0.1175,0.361686383136169);

\path [draw=color2, line width=0.7000000000000001pt] (axis cs:0.119583333333333,0.369016309836902)
--(axis cs:0.119583333333333,0.37689623103769);

\path [draw=color2, line width=0.7000000000000001pt] (axis cs:0.121666666666667,0.382966170338297)
--(axis cs:0.121666666666667,0.390906090939091);

\path [draw=color2, line width=0.7000000000000001pt] (axis cs:0.12375,0.401585984140159)
--(axis cs:0.12375,0.409585904140959);

\path [draw=color2, line width=0.7000000000000001pt] (axis cs:0.125833333333333,0.414255857441426)
--(axis cs:0.125833333333333,0.42229577704223);

\path [draw=color2, line width=0.7000000000000001pt] (axis cs:0.127916666666667,0.426665733342667)
--(axis cs:0.127916666666667,0.434725652743473);

\path [draw=color2, line width=0.7000000000000001pt] (axis cs:0.13,0.442025579744203)
--(axis cs:0.13,0.450125498745013);

\path [draw=color3, line width=0.7000000000000001pt] (axis cs:0.08,0.0852291477085229)
--(axis cs:0.08,0.0898291017089829);

\path [draw=color3, line width=0.7000000000000001pt] (axis cs:0.0820833333333333,0.0964790352096479)
--(axis cs:0.0820833333333333,0.101338986610134);

\path [draw=color3, line width=0.7000000000000001pt] (axis cs:0.0841666666666667,0.111168888311117)
--(axis cs:0.0841666666666667,0.116338836611634);

\path [draw=color3, line width=0.7000000000000001pt] (axis cs:0.08625,0.123708762912371)
--(axis cs:0.08625,0.129118708812912);

\path [draw=color3, line width=0.7000000000000001pt] (axis cs:0.0883333333333333,0.137068629313707)
--(axis cs:0.0883333333333333,0.142718572814272);

\path [draw=color3, line width=0.7000000000000001pt] (axis cs:0.0904166666666667,0.151558484415156)
--(axis cs:0.0904166666666667,0.157448425515745);

\path [draw=color3, line width=0.7000000000000001pt] (axis cs:0.0925,0.166718332816672)
--(axis cs:0.0925,0.172828271717283);

\path [draw=color3, line width=0.7000000000000001pt] (axis cs:0.0945833333333333,0.18279817201828)
--(axis cs:0.0945833333333333,0.189138108618914);

\path [draw=color3, line width=0.7000000000000001pt] (axis cs:0.0966666666666667,0.199788002119979)
--(axis cs:0.0966666666666667,0.206337936620634);

\path [draw=color3, line width=0.7000000000000001pt] (axis cs:0.09875,0.216217837821622)
--(axis cs:0.09875,0.222967770322297);

\path [draw=color3, line width=0.7000000000000001pt] (axis cs:0.100833333333333,0.233647663523365)
--(axis cs:0.100833333333333,0.240577594224058);

\path [draw=color3, line width=0.7000000000000001pt] (axis cs:0.102916666666667,0.253957460425396)
--(axis cs:0.102916666666667,0.261087389126109);

\path [draw=color3, line width=0.7000000000000001pt] (axis cs:0.105,0.270877291227088)
--(axis cs:0.105,0.278147218527815);

\path [draw=color3, line width=0.7000000000000001pt] (axis cs:0.107083333333333,0.289257107428926)
--(axis cs:0.107083333333333,0.296667033329667);

\path [draw=color3, line width=0.7000000000000001pt] (axis cs:0.109166666666667,0.306806931930681)
--(axis cs:0.109166666666667,0.314346856531435);

\path [draw=color3, line width=0.7000000000000001pt] (axis cs:0.11125,0.322876771232288)
--(axis cs:0.11125,0.330516694833052);

\path [draw=color3, line width=0.7000000000000001pt] (axis cs:0.113333333333333,0.338926610733893)
--(axis cs:0.113333333333333,0.346656533434666);

\path [draw=color3, line width=0.7000000000000001pt] (axis cs:0.115416666666667,0.358036419635804)
--(axis cs:0.115416666666667,0.365866341336587);

\path [draw=color3, line width=0.7000000000000001pt] (axis cs:0.1175,0.374756252437476)
--(axis cs:0.1175,0.382656173438266);

\path [draw=color3, line width=0.7000000000000001pt] (axis cs:0.119583333333333,0.391936080639194)
--(axis cs:0.119583333333333,0.399906000939991);

\path [draw=color3, line width=0.7000000000000001pt] (axis cs:0.121666666666667,0.414745852541475)
--(axis cs:0.121666666666667,0.422785772142279);

\path [draw=color3, line width=0.7000000000000001pt] (axis cs:0.12375,0.428585714142859)
--(axis cs:0.12375,0.436655633443666);

\path [draw=color3, line width=0.7000000000000001pt] (axis cs:0.125833333333333,0.449935500644994)
--(axis cs:0.125833333333333,0.458045419545805);

\path [draw=color3, line width=0.7000000000000001pt] (axis cs:0.127916666666667,0.462915370846292)
--(axis cs:0.127916666666667,0.471035289647104);

\path [draw=color3, line width=0.7000000000000001pt] (axis cs:0.13,0.47930520694793)
--(axis cs:0.13,0.487445125548745);

\path [draw=color4, line width=0.7000000000000001pt] (axis cs:0.08,0.0686393136068639)
--(axis cs:0.08,0.0728192718072819);

\path [draw=color4, line width=0.7000000000000001pt] (axis cs:0.0820833333333333,0.0794792052079479)
--(axis cs:0.0820833333333333,0.0839391606083939);

\path [draw=color4, line width=0.7000000000000001pt] (axis cs:0.0841666666666667,0.0906590934090659)
--(axis cs:0.0841666666666667,0.0953890461095389);

\path [draw=color4, line width=0.7000000000000001pt] (axis cs:0.08625,0.10489895101049)
--(axis cs:0.08625,0.109938900610994);

\path [draw=color4, line width=0.7000000000000001pt] (axis cs:0.0883333333333333,0.121678783212168)
--(axis cs:0.0883333333333333,0.127048729512705);

\path [draw=color4, line width=0.7000000000000001pt] (axis cs:0.0904166666666667,0.138078619213808)
--(axis cs:0.0904166666666667,0.143748562514375);

\path [draw=color4, line width=0.7000000000000001pt] (axis cs:0.0925,0.15429845701543)
--(axis cs:0.0925,0.160228397716023);

\path [draw=color4, line width=0.7000000000000001pt] (axis cs:0.0945833333333333,0.172718272817272)
--(axis cs:0.0945833333333333,0.178918210817892);

\path [draw=color4, line width=0.7000000000000001pt] (axis cs:0.0966666666666667,0.189188108118919)
--(axis cs:0.0966666666666667,0.195618043819562);

\path [draw=color4, line width=0.7000000000000001pt] (axis cs:0.09875,0.208747912520875)
--(axis cs:0.09875,0.215407845921541);

\path [draw=color4, line width=0.7000000000000001pt] (axis cs:0.100833333333333,0.228817711822882)
--(axis cs:0.100833333333333,0.235687643123569);

\path [draw=color4, line width=0.7000000000000001pt] (axis cs:0.102916666666667,0.250887491125089)
--(axis cs:0.102916666666667,0.257977420225798);

\path [draw=color4, line width=0.7000000000000001pt] (axis cs:0.105,0.274117258827412)
--(axis cs:0.105,0.281407185928141);

\path [draw=color4, line width=0.7000000000000001pt] (axis cs:0.107083333333333,0.290507094929051)
--(axis cs:0.107083333333333,0.297937020629794);

\path [draw=color4, line width=0.7000000000000001pt] (axis cs:0.109166666666667,0.310566894331057)
--(axis cs:0.109166666666667,0.318136818631814);

\path [draw=color4, line width=0.7000000000000001pt] (axis cs:0.11125,0.33169668303317)
--(axis cs:0.11125,0.339386606133939);

\path [draw=color4, line width=0.7000000000000001pt] (axis cs:0.113333333333333,0.354446455535445)
--(axis cs:0.113333333333333,0.362256377436226);

\path [draw=color4, line width=0.7000000000000001pt] (axis cs:0.115416666666667,0.373016269837302)
--(axis cs:0.115416666666667,0.380916190838092);

\path [draw=color4, line width=0.7000000000000001pt] (axis cs:0.1175,0.391136088639114)
--(axis cs:0.1175,0.39909600903991);

\path [draw=color4, line width=0.7000000000000001pt] (axis cs:0.119583333333333,0.415225847741523)
--(axis cs:0.119583333333333,0.423265767342327);

\path [draw=color4, line width=0.7000000000000001pt] (axis cs:0.121666666666667,0.437765622343777)
--(axis cs:0.121666666666667,0.445855541444586);

\path [draw=color4, line width=0.7000000000000001pt] (axis cs:0.12375,0.454405455945441)
--(axis cs:0.12375,0.462525374746253);

\path [draw=color4, line width=0.7000000000000001pt] (axis cs:0.125833333333333,0.473885261147389)
--(axis cs:0.125833333333333,0.482025179748203);

\path [draw=color4, line width=0.7000000000000001pt] (axis cs:0.127916666666667,0.494075059249408)
--(axis cs:0.127916666666667,0.502214977850222);

\path [draw=color4, line width=0.7000000000000001pt] (axis cs:0.13,0.507734922650773)
--(axis cs:0.13,0.515874841251587);

\path [draw=color5, line width=0.7000000000000001pt] (axis cs:0.08,0.0536394636053639)
--(axis cs:0.08,0.0573694263057369);

\path [draw=color5, line width=0.7000000000000001pt] (axis cs:0.0820833333333333,0.0647393526064739)
--(axis cs:0.0820833333333333,0.0687993120068799);

\path [draw=color5, line width=0.7000000000000001pt] (axis cs:0.0841666666666667,0.0750492495075049)
--(axis cs:0.0841666666666667,0.0793992060079399);

\path [draw=color5, line width=0.7000000000000001pt] (axis cs:0.08625,0.0913590864091359)
--(axis cs:0.08625,0.0961090389096109);

\path [draw=color5, line width=0.7000000000000001pt] (axis cs:0.0883333333333333,0.107118928810712)
--(axis cs:0.0883333333333333,0.112208877911221);

\path [draw=color5, line width=0.7000000000000001pt] (axis cs:0.0904166666666667,0.122128778712213)
--(axis cs:0.0904166666666667,0.127518724812752);

\path [draw=color5, line width=0.7000000000000001pt] (axis cs:0.0925,0.14049859501405)
--(axis cs:0.0925,0.146208537914621);

\path [draw=color5, line width=0.7000000000000001pt] (axis cs:0.0945833333333333,0.160658393416066)
--(axis cs:0.0945833333333333,0.166688333116669);

\path [draw=color5, line width=0.7000000000000001pt] (axis cs:0.0966666666666667,0.180678193218068)
--(axis cs:0.0966666666666667,0.186988130118699);

\path [draw=color5, line width=0.7000000000000001pt] (axis cs:0.09875,0.201517984820152)
--(axis cs:0.09875,0.208087919120809);

\path [draw=color5, line width=0.7000000000000001pt] (axis cs:0.100833333333333,0.223717762822372)
--(axis cs:0.100833333333333,0.230547694523055);

\path [draw=color5, line width=0.7000000000000001pt] (axis cs:0.102916666666667,0.246557534424656)
--(axis cs:0.102916666666667,0.253607463925361);

\path [draw=color5, line width=0.7000000000000001pt] (axis cs:0.105,0.268437315626844)
--(axis cs:0.105,0.275687243127569);

\path [draw=color5, line width=0.7000000000000001pt] (axis cs:0.107083333333333,0.29179708202918)
--(axis cs:0.107083333333333,0.299227007729923);

\path [draw=color5, line width=0.7000000000000001pt] (axis cs:0.109166666666667,0.318556814431856)
--(axis cs:0.109166666666667,0.326166738332617);

\path [draw=color5, line width=0.7000000000000001pt] (axis cs:0.11125,0.340546594534055)
--(axis cs:0.11125,0.348286517134829);

\path [draw=color5, line width=0.7000000000000001pt] (axis cs:0.113333333333333,0.362156378436216)
--(axis cs:0.113333333333333,0.370006299937001);

\path [draw=color5, line width=0.7000000000000001pt] (axis cs:0.115416666666667,0.388676113238868)
--(axis cs:0.115416666666667,0.396636033639664);

\path [draw=color5, line width=0.7000000000000001pt] (axis cs:0.1175,0.412535874641254)
--(axis cs:0.1175,0.420565794342057);

\path [draw=color5, line width=0.7000000000000001pt] (axis cs:0.119583333333333,0.431375686243138)
--(axis cs:0.119583333333333,0.439455605443946);

\path [draw=color5, line width=0.7000000000000001pt] (axis cs:0.121666666666667,0.455405445945541)
--(axis cs:0.121666666666667,0.463525364746353);

\path [draw=color5, line width=0.7000000000000001pt] (axis cs:0.12375,0.476685233147669)
--(axis cs:0.12375,0.484825151748483);

\path [draw=color5, line width=0.7000000000000001pt] (axis cs:0.125833333333333,0.500024999750002)
--(axis cs:0.125833333333333,0.508164918350816);

\path [draw=color5, line width=0.7000000000000001pt] (axis cs:0.127916666666667,0.514244857551424)
--(axis cs:0.127916666666667,0.522384776152238);

\path [draw=color5, line width=0.7000000000000001pt] (axis cs:0.13,0.536064639353606)
--(axis cs:0.13,0.544184558154418);

\addplot [line width=0.7000000000000001pt, color1]
table {%
0.095617288756 0.217830126132605
0.095764595153303 0.218593099725736
0.0959119015506061 0.219356145847068
0.0960592079479091 0.220119264496602
0.0962065143452121 0.220882455674338
0.0963538207425151 0.221645719380276
0.0965011271398182 0.222409055614417
0.0966484335371212 0.223172464376759
0.0967957399344242 0.223935945667303
0.0969430463317273 0.224699499486049
0.0970903527290303 0.225463125832997
0.0972376591263333 0.226226824708147
0.0973849655236364 0.226990596111499
0.0975322719209394 0.227754440043053
0.0976795783182424 0.228518356502809
0.0978268847155454 0.229282345490767
0.0979741911128485 0.230046407006927
0.0981214975101515 0.230810541051289
0.0982688039074545 0.231574747623853
0.0984161103047576 0.232339026724619
0.0985634167020606 0.233103378353587
0.0987107230993636 0.233867802510756
0.0988580294966667 0.234632299196128
0.0990053358939697 0.235396868409702
0.0991526422912727 0.236161510151478
0.0992999486885758 0.236926224421455
0.0994472550858788 0.237691011219635
0.0995945614831818 0.238455870546017
0.0997418678804848 0.2392208024006
0.0998891742777879 0.239985806783386
0.100036480675091 0.240750883694374
0.100183787072394 0.241516033133563
0.100331093469697 0.242281255100955
0.100478399867 0.243046549596548
0.100625706264303 0.243811916620344
0.100773012661606 0.244577356172341
0.100920319058909 0.245342868252541
0.101067625456212 0.246108452860942
0.101214931853515 0.246874109997545
0.101362238250818 0.247639839662351
0.101509544648121 0.248405641855358
0.101656851045424 0.249171516576567
0.101804157442727 0.249937463825979
0.10195146384003 0.250703483603592
0.102098770237333 0.251469575909407
0.102246076634636 0.252235740743424
0.102393383031939 0.253001978105644
0.102540689429242 0.253768287996065
0.102687995826545 0.254534670414688
0.102835302223848 0.255301125361513
0.102982608621152 0.25606765283654
0.103129915018455 0.256834252839769
0.103277221415758 0.2576009253712
0.103424527813061 0.258367670430833
0.103571834210364 0.259134488018668
0.103719140607667 0.259901378134705
0.10386644700497 0.260668340778944
0.104013753402273 0.261435375951385
0.104161059799576 0.262202483652028
0.104308366196879 0.262969663880873
0.104455672594182 0.26373691663792
0.104602978991485 0.264504241923169
0.104750285388788 0.26527163973662
0.104897591786091 0.266039110078272
0.105044898183394 0.266806652948127
0.105192204580697 0.267574268346184
0.105339510978 0.268341956272443
0.105486817375303 0.269109716726903
0.105634123772606 0.269877549709566
0.105781430169909 0.270645455220431
0.105928736567212 0.271413433259497
0.106076042964515 0.272181483826766
0.106223349361818 0.272949606922236
0.106370655759121 0.273717802545909
0.106517962156424 0.274486070697783
0.106665268553727 0.27525441137786
0.10681257495103 0.276022824586138
0.106959881348333 0.276791310322619
0.107107187745636 0.277559868587301
0.107254494142939 0.278328499380186
0.107401800540242 0.279097202701272
0.107549106937545 0.27986597855056
0.107696413334848 0.280634826928051
0.107843719732152 0.281403747833743
0.107991026129455 0.282172741267637
0.108138332526758 0.282941807229733
0.108285638924061 0.283710945720032
0.108432945321364 0.284480156738532
0.108580251718667 0.285249440285234
0.10872755811597 0.286018796360138
0.108874864513273 0.286788224963244
0.109022170910576 0.287557726094552
0.109169477307879 0.288327299754063
0.109316783705182 0.289096945941775
0.109464090102485 0.289866664657689
0.109611396499788 0.290636455901805
0.109758702897091 0.291406319674123
0.109906009294394 0.292176255974643
0.110053315691697 0.292946264803365
0.110200622089 0.293716346160289
};
\addplot [line width=0.7000000000000001pt, color2]
table {%
0.095617288756 0.204430469106776
0.095764595153303 0.205445997673898
0.0959119015506061 0.206461655158259
0.0960592079479091 0.207477441559858
0.0962065143452121 0.208493356878697
0.0963538207425151 0.209509401114774
0.0965011271398182 0.21052557426809
0.0966484335371212 0.211541876338645
0.0967957399344242 0.21255830732644
0.0969430463317273 0.213574867231473
0.0970903527290303 0.214591556053745
0.0972376591263333 0.215608373793256
0.0973849655236364 0.216625320450006
0.0975322719209394 0.217642396023995
0.0976795783182424 0.218659600515222
0.0978268847155454 0.219676933923689
0.0979741911128485 0.220694396249395
0.0981214975101515 0.221711987492339
0.0982688039074545 0.222729707652523
0.0984161103047576 0.223747556729945
0.0985634167020606 0.224765534724606
0.0987107230993636 0.225783641636507
0.0988580294966667 0.226801877465646
0.0990053358939697 0.227820242212024
0.0991526422912727 0.228838735875641
0.0992999486885758 0.229857358456497
0.0994472550858788 0.230876109954592
0.0995945614831818 0.231894990369926
0.0997418678804848 0.232913999702499
0.0998891742777879 0.23393313795231
0.100036480675091 0.234952405119361
0.100183787072394 0.235971801203651
0.100331093469697 0.236991326205179
0.100478399867 0.238010980123947
0.100625706264303 0.239030762959953
0.100773012661606 0.240050674713198
0.100920319058909 0.241070715383682
0.101067625456212 0.242090884971406
0.101214931853515 0.243111183476368
0.101362238250818 0.244131610898569
0.101509544648121 0.245152167238009
0.101656851045424 0.246172852494687
0.101804157442727 0.247193666668605
0.10195146384003 0.248214609759762
0.102098770237333 0.249235681768158
0.102246076634636 0.250256882693792
0.102393383031939 0.251278212536666
0.102540689429242 0.252299671296778
0.102687995826545 0.25332125897413
0.102835302223848 0.25434297556872
0.102982608621152 0.255364821080549
0.103129915018455 0.256386795509617
0.103277221415758 0.257408898855924
0.103424527813061 0.25843113111947
0.103571834210364 0.259453492300255
0.103719140607667 0.260475982398279
0.10386644700497 0.261498601413542
0.104013753402273 0.262521349346044
0.104161059799576 0.263544226195785
0.104308366196879 0.264567231962764
0.104455672594182 0.265590366646983
0.104602978991485 0.26661363024844
0.104750285388788 0.267637022767137
0.104897591786091 0.268660544203072
0.105044898183394 0.269684194556246
0.105192204580697 0.270707973826659
0.105339510978 0.271731882014311
0.105486817375303 0.272755919119202
0.105634123772606 0.273780085141333
0.105781430169909 0.274804380080701
0.105928736567212 0.275828803937309
0.106076042964515 0.276853356711156
0.106223349361818 0.277878038402242
0.106370655759121 0.278902849010566
0.106517962156424 0.27992778853613
0.106665268553727 0.280952856978932
0.10681257495103 0.281978054338974
0.106959881348333 0.283003380616254
0.107107187745636 0.284028835810773
0.107254494142939 0.285054419922532
0.107401800540242 0.286080132951529
0.107549106937545 0.287105974897765
0.107696413334848 0.28813194576124
0.107843719732152 0.289158045541954
0.107991026129455 0.290184274239907
0.108138332526758 0.291210631855098
0.108285638924061 0.292237118387529
0.108432945321364 0.293263733837199
0.108580251718667 0.294290478204107
0.10872755811597 0.295317351488255
0.108874864513273 0.296344353689641
0.109022170910576 0.297371484808267
0.109169477307879 0.298398744844131
0.109316783705182 0.299426133797234
0.109464090102485 0.300453651667576
0.109611396499788 0.301481298455157
0.109758702897091 0.302509074159977
0.109906009294394 0.303536978782036
0.110053315691697 0.304565012321334
0.110200622089 0.305593174777871
};
\addplot [line width=0.7000000000000001pt, color3]
table {%
0.095617288756 0.192752114430375
0.095764595153303 0.193987412413639
0.0959119015506061 0.195222901702674
0.0960592079479091 0.196458582297479
0.0962065143452121 0.197694454198055
0.0963538207425151 0.198930517404402
0.0965011271398182 0.200166771916519
0.0966484335371212 0.201403217734407
0.0967957399344242 0.202639854858066
0.0969430463317273 0.203876683287495
0.0970903527290303 0.205113703022695
0.0972376591263333 0.206350914063665
0.0973849655236364 0.207588316410406
0.0975322719209394 0.208825910062918
0.0976795783182424 0.210063695021201
0.0978268847155454 0.211301671285254
0.0979741911128485 0.212539838855078
0.0981214975101515 0.213778197730672
0.0982688039074545 0.215016747912037
0.0984161103047576 0.216255489399173
0.0985634167020606 0.21749442219208
0.0987107230993636 0.218733546290757
0.0988580294966667 0.219972861695204
0.0990053358939697 0.221212368405423
0.0991526422912727 0.222452066421412
0.0992999486885758 0.223691955743171
0.0994472550858788 0.224932036370702
0.0995945614831818 0.226172308304003
0.0997418678804848 0.227412771543074
0.0998891742777879 0.228653426087917
0.100036480675091 0.22989427193853
0.100183787072394 0.231135309094913
0.100331093469697 0.232376537557068
0.100478399867 0.233617957324993
0.100625706264303 0.234859568398688
0.100773012661606 0.236101370778154
0.100920319058909 0.237343364463391
0.101067625456212 0.238585549454399
0.101214931853515 0.239827925751177
0.101362238250818 0.241070493353726
0.101509544648121 0.242313252262045
0.101656851045424 0.243556202476135
0.101804157442727 0.244799343995996
0.10195146384003 0.246042676821628
0.102098770237333 0.24728620095303
0.102246076634636 0.248529916390203
0.102393383031939 0.249773823133146
0.102540689429242 0.25101792118186
0.102687995826545 0.252262210536345
0.102835302223848 0.2535066911966
0.102982608621152 0.254751363162626
0.103129915018455 0.255996226434423
0.103277221415758 0.25724128101199
0.103424527813061 0.258486526895328
0.103571834210364 0.259731964084437
0.103719140607667 0.260977592579316
0.10386644700497 0.262223412379966
0.104013753402273 0.263469423486387
0.104161059799576 0.264715625898578
0.104308366196879 0.26596201961654
0.104455672594182 0.267208604640273
0.104602978991485 0.268455380969776
0.104750285388788 0.26970234860505
0.104897591786091 0.270949507546094
0.105044898183394 0.27219685779291
0.105192204580697 0.273444399345495
0.105339510978 0.274692132203852
0.105486817375303 0.275940056367979
0.105634123772606 0.277188171837877
0.105781430169909 0.278436478613545
0.105928736567212 0.279684976694984
0.106076042964515 0.280933666082194
0.106223349361818 0.282182546775175
0.106370655759121 0.283431618773926
0.106517962156424 0.284680882078448
0.106665268553727 0.28593033668874
0.10681257495103 0.287179982604803
0.106959881348333 0.288429819826637
0.107107187745636 0.289679848354241
0.107254494142939 0.290930068187616
0.107401800540242 0.292180479326762
0.107549106937545 0.293431081771678
0.107696413334848 0.294681875522365
0.107843719732152 0.295932860578823
0.107991026129455 0.297184036941051
0.108138332526758 0.29843540460905
0.108285638924061 0.299686963582819
0.108432945321364 0.30093871386236
0.108580251718667 0.30219065544767
0.10872755811597 0.303442788338752
0.108874864513273 0.304695112535604
0.109022170910576 0.305947628038227
0.109169477307879 0.307200334846621
0.109316783705182 0.308453232960785
0.109464090102485 0.309706322380719
0.109611396499788 0.310959603106425
0.109758702897091 0.312213075137901
0.109906009294394 0.313466738475148
0.110053315691697 0.314720593118165
0.110200622089 0.315974639066953
};
\addplot [line width=0.7000000000000001pt, color4]
table {%
0.095617288756 0.182182840566572
0.095764595153303 0.183616760338882
0.0959119015506061 0.185050938560822
0.0960592079479091 0.186485375232392
0.0962065143452121 0.187920070353593
0.0963538207425151 0.189355023924424
0.0965011271398182 0.190790235944886
0.0966484335371212 0.192225706414977
0.0967957399344242 0.193661435334699
0.0969430463317273 0.195097422704051
0.0970903527290303 0.196533668523033
0.0972376591263333 0.197970172791646
0.0973849655236364 0.199406935509889
0.0975322719209394 0.200843956677762
0.0976795783182424 0.202281236295266
0.0978268847155454 0.203718774362399
0.0979741911128485 0.205156570879164
0.0981214975101515 0.206594625845558
0.0982688039074545 0.208032939261583
0.0984161103047576 0.209471511127237
0.0985634167020606 0.210910341442523
0.0987107230993636 0.212349430207438
0.0988580294966667 0.213788777421984
0.0990053358939697 0.21522838308616
0.0991526422912727 0.216668247199966
0.0992999486885758 0.218108369763403
0.0994472550858788 0.21954875077647
0.0995945614831818 0.220989390239167
0.0997418678804848 0.222430288151494
0.0998891742777879 0.223871444513452
0.100036480675091 0.22531285932504
0.100183787072394 0.226754532586258
0.100331093469697 0.228196464297106
0.100478399867 0.229638654457585
0.100625706264303 0.231081103067694
0.100773012661606 0.232523810127434
0.100920319058909 0.233966775636803
0.101067625456212 0.235409999595803
0.101214931853515 0.236853482004433
0.101362238250818 0.238297222862694
0.101509544648121 0.239741222170585
0.101656851045424 0.241185479928106
0.101804157442727 0.242629996135257
0.10195146384003 0.244074770792039
0.102098770237333 0.24551980389845
0.102246076634636 0.246965095454493
0.102393383031939 0.248410645460165
0.102540689429242 0.249856453915468
0.102687995826545 0.251302520820401
0.102835302223848 0.252748846174964
0.102982608621152 0.254195429979158
0.103129915018455 0.255642272232981
0.103277221415758 0.257089372936435
0.103424527813061 0.25853673208952
0.103571834210364 0.259984349692234
0.103719140607667 0.26143222574458
0.10386644700497 0.262880360246555
0.104013753402273 0.26432875319816
0.104161059799576 0.265777404599396
0.104308366196879 0.267226314450262
0.104455672594182 0.268675482750758
0.104602978991485 0.270124909500885
0.104750285388788 0.271574594700642
0.104897591786091 0.273024538350029
0.105044898183394 0.274474740449046
0.105192204580697 0.275925200997694
0.105339510978 0.277375919995972
0.105486817375303 0.27882689744388
0.105634123772606 0.280278133341419
0.105781430169909 0.281729627688588
0.105928736567212 0.283181380485387
0.106076042964515 0.284633391731816
0.106223349361818 0.286085661427876
0.106370655759121 0.287538189573566
0.106517962156424 0.288990976168886
0.106665268553727 0.290444021213837
0.10681257495103 0.291897324708418
0.106959881348333 0.293350886652629
0.107107187745636 0.29480470704647
0.107254494142939 0.296258785889942
0.107401800540242 0.297713123183044
0.107549106937545 0.299167718925776
0.107696413334848 0.300622573118138
0.107843719732152 0.302077685760131
0.107991026129455 0.303533056851754
0.108138332526758 0.304988686393007
0.108285638924061 0.306444574383891
0.108432945321364 0.307900720824405
0.108580251718667 0.309357125714549
0.10872755811597 0.310813789054323
0.108874864513273 0.312270710843728
0.109022170910576 0.313727891082763
0.109169477307879 0.315185329771428
0.109316783705182 0.316643026909724
0.109464090102485 0.31810098249765
0.109611396499788 0.319559196535206
0.109758702897091 0.321017669022392
0.109906009294394 0.322476399959209
0.110053315691697 0.323935389345656
0.110200622089 0.325394637181733
};
\addplot [line width=0.7000000000000001pt, color5]
table {%
0.095617288756 0.172411813068058
0.095764595153303 0.174029119155591
0.0959119015506061 0.1756467548291
0.0960592079479091 0.177264720088585
0.0962065143452121 0.178883014934045
0.0963538207425151 0.180501639365481
0.0965011271398182 0.182120593382892
0.0966484335371212 0.183739876986279
0.0967957399344242 0.185359490175641
0.0969430463317273 0.186979432950979
0.0970903527290303 0.188599705312292
0.0972376591263333 0.190220307259581
0.0973849655236364 0.191841238792846
0.0975322719209394 0.193462499912085
0.0976795783182424 0.195084090617301
0.0978268847155454 0.196706010908492
0.0979741911128485 0.198328260785659
0.0981214975101515 0.199950840248801
0.0982688039074545 0.201573749297919
0.0984161103047576 0.203196987933012
0.0985634167020606 0.204820556154081
0.0987107230993636 0.206444453961125
0.0988580294966667 0.208068681354145
0.0990053358939697 0.20969323833314
0.0991526422912727 0.211318124898111
0.0992999486885758 0.212943341049058
0.0994472550858788 0.21456888678598
0.0995945614831818 0.216194762108877
0.0997418678804848 0.217820967017751
0.0998891742777879 0.219447501512599
0.100036480675091 0.221074365593423
0.100183787072394 0.222701559260223
0.100331093469697 0.224329082512998
0.100478399867 0.225956935351749
0.100625706264303 0.227585117776476
0.100773012661606 0.229213629787178
0.100920319058909 0.230842471383855
0.101067625456212 0.232471642566508
0.101214931853515 0.234101143335137
0.101362238250818 0.235730973689741
0.101509544648121 0.23736113363032
0.101656851045424 0.238991623156876
0.101804157442727 0.240622442269406
0.10195146384003 0.242253590967913
0.102098770237333 0.243885069252394
0.102246076634636 0.245516877122852
0.102393383031939 0.247149014579284
0.102540689429242 0.248781481621693
0.102687995826545 0.250414278250077
0.102835302223848 0.252047404464436
0.102982608621152 0.253680860264771
0.103129915018455 0.255314645651082
0.103277221415758 0.256948760623368
0.103424527813061 0.25858320518163
0.103571834210364 0.260217979325867
0.103719140607667 0.26185308305608
0.10386644700497 0.263488516372268
0.104013753402273 0.265124279274432
0.104161059799576 0.266760371762571
0.104308366196879 0.268396793836686
0.104455672594182 0.270033545496777
0.104602978991485 0.271670626742843
0.104750285388788 0.273308037574884
0.104897591786091 0.274945777992901
0.105044898183394 0.276583847996894
0.105192204580697 0.278222247586862
0.105339510978 0.279860976762806
0.105486817375303 0.281500035524725
0.105634123772606 0.28313942387262
0.105781430169909 0.28477914180649
0.105928736567212 0.286419189326336
0.106076042964515 0.288059566432158
0.106223349361818 0.289700273123955
0.106370655759121 0.291341309401727
0.106517962156424 0.292982675265475
0.106665268553727 0.294624370715199
0.10681257495103 0.296266395750898
0.106959881348333 0.297908750372573
0.107107187745636 0.299551434580223
0.107254494142939 0.301194448373849
0.107401800540242 0.30283779175345
0.107549106937545 0.304481464719027
0.107696413334848 0.306125467270579
0.107843719732152 0.307769799408107
0.107991026129455 0.309414461131611
0.108138332526758 0.31105945244109
0.108285638924061 0.312704773336544
0.108432945321364 0.314350423817974
0.108580251718667 0.31599640388538
0.10872755811597 0.317642713538761
0.108874864513273 0.319289352778118
0.109022170910576 0.32093632160345
0.109169477307879 0.322583620014758
0.109316783705182 0.324231248012041
0.109464090102485 0.3258792055953
0.109611396499788 0.327527492764535
0.109758702897091 0.329176109519745
0.109906009294394 0.33082505586093
0.110053315691697 0.332474331788091
0.110200622089 0.334123937301228
};
\path [draw=white, fill opacity=0] (axis cs:0,0.0291122088779112)
--(axis cs:0,0.568711812881871);

\path [draw=white, fill opacity=0] (axis cs:1,0.0291122088779112)
--(axis cs:1,0.568711812881871);

\path [draw=white, fill opacity=0] (axis cs:0.0775,0)
--(axis cs:0.1325,0);

\path [draw=white, fill opacity=0] (axis cs:0.0775,1)
--(axis cs:0.1325,1);

\end{axis}

\end{tikzpicture}

%% file: tanner_graph.tex
\begin{tikzpicture}[x = 12pt, y = 12pt, gridline/.style = {black, line width = 2pt, line join = round, line cap = round}]
\begin{scope}[shift = {(0, 0)}, opacity=0.5]
\foreach \x/\y in {1/1, 1/5, 3/3, 3/7, 5/1, 5/5, 7/3, 7/7}{
    \fill[white] (\x, \y) rectangle +(2,2);
}
\foreach \x/\y in {1/3, 1/7, 3/1, 3/5, 5/3, 5/7, 7/1, 7/5}{
    \fill[quantumgray!50!white] (\x, \y) rectangle +(2,2);
}
\draw[gridline, step=2, shift = {(-1,-1)}] (2, 2) grid (10, 10);
\foreach \x/\y in {3/9, 7/9}{
    \filldraw[gridline, fill = white] (\x, \y) arc (0:180:1) -- cycle;
}
\foreach \x/\y in {9/3, 9/7}{
    \filldraw[gridline, fill = quantumgray!50!white] (\x, \y) arc (-90:90:1) -- cycle;
}
\foreach \x/\y in {1/3, 1/7}{
    \filldraw[gridline, fill = quantumgray!50!white] (\x, \y) arc (90:270:1) -- cycle;
}
\foreach \x/\y in {3/1, 7/1}{
    \filldraw[gridline, fill = white] (\x, \y) arc (180:360:1) -- cycle;
}
\end{scope}

\foreach \x in {1, 3, ..., 9}{
    \foreach \y in {1, 3, ..., 9}{
    \node (qubit_\x_\y) at (\x, \y) [circle, line width = 2pt, draw = black, fill=black, minimum size = 6 pt, inner sep = 0 pt] {};
    }
}

\foreach \x in {2, 4, 6, 8}{
    \foreach \y in {2, 4, 6, 8}{
        \node (check_\x_\y) at (\x, \y) [circle, line width = 2pt, draw = black, fill=white, minimum size = 6 pt, inner sep = 0 pt] {};
        \foreach \sx/\sy in {1/1, -1/1, 1/-1, -1/-1} {
            \draw[gridline] (check_\x_\y) --++ (\sx, \sy); 
        }
    }
}

\foreach \x/\y/\sxa/\sya/\sxb/\syb in {0/2/1/1/1/-1, 10/4/-1/1/-1/-1, 0/6/1/1/1/-1, 10/8/-1/1/-1/-1, 4/0/-1/1/1/1, 8/0/-1/1/1/1, 2/10/1/-1/-1/-1, 6/10/1/-1/-1/-1}{
    \node (check_\x_\y) at (\x, \y) [circle, line width = 2pt, draw = black, fill=white, minimum size = 6 pt, inner sep = 0 pt] {};
    \draw[gridline] (check_\x_\y) --++ (\sxa, \sya);
    \draw[gridline] (check_\x_\y) --++ (\sxb, \syb);
}

\end{tikzpicture}

%% file: x_upper_left.tex
\begin{tikzpicture}[x = 12pt, y = 12pt, gridline/.style = {black, line width = 2pt, line join = round, line cap = round}]
\begin{scope}[shift = {(0, 0)}]
\foreach \x/\y in {1/1, 3/3}{
    \fill[white] (\x, \y) rectangle +(2,2);
}
\foreach \x/\y in {1/3, 3/1}{
    \fill[quantumgray!50!white] (\x, \y) rectangle +(2,2);
}
\draw[gridline, step=2, shift = {(-1,-1)}] (2, 2) grid (6, 6);
\foreach \x/\y in {3/5}{
    \filldraw[gridline, fill = white] (\x, \y) arc (0:180:1) -- cycle;
}
\foreach \x/\y in {5/3}{
    \filldraw[gridline, fill = quantumgray!50!white] (\x, \y) arc (-90:90:1) -- cycle;
}
\foreach \x/\y in {1/3}{
    \filldraw[gridline, fill = quantumgray!50!white] (\x, \y) arc (90:270:1) -- cycle;
}
\foreach \x/\y in {3/1}{
    \filldraw[gridline, fill = white] (\x, \y) arc (180:360:1) -- cycle;
}
\end{scope}

\draw[black, line width=14pt, line cap=round, line join=round] (1, 5) -- ++(0, 0)  ;
\draw[white, line width=12pt, line cap=round, line join=round] (1, 5) node[black]{$X$} -- ++(0, 0);

\filldraw[draw=white, fill=quantumviolet] (2, 4) circle (7pt) node[white]{$X$};

\end{tikzpicture}

%% file: x_top_middle.tex
\begin{tikzpicture}[x = 12pt, y = 12pt, gridline/.style = {black, line width = 2pt, line join = round, line cap = round}]
\begin{scope}[shift = {(0, 0)}]
\foreach \x/\y in {1/1, 3/3}{
    \fill[white] (\x, \y) rectangle +(2,2);
}
\foreach \x/\y in {1/3, 3/1}{
    \fill[quantumgray!50!white] (\x, \y) rectangle +(2,2);
}
\draw[gridline, step=2, shift = {(-1,-1)}] (2, 2) grid (6, 6);
\foreach \x/\y in {3/5}{
    \filldraw[gridline, fill = white] (\x, \y) arc (0:180:1) -- cycle;
}
\foreach \x/\y in {5/3}{
    \filldraw[gridline, fill = quantumgray!50!white] (\x, \y) arc (-90:90:1) -- cycle;
}
\foreach \x/\y in {1/3}{
    \filldraw[gridline, fill = quantumgray!50!white] (\x, \y) arc (90:270:1) -- cycle;
}
\foreach \x/\y in {3/1}{
    \filldraw[gridline, fill = white] (\x, \y) arc (180:360:1) -- cycle;
}
\end{scope}

\draw[black, line width=14pt, line cap=round, line join=round] (3, 5) -- ++(0, 0)  ;
\draw[white, line width=12pt, line cap=round, line join=round] (3, 5) node[black]{$X$} -- ++(0, 0);

\filldraw[draw=white, fill=quantumviolet] (2, 4) circle (7pt) node[white]{$X$};

\end{tikzpicture}

%% file: failed_beliefs.tex
\begin{tikzpicture}

\definecolor{color1}{rgb}{0.298039215686275,0.447058823529412,0.690196078431373}
\definecolor{color0}{rgb}{0.917647058823529,0.917647058823529,0.949019607843137}
\definecolor{color2}{rgb}{0.333333333333333,0.658823529411765,0.407843137254902}

\begin{axis}[
title={Beliefs, Upper-Left Qubit},
xlabel={number of iterations},
ylabel={marginal probability},
xmin=-0.5, xmax=10.5,
ymin=-0.028, ymax=1.028,
width=\figurewidth,
height=\figureheight,
ytick={-0.2,0,0.2,0.4,0.6,0.8,1,1.2},
yticklabels={,0.0,0.2,0.4,0.6,0.8,1.0,},
tick align=outside,
tick pos=left,
xmajorgrids,
x grid style={white},
ymajorgrids,
y grid style={white},
axis line style={white},
axis background/.style={fill=color0},
legend pos = north east,
legend entries = {$p_{\id} + p_{Z}$, $p_{X} + p_{Y}$}
]
\addplot [line width=0.7pt, quantumviolet!80, mark=*, mark size=2, mark options={solid}, only marks]
table {%
0 0.98
1 0.853259847347793
2 0.517909308447022
3 0.506101668733812
4 0.507597348502516
5 0.507694802413582
6 0.507707564920307
7 0.507721528566171
8 0.507722170920779
9 0.507722092505057
10 0.507722092274707
};
\addplot [line width=0.7pt, quantumgray!80, mark=*, mark size=2, mark options={solid}, only marks]
table {%
0 0.02
1 0.146740152652207
2 0.482090691552978
3 0.493898331266188
4 0.492402651497484
5 0.492305197586418
6 0.492292435079693
7 0.492278471433829
8 0.492277829079221
9 0.492277907494943
10 0.492277907725293
};
\path [draw=white, fill opacity=0] (axis cs:0,-0.028)
--(axis cs:0,1.028);

\path [draw=white, fill opacity=0] (axis cs:1,-0.028)
--(axis cs:1,1.028);

\path [draw=white, fill opacity=0] (axis cs:-0.5,0)
--(axis cs:10.5,0);

\path [draw=white, fill opacity=0] (axis cs:-0.5,1)
--(axis cs:10.5,1);

\end{axis}

\end{tikzpicture}

%% file: paper_dep_no_bp.tex
\begin{tikzpicture}

\definecolor{color1}{rgb}{0.967797559291991,0.441274560091574,0.53581031550587}
\definecolor{color0}{rgb}{0.917647058823529,0.917647058823529,0.949019607843137}
\definecolor{color3}{rgb}{0.201253172212011,0.690792081537903,0.479667611892753}
\definecolor{color2}{rgb}{0.680418912779335,0.615149751467757,0.194054521114453}
\definecolor{color5}{rgb}{0.800493618642396,0.477033635337372,0.957954719600752}
\definecolor{color4}{rgb}{0.219799566082832,0.662515787685034,0.773209315931721}

\begin{axis}[
title={Depolarizing Error Model, without BP/multi-path summation},
xlabel={physical error probability},
ylabel={logical error probability},
xmin=0.1275, xmax=0.1825,
ymin=0.0830610557808133, ymax=0.489233555091918,
width=\figurewidth,
height=\figureheight,
xtick={0.12,0.13,0.14,0.15,0.16,0.17,0.18,0.19},
xticklabels={,0.13,0.14,0.15,0.16,0.17,0.18,},
ytick={0.05,0.1,0.15,0.2,0.25,0.3,0.35,0.4,0.45,0.5},
yticklabels={,0.10,0.15,0.20,0.25,0.30,0.35,0.40,0.45,},
tick align=outside,
tick pos=left,
xmajorgrids,
x grid style={white},
ymajorgrids,
y grid style={white},
axis line style={white},
axis background/.style={fill=color0},
legend pos=north west,
legend entries={$d = 7$, $d = 11$, $d = 15$, $d = 19$, $d = 23$}
]
\path [draw=color1, line width=0.7000000000000001pt] (axis cs:0.13,0.159902373735386)
--(axis cs:0.13,0.165876077628627);

\path [draw=color1, line width=0.7000000000000001pt] (axis cs:0.132083333333333,0.166318938708027)
--(axis cs:0.132083333333333,0.172381214817148);

\path [draw=color1, line width=0.7000000000000001pt] (axis cs:0.134166666666667,0.170934535291107)
--(axis cs:0.134166666666667,0.177065700901468);

\path [draw=color1, line width=0.7000000000000001pt] (axis cs:0.13625,0.181071133330709)
--(axis cs:0.13625,0.18734007794355);

\path [draw=color1, line width=0.7000000000000001pt] (axis cs:0.138333333333333,0.18703499586663)
--(axis cs:0.138333333333333,0.193382671338031);

\path [draw=color1, line width=0.7000000000000001pt] (axis cs:0.140416666666667,0.193825532417431)
--(axis cs:0.140416666666667,0.200251938747392);

\path [draw=color1, line width=0.7000000000000001pt] (axis cs:0.1425,0.200517655395032)
--(axis cs:0.1425,0.207022792583553);

\path [draw=color1, line width=0.7000000000000001pt] (axis cs:0.144583333333333,0.209148525764673)
--(axis cs:0.144583333333333,0.215752076526395);

\path [draw=color1, line width=0.7000000000000001pt] (axis cs:0.146666666666667,0.214718734007794)
--(axis cs:0.146666666666667,0.221401015628075);

\path [draw=color1, line width=0.7000000000000001pt] (axis cs:0.14875,0.226931858441916)
--(axis cs:0.14875,0.233732236350037);

\path [draw=color1, line width=0.7000000000000001pt] (axis cs:0.150833333333333,0.232226508680077)
--(axis cs:0.150833333333333,0.239085934732118);

\path [draw=color1, line width=0.7000000000000001pt] (axis cs:0.152916666666667,0.238091957642798)
--(axis cs:0.152916666666667,0.245010431838759);

\path [draw=color1, line width=0.7000000000000001pt] (axis cs:0.155,0.2471460063772)
--(axis cs:0.155,0.254143211431721);

\path [draw=color1, line width=0.7000000000000001pt] (axis cs:0.157083333333333,0.253513364563241)
--(axis cs:0.157083333333333,0.260569617761682);

\path [draw=color1, line width=0.7000000000000001pt] (axis cs:0.159166666666667,0.261593118922962)
--(axis cs:0.159166666666667,0.268728102979963);

\path [draw=color1, line width=0.7000000000000001pt] (axis cs:0.16125,0.265942998858403)
--(axis cs:0.16125,0.273117348344684);

\path [draw=color1, line width=0.7000000000000001pt] (axis cs:0.163333333333333,0.275410384600244)
--(axis cs:0.163333333333333,0.282663464945085);

\path [draw=color1, line width=0.7000000000000001pt] (axis cs:0.165416666666667,0.284562846907846)
--(axis cs:0.165416666666667,0.291874975396607);

\path [draw=color1, line width=0.7000000000000001pt] (axis cs:0.1675,0.292770538912727)
--(axis cs:0.1675,0.300151556902728);

\path [draw=color1, line width=0.7000000000000001pt] (axis cs:0.169583333333333,0.296943274416407)
--(axis cs:0.169583333333333,0.304353816478369);

\path [draw=color1, line width=0.7000000000000001pt] (axis cs:0.171666666666667,0.306154784867929)
--(axis cs:0.171666666666667,0.31362437507381);

\path [draw=color1, line width=0.7000000000000001pt] (axis cs:0.17375,0.31389993307877)
--(axis cs:0.17375,0.321428571428571);

\path [draw=color1, line width=0.7000000000000001pt] (axis cs:0.175833333333333,0.320779041845451)
--(axis cs:0.175833333333333,0.328347045624533);

\path [draw=color1, line width=0.7000000000000001pt] (axis cs:0.177916666666667,0.329173719639413)
--(axis cs:0.177916666666667,0.336790930205094);

\path [draw=color1, line width=0.7000000000000001pt] (axis cs:0.18,0.337253473999134)
--(axis cs:0.18,0.344919891351415);

\path [draw=color2, line width=0.7000000000000001pt] (axis cs:0.13,0.146163838916663)
--(axis cs:0.13,0.151921032948864);

\path [draw=color2, line width=0.7000000000000001pt] (axis cs:0.132083333333333,0.155385190725505)
--(axis cs:0.132083333333333,0.161280163760186);

\path [draw=color2, line width=0.7000000000000001pt] (axis cs:0.134166666666667,0.163110656221706)
--(axis cs:0.134166666666667,0.169123725544227);

\path [draw=color2, line width=0.7000000000000001pt] (axis cs:0.13625,0.170442467425107)
--(axis cs:0.13625,0.176563791678148);

\path [draw=color2, line width=0.7000000000000001pt] (axis cs:0.138333333333333,0.179053655080109)
--(axis cs:0.138333333333333,0.18529307562099);

\path [draw=color2, line width=0.7000000000000001pt] (axis cs:0.140416666666667,0.18945596976735)
--(axis cs:0.140416666666667,0.195823327953391);

\path [draw=color2, line width=0.7000000000000001pt] (axis cs:0.1425,0.198175412352872)
--(axis cs:0.1425,0.204660866826753);

\path [draw=color2, line width=0.7000000000000001pt] (axis cs:0.144583333333333,0.209630752273354)
--(axis cs:0.144583333333333,0.216253985749715);

\path [draw=color2, line width=0.7000000000000001pt] (axis cs:0.146666666666667,0.217897492422155)
--(axis cs:0.146666666666667,0.224599456757076);

\path [draw=color2, line width=0.7000000000000001pt] (axis cs:0.14875,0.230071251426997)
--(axis cs:0.14875,0.236910994764398);

\path [draw=color2, line width=0.7000000000000001pt] (axis cs:0.150833333333333,0.239164665590678)
--(axis cs:0.150833333333333,0.246083139786639);

\path [draw=color2, line width=0.7000000000000001pt] (axis cs:0.152916666666667,0.245020273196079)
--(axis cs:0.152916666666667,0.25199779553596);

\path [draw=color2, line width=0.7000000000000001pt] (axis cs:0.155,0.256160689682321)
--(axis cs:0.155,0.263246466952722);

\path [draw=color2, line width=0.7000000000000001pt] (axis cs:0.157083333333333,0.264781718694642)
--(axis cs:0.157083333333333,0.271936385466284);

\path [draw=color2, line width=0.7000000000000001pt] (axis cs:0.159166666666667,0.277969137503444)
--(axis cs:0.159166666666667,0.285241900562926);

\path [draw=color2, line width=0.7000000000000001pt] (axis cs:0.16125,0.288204149116246)
--(axis cs:0.16125,0.295555643034287);

\path [draw=color2, line width=0.7000000000000001pt] (axis cs:0.163333333333333,0.299265834743928)
--(axis cs:0.163333333333333,0.306696059520529);

\path [draw=color2, line width=0.7000000000000001pt] (axis cs:0.165416666666667,0.305465889855529)
--(axis cs:0.165416666666667,0.31293548006141);

\path [draw=color2, line width=0.7000000000000001pt] (axis cs:0.1675,0.318318702515451)
--(axis cs:0.1675,0.325867023579892);

\path [draw=color2, line width=0.7000000000000001pt] (axis cs:0.169583333333333,0.328543872770933)
--(axis cs:0.169583333333333,0.336151241979294);

\path [draw=color2, line width=0.7000000000000001pt] (axis cs:0.171666666666667,0.335816635830414)
--(axis cs:0.171666666666667,0.343473211825375);

\path [draw=color2, line width=0.7000000000000001pt] (axis cs:0.17375,0.348088808408456)
--(axis cs:0.17375,0.355804432547337);

\path [draw=color2, line width=0.7000000000000001pt] (axis cs:0.175833333333333,0.357379049718537)
--(axis cs:0.175833333333333,0.365143880644018);

\path [draw=color2, line width=0.7000000000000001pt] (axis cs:0.177916666666667,0.369434712435539)
--(axis cs:0.177916666666667,0.37724875014762);

\path [draw=color2, line width=0.7000000000000001pt] (axis cs:0.18,0.37771129394166)
--(axis cs:0.18,0.385564697083022);

\path [draw=color3, line width=0.7000000000000001pt] (axis cs:0.13,0.129758296264221)
--(axis cs:0.13,0.135239932291462);

\path [draw=color3, line width=0.7000000000000001pt] (axis cs:0.132083333333333,0.139924418375782)
--(axis cs:0.132083333333333,0.145583198834783);

\path [draw=color3, line width=0.7000000000000001pt] (axis cs:0.134166666666667,0.149224501043184)
--(axis cs:0.134166666666667,0.155030901861985);

\path [draw=color3, line width=0.7000000000000001pt] (axis cs:0.13625,0.158288391134905)
--(axis cs:0.13625,0.164232570956186);

\path [draw=color3, line width=0.7000000000000001pt] (axis cs:0.138333333333333,0.170845963075227)
--(axis cs:0.138333333333333,0.176977128685588);

\path [draw=color3, line width=0.7000000000000001pt] (axis cs:0.140416666666667,0.184043223241349)
--(axis cs:0.140416666666667,0.19035153328347);

\path [draw=color3, line width=0.7000000000000001pt] (axis cs:0.1425,0.194819509506751)
--(axis cs:0.1425,0.201265598551352);

\path [draw=color3, line width=0.7000000000000001pt] (axis cs:0.144583333333333,0.203470062591033)
--(axis cs:0.144583333333333,0.210014565208834);

\path [draw=color3, line width=0.7000000000000001pt] (axis cs:0.146666666666667,0.217661299846475)
--(axis cs:0.146666666666667,0.224363264181396);

\path [draw=color3, line width=0.7000000000000001pt] (axis cs:0.14875,0.227758532456796)
--(axis cs:0.14875,0.234578593079558);

\path [draw=color3, line width=0.7000000000000001pt] (axis cs:0.150833333333333,0.240739282761879)
--(axis cs:0.150833333333333,0.24767743967248);

\path [draw=color3, line width=0.7000000000000001pt] (axis cs:0.152916666666667,0.25192890603472)
--(axis cs:0.152916666666667,0.258985159233161);

\path [draw=color3, line width=0.7000000000000001pt] (axis cs:0.155,0.264456953903082)
--(axis cs:0.155,0.271611620674723);

\path [draw=color3, line width=0.7000000000000001pt] (axis cs:0.157083333333333,0.276729126481124)
--(axis cs:0.157083333333333,0.283982206825965);

\path [draw=color3, line width=0.7000000000000001pt] (axis cs:0.159166666666667,0.288489548478526)
--(axis cs:0.159166666666667,0.295841042396567);

\path [draw=color3, line width=0.7000000000000001pt] (axis cs:0.16125,0.300997913632248)
--(axis cs:0.16125,0.308437979766169);

\path [draw=color3, line width=0.7000000000000001pt] (axis cs:0.163333333333333,0.31514978545841)
--(axis cs:0.163333333333333,0.322678423808212);

\path [draw=color3, line width=0.7000000000000001pt] (axis cs:0.165416666666667,0.326359091445892)
--(axis cs:0.165416666666667,0.333956619296933);

\path [draw=color3, line width=0.7000000000000001pt] (axis cs:0.1675,0.339566192969334)
--(axis cs:0.1675,0.347242451678936);

\path [draw=color3, line width=0.7000000000000001pt] (axis cs:0.169583333333333,0.349742156438216)
--(axis cs:0.169583333333333,0.357467621934417);

\path [draw=color3, line width=0.7000000000000001pt] (axis cs:0.171666666666667,0.364582923276778)
--(axis cs:0.171666666666667,0.37237727827422);

\path [draw=color3, line width=0.7000000000000001pt] (axis cs:0.17375,0.37644175884738)
--(axis cs:0.17375,0.384285320631422);

\path [draw=color3, line width=0.7000000000000001pt] (axis cs:0.175833333333333,0.390180293666102)
--(axis cs:0.175833333333333,0.398073062236744);

\path [draw=color3, line width=0.7000000000000001pt] (axis cs:0.177916666666667,0.398801322678424)
--(axis cs:0.177916666666667,0.406733456678345);

\path [draw=color3, line width=0.7000000000000001pt] (axis cs:0.18,0.411910010628666)
--(axis cs:0.18,0.419881510057867);

\path [draw=color4, line width=0.7000000000000001pt] (axis cs:0.13,0.116561036098099)
--(axis cs:0.13,0.121796638192339);

\path [draw=color4, line width=0.7000000000000001pt] (axis cs:0.132083333333333,0.12711097114514)
--(axis cs:0.132083333333333,0.132543400385781);

\path [draw=color4, line width=0.7000000000000001pt] (axis cs:0.134166666666667,0.135436759437862)
--(axis cs:0.134166666666667,0.141016809038303);

\path [draw=color4, line width=0.7000000000000001pt] (axis cs:0.13625,0.148319096169744)
--(axis cs:0.13625,0.154115655631225);

\path [draw=color4, line width=0.7000000000000001pt] (axis cs:0.138333333333333,0.160837302680786)
--(axis cs:0.138333333333333,0.166820847931347);

\path [draw=color4, line width=0.7000000000000001pt] (axis cs:0.140416666666667,0.172794551824588)
--(axis cs:0.140416666666667,0.178945400149589);

\path [draw=color4, line width=0.7000000000000001pt] (axis cs:0.1425,0.18458449789395)
--(axis cs:0.1425,0.190892807936071);

\path [draw=color4, line width=0.7000000000000001pt] (axis cs:0.144583333333333,0.200960516474432)
--(axis cs:0.144583333333333,0.207475495020273);

\path [draw=color4, line width=0.7000000000000001pt] (axis cs:0.146666666666667,0.211441562020234)
--(axis cs:0.146666666666667,0.218084478211235);

\path [draw=color4, line width=0.7000000000000001pt] (axis cs:0.14875,0.226006770853836)
--(axis cs:0.14875,0.232807148761957);

\path [draw=color4, line width=0.7000000000000001pt] (axis cs:0.150833333333333,0.242392630791639)
--(axis cs:0.150833333333333,0.24935047041688);

\path [draw=color4, line width=0.7000000000000001pt] (axis cs:0.152916666666667,0.253218123843641)
--(axis cs:0.152916666666667,0.260274377042082);

\path [draw=color4, line width=0.7000000000000001pt] (axis cs:0.155,0.266041412431603)
--(axis cs:0.155,0.273215761917884);

\path [draw=color4, line width=0.7000000000000001pt] (axis cs:0.157083333333333,0.282397748297445)
--(axis cs:0.157083333333333,0.289700035428886);

\path [draw=color4, line width=0.7000000000000001pt] (axis cs:0.159166666666667,0.292524504979727)
--(axis cs:0.159166666666667,0.299905522969728);

\path [draw=color4, line width=0.7000000000000001pt] (axis cs:0.16125,0.31106562217061)
--(axis cs:0.16125,0.318574577805771);

\path [draw=color4, line width=0.7000000000000001pt] (axis cs:0.163333333333333,0.324882887847892)
--(axis cs:0.163333333333333,0.332470574341613);

\path [draw=color4, line width=0.7000000000000001pt] (axis cs:0.165416666666667,0.339369365822934)
--(axis cs:0.165416666666667,0.347045624532536);

\path [draw=color4, line width=0.7000000000000001pt] (axis cs:0.1675,0.353137424713616)
--(axis cs:0.1675,0.360882572924458);

\path [draw=color4, line width=0.7000000000000001pt] (axis cs:0.169583333333333,0.369493760579459)
--(axis cs:0.169583333333333,0.37730779829154);

\path [draw=color4, line width=0.7000000000000001pt] (axis cs:0.171666666666667,0.385692634728182)
--(axis cs:0.171666666666667,0.393575561941503);

\path [draw=color4, line width=0.7000000000000001pt] (axis cs:0.17375,0.394894303822383)
--(axis cs:0.17375,0.402806755107664);

\path [draw=color4, line width=0.7000000000000001pt] (axis cs:0.175833333333333,0.412657953784986)
--(axis cs:0.175833333333333,0.420629453214187);

\path [draw=color4, line width=0.7000000000000001pt] (axis cs:0.177916666666667,0.428797779789789)
--(axis cs:0.177916666666667,0.43680864464827);

\path [draw=color4, line width=0.7000000000000001pt] (axis cs:0.18,0.441404558516711)
--(axis cs:0.18,0.449435106089832);

\path [draw=color5, line width=0.7000000000000001pt] (axis cs:0.13,0.101523442113136)
--(axis cs:0.13,0.106453962130457);

\path [draw=color5, line width=0.7000000000000001pt] (axis cs:0.132083333333333,0.114907688068338)
--(axis cs:0.132083333333333,0.120113766090619);

\path [draw=color5, line width=0.7000000000000001pt] (axis cs:0.134166666666667,0.12705192300122)
--(axis cs:0.134166666666667,0.132484352241861);

\path [draw=color5, line width=0.7000000000000001pt] (axis cs:0.13625,0.137582175333622)
--(axis cs:0.13625,0.143201590363343);

\path [draw=color5, line width=0.7000000000000001pt] (axis cs:0.138333333333333,0.151241979293784)
--(axis cs:0.138333333333333,0.157077904184545);

\path [draw=color5, line width=0.7000000000000001pt] (axis cs:0.140416666666667,0.165748139983467)
--(axis cs:0.140416666666667,0.171800574735267);

\path [draw=color5, line width=0.7000000000000001pt] (axis cs:0.1425,0.181819076487029)
--(axis cs:0.1425,0.18808802109987);

\path [draw=color5, line width=0.7000000000000001pt] (axis cs:0.144583333333333,0.191512813447231)
--(axis cs:0.144583333333333,0.197919537062552);

\path [draw=color5, line width=0.7000000000000001pt] (axis cs:0.146666666666667,0.210427902216274)
--(axis cs:0.146666666666667,0.217051135692635);

\path [draw=color5, line width=0.7000000000000001pt] (axis cs:0.14875,0.222473723575956)
--(axis cs:0.14875,0.229234736054797);

\path [draw=color5, line width=0.7000000000000001pt] (axis cs:0.150833333333333,0.237796716923198)
--(axis cs:0.150833333333333,0.244715191119159);

\path [draw=color5, line width=0.7000000000000001pt] (axis cs:0.152916666666667,0.257892768570641)
--(axis cs:0.152916666666667,0.264988387198362);

\path [draw=color5, line width=0.7000000000000001pt] (axis cs:0.155,0.273255127347164)
--(axis cs:0.155,0.280488524977365);

\path [draw=color5, line width=0.7000000000000001pt] (axis cs:0.157083333333333,0.287534936818486)
--(axis cs:0.157083333333333,0.294876589379207);

\path [draw=color5, line width=0.7000000000000001pt] (axis cs:0.159166666666667,0.302139511081368)
--(axis cs:0.159166666666667,0.30958941857261);

\path [draw=color5, line width=0.7000000000000001pt] (axis cs:0.16125,0.317954572294611)
--(axis cs:0.16125,0.325502893359052);

\path [draw=color5, line width=0.7000000000000001pt] (axis cs:0.163333333333333,0.338001417155454)
--(axis cs:0.163333333333333,0.345667834507735);

\path [draw=color5, line width=0.7000000000000001pt] (axis cs:0.165416666666667,0.353285045073417)
--(axis cs:0.165416666666667,0.361030193284258);

\path [draw=color5, line width=0.7000000000000001pt] (axis cs:0.1675,0.371589969688619)
--(axis cs:0.1675,0.379413848758021);

\path [draw=color5, line width=0.7000000000000001pt] (axis cs:0.169583333333333,0.386292957524702)
--(axis cs:0.169583333333333,0.394175884738023);

\path [draw=color5, line width=0.7000000000000001pt] (axis cs:0.171666666666667,0.403672794551825)
--(axis cs:0.171666666666667,0.411614769909066);

\path [draw=color5, line width=0.7000000000000001pt] (axis cs:0.17375,0.416407510923907)
--(axis cs:0.17375,0.424379010353108);

\path [draw=color5, line width=0.7000000000000001pt] (axis cs:0.175833333333333,0.43665118293115)
--(axis cs:0.175833333333333,0.444681730504271);

\path [draw=color5, line width=0.7000000000000001pt] (axis cs:0.177916666666667,0.449130024012912)
--(axis cs:0.177916666666667,0.457180254300673);

\path [draw=color5, line width=0.7000000000000001pt] (axis cs:0.18,0.462701255757194)
--(axis cs:0.18,0.470771168759595);

\addplot [line width=0.7000000000000001pt, color1]
table {%
0.140970998597 0.202116329324367
0.141118304994303 0.202629032467611
0.141265611391606 0.203141994003482
0.141412917788909 0.203655213931983
0.141560224186212 0.204168692253112
0.141707530583515 0.20468242896687
0.141854836980818 0.205196424073257
0.142002143378121 0.205710677572272
0.142149449775424 0.206225189463916
0.142296756172727 0.206739959748188
0.14244406257003 0.207254988425089
0.142591368967333 0.207770275494619
0.142738675364636 0.208285820956777
0.142885981761939 0.208801624811564
0.143033288159242 0.20931768705898
0.143180594556545 0.209834007699024
0.143327900953848 0.210350586731697
0.143475207351152 0.210867424156998
0.143622513748455 0.211384519974928
0.143769820145758 0.211901874185487
0.143917126543061 0.212419486788675
0.144064432940364 0.212937357784491
0.144211739337667 0.213455487172935
0.14435904573497 0.213973874954009
0.144506352132273 0.214492521127711
0.144653658529576 0.215011425694041
0.144800964926879 0.215530588653001
0.144948271324182 0.216050010004588
0.145095577721485 0.216569689748805
0.145242884118788 0.21708962788565
0.145390190516091 0.217609824415124
0.145537496913394 0.218130279337226
0.145684803310697 0.218650992651957
0.145832109708 0.219171964359317
0.145979416105303 0.219693194459305
0.146126722502606 0.220214682951922
0.146274028899909 0.220736429837168
0.146421335297212 0.221258435115042
0.146568641694515 0.221780698785545
0.146715948091818 0.222303220848677
0.146863254489121 0.222826001304437
0.147010560886424 0.223349040152826
0.147157867283727 0.223872337393843
0.14730517368103 0.224395893027489
0.147452480078333 0.224919707053764
0.147599786475636 0.225443779472667
0.147747092872939 0.225968110284199
0.147894399270242 0.22649269948836
0.148041705667545 0.227017547085149
0.148189012064848 0.227542653074567
0.148336318462152 0.228068017456613
0.148483624859455 0.228593640231289
0.148630931256758 0.229119521398592
0.148778237654061 0.229645660958525
0.148925544051364 0.230172058911086
0.149072850448667 0.230698715256275
0.14922015684597 0.231225629994094
0.149367463243273 0.231752803124541
0.149514769640576 0.232280234647616
0.149662076037879 0.232807924563321
0.149809382435182 0.233335872871654
0.149956688832485 0.233864079572615
0.150103995229788 0.234392544666205
0.150251301627091 0.234921268152424
0.150398608024394 0.235450250031271
0.150545914421697 0.235979490302748
0.150693220819 0.236508988966852
0.150840527216303 0.237038746023586
0.150987833613606 0.237568761472948
0.151135140010909 0.238099035314938
0.151282446408212 0.238629567549557
0.151429752805515 0.239160358176805
0.151577059202818 0.239691407196682
0.151724365600121 0.240222714609187
0.151871671997424 0.240754280414321
0.152018978394727 0.241286104612083
0.15216628479203 0.241818187202474
0.152313591189333 0.242350528185494
0.152460897586636 0.242883127561142
0.152608203983939 0.243415985329419
0.152755510381242 0.243949101490325
0.152902816778545 0.244482476043859
0.153050123175848 0.245016108990022
0.153197429573152 0.245550000328814
0.153344735970455 0.246084150060234
0.153492042367758 0.246618558184283
0.153639348765061 0.24715322470096
0.153786655162364 0.247688149610266
0.153933961559667 0.248223332912201
0.15408126795697 0.248758774606764
0.154228574354273 0.249294474693956
0.154375880751576 0.249830433173777
0.154523187148879 0.250366650046226
0.154670493546182 0.250903125311304
0.154817799943485 0.25143985896901
0.154965106340788 0.251976851019346
0.155112412738091 0.252514101462309
0.155259719135394 0.253051610297902
0.155407025532697 0.253589377526123
0.15555433193 0.254127403146973
};
\addplot [line width=0.7000000000000001pt, color2]
table {%
0.140970998597 0.193520676513752
0.141118304994303 0.194191387343286
0.141265611391606 0.194862547902969
0.141412917788909 0.195534158192802
0.141560224186212 0.196206218212785
0.141707530583515 0.196878727962917
0.141854836980818 0.197551687443199
0.142002143378121 0.19822509665363
0.142149449775424 0.198898955594211
0.142296756172727 0.199573264264941
0.14244406257003 0.200248022665821
0.142591368967333 0.20092323079685
0.142738675364636 0.201598888658029
0.142885981761939 0.202274996249357
0.143033288159242 0.202951553570835
0.143180594556545 0.203628560622462
0.143327900953848 0.204306017404239
0.143475207351152 0.204983923916166
0.143622513748455 0.205662280158242
0.143769820145758 0.206341086130467
0.143917126543061 0.207020341832843
0.144064432940364 0.207700047265367
0.144211739337667 0.208380202428041
0.14435904573497 0.209060807320865
0.144506352132273 0.209741861943838
0.144653658529576 0.210423366296961
0.144800964926879 0.211105320380233
0.144948271324182 0.211787724193655
0.145095577721485 0.212470577737226
0.145242884118788 0.213153881010947
0.145390190516091 0.213837634014818
0.145537496913394 0.214521836748838
0.145684803310697 0.215206489213007
0.145832109708 0.215891591407326
0.145979416105303 0.216577143331794
0.146126722502606 0.217263144986413
0.146274028899909 0.21794959637118
0.146421335297212 0.218636497486097
0.146568641694515 0.219323848331164
0.146715948091818 0.22001164890638
0.146863254489121 0.220699899211746
0.147010560886424 0.221388599247261
0.147157867283727 0.222077749012926
0.14730517368103 0.22276734850874
0.147452480078333 0.223457397734704
0.147599786475636 0.224147896690817
0.147747092872939 0.22483884537708
0.147894399270242 0.225530243793493
0.148041705667545 0.226222091940055
0.148189012064848 0.226914389816766
0.148336318462152 0.227607137423627
0.148483624859455 0.228300334760638
0.148630931256758 0.228993981827798
0.148778237654061 0.229688078625107
0.148925544051364 0.230382625152566
0.149072850448667 0.231077621410175
0.14922015684597 0.231773067397933
0.149367463243273 0.232468963115841
0.149514769640576 0.233165308563898
0.149662076037879 0.233862103742105
0.149809382435182 0.234559348650461
0.149956688832485 0.235257043288967
0.150103995229788 0.235955187657623
0.150251301627091 0.236653781756428
0.150398608024394 0.237352825585382
0.150545914421697 0.238052319144486
0.150693220819 0.23875226243374
0.150840527216303 0.239452655453143
0.150987833613606 0.240153498202695
0.151135140010909 0.240854790682397
0.151282446408212 0.241556532892249
0.151429752805515 0.24225872483225
0.151577059202818 0.242961366502401
0.151724365600121 0.243664457902701
0.151871671997424 0.244367999033151
0.152018978394727 0.24507198989375
0.15216628479203 0.245776430484499
0.152313591189333 0.246481320805397
0.152460897586636 0.247186660856445
0.152608203983939 0.247892450637643
0.152755510381242 0.24859869014899
0.152902816778545 0.249305379390486
0.153050123175848 0.250012518362132
0.153197429573152 0.250720107063928
0.153344735970455 0.251428145495873
0.153492042367758 0.252136633657967
0.153639348765061 0.252845571550211
0.153786655162364 0.253554959172605
0.153933961559667 0.254264796525148
0.15408126795697 0.254975083607841
0.154228574354273 0.255685820420683
0.154375880751576 0.256397006963675
0.154523187148879 0.257108643236816
0.154670493546182 0.257820729240107
0.154817799943485 0.258533264973548
0.154965106340788 0.259246250437138
0.155112412738091 0.259959685630877
0.155259719135394 0.260673570554766
0.155407025532697 0.261387905208805
0.15555433193 0.262102689592993
};
\addplot [line width=0.7000000000000001pt, color3]
table {%
0.140970998597 0.186141848649413
0.141118304994303 0.186947065058314
0.141265611391606 0.187752939280296
0.141412917788909 0.188559471315362
0.141560224186212 0.189366661163511
0.141707530583515 0.190174508824742
0.141854836980818 0.190983014299057
0.142002143378121 0.191792177586454
0.142149449775424 0.192601998686934
0.142296756172727 0.193412477600496
0.14244406257003 0.194223614327141
0.142591368967333 0.19503540886687
0.142738675364636 0.195847861219681
0.142885981761939 0.196660971385575
0.143033288159242 0.197474739364551
0.143180594556545 0.198289165156611
0.143327900953848 0.199104248761753
0.143475207351152 0.199919990179978
0.143622513748455 0.200736389411286
0.143769820145758 0.201553446455676
0.143917126543061 0.20237116131315
0.144064432940364 0.203189533983706
0.144211739337667 0.204008564467345
0.14435904573497 0.204828252764067
0.144506352132273 0.205648598873871
0.144653658529576 0.206469602796759
0.144800964926879 0.207291264532729
0.144948271324182 0.208113584081782
0.145095577721485 0.208936561443917
0.145242884118788 0.209760196619136
0.145390190516091 0.210584489607437
0.145537496913394 0.211409440408821
0.145684803310697 0.212235049023289
0.145832109708 0.213061315450838
0.145979416105303 0.213888239691471
0.146126722502606 0.214715821745186
0.146274028899909 0.215544061611984
0.146421335297212 0.216372959291865
0.146568641694515 0.217202514784829
0.146715948091818 0.218032728090876
0.146863254489121 0.218863599210005
0.147010560886424 0.219695128142217
0.147157867283727 0.220527314887512
0.14730517368103 0.22136015944589
0.147452480078333 0.22219366181735
0.147599786475636 0.223027822001894
0.147747092872939 0.22386263999952
0.147894399270242 0.224698115810229
0.148041705667545 0.22553424943402
0.148189012064848 0.226371040870895
0.148336318462152 0.227208490120852
0.148483624859455 0.228046597183892
0.148630931256758 0.228885362060015
0.148778237654061 0.229724784749221
0.148925544051364 0.230564865251509
0.149072850448667 0.231405603566881
0.14922015684597 0.232246999695335
0.149367463243273 0.233089053636872
0.149514769640576 0.233931765391491
0.149662076037879 0.234775134959194
0.149809382435182 0.235619162339979
0.149956688832485 0.236463847533847
0.150103995229788 0.237309190540798
0.150251301627091 0.238155191360832
0.150398608024394 0.239001849993948
0.150545914421697 0.239849166440148
0.150693220819 0.24069714069943
0.150840527216303 0.241545772771795
0.150987833613606 0.242395062657242
0.151135140010909 0.243245010355773
0.151282446408212 0.244095615867386
0.151429752805515 0.244946879192082
0.151577059202818 0.245798800329861
0.151724365600121 0.246651379280723
0.151871671997424 0.247504616044667
0.152018978394727 0.248358510621694
0.15216628479203 0.249213063011805
0.152313591189333 0.250068273214997
0.152460897586636 0.250924141231273
0.152608203983939 0.251780667060631
0.152755510381242 0.252637850703073
0.152902816778545 0.253495692158597
0.153050123175848 0.254354191427204
0.153197429573152 0.255213348508893
0.153344735970455 0.256073163403666
0.153492042367758 0.256933636111521
0.153639348765061 0.257794766632459
0.153786655162364 0.25865655496648
0.153933961559667 0.259519001113584
0.15408126795697 0.26038210507377
0.154228574354273 0.261245866847039
0.154375880751576 0.262110286433391
0.154523187148879 0.262975363832826
0.154670493546182 0.263841099045344
0.154817799943485 0.264707492070944
0.154965106340788 0.265574542909627
0.155112412738091 0.266442251561393
0.155259719135394 0.267310618026242
0.155407025532697 0.268179642304174
0.15555433193 0.269049324395188
};
\addplot [line width=0.7000000000000001pt, color4]
table {%
0.140970998597 0.179543474503694
0.141118304994303 0.180468062194554
0.141265611391606 0.181393528855711
0.141412917788909 0.182319874487164
0.141560224186212 0.183247099088913
0.141707530583515 0.184175202660958
0.141854836980818 0.1851041852033
0.142002143378121 0.186034046715937
0.142149449775424 0.186964787198871
0.142296756172727 0.187896406652102
0.14244406257003 0.188828905075628
0.142591368967333 0.189762282469451
0.142738675364636 0.19069653883357
0.142885981761939 0.191631674167985
0.143033288159242 0.192567688472697
0.143180594556545 0.193504581747705
0.143327900953848 0.194442353993009
0.143475207351152 0.195381005208609
0.143622513748455 0.196320535394505
0.143769820145758 0.197260944550698
0.143917126543061 0.198202232677187
0.144064432940364 0.199144399773972
0.144211739337667 0.200087445841054
0.14435904573497 0.201031370878432
0.144506352132273 0.201976174886106
0.144653658529576 0.202921857864076
0.144800964926879 0.203868419812342
0.144948271324182 0.204815860730905
0.145095577721485 0.205764180619764
0.145242884118788 0.206713379478919
0.145390190516091 0.207663457308371
0.145537496913394 0.208614414108118
0.145684803310697 0.209566249878162
0.145832109708 0.210518964618503
0.145979416105303 0.211472558329139
0.146126722502606 0.212427031010072
0.146274028899909 0.213382382661301
0.146421335297212 0.214338613282826
0.146568641694515 0.215295722874647
0.146715948091818 0.216253711436765
0.146863254489121 0.217212578969179
0.147010560886424 0.218172325471889
0.147157867283727 0.219132950944895
0.14730517368103 0.220094455388198
0.147452480078333 0.221056838801797
0.147599786475636 0.222020101185692
0.147747092872939 0.222984242539883
0.147894399270242 0.223949262864371
0.148041705667545 0.224915162159155
0.148189012064848 0.225881940424235
0.148336318462152 0.226849597659611
0.148483624859455 0.227818133865284
0.148630931256758 0.228787549041253
0.148778237654061 0.229757843187518
0.148925544051364 0.230729016304079
0.149072850448667 0.231701068390937
0.14922015684597 0.232673999448091
0.149367463243273 0.233647809475541
0.149514769640576 0.234622498473287
0.149662076037879 0.23559806644133
0.149809382435182 0.236574513379669
0.149956688832485 0.237551839288304
0.150103995229788 0.238530044167235
0.150251301627091 0.239509128016463
0.150398608024394 0.240489090835987
0.150545914421697 0.241469932625807
0.150693220819 0.242451653385923
0.150840527216303 0.243434253116335
0.150987833613606 0.244417731817044
0.151135140010909 0.245402089488049
0.151282446408212 0.246387326129351
0.151429752805515 0.247373441740948
0.151577059202818 0.248360436322842
0.151724365600121 0.249348309875032
0.151871671997424 0.250337062397518
0.152018978394727 0.251326693890301
0.15216628479203 0.25231720435338
0.152313591189333 0.253308593786755
0.152460897586636 0.254300862190426
0.152608203983939 0.255294009564393
0.152755510381242 0.256288035908657
0.152902816778545 0.257282941223217
0.153050123175848 0.258278725508073
0.153197429573152 0.259275388763226
0.153344735970455 0.260272930988675
0.153492042367758 0.26127135218442
0.153639348765061 0.262270652350461
0.153786655162364 0.263270831486798
0.153933961559667 0.264271889593432
0.15408126795697 0.265273826670362
0.154228574354273 0.266276642717588
0.154375880751576 0.267280337735111
0.154523187148879 0.26828491172293
0.154670493546182 0.269290364681045
0.154817799943485 0.270296696609456
0.154965106340788 0.271303907508163
0.155112412738091 0.272311997377167
0.155259719135394 0.273320966216467
0.155407025532697 0.274330814026063
0.15555433193 0.275341540805955
};
\addplot [line width=0.7000000000000001pt, color5]
table {%
0.140970998597 0.173505164675499
0.141118304994303 0.174538224991427
0.141265611391606 0.175572396288794
0.141412917788909 0.176607678567598
0.141560224186212 0.17764407182784
0.141707530583515 0.178681576069521
0.141854836980818 0.179720191292639
0.142002143378121 0.180759917497195
0.142149449775424 0.181800754683189
0.142296756172727 0.182842702850622
0.14244406257003 0.183885761999492
0.142591368967333 0.1849299321298
0.142738675364636 0.185975213241546
0.142885981761939 0.18702160533473
0.143033288159242 0.188069108409352
0.143180594556545 0.189117722465413
0.143327900953848 0.190167447502911
0.143475207351152 0.191218283521847
0.143622513748455 0.192270230522221
0.143769820145758 0.193323288504033
0.143917126543061 0.194377457467283
0.144064432940364 0.195432737411971
0.144211739337667 0.196489128338097
0.14435904573497 0.197546630245661
0.144506352132273 0.198605243134663
0.144653658529576 0.199664967005103
0.144800964926879 0.200725801856981
0.144948271324182 0.201787747690297
0.145095577721485 0.202850804505051
0.145242884118788 0.203914972301243
0.145390190516091 0.204980251078873
0.145537496913394 0.206046640837941
0.145684803310697 0.207114141578447
0.145832109708 0.208182753300391
0.145979416105303 0.209252476003773
0.146126722502606 0.210323309688593
0.146274028899909 0.211395254354851
0.146421335297212 0.212468310002546
0.146568641694515 0.21354247663168
0.146715948091818 0.214617754242252
0.146863254489121 0.215694142834262
0.147010560886424 0.21677164240771
0.147157867283727 0.217850252962595
0.14730517368103 0.218929974498919
0.147452480078333 0.220010807016681
0.147599786475636 0.221092750515881
0.147747092872939 0.222175804996518
0.147894399270242 0.223259970458594
0.148041705667545 0.224345246902108
0.148189012064848 0.22543163432706
0.148336318462152 0.226519132733449
0.148483624859455 0.227607742121277
0.148630931256758 0.228697462490543
0.148778237654061 0.229788293841246
0.148925544051364 0.230880236173388
0.149072850448667 0.231973289486967
0.14922015684597 0.233067453781985
0.149367463243273 0.234162729058441
0.149514769640576 0.235259115316334
0.149662076037879 0.236356612555666
0.149809382435182 0.237455220776435
0.149956688832485 0.238554939978643
0.150103995229788 0.239655770162288
0.150251301627091 0.240757711327372
0.150398608024394 0.241860763473893
0.150545914421697 0.242964926601853
0.150693220819 0.24407020071125
0.150840527216303 0.245176585802086
0.150987833613606 0.246284081874359
0.151135140010909 0.247392688928071
0.151282446408212 0.24850240696322
0.151429752805515 0.249613235979808
0.151577059202818 0.250725175977833
0.151724365600121 0.251838226957296
0.151871671997424 0.252952388918198
0.152018978394727 0.254067661860537
0.15216628479203 0.255184045784315
0.152313591189333 0.25630154068953
0.152460897586636 0.257420146576183
0.152608203983939 0.258539863444274
0.152755510381242 0.259660691293804
0.152902816778545 0.260782630124771
0.153050123175848 0.261905679937176
0.153197429573152 0.263029840731019
0.153344735970455 0.264155112506301
0.153492042367758 0.26528149526302
0.153639348765061 0.266408989001177
0.153786655162364 0.267537593720772
0.153933961559667 0.268667309421806
0.15408126795697 0.269798136104277
0.154228574354273 0.270930073768186
0.154375880751576 0.272063122413533
0.154523187148879 0.273197282040318
0.154670493546182 0.274332552648541
0.154817799943485 0.275468934238202
0.154965106340788 0.276606426809302
0.155112412738091 0.277745030361839
0.155259719135394 0.278884744895814
0.155407025532697 0.280025570411227
0.15555433193 0.281167506908078
};
\path [draw=white, fill opacity=0] (axis cs:0,0.0830610557808133)
--(axis cs:0,0.489233555091918);

\path [draw=white, fill opacity=0] (axis cs:1,0.0830610557808133)
--(axis cs:1,0.489233555091918);

\path [draw=white, fill opacity=0] (axis cs:0.1275,0)
--(axis cs:0.1825,0);

\path [draw=white, fill opacity=0] (axis cs:0.1275,1)
--(axis cs:0.1825,1);

\end{axis}

\end{tikzpicture}

%% file: paper_dep.tex
\begin{tikzpicture}

\definecolor{color1}{rgb}{0.967797559291991,0.441274560091574,0.53581031550587}
\definecolor{color0}{rgb}{0.917647058823529,0.917647058823529,0.949019607843137}
\definecolor{color3}{rgb}{0.201253172212011,0.690792081537903,0.479667611892753}
\definecolor{color2}{rgb}{0.680418912779335,0.615149751467757,0.194054521114453}
\definecolor{color5}{rgb}{0.800493618642396,0.477033635337372,0.957954719600752}
\definecolor{color4}{rgb}{0.219799566082832,0.662515787685034,0.773209315931721}

\begin{axis}[
title={Depolarizing Error Model, with BP/multi-path summation},
xlabel={physical error probability},
ylabel={logical error probability},
xmin=0.1475, xmax=0.2025,
ymin=0.0785084325396825, ymax=0.464021329365079,
width=\figurewidth,
height=\figureheight,
xtick={0.14,0.15,0.16,0.17,0.18,0.19,0.2,0.21},
xticklabels={,0.15,0.16,0.17,0.18,0.19,0.20,},
ytick={0.05,0.1,0.15,0.2,0.25,0.3,0.35,0.4,0.45,0.5},
yticklabels={,0.10,0.15,0.20,0.25,0.30,0.35,0.40,0.45,},
tick align=outside,
tick pos=left,
xmajorgrids,
x grid style={white},
ymajorgrids,
y grid style={white},
axis line style={white},
axis background/.style={fill=color0},
legend pos=north west,
legend entries={$d = 7$, $d = 11$, $d = 15$, $d = 19$, $d = 23$}
]
\path [draw=color1, line width=0.7000000000000001pt] (axis cs:0.15,0.173373015873016)
--(axis cs:0.15,0.179553571428571);

\path [draw=color1, line width=0.7000000000000001pt] (axis cs:0.152083333333333,0.179027777777778)
--(axis cs:0.152083333333333,0.185287698412698);

\path [draw=color1, line width=0.7000000000000001pt] (axis cs:0.154166666666667,0.187857142857143)
--(axis cs:0.154166666666667,0.194236111111111);

\path [draw=color1, line width=0.7000000000000001pt] (axis cs:0.15625,0.192668650793651)
--(axis cs:0.15625,0.199107142857143);

\path [draw=color1, line width=0.7000000000000001pt] (axis cs:0.158333333333333,0.200545634920635)
--(axis cs:0.158333333333333,0.207083333333333);

\path [draw=color1, line width=0.7000000000000001pt] (axis cs:0.160416666666667,0.206329365079365)
--(axis cs:0.160416666666667,0.212926587301587);

\path [draw=color1, line width=0.7000000000000001pt] (axis cs:0.1625,0.213105158730159)
--(axis cs:0.1625,0.219781746031746);

\path [draw=color1, line width=0.7000000000000001pt] (axis cs:0.164583333333333,0.218055555555556)
--(axis cs:0.164583333333333,0.224791666666667);

\path [draw=color1, line width=0.7000000000000001pt] (axis cs:0.166666666666667,0.227242063492063)
--(axis cs:0.166666666666667,0.234077380952381);

\path [draw=color1, line width=0.7000000000000001pt] (axis cs:0.16875,0.235823412698413)
--(axis cs:0.16875,0.242738095238095);

\path [draw=color1, line width=0.7000000000000001pt] (axis cs:0.170833333333333,0.241924603174603)
--(axis cs:0.170833333333333,0.24889880952381);

\path [draw=color1, line width=0.7000000000000001pt] (axis cs:0.172916666666667,0.248214285714286)
--(axis cs:0.172916666666667,0.255248015873016);

\path [draw=color1, line width=0.7000000000000001pt] (axis cs:0.175,0.254900793650794)
--(axis cs:0.175,0.262003968253968);

\path [draw=color1, line width=0.7000000000000001pt] (axis cs:0.177083333333333,0.26343253968254)
--(axis cs:0.177083333333333,0.270605158730159);

\path [draw=color1, line width=0.7000000000000001pt] (axis cs:0.179166666666667,0.269930555555556)
--(axis cs:0.179166666666667,0.277162698412698);

\path [draw=color1, line width=0.7000000000000001pt] (axis cs:0.18125,0.279484126984127)
--(axis cs:0.18125,0.286795634920635);

\path [draw=color1, line width=0.7000000000000001pt] (axis cs:0.183333333333333,0.284503968253968)
--(axis cs:0.183333333333333,0.291855158730159);

\path [draw=color1, line width=0.7000000000000001pt] (axis cs:0.185416666666667,0.290684523809524)
--(axis cs:0.185416666666667,0.298075396825397);

\path [draw=color1, line width=0.7000000000000001pt] (axis cs:0.1875,0.299246031746032)
--(axis cs:0.1875,0.306696428571429);

\path [draw=color1, line width=0.7000000000000001pt] (axis cs:0.189583333333333,0.305823412698413)
--(axis cs:0.189583333333333,0.313323412698413);

\path [draw=color1, line width=0.7000000000000001pt] (axis cs:0.191666666666667,0.313710317460317)
--(axis cs:0.191666666666667,0.321259920634921);

\path [draw=color1, line width=0.7000000000000001pt] (axis cs:0.19375,0.319811507936508)
--(axis cs:0.19375,0.327400793650794);

\path [draw=color1, line width=0.7000000000000001pt] (axis cs:0.195833333333333,0.330962301587302)
--(axis cs:0.195833333333333,0.338621031746032);

\path [draw=color1, line width=0.7000000000000001pt] (axis cs:0.197916666666667,0.338581349206349)
--(axis cs:0.197916666666667,0.346279761904762);

\path [draw=color1, line width=0.7000000000000001pt] (axis cs:0.2,0.347043650793651)
--(axis cs:0.2,0.354791666666667);

\path [draw=color2, line width=0.7000000000000001pt] (axis cs:0.15,0.151200396825397)
--(axis cs:0.15,0.157063492063492);

\path [draw=color2, line width=0.7000000000000001pt] (axis cs:0.152083333333333,0.154464285714286)
--(axis cs:0.152083333333333,0.160367063492064);

\path [draw=color2, line width=0.7000000000000001pt] (axis cs:0.154166666666667,0.16468253968254)
--(axis cs:0.154166666666667,0.170744047619048);

\path [draw=color2, line width=0.7000000000000001pt] (axis cs:0.15625,0.171676587301587)
--(axis cs:0.15625,0.177837301587302);

\path [draw=color2, line width=0.7000000000000001pt] (axis cs:0.158333333333333,0.182222222222222)
--(axis cs:0.158333333333333,0.188521825396825);

\path [draw=color2, line width=0.7000000000000001pt] (axis cs:0.160416666666667,0.188918650793651)
--(axis cs:0.160416666666667,0.19531746031746);

\path [draw=color2, line width=0.7000000000000001pt] (axis cs:0.1625,0.199246031746032)
--(axis cs:0.1625,0.205763888888889);

\path [draw=color2, line width=0.7000000000000001pt] (axis cs:0.164583333333333,0.206656746031746)
--(axis cs:0.164583333333333,0.21327380952381);

\path [draw=color2, line width=0.7000000000000001pt] (axis cs:0.166666666666667,0.218363095238095)
--(axis cs:0.166666666666667,0.225099206349206);

\path [draw=color2, line width=0.7000000000000001pt] (axis cs:0.16875,0.224553571428571)
--(axis cs:0.16875,0.231369047619048);

\path [draw=color2, line width=0.7000000000000001pt] (axis cs:0.170833333333333,0.236825396825397)
--(axis cs:0.170833333333333,0.243759920634921);

\path [draw=color2, line width=0.7000000000000001pt] (axis cs:0.172916666666667,0.24860119047619)
--(axis cs:0.172916666666667,0.255654761904762);

\path [draw=color2, line width=0.7000000000000001pt] (axis cs:0.175,0.255049603174603)
--(axis cs:0.175,0.262162698412698);

\path [draw=color2, line width=0.7000000000000001pt] (axis cs:0.177083333333333,0.264444444444444)
--(axis cs:0.177083333333333,0.271636904761905);

\path [draw=color2, line width=0.7000000000000001pt] (axis cs:0.179166666666667,0.273075396825397)
--(axis cs:0.179166666666667,0.280327380952381);

\path [draw=color2, line width=0.7000000000000001pt] (axis cs:0.18125,0.28531746031746)
--(axis cs:0.18125,0.292668650793651);

\path [draw=color2, line width=0.7000000000000001pt] (axis cs:0.183333333333333,0.294652777777778)
--(axis cs:0.183333333333333,0.302083333333333);

\path [draw=color2, line width=0.7000000000000001pt] (axis cs:0.185416666666667,0.307113095238095)
--(axis cs:0.185416666666667,0.314623015873016);

\path [draw=color2, line width=0.7000000000000001pt] (axis cs:0.1875,0.31469246031746)
--(axis cs:0.1875,0.322251984126984);

\path [draw=color2, line width=0.7000000000000001pt] (axis cs:0.189583333333333,0.325178571428571)
--(axis cs:0.189583333333333,0.33280753968254);

\path [draw=color2, line width=0.7000000000000001pt] (axis cs:0.191666666666667,0.339672619047619)
--(axis cs:0.191666666666667,0.347380952380952);

\path [draw=color2, line width=0.7000000000000001pt] (axis cs:0.19375,0.344474206349206)
--(axis cs:0.19375,0.352202380952381);

\path [draw=color2, line width=0.7000000000000001pt] (axis cs:0.195833333333333,0.352599206349206)
--(axis cs:0.195833333333333,0.360367063492063);

\path [draw=color2, line width=0.7000000000000001pt] (axis cs:0.197916666666667,0.363839285714286)
--(axis cs:0.197916666666667,0.371666666666667);

\path [draw=color2, line width=0.7000000000000001pt] (axis cs:0.2,0.374553571428571)
--(axis cs:0.2,0.382420634920635);

\path [draw=color3, line width=0.7000000000000001pt] (axis cs:0.15,0.125763888888889)
--(axis cs:0.15,0.131190476190476);

\path [draw=color3, line width=0.7000000000000001pt] (axis cs:0.152083333333333,0.13593253968254)
--(axis cs:0.152083333333333,0.141537698412698);

\path [draw=color3, line width=0.7000000000000001pt] (axis cs:0.154166666666667,0.146240079365079)
--(axis cs:0.154166666666667,0.15202380952381);

\path [draw=color3, line width=0.7000000000000001pt] (axis cs:0.15625,0.154454365079365)
--(axis cs:0.15625,0.160357142857143);

\path [draw=color3, line width=0.7000000000000001pt] (axis cs:0.158333333333333,0.16218253968254)
--(axis cs:0.158333333333333,0.168204365079365);

\path [draw=color3, line width=0.7000000000000001pt] (axis cs:0.160416666666667,0.174007936507937)
--(axis cs:0.160416666666667,0.180208333333333);

\path [draw=color3, line width=0.7000000000000001pt] (axis cs:0.1625,0.184930555555556)
--(axis cs:0.1625,0.191269841269841);

\path [draw=color3, line width=0.7000000000000001pt] (axis cs:0.164583333333333,0.192777777777778)
--(axis cs:0.164583333333333,0.19921626984127);

\path [draw=color3, line width=0.7000000000000001pt] (axis cs:0.166666666666667,0.206230158730159)
--(axis cs:0.166666666666667,0.212827380952381);

\path [draw=color3, line width=0.7000000000000001pt] (axis cs:0.16875,0.216458333333333)
--(axis cs:0.16875,0.223174603174603);

\path [draw=color3, line width=0.7000000000000001pt] (axis cs:0.170833333333333,0.229712301587302)
--(axis cs:0.170833333333333,0.23656746031746);

\path [draw=color3, line width=0.7000000000000001pt] (axis cs:0.172916666666667,0.241537698412698)
--(axis cs:0.172916666666667,0.248511904761905);

\path [draw=color3, line width=0.7000000000000001pt] (axis cs:0.175,0.253650793650794)
--(axis cs:0.175,0.260744047619048);

\path [draw=color3, line width=0.7000000000000001pt] (axis cs:0.177083333333333,0.26421626984127)
--(axis cs:0.177083333333333,0.27140873015873);

\path [draw=color3, line width=0.7000000000000001pt] (axis cs:0.179166666666667,0.275188492063492)
--(axis cs:0.179166666666667,0.282460317460317);

\path [draw=color3, line width=0.7000000000000001pt] (axis cs:0.18125,0.288244047619048)
--(axis cs:0.18125,0.295615079365079);

\path [draw=color3, line width=0.7000000000000001pt] (axis cs:0.183333333333333,0.300922619047619)
--(axis cs:0.183333333333333,0.308392857142857);

\path [draw=color3, line width=0.7000000000000001pt] (axis cs:0.185416666666667,0.313214285714286)
--(axis cs:0.185416666666667,0.320763888888889);

\path [draw=color3, line width=0.7000000000000001pt] (axis cs:0.1875,0.323799603174603)
--(axis cs:0.1875,0.33140873015873);

\path [draw=color3, line width=0.7000000000000001pt] (axis cs:0.189583333333333,0.337599206349206)
--(axis cs:0.189583333333333,0.345287698412698);

\path [draw=color3, line width=0.7000000000000001pt] (axis cs:0.191666666666667,0.350416666666667)
--(axis cs:0.191666666666667,0.358174603174603);

\path [draw=color3, line width=0.7000000000000001pt] (axis cs:0.19375,0.361617063492064)
--(axis cs:0.19375,0.369434523809524);

\path [draw=color3, line width=0.7000000000000001pt] (axis cs:0.195833333333333,0.374136904761905)
--(axis cs:0.195833333333333,0.382003968253968);

\path [draw=color3, line width=0.7000000000000001pt] (axis cs:0.197916666666667,0.386904761904762)
--(axis cs:0.197916666666667,0.394821428571429);

\path [draw=color3, line width=0.7000000000000001pt] (axis cs:0.2,0.39797619047619)
--(axis cs:0.2,0.40593253968254);

\path [draw=color4, line width=0.7000000000000001pt] (axis cs:0.15,0.108234126984127)
--(axis cs:0.15,0.113323412698413);

\path [draw=color4, line width=0.7000000000000001pt] (axis cs:0.152083333333333,0.119087301587302)
--(axis cs:0.152083333333333,0.124394841269841);

\path [draw=color4, line width=0.7000000000000001pt] (axis cs:0.154166666666667,0.128234126984127)
--(axis cs:0.154166666666667,0.133710317460317);

\path [draw=color4, line width=0.7000000000000001pt] (axis cs:0.15625,0.139434523809524)
--(axis cs:0.15625,0.145099206349206);

\path [draw=color4, line width=0.7000000000000001pt] (axis cs:0.158333333333333,0.149355158730159)
--(axis cs:0.158333333333333,0.155178571428571);

\path [draw=color4, line width=0.7000000000000001pt] (axis cs:0.160416666666667,0.161061507936508)
--(axis cs:0.160416666666667,0.167063492063492);

\path [draw=color4, line width=0.7000000000000001pt] (axis cs:0.1625,0.171914682539683)
--(axis cs:0.1625,0.178075396825397);

\path [draw=color4, line width=0.7000000000000001pt] (axis cs:0.164583333333333,0.184156746031746)
--(axis cs:0.164583333333333,0.190486111111111);

\path [draw=color4, line width=0.7000000000000001pt] (axis cs:0.166666666666667,0.196865079365079)
--(axis cs:0.166666666666667,0.203363095238095);

\path [draw=color4, line width=0.7000000000000001pt] (axis cs:0.16875,0.209275793650794)
--(axis cs:0.16875,0.215912698412698);

\path [draw=color4, line width=0.7000000000000001pt] (axis cs:0.170833333333333,0.223779761904762)
--(axis cs:0.170833333333333,0.230575396825397);

\path [draw=color4, line width=0.7000000000000001pt] (axis cs:0.172916666666667,0.236200396825397)
--(axis cs:0.172916666666667,0.243134920634921);

\path [draw=color4, line width=0.7000000000000001pt] (axis cs:0.175,0.247767857142857)
--(axis cs:0.175,0.254801587301587);

\path [draw=color4, line width=0.7000000000000001pt] (axis cs:0.177083333333333,0.262559523809524)
--(axis cs:0.177083333333333,0.269732142857143);

\path [draw=color4, line width=0.7000000000000001pt] (axis cs:0.179166666666667,0.277628968253968)
--(axis cs:0.179166666666667,0.284920634920635);

\path [draw=color4, line width=0.7000000000000001pt] (axis cs:0.18125,0.293035714285714)
--(axis cs:0.18125,0.300446428571429);

\path [draw=color4, line width=0.7000000000000001pt] (axis cs:0.183333333333333,0.304831349206349)
--(axis cs:0.183333333333333,0.312321428571429);

\path [draw=color4, line width=0.7000000000000001pt] (axis cs:0.185416666666667,0.321309523809524)
--(axis cs:0.185416666666667,0.32890873015873);

\path [draw=color4, line width=0.7000000000000001pt] (axis cs:0.1875,0.333690476190476)
--(axis cs:0.1875,0.341359126984127);

\path [draw=color4, line width=0.7000000000000001pt] (axis cs:0.189583333333333,0.348809523809524)
--(axis cs:0.189583333333333,0.35655753968254);

\path [draw=color4, line width=0.7000000000000001pt] (axis cs:0.191666666666667,0.363839285714286)
--(axis cs:0.191666666666667,0.371666666666667);

\path [draw=color4, line width=0.7000000000000001pt] (axis cs:0.19375,0.378065476190476)
--(axis cs:0.19375,0.385952380952381);

\path [draw=color4, line width=0.7000000000000001pt] (axis cs:0.195833333333333,0.392440476190476)
--(axis cs:0.195833333333333,0.400376984126984);

\path [draw=color4, line width=0.7000000000000001pt] (axis cs:0.197916666666667,0.409077380952381)
--(axis cs:0.197916666666667,0.417073412698413);

\path [draw=color4, line width=0.7000000000000001pt] (axis cs:0.2,0.421884920634921)
--(axis cs:0.2,0.429910714285714);

\path [draw=color5, line width=0.7000000000000001pt] (axis cs:0.15,0.096031746031746)
--(axis cs:0.15,0.100863095238095);

\path [draw=color5, line width=0.7000000000000001pt] (axis cs:0.152083333333333,0.104236111111111)
--(axis cs:0.152083333333333,0.109246031746032);

\path [draw=color5, line width=0.7000000000000001pt] (axis cs:0.154166666666667,0.112361111111111)
--(axis cs:0.154166666666667,0.117539682539683);

\path [draw=color5, line width=0.7000000000000001pt] (axis cs:0.15625,0.125089285714286)
--(axis cs:0.15625,0.130505952380952);

\path [draw=color5, line width=0.7000000000000001pt] (axis cs:0.158333333333333,0.134930555555556)
--(axis cs:0.158333333333333,0.140515873015873);

\path [draw=color5, line width=0.7000000000000001pt] (axis cs:0.160416666666667,0.14984126984127)
--(axis cs:0.160416666666667,0.155684523809524);

\path [draw=color5, line width=0.7000000000000001pt] (axis cs:0.1625,0.159563492063492)
--(axis cs:0.1625,0.165545634920635);

\path [draw=color5, line width=0.7000000000000001pt] (axis cs:0.164583333333333,0.173313492063492)
--(axis cs:0.164583333333333,0.179494047619048);

\path [draw=color5, line width=0.7000000000000001pt] (axis cs:0.166666666666667,0.187371031746032)
--(axis cs:0.166666666666667,0.19375);

\path [draw=color5, line width=0.7000000000000001pt] (axis cs:0.16875,0.201755952380952)
--(axis cs:0.16875,0.208313492063492);

\path [draw=color5, line width=0.7000000000000001pt] (axis cs:0.170833333333333,0.216329365079365)
--(axis cs:0.170833333333333,0.223045634920635);

\path [draw=color5, line width=0.7000000000000001pt] (axis cs:0.172916666666667,0.23343253968254)
--(axis cs:0.172916666666667,0.240327380952381);

\path [draw=color5, line width=0.7000000000000001pt] (axis cs:0.175,0.24625)
--(axis cs:0.175,0.25327380952381);

\path [draw=color5, line width=0.7000000000000001pt] (axis cs:0.177083333333333,0.259315476190476)
--(axis cs:0.177083333333333,0.266458333333333);

\path [draw=color5, line width=0.7000000000000001pt] (axis cs:0.179166666666667,0.278611111111111)
--(axis cs:0.179166666666667,0.285912698412698);

\path [draw=color5, line width=0.7000000000000001pt] (axis cs:0.18125,0.291924603174603)
--(axis cs:0.18125,0.299335317460317);

\path [draw=color5, line width=0.7000000000000001pt] (axis cs:0.183333333333333,0.308333333333333)
--(axis cs:0.183333333333333,0.315853174603175);

\path [draw=color5, line width=0.7000000000000001pt] (axis cs:0.185416666666667,0.325287698412698)
--(axis cs:0.185416666666667,0.332916666666667);

\path [draw=color5, line width=0.7000000000000001pt] (axis cs:0.1875,0.34264880952381)
--(axis cs:0.1875,0.350367063492063);

\path [draw=color5, line width=0.7000000000000001pt] (axis cs:0.189583333333333,0.359384920634921)
--(axis cs:0.189583333333333,0.36719246031746);

\path [draw=color5, line width=0.7000000000000001pt] (axis cs:0.191666666666667,0.374543650793651)
--(axis cs:0.191666666666667,0.382410714285714);

\path [draw=color5, line width=0.7000000000000001pt] (axis cs:0.19375,0.392440476190476)
--(axis cs:0.19375,0.400376984126984);

\path [draw=color5, line width=0.7000000000000001pt] (axis cs:0.195833333333333,0.408640873015873)
--(axis cs:0.195833333333333,0.416626984126984);

\path [draw=color5, line width=0.7000000000000001pt] (axis cs:0.197916666666667,0.42235119047619)
--(axis cs:0.197916666666667,0.430376984126984);

\path [draw=color5, line width=0.7000000000000001pt] (axis cs:0.2,0.43844246031746)
--(axis cs:0.2,0.446498015873016);

\addplot [line width=0.7000000000000001pt, color1]
table {%
0.170225147398 0.24453453461793
0.170372453795313 0.245030517416965
0.170519760192626 0.245526893722437
0.170667066589939 0.246023663534346
0.170814372987253 0.24652082685269
0.170961679384566 0.247018383677472
0.171108985781879 0.24751633400869
0.171256292179192 0.248014677846345
0.171403598576505 0.248513415190436
0.171550904973818 0.249012546040964
0.171698211371131 0.249512070397929
0.171845517768444 0.25001198826133
0.171992824165758 0.250512299631168
0.172140130563071 0.251013004507443
0.172287436960384 0.251514102890154
0.172434743357697 0.252015594779302
0.17258204975501 0.252517480174886
0.172729356152323 0.253019759076907
0.172876662549636 0.253522431485364
0.173023968946949 0.254025497400259
0.173171275344263 0.254528956821589
0.173318581741576 0.255032809749357
0.173465888138889 0.255537056183561
0.173613194536202 0.256041696124201
0.173760500933515 0.256546729571278
0.173907807330828 0.257052156524792
0.174055113728141 0.257557976984743
0.174202420125455 0.25806419095113
0.174349726522768 0.258570798423953
0.174497032920081 0.259077799403214
0.174644339317394 0.259585193888911
0.174791645714707 0.260092981881044
0.17493895211202 0.260601163379614
0.175086258509333 0.261109738384621
0.175233564906646 0.261618706896064
0.17538087130396 0.262128068913944
0.175528177701273 0.262637824438261
0.175675484098586 0.263147973469014
0.175822790495899 0.263658516006204
0.175970096893212 0.26416945204983
0.176117403290525 0.264680781599893
0.176264709687838 0.265192504656393
0.176412016085152 0.265704621219329
0.176559322482465 0.266217131288702
0.176706628879778 0.266730034864511
0.176853935277091 0.267243331946757
0.177001241674404 0.26775702253544
0.177148548071717 0.268271106630559
0.17729585446903 0.268785584232115
0.177443160866343 0.269300455340108
0.177590467263657 0.269815719954537
0.17773777366097 0.270331378075403
0.177885080058283 0.270847429702705
0.178032386455596 0.271363874836444
0.178179692852909 0.27188071347662
0.178326999250222 0.272397945623232
0.178474305647535 0.272915571276281
0.178621612044848 0.273433590435766
0.178768918442162 0.273952003101688
0.178916224839475 0.274470809274047
0.179063531236788 0.274990008952842
0.179210837634101 0.275509602138074
0.179358144031414 0.276029588829742
0.179505450428727 0.276549969027848
0.17965275682604 0.277070742732389
0.179800063223354 0.277591909943368
0.179947369620667 0.278113470660783
0.18009467601798 0.278635424884634
0.180241982415293 0.279157772614922
0.180389288812606 0.279680513851647
0.180536595209919 0.280203648594808
0.180683901607232 0.280727176844406
0.180831208004545 0.281251098600441
0.180978514401859 0.281775413862912
0.181125820799172 0.28230012263182
0.181273127196485 0.282825224907165
0.181420433593798 0.283350720688946
0.181567739991111 0.283876609977163
0.181715046388424 0.284402892771818
0.181862352785737 0.284929569072908
0.182009659183051 0.285456638880436
0.182156965580364 0.2859841021944
0.182304271977677 0.286511959014801
0.18245157837499 0.287040209341638
0.182598884772303 0.287568853174912
0.182746191169616 0.288097890514623
0.182893497566929 0.28862732136077
0.183040803964242 0.289157145713354
0.183188110361556 0.289687363572374
0.183335416758869 0.290217974937831
0.183482723156182 0.290748979809725
0.183630029553495 0.291280378188055
0.183777335950808 0.291812170072822
0.183924642348121 0.292344355464025
0.184071948745434 0.292876934361665
0.184219255142747 0.293409906765742
0.184366561540061 0.293943272676255
0.184513867937374 0.294477032093205
0.184661174334687 0.295011185016592
0.184808480732 0.295545731446415
};
\addplot [line width=0.7000000000000001pt, color2]
table {%
0.170225147398 0.236360862053073
0.170372453795313 0.237012669461572
0.170519760192626 0.237665174546223
0.170667066589939 0.238318377307026
0.170814372987253 0.238972277743981
0.170961679384566 0.239626875857088
0.171108985781879 0.240282171646347
0.171256292179192 0.240938165111758
0.171403598576505 0.241594856253322
0.171550904973818 0.242252245071037
0.171698211371131 0.242910331564904
0.171845517768444 0.243569115734923
0.171992824165758 0.244228597581094
0.172140130563071 0.244888777103418
0.172287436960384 0.245549654301893
0.172434743357697 0.24621122917652
0.17258204975501 0.246873501727299
0.172729356152323 0.247536471954231
0.172876662549636 0.248200139857314
0.173023968946949 0.248864505436549
0.173171275344263 0.249529568691937
0.173318581741576 0.250195329623476
0.173465888138889 0.250861788231167
0.173613194536202 0.251528944515011
0.173760500933515 0.252196798475006
0.173907807330828 0.252865350111153
0.174055113728141 0.253534599423453
0.174202420125455 0.254204546411904
0.174349726522768 0.254875191076507
0.174497032920081 0.255546533417263
0.174644339317394 0.25621857343417
0.174791645714707 0.25689131112723
0.17493895211202 0.257564746496441
0.175086258509333 0.258238879541805
0.175233564906646 0.25891371026332
0.17538087130396 0.259589238660988
0.175528177701273 0.260265464734807
0.175675484098586 0.260942388484779
0.175822790495899 0.261620009910902
0.175970096893212 0.262298329013178
0.176117403290525 0.262977345791605
0.176264709687838 0.263657060246185
0.176412016085152 0.264337472376916
0.176559322482465 0.2650185821838
0.176706628879778 0.265700389666836
0.176853935277091 0.266382894826023
0.177001241674404 0.267066097661363
0.177148548071717 0.267749998172855
0.17729585446903 0.268434596360498
0.177443160866343 0.269119892224294
0.177590467263657 0.269805885764242
0.17773777366097 0.270492576980341
0.177885080058283 0.271179965872593
0.178032386455596 0.271868052440997
0.178179692852909 0.272556836685553
0.178326999250222 0.27324631860626
0.178474305647535 0.27393649820312
0.178621612044848 0.274627375476132
0.178768918442162 0.275318950425295
0.178916224839475 0.276011223050611
0.179063531236788 0.276704193352079
0.179210837634101 0.277397861329699
0.179358144031414 0.278092226983471
0.179505450428727 0.278787290313395
0.17965275682604 0.27948305131947
0.179800063223354 0.280179510001698
0.179947369620667 0.280876666360078
0.18009467601798 0.28157452039461
0.180241982415293 0.282273072105294
0.180389288812606 0.28297232149213
0.180536595209919 0.283672268555118
0.180683901607232 0.284372913294258
0.180831208004545 0.28507425570955
0.180978514401859 0.285776295800994
0.181125820799172 0.28647903356859
0.181273127196485 0.287182469012338
0.181420433593798 0.287886602132238
0.181567739991111 0.28859143292829
0.181715046388424 0.289296961400494
0.181862352785737 0.29000318754885
0.182009659183051 0.290710111373358
0.182156965580364 0.291417732874018
0.182304271977677 0.29212605205083
0.18245157837499 0.292835068903794
0.182598884772303 0.29354478343291
0.182746191169616 0.294255195638178
0.182893497566929 0.294966305519598
0.183040803964242 0.295678113077171
0.183188110361556 0.296390618310895
0.183335416758869 0.297103821220771
0.183482723156182 0.297817721806799
0.183630029553495 0.298532320068979
0.183777335950808 0.299247616007311
0.183924642348121 0.299963609621796
0.184071948745434 0.300680300912432
0.184219255142747 0.30139768987922
0.184366561540061 0.30211577652216
0.184513867937374 0.302834560841253
0.184661174334687 0.303554042836497
0.184808480732 0.304274222507893
};
\addplot [line width=0.7000000000000001pt, color3]
table {%
0.170225147398 0.229321963208535
0.170372453795313 0.230106149158574
0.170519760192626 0.230891368618706
0.170667066589939 0.231677621588931
0.170814372987253 0.232464908069248
0.170961679384566 0.233253228059658
0.171108985781879 0.234042581560161
0.171256292179192 0.234832968570756
0.171403598576505 0.235624389091444
0.171550904973818 0.236416843122225
0.171698211371131 0.237210330663099
0.171845517768444 0.238004851714065
0.171992824165758 0.238800406275124
0.172140130563071 0.239596994346276
0.172287436960384 0.24039461592752
0.172434743357697 0.241193271018857
0.17258204975501 0.241992959620287
0.172729356152323 0.242793681731809
0.172876662549636 0.243595437353424
0.173023968946949 0.244398226485132
0.173171275344263 0.245202049126933
0.173318581741576 0.246006905278826
0.173465888138889 0.246812794940812
0.173613194536202 0.24761971811289
0.173760500933515 0.248427674795061
0.173907807330828 0.249236664987325
0.174055113728141 0.250046688689682
0.174202420125455 0.250857745902132
0.174349726522768 0.251669836624674
0.174497032920081 0.252482960857308
0.174644339317394 0.253297118600036
0.174791645714707 0.254112309852856
0.17493895211202 0.254928534615769
0.175086258509333 0.255745792888774
0.175233564906646 0.256564084671873
0.17538087130396 0.257383409965064
0.175528177701273 0.258203768768347
0.175675484098586 0.259025161081724
0.175822790495899 0.259847586905193
0.175970096893212 0.260671046238755
0.176117403290525 0.261495539082409
0.176264709687838 0.262321065436156
0.176412016085152 0.263147625299996
0.176559322482465 0.263975218673929
0.176706628879778 0.264803845557954
0.176853935277091 0.265633505952072
0.177001241674404 0.266464199856283
0.177148548071717 0.267295927270586
0.17729585446903 0.268128688194982
0.177443160866343 0.268962482629471
0.177590467263657 0.269797310574052
0.17773777366097 0.270633172028726
0.177885080058283 0.271470066993493
0.178032386455596 0.272307995468353
0.178179692852909 0.273146957453305
0.178326999250222 0.27398695294835
0.178474305647535 0.274827981953487
0.178621612044848 0.275670044468718
0.178768918442162 0.276513140494041
0.178916224839475 0.277357270029456
0.179063531236788 0.278202433074965
0.179210837634101 0.279048629630566
0.179358144031414 0.27989585969626
0.179505450428727 0.280744123272046
0.17965275682604 0.281593420357925
0.179800063223354 0.282443750953897
0.179947369620667 0.283295115059962
0.18009467601798 0.284147512676119
0.180241982415293 0.285000943802369
0.180389288812606 0.285855408438712
0.180536595209919 0.286710906585147
0.180683901607232 0.287567438241675
0.180831208004545 0.288425003408296
0.180978514401859 0.28928360208501
0.181125820799172 0.290143234271816
0.181273127196485 0.291003899968715
0.181420433593798 0.291865599175706
0.181567739991111 0.292728331892791
0.181715046388424 0.293592098119968
0.181862352785737 0.294456897857237
0.182009659183051 0.2953227311046
0.182156965580364 0.296189597862055
0.182304271977677 0.297057498129603
0.18245157837499 0.297926431907243
0.182598884772303 0.298796399194976
0.182746191169616 0.299667399992802
0.182893497566929 0.30053943430072
0.183040803964242 0.301412502118732
0.183188110361556 0.302286603446836
0.183335416758869 0.303161738285032
0.183482723156182 0.304037906633322
0.183630029553495 0.304915108491704
0.183777335950808 0.305793343860178
0.183924642348121 0.306672612738746
0.184071948745434 0.307552915127406
0.184219255142747 0.308434251026159
0.184366561540061 0.309316620435004
0.184513867937374 0.310200023353943
0.184661174334687 0.311084459782973
0.184808480732 0.311969929722097
};
\addplot [line width=0.7000000000000001pt, color4]
table {%
0.170225147398 0.223018609121298
0.170372453795313 0.223919865130352
0.170519760192626 0.22482251553562
0.170667066589939 0.225726560337103
0.170814372987253 0.226631999534799
0.170961679384566 0.22753883312871
0.171108985781879 0.228447061118835
0.171256292179192 0.229356683505174
0.171403598576505 0.230267700287727
0.171550904973818 0.231180111466494
0.171698211371131 0.232093917041476
0.171845517768444 0.233009117012671
0.171992824165758 0.233925711380081
0.172140130563071 0.234843700143705
0.172287436960384 0.235763083303543
0.172434743357697 0.236683860859595
0.17258204975501 0.237606032811861
0.172729356152323 0.238529599160342
0.172876662549636 0.239454559905037
0.173023968946949 0.240380915045945
0.173171275344263 0.241308664583068
0.173318581741576 0.242237808516405
0.173465888138889 0.243168346845957
0.173613194536202 0.244100279571722
0.173760500933515 0.245033606693701
0.173907807330828 0.245968328211895
0.174055113728141 0.246904444126303
0.174202420125455 0.247841954436925
0.174349726522768 0.248780859143761
0.174497032920081 0.249721158246811
0.174644339317394 0.250662851746076
0.174791645714707 0.251605939641554
0.17493895211202 0.252550421933247
0.175086258509333 0.253496298621154
0.175233564906646 0.254443569705275
0.17538087130396 0.25539223518561
0.175528177701273 0.256342295062159
0.175675484098586 0.257293749334923
0.175822790495899 0.2582465980039
0.175970096893212 0.259200841069092
0.176117403290525 0.260156478530498
0.176264709687838 0.261113510388118
0.176412016085152 0.262071936641952
0.176559322482465 0.263031757292001
0.176706628879778 0.263992972338263
0.176853935277091 0.26495558178074
0.177001241674404 0.265919585619431
0.177148548071717 0.266884983854336
0.17729585446903 0.267851776485455
0.177443160866343 0.268819963512788
0.177590467263657 0.269789544936335
0.17773777366097 0.270760520756097
0.177885080058283 0.271732890972073
0.178032386455596 0.272706655584262
0.178179692852909 0.273681814592666
0.178326999250222 0.274658367997284
0.178474305647535 0.275636315798117
0.178621612044848 0.276615657995163
0.178768918442162 0.277596394588424
0.178916224839475 0.278578525577899
0.179063531236788 0.279562050963588
0.179210837634101 0.28054697074549
0.179358144031414 0.281533284923608
0.179505450428727 0.282520993497939
0.17965275682604 0.283510096468484
0.179800063223354 0.284500593835244
0.179947369620667 0.285492485598218
0.18009467601798 0.286485771757406
0.180241982415293 0.287480452312808
0.180389288812606 0.288476527264424
0.180536595209919 0.289473996612254
0.180683901607232 0.290472860356299
0.180831208004545 0.291473118496558
0.180978514401859 0.29247477103303
0.181125820799172 0.293477817965717
0.181273127196485 0.294482259294619
0.181420433593798 0.295488095019734
0.181567739991111 0.296495325141063
0.181715046388424 0.297503949658607
0.181862352785737 0.298513968572364
0.182009659183051 0.299525381882336
0.182156965580364 0.300538189588522
0.182304271977677 0.301552391690922
0.18245157837499 0.302567988189537
0.182598884772303 0.303584979084365
0.182746191169616 0.304603364375408
0.182893497566929 0.305623144062664
0.183040803964242 0.306644318146135
0.183188110361556 0.307666886625821
0.183335416758869 0.30869084950172
0.183482723156182 0.309716206773833
0.183630029553495 0.31074295844216
0.183777335950808 0.311771104506702
0.183924642348121 0.312800644967458
0.184071948745434 0.313831579824428
0.184219255142747 0.314863909077612
0.184366561540061 0.31589763272701
0.184513867937374 0.316932750772623
0.184661174334687 0.317969263214449
0.184808480732 0.31900717005249
};
\addplot [line width=0.7000000000000001pt, color5]
table {%
0.170225147398 0.217247524749906
0.170372453795313 0.218254702440324
0.170519760192626 0.219263656417856
0.170667066589939 0.220274386682503
0.170814372987253 0.221286893234263
0.170961679384566 0.222301176073138
0.171108985781879 0.223317235199126
0.171256292179192 0.224335070612229
0.171403598576505 0.225354682312446
0.171550904973818 0.226376070299777
0.171698211371131 0.227399234574223
0.171845517768444 0.228424175135782
0.171992824165758 0.229450891984456
0.172140130563071 0.230479385120244
0.172287436960384 0.231509654543146
0.172434743357697 0.232541700253162
0.17258204975501 0.233575522250292
0.172729356152323 0.234611120534536
0.172876662549636 0.235648495105895
0.173023968946949 0.236687645964368
0.173171275344263 0.237728573109954
0.173318581741576 0.238771276542655
0.173465888138889 0.23981575626247
0.173613194536202 0.2408620122694
0.173760500933515 0.241910044563443
0.173907807330828 0.242959853144601
0.174055113728141 0.244011438012872
0.174202420125455 0.245064799168258
0.174349726522768 0.246119936610758
0.174497032920081 0.247176850340373
0.174644339317394 0.248235540357101
0.174791645714707 0.249296006660943
0.17493895211202 0.2503582492519
0.175086258509333 0.251422268129971
0.175233564906646 0.252488063295156
0.17538087130396 0.253555634747455
0.175528177701273 0.254624982486868
0.175675484098586 0.255696106513395
0.175822790495899 0.256769006827037
0.175970096893212 0.257843683427793
0.176117403290525 0.258920136315662
0.176264709687838 0.259998365490647
0.176412016085152 0.261078370952744
0.176559322482465 0.262160152701957
0.176706628879778 0.263243710738283
0.176853935277091 0.264329045061724
0.177001241674404 0.265416155672279
0.177148548071717 0.266505042569948
0.17729585446903 0.267595705754731
0.177443160866343 0.268688145226628
0.177590467263657 0.269782360985639
0.17773777366097 0.270878353031765
0.177885080058283 0.271976121365004
0.178032386455596 0.273075665985358
0.178179692852909 0.274176986892826
0.178326999250222 0.275280084087408
0.178474305647535 0.276384957569105
0.178621612044848 0.277491607337915
0.178768918442162 0.278600033393839
0.178916224839475 0.279710235736878
0.179063531236788 0.280822214367031
0.179210837634101 0.281935969284298
0.179358144031414 0.283051500488679
0.179505450428727 0.284168807980175
0.17965275682604 0.285287891758784
0.179800063223354 0.286408751824508
0.179947369620667 0.287531388177345
0.18009467601798 0.288655800817297
0.180241982415293 0.289781989744363
0.180389288812606 0.290909954958544
0.180536595209919 0.292039696459838
0.180683901607232 0.293171214248246
0.180831208004545 0.294304508323769
0.180978514401859 0.295439578686406
0.181125820799172 0.296576425336157
0.181273127196485 0.297715048273022
0.181420433593798 0.298855447497001
0.181567739991111 0.299997623008095
0.181715046388424 0.301141574806302
0.181862352785737 0.302287302891624
0.182009659183051 0.30343480726406
0.182156965580364 0.30458408792361
0.182304271977677 0.305735144870274
0.18245157837499 0.306887978104052
0.182598884772303 0.308042587624945
0.182746191169616 0.309198973432951
0.182893497566929 0.310357135528072
0.183040803964242 0.311517073910307
0.183188110361556 0.312678788579656
0.183335416758869 0.31384227953612
0.183482723156182 0.315007546779697
0.183630029553495 0.316174590310389
0.183777335950808 0.317343410128194
0.183924642348121 0.318514006233114
0.184071948745434 0.319686378625148
0.184219255142747 0.320860527304296
0.184366561540061 0.322036452270559
0.184513867937374 0.323214153523935
0.184661174334687 0.324393631064425
0.184808480732 0.32557488489203
};
\path [draw=white, fill opacity=0] (axis cs:0,0.0785084325396825)
--(axis cs:0,0.464021329365079);

\path [draw=white, fill opacity=0] (axis cs:1,0.0785084325396825)
--(axis cs:1,0.464021329365079);

\path [draw=white, fill opacity=0] (axis cs:0.1475,0)
--(axis cs:0.2025,0);

\path [draw=white, fill opacity=0] (axis cs:0.1475,1)
--(axis cs:0.2025,1);

\end{axis}

\end{tikzpicture}

%% file: paper_planar_iidxz_no_mp.tex
\begin{tikzpicture}

\definecolor{color4}{rgb}{0.219799566082832,0.662515787685034,0.773209315931721}
\definecolor{color2}{rgb}{0.680418912779335,0.615149751467757,0.194054521114453}
\definecolor{color5}{rgb}{0.800493618642396,0.477033635337372,0.957954719600752}
\definecolor{color3}{rgb}{0.201253172212011,0.690792081537903,0.479667611892753}
\definecolor{color0}{rgb}{0.917647058823529,0.917647058823529,0.949019607843137}
\definecolor{color1}{rgb}{0.967797559291991,0.441274560091574,0.53581031550587}

\begin{axis}[
title={IID $X$/$Z$ Error Model, without multi-path summation},
xlabel={physical error probability},
ylabel={logical error probability},
xmin=0.078, xmax=0.122,
ymin=0.016519, ymax=0.551361,
width=\figurewidth,
height=\figureheight,
xtick={0.075,0.08,0.085,0.09,0.095,0.1,0.105,0.11,0.115,0.12,0.125},
xticklabels={,0.080,0.085,0.090,0.095,0.100,0.105,0.110,0.115,0.120,},
ytick={0,0.1,0.2,0.3,0.4,0.5,0.6},
yticklabels={,0.1,0.2,0.3,0.4,0.5,},
tick align=outside,
tick pos=left,
xmajorgrids,
x grid style={white},
ymajorgrids,
y grid style={white},
axis line style={white},
axis background/.style={fill=color0},
legend pos=north west,
legend entries={$d = 7$, $d = 11$, $d = 15$, $d = 19$, $d = 23$}
]
\path [draw=color1, line width=0.7000000000000001pt] (axis cs:0.08,0.13826)
--(axis cs:0.08,0.14393);

\path [draw=color1, line width=0.7000000000000001pt] (axis cs:0.08166667,0.14713)
--(axis cs:0.08166667,0.15295);

\path [draw=color1, line width=0.7000000000000001pt] (axis cs:0.08333333,0.15742)
--(axis cs:0.08333333,0.1634);

\path [draw=color1, line width=0.7000000000000001pt] (axis cs:0.085,0.16498)
--(axis cs:0.085,0.17107);

\path [draw=color1, line width=0.7000000000000001pt] (axis cs:0.08666667,0.17684)
--(axis cs:0.08666667,0.1831);

\path [draw=color1, line width=0.7000000000000001pt] (axis cs:0.08833333,0.1868)
--(axis cs:0.08833333,0.19319);

\path [draw=color1, line width=0.7000000000000001pt] (axis cs:0.09,0.19751)
--(axis cs:0.09,0.20404);

\path [draw=color1, line width=0.7000000000000001pt] (axis cs:0.09166667,0.20804)
--(axis cs:0.09166667,0.21469);

\path [draw=color1, line width=0.7000000000000001pt] (axis cs:0.09333333,0.21607)
--(axis cs:0.09333333,0.22282);

\path [draw=color1, line width=0.7000000000000001pt] (axis cs:0.095,0.23036)
--(axis cs:0.095,0.23725);

\path [draw=color1, line width=0.7000000000000001pt] (axis cs:0.09666667,0.23728)
--(axis cs:0.09666667,0.24425);

\path [draw=color1, line width=0.7000000000000001pt] (axis cs:0.09833333,0.24757)
--(axis cs:0.09833333,0.25464);

\path [draw=color1, line width=0.7000000000000001pt] (axis cs:0.1,0.25929)
--(axis cs:0.1,0.26646);

\path [draw=color1, line width=0.7000000000000001pt] (axis cs:0.10166667,0.26911)
--(axis cs:0.10166667,0.27636);

\path [draw=color1, line width=0.7000000000000001pt] (axis cs:0.10333333,0.28036)
--(axis cs:0.10333333,0.28771);

\path [draw=color1, line width=0.7000000000000001pt] (axis cs:0.105,0.29269)
--(axis cs:0.105,0.30013);

\path [draw=color1, line width=0.7000000000000001pt] (axis cs:0.10666667,0.30225)
--(axis cs:0.10666667,0.30976);

\path [draw=color1, line width=0.7000000000000001pt] (axis cs:0.10833333,0.31347)
--(axis cs:0.10833333,0.32105);

\path [draw=color1, line width=0.7000000000000001pt] (axis cs:0.11,0.32552)
--(axis cs:0.11,0.33317);

\path [draw=color1, line width=0.7000000000000001pt] (axis cs:0.11166667,0.33281)
--(axis cs:0.11166667,0.34051);

\path [draw=color1, line width=0.7000000000000001pt] (axis cs:0.11333333,0.34618)
--(axis cs:0.11333333,0.35395);

\path [draw=color1, line width=0.7000000000000001pt] (axis cs:0.115,0.35551)
--(axis cs:0.115,0.36333);

\path [draw=color1, line width=0.7000000000000001pt] (axis cs:0.11666667,0.36878)
--(axis cs:0.11666667,0.37666);

\path [draw=color1, line width=0.7000000000000001pt] (axis cs:0.11833333,0.37648)
--(axis cs:0.11833333,0.38439);

\path [draw=color1, line width=0.7000000000000001pt] (axis cs:0.12,0.38651)
--(axis cs:0.12,0.39446);

\path [draw=color2, line width=0.7000000000000001pt] (axis cs:0.08,0.10219)
--(axis cs:0.08,0.10718);

\path [draw=color2, line width=0.7000000000000001pt] (axis cs:0.08166667,0.11099)
--(axis cs:0.08166667,0.11616);

\path [draw=color2, line width=0.7000000000000001pt] (axis cs:0.08333333,0.12254)
--(axis cs:0.08333333,0.12793);

\path [draw=color2, line width=0.7000000000000001pt] (axis cs:0.085,0.13369)
--(axis cs:0.085,0.13928);

\path [draw=color2, line width=0.7000000000000001pt] (axis cs:0.08666667,0.14427)
--(axis cs:0.08666667,0.15004);

\path [draw=color2, line width=0.7000000000000001pt] (axis cs:0.08833333,0.15639)
--(axis cs:0.08833333,0.16235);

\path [draw=color2, line width=0.7000000000000001pt] (axis cs:0.09,0.16971)
--(axis cs:0.09,0.17587);

\path [draw=color2, line width=0.7000000000000001pt] (axis cs:0.09166667,0.18183)
--(axis cs:0.09166667,0.18816);

\path [draw=color2, line width=0.7000000000000001pt] (axis cs:0.09333333,0.1957)
--(axis cs:0.09333333,0.20221);

\path [draw=color2, line width=0.7000000000000001pt] (axis cs:0.095,0.20828)
--(axis cs:0.095,0.21493);

\path [draw=color2, line width=0.7000000000000001pt] (axis cs:0.09666667,0.22358)
--(axis cs:0.09666667,0.23041);

\path [draw=color2, line width=0.7000000000000001pt] (axis cs:0.09833333,0.2364)
--(axis cs:0.09833333,0.24335);

\path [draw=color2, line width=0.7000000000000001pt] (axis cs:0.1,0.25094)
--(axis cs:0.1,0.25803);

\path [draw=color2, line width=0.7000000000000001pt] (axis cs:0.10166667,0.26563)
--(axis cs:0.10166667,0.27286);

\path [draw=color2, line width=0.7000000000000001pt] (axis cs:0.10333333,0.27965)
--(axis cs:0.10333333,0.28699);

\path [draw=color2, line width=0.7000000000000001pt] (axis cs:0.105,0.29311)
--(axis cs:0.105,0.30056);

\path [draw=color2, line width=0.7000000000000001pt] (axis cs:0.10666667,0.31112)
--(axis cs:0.10666667,0.31869);

\path [draw=color2, line width=0.7000000000000001pt] (axis cs:0.10833333,0.32561)
--(axis cs:0.10833333,0.33327);

\path [draw=color2, line width=0.7000000000000001pt] (axis cs:0.11,0.33931)
--(axis cs:0.11,0.34704);

\path [draw=color2, line width=0.7000000000000001pt] (axis cs:0.11166667,0.35572)
--(axis cs:0.11166667,0.36354);

\path [draw=color2, line width=0.7000000000000001pt] (axis cs:0.11333333,0.36864)
--(axis cs:0.11333333,0.37652);

\path [draw=color2, line width=0.7000000000000001pt] (axis cs:0.115,0.38126)
--(axis cs:0.115,0.38919);

\path [draw=color2, line width=0.7000000000000001pt] (axis cs:0.11666667,0.3975)
--(axis cs:0.11666667,0.40548);

\path [draw=color2, line width=0.7000000000000001pt] (axis cs:0.11833333,0.41272)
--(axis cs:0.11833333,0.42075);

\path [draw=color2, line width=0.7000000000000001pt] (axis cs:0.12,0.42393)
--(axis cs:0.12,0.43199);

\path [draw=color3, line width=0.7000000000000001pt] (axis cs:0.08,0.07409)
--(axis cs:0.08,0.07841);

\path [draw=color3, line width=0.7000000000000001pt] (axis cs:0.08166667,0.08294)
--(axis cs:0.08166667,0.08749);

\path [draw=color3, line width=0.7000000000000001pt] (axis cs:0.08333333,0.09548)
--(axis cs:0.08333333,0.10032);

\path [draw=color3, line width=0.7000000000000001pt] (axis cs:0.085,0.10577)
--(axis cs:0.085,0.11083);

\path [draw=color3, line width=0.7000000000000001pt] (axis cs:0.08666667,0.12031)
--(axis cs:0.08666667,0.12566);

\path [draw=color3, line width=0.7000000000000001pt] (axis cs:0.08833333,0.13361)
--(axis cs:0.08833333,0.1392);

\path [draw=color3, line width=0.7000000000000001pt] (axis cs:0.09,0.1469)
--(axis cs:0.09,0.15271);

\path [draw=color3, line width=0.7000000000000001pt] (axis cs:0.09166667,0.16114)
--(axis cs:0.09166667,0.16717);

\path [draw=color3, line width=0.7000000000000001pt] (axis cs:0.09333333,0.17731)
--(axis cs:0.09333333,0.18358);

\path [draw=color3, line width=0.7000000000000001pt] (axis cs:0.095,0.19328)
--(axis cs:0.095,0.19975);

\path [draw=color3, line width=0.7000000000000001pt] (axis cs:0.09666667,0.20869)
--(axis cs:0.09666667,0.21534);

\path [draw=color3, line width=0.7000000000000001pt] (axis cs:0.09833333,0.227)
--(axis cs:0.09833333,0.23386);

\path [draw=color3, line width=0.7000000000000001pt] (axis cs:0.1,0.24514)
--(axis cs:0.1,0.25219);

\path [draw=color3, line width=0.7000000000000001pt] (axis cs:0.10166667,0.26247)
--(axis cs:0.10166667,0.26967);

\path [draw=color3, line width=0.7000000000000001pt] (axis cs:0.10333333,0.28015)
--(axis cs:0.10333333,0.2875);

\path [draw=color3, line width=0.7000000000000001pt] (axis cs:0.105,0.30174)
--(axis cs:0.105,0.30925);

\path [draw=color3, line width=0.7000000000000001pt] (axis cs:0.10666667,0.31609)
--(axis cs:0.10666667,0.32369);

\path [draw=color3, line width=0.7000000000000001pt] (axis cs:0.10833333,0.33523)
--(axis cs:0.10833333,0.34294);

\path [draw=color3, line width=0.7000000000000001pt] (axis cs:0.11,0.35588)
--(axis cs:0.11,0.3637);

\path [draw=color3, line width=0.7000000000000001pt] (axis cs:0.11166667,0.37032)
--(axis cs:0.11166667,0.3782);

\path [draw=color3, line width=0.7000000000000001pt] (axis cs:0.11333333,0.39016)
--(axis cs:0.11333333,0.39812);

\path [draw=color3, line width=0.7000000000000001pt] (axis cs:0.115,0.40641)
--(axis cs:0.115,0.41443);

\path [draw=color3, line width=0.7000000000000001pt] (axis cs:0.11666667,0.42611)
--(axis cs:0.11666667,0.43417);

\path [draw=color3, line width=0.7000000000000001pt] (axis cs:0.11833333,0.44267)
--(axis cs:0.11833333,0.45077);

\path [draw=color3, line width=0.7000000000000001pt] (axis cs:0.12,0.4593)
--(axis cs:0.12,0.46742);

\path [draw=color4, line width=0.7000000000000001pt] (axis cs:0.08,0.05285)
--(axis cs:0.08,0.05655);

\path [draw=color4, line width=0.7000000000000001pt] (axis cs:0.08166667,0.06418)
--(axis cs:0.08166667,0.06823);

\path [draw=color4, line width=0.7000000000000001pt] (axis cs:0.08333333,0.0735)
--(axis cs:0.08333333,0.07781);

\path [draw=color4, line width=0.7000000000000001pt] (axis cs:0.085,0.0864)
--(axis cs:0.085,0.09103);

\path [draw=color4, line width=0.7000000000000001pt] (axis cs:0.08666667,0.10026)
--(axis cs:0.08666667,0.10521);

\path [draw=color4, line width=0.7000000000000001pt] (axis cs:0.08833333,0.11507)
--(axis cs:0.08833333,0.12032);

\path [draw=color4, line width=0.7000000000000001pt] (axis cs:0.09,0.12839)
--(axis cs:0.09,0.13389);

\path [draw=color4, line width=0.7000000000000001pt] (axis cs:0.09166667,0.14488)
--(axis cs:0.09166667,0.15066);

\path [draw=color4, line width=0.7000000000000001pt] (axis cs:0.09333333,0.16189)
--(axis cs:0.09333333,0.16794);

\path [draw=color4, line width=0.7000000000000001pt] (axis cs:0.095,0.18079)
--(axis cs:0.095,0.1871);

\path [draw=color4, line width=0.7000000000000001pt] (axis cs:0.09666667,0.20037)
--(axis cs:0.09666667,0.20693);

\path [draw=color4, line width=0.7000000000000001pt] (axis cs:0.09833333,0.21693)
--(axis cs:0.09833333,0.22368);

\path [draw=color4, line width=0.7000000000000001pt] (axis cs:0.1,0.23784)
--(axis cs:0.1,0.24481);

\path [draw=color4, line width=0.7000000000000001pt] (axis cs:0.10166667,0.26217)
--(axis cs:0.10166667,0.26936);

\path [draw=color4, line width=0.7000000000000001pt] (axis cs:0.10333333,0.28346)
--(axis cs:0.10333333,0.29083);

\path [draw=color4, line width=0.7000000000000001pt] (axis cs:0.105,0.3074)
--(axis cs:0.105,0.31494);

\path [draw=color4, line width=0.7000000000000001pt] (axis cs:0.10666667,0.322)
--(axis cs:0.10666667,0.32963);

\path [draw=color4, line width=0.7000000000000001pt] (axis cs:0.10833333,0.34478)
--(axis cs:0.10833333,0.35254);

\path [draw=color4, line width=0.7000000000000001pt] (axis cs:0.11,0.37093)
--(axis cs:0.11,0.37882);

\path [draw=color4, line width=0.7000000000000001pt] (axis cs:0.11166667,0.38745)
--(axis cs:0.11166667,0.3954);

\path [draw=color4, line width=0.7000000000000001pt] (axis cs:0.11333333,0.41052)
--(axis cs:0.11333333,0.41854);

\path [draw=color4, line width=0.7000000000000001pt] (axis cs:0.115,0.42907)
--(axis cs:0.115,0.43714);

\path [draw=color4, line width=0.7000000000000001pt] (axis cs:0.11666667,0.44899)
--(axis cs:0.11666667,0.4571);

\path [draw=color4, line width=0.7000000000000001pt] (axis cs:0.11833333,0.47144)
--(axis cs:0.11833333,0.47958);

\path [draw=color4, line width=0.7000000000000001pt] (axis cs:0.12,0.4894)
--(axis cs:0.12,0.49754);

\path [draw=color5, line width=0.7000000000000001pt] (axis cs:0.08,0.04083)
--(axis cs:0.08,0.04411);

\path [draw=color5, line width=0.7000000000000001pt] (axis cs:0.08166667,0.05033)
--(axis cs:0.08166667,0.05395);

\path [draw=color5, line width=0.7000000000000001pt] (axis cs:0.08333333,0.05929)
--(axis cs:0.08333333,0.06319);

\path [draw=color5, line width=0.7000000000000001pt] (axis cs:0.085,0.07063)
--(axis cs:0.085,0.07486);

\path [draw=color5, line width=0.7000000000000001pt] (axis cs:0.08666667,0.0833)
--(axis cs:0.08666667,0.08786);

\path [draw=color5, line width=0.7000000000000001pt] (axis cs:0.08833333,0.09686)
--(axis cs:0.08833333,0.10173);

\path [draw=color5, line width=0.7000000000000001pt] (axis cs:0.09,0.11333)
--(axis cs:0.09,0.11854);

\path [draw=color5, line width=0.7000000000000001pt] (axis cs:0.09166667,0.12832)
--(axis cs:0.09166667,0.13382);

\path [draw=color5, line width=0.7000000000000001pt] (axis cs:0.09333333,0.14782)
--(axis cs:0.09333333,0.15365);

\path [draw=color5, line width=0.7000000000000001pt] (axis cs:0.095,0.17013)
--(axis cs:0.095,0.1763);

\path [draw=color5, line width=0.7000000000000001pt] (axis cs:0.09666667,0.18985)
--(axis cs:0.09666667,0.19628);

\path [draw=color5, line width=0.7000000000000001pt] (axis cs:0.09833333,0.21047)
--(axis cs:0.09833333,0.21714);

\path [draw=color5, line width=0.7000000000000001pt] (axis cs:0.1,0.23401)
--(axis cs:0.1,0.24094);

\path [draw=color5, line width=0.7000000000000001pt] (axis cs:0.10166667,0.25891)
--(axis cs:0.10166667,0.26608);

\path [draw=color5, line width=0.7000000000000001pt] (axis cs:0.10333333,0.28193)
--(axis cs:0.10333333,0.28929);

\path [draw=color5, line width=0.7000000000000001pt] (axis cs:0.105,0.30844)
--(axis cs:0.105,0.31599);

\path [draw=color5, line width=0.7000000000000001pt] (axis cs:0.10666667,0.32853)
--(axis cs:0.10666667,0.3362);

\path [draw=color5, line width=0.7000000000000001pt] (axis cs:0.10833333,0.35592)
--(axis cs:0.10833333,0.36374);

\path [draw=color5, line width=0.7000000000000001pt] (axis cs:0.11,0.38213)
--(axis cs:0.11,0.39006);

\path [draw=color5, line width=0.7000000000000001pt] (axis cs:0.11166667,0.40538)
--(axis cs:0.11166667,0.41339);

\path [draw=color5, line width=0.7000000000000001pt] (axis cs:0.11333333,0.42852)
--(axis cs:0.11333333,0.43659);

\path [draw=color5, line width=0.7000000000000001pt] (axis cs:0.115,0.45169)
--(axis cs:0.115,0.45981);

\path [draw=color5, line width=0.7000000000000001pt] (axis cs:0.11666667,0.474)
--(axis cs:0.11666667,0.48214);

\path [draw=color5, line width=0.7000000000000001pt] (axis cs:0.11833333,0.49416)
--(axis cs:0.11833333,0.5023);

\path [draw=color5, line width=0.7000000000000001pt] (axis cs:0.12,0.51891)
--(axis cs:0.12,0.52705);

\addplot [line width=0.7000000000000001pt, color1]
table {%
0.0976692028103785 0.246282648970682
0.0977870481639138 0.247071821289228
0.0979048935174492 0.24786076654306
0.0980227388709845 0.248649484732181
0.0981405842245199 0.249437975856588
0.0982584295780552 0.250226239916284
0.0983762749315906 0.251014276911266
0.0984941202851259 0.251802086841536
0.0986119656386613 0.252589669707093
0.0987298109921966 0.253377025507938
0.098847656345732 0.25416415424407
0.0989655016992673 0.254951055915489
0.0990833470528027 0.255737730522196
0.0992011924063381 0.256524178064191
0.0993190377598734 0.257310398541472
0.0994368831134088 0.258096391954041
0.0995547284669441 0.258882158301898
0.0996725738204795 0.259667697585042
0.0997904191740148 0.260453009803473
0.0999082645275502 0.261238094957192
0.100026109881086 0.262022953046198
0.100143955234621 0.262807584070491
0.100261800588156 0.263591988030072
0.100379645941692 0.26437616492494
0.100497491295227 0.265160114755096
0.100615336648762 0.265943837520539
0.100733182002298 0.266727333221269
0.100851027355833 0.267510601857287
0.100968872709368 0.268293643428592
0.101086718062904 0.269076457935185
0.101204563416439 0.269859045377065
0.101322408769974 0.270641405754232
0.10144025412351 0.271423539066687
0.101558099477045 0.272205445314429
0.10167594483058 0.272987124497459
0.101793790184116 0.273768576615776
0.101911635537651 0.27454980166938
0.102029480891187 0.275330799658272
0.102147326244722 0.276111570582451
0.102265171598257 0.276892114441918
0.102383016951793 0.277672431236672
0.102500862305328 0.278452520966713
0.102618707658863 0.279232383632042
0.102736553012399 0.280012019232658
0.102854398365934 0.280791427768562
0.102972243719469 0.281570609239753
0.103090089073005 0.282349563646231
0.10320793442654 0.283128290987997
0.103325779780075 0.28390679126505
0.103443625133611 0.284685064477391
0.103561470487146 0.285463110625018
0.103679315840681 0.286240929707934
0.103797161194217 0.287018521726136
0.103915006547752 0.287795886679627
0.104032851901288 0.288573024568404
0.104150697254823 0.289349935392469
0.104268542608358 0.290126619151821
0.104386387961894 0.290903075846461
0.104504233315429 0.291679305476388
0.104622078668964 0.292455308041603
0.1047399240225 0.293231083542105
0.104857769376035 0.294006631977894
0.10497561472957 0.294781953348971
0.105093460083106 0.295557047655335
0.105211305436641 0.296331914896986
0.105329150790176 0.297106555073925
0.105446996143712 0.297880968186151
0.105564841497247 0.298655154233665
0.105682686850782 0.299429113216466
0.105800532204318 0.300202845134554
0.105918377557853 0.30097634998793
0.106036222911389 0.301749627776593
0.106154068264924 0.302522678500544
0.106271913618459 0.303295502159782
0.106389758971995 0.304068098754308
0.10650760432553 0.30484046828412
0.106625449679065 0.305612610749221
0.106743295032601 0.306384526149608
0.106861140386136 0.307156214485283
0.106978985739671 0.307927675756246
0.107096831093207 0.308698909962496
0.107214676446742 0.309469917104033
0.107332521800277 0.310240697180857
0.107450367153813 0.311011250192969
0.107568212507348 0.311781576140369
0.107686057860884 0.312551675023055
0.107803903214419 0.31332154684103
0.107921748567954 0.314091191594291
0.10803959392149 0.31486060928284
0.108157439275025 0.315629799906677
0.10827528462856 0.3163987634658
0.108393129982096 0.317167499960211
0.108510975335631 0.31793600938991
0.108628820689166 0.318704291754896
0.108746666042702 0.319472347055169
0.108864511396237 0.32024017529073
0.108982356749772 0.321007776461578
0.109100202103308 0.321775150567714
0.109218047456843 0.322542297609137
0.109335892810378 0.323309217585847
};
\addplot [line width=0.7000000000000001pt, color2]
table {%
0.0976692028103785 0.233734347980203
0.0977870481639138 0.234804293793162
0.0979048935174492 0.235873825970558
0.0980227388709845 0.236942944512394
0.0981405842245199 0.238011649418668
0.0982584295780552 0.239079940689381
0.0983762749315906 0.240147818324533
0.0984941202851259 0.241215282324123
0.0986119656386613 0.242282332688152
0.0987298109921966 0.243348969416619
0.098847656345732 0.244415192509525
0.0989655016992673 0.24548100196687
0.0990833470528027 0.246546397788654
0.0992011924063381 0.247611379974876
0.0993190377598734 0.248675948525537
0.0994368831134088 0.249740103440636
0.0995547284669441 0.250803844720174
0.0996725738204795 0.251867172364151
0.0997904191740148 0.252930086372567
0.0999082645275502 0.253992586745421
0.100026109881086 0.255054673482714
0.100143955234621 0.256116346584445
0.100261800588156 0.257177606050615
0.100379645941692 0.258238451881224
0.100497491295227 0.259298884076272
0.100615336648762 0.260358902635758
0.100733182002298 0.261418507559683
0.100851027355833 0.262477698848046
0.100968872709368 0.263536476500848
0.101086718062904 0.264594840518089
0.101204563416439 0.265652790899769
0.101322408769974 0.266710327645887
0.10144025412351 0.267767450756444
0.101558099477045 0.268824160231439
0.10167594483058 0.269880456070873
0.101793790184116 0.270936338274746
0.101911635537651 0.271991806843057
0.102029480891187 0.273046861775807
0.102147326244722 0.274101503072996
0.102265171598257 0.275155730734624
0.102383016951793 0.27620954476069
0.102500862305328 0.277262945151195
0.102618707658863 0.278315931906138
0.102736553012399 0.27936850502552
0.102854398365934 0.280420664509341
0.102972243719469 0.2814724103576
0.103090089073005 0.282523742570298
0.10320793442654 0.283574661147435
0.103325779780075 0.284625166089011
0.103443625133611 0.285675257395025
0.103561470487146 0.286724935065477
0.103679315840681 0.287774199100369
0.103797161194217 0.288823049499699
0.103915006547752 0.289871486263468
0.104032851901288 0.290919509391675
0.104150697254823 0.291967118884321
0.104268542608358 0.293014314741406
0.104386387961894 0.294061096962929
0.104504233315429 0.295107465548891
0.104622078668964 0.296153420499292
0.1047399240225 0.297198961814131
0.104857769376035 0.298244089493409
0.10497561472957 0.299288803537126
0.105093460083106 0.300333103945281
0.105211305436641 0.301376990717876
0.105329150790176 0.302420463854908
0.105446996143712 0.30346352335638
0.105564841497247 0.30450616922229
0.105682686850782 0.305548401452638
0.105800532204318 0.306590220047426
0.105918377557853 0.307631625006652
0.106036222911389 0.308672616330316
0.106154068264924 0.30971319401842
0.106271913618459 0.310753358070962
0.106389758971995 0.311793108487942
0.10650760432553 0.312832445269362
0.106625449679065 0.31387136841522
0.106743295032601 0.314909877925516
0.106861140386136 0.315947973800252
0.106978985739671 0.316985656039426
0.107096831093207 0.318022924643038
0.107214676446742 0.31905977961109
0.107332521800277 0.32009622094358
0.107450367153813 0.321132248640508
0.107568212507348 0.322167862701876
0.107686057860884 0.323203063127682
0.107803903214419 0.324237849917926
0.107921748567954 0.32527222307261
0.10803959392149 0.326306182591732
0.108157439275025 0.327339728475292
0.10827528462856 0.328372860723292
0.108393129982096 0.32940557933573
0.108510975335631 0.330437884312606
0.108628820689166 0.331469775653921
0.108746666042702 0.332501253359675
0.108864511396237 0.333532317429868
0.108982356749772 0.334562967864499
0.109100202103308 0.335593204663569
0.109218047456843 0.336623027827078
0.109335892810378 0.337652437355025
};
\addplot [line width=0.7000000000000001pt, color3]
table {%
0.0976692028103785 0.222621210375065
0.0977870481639138 0.223940823695714
0.0979048935174492 0.225259812774715
0.0980227388709845 0.22657817761207
0.0981405842245199 0.227895918207777
0.0982584295780552 0.229213034561836
0.0983762749315906 0.230529526674249
0.0984941202851259 0.231845394545014
0.0986119656386613 0.233160638174132
0.0987298109921966 0.234475257561602
0.098847656345732 0.235789252707426
0.0989655016992673 0.237102623611601
0.0990833470528027 0.23841537027413
0.0992011924063381 0.239727492695012
0.0993190377598734 0.241038990874246
0.0994368831134088 0.242349864811832
0.0995547284669441 0.243660114507772
0.0996725738204795 0.244969739962064
0.0997904191740148 0.246278741174709
0.0999082645275502 0.247587118145706
0.100026109881086 0.248894870875056
0.100143955234621 0.250201999362759
0.100261800588156 0.251508503608815
0.100379645941692 0.252814383613223
0.100497491295227 0.254119639375984
0.100615336648762 0.255424270897098
0.100733182002298 0.256728278176564
0.100851027355833 0.258031661214383
0.100968872709368 0.259334420010555
0.101086718062904 0.26063655456508
0.101204563416439 0.261938064877957
0.101322408769974 0.263238950949187
0.10144025412351 0.264539212778769
0.101558099477045 0.265838850366705
0.10167594483058 0.267137863712993
0.101793790184116 0.268436252817633
0.101911635537651 0.269734017680627
0.102029480891187 0.271031158301973
0.102147326244722 0.272327674681671
0.102265171598257 0.273623566819723
0.102383016951793 0.274918834716127
0.102500862305328 0.276213478370884
0.102618707658863 0.277507497783994
0.102736553012399 0.278800892955456
0.102854398365934 0.280093663885271
0.102972243719469 0.281385810573438
0.103090089073005 0.282677333019959
0.10320793442654 0.283968231224832
0.103325779780075 0.285258505188057
0.103443625133611 0.286548154909636
0.103561470487146 0.287837180389567
0.103679315840681 0.289125581627851
0.103797161194217 0.290413358624487
0.103915006547752 0.291700511379476
0.104032851901288 0.292987039892818
0.104150697254823 0.294272944164513
0.104268542608358 0.29555822419456
0.104386387961894 0.29684287998296
0.104504233315429 0.298126911529713
0.104622078668964 0.299410318834818
0.1047399240225 0.300693101898276
0.104857769376035 0.301975260720087
0.10497561472957 0.303256795300251
0.105093460083106 0.304537705638767
0.105211305436641 0.305817991735636
0.105329150790176 0.307097653590857
0.105446996143712 0.308376691204432
0.105564841497247 0.309655104576358
0.105682686850782 0.310932893706638
0.105800532204318 0.312210058595271
0.105918377557853 0.313486599242256
0.106036222911389 0.314762515647593
0.106154068264924 0.316037807811284
0.106271913618459 0.317312475733327
0.106389758971995 0.318586519413723
0.10650760432553 0.319859938852471
0.106625449679065 0.321132734049573
0.106743295032601 0.322404905005026
0.106861140386136 0.323676451718833
0.106978985739671 0.324947374190992
0.107096831093207 0.326217672421505
0.107214676446742 0.327487346410369
0.107332521800277 0.328756396157587
0.107450367153813 0.330024821663157
0.107568212507348 0.33129262292708
0.107686057860884 0.332559799949355
0.107803903214419 0.333826352729983
0.107921748567954 0.335092281268964
0.10803959392149 0.336357585566298
0.108157439275025 0.337622265621984
0.10827528462856 0.338886321436023
0.108393129982096 0.340149753008415
0.108510975335631 0.341412560339159
0.108628820689166 0.342674743428256
0.108746666042702 0.343936302275706
0.108864511396237 0.345197236881509
0.108982356749772 0.346457547245664
0.109100202103308 0.347717233368172
0.109218047456843 0.348976295249032
0.109335892810378 0.350234732888245
};
\addplot [line width=0.7000000000000001pt, color4]
table {%
0.0976692028103785 0.212436200319313
0.0977870481639138 0.213985450022134
0.0979048935174492 0.215533845488813
0.0980227388709845 0.217081386719351
0.0981405842245199 0.218628073713748
0.0982584295780552 0.220173906472003
0.0983762749315906 0.221718884994117
0.0984941202851259 0.22326300928009
0.0986119656386613 0.224806279329921
0.0987298109921966 0.22634869514361
0.098847656345732 0.227890256721159
0.0989655016992673 0.229430964062565
0.0990833470528027 0.230970817167831
0.0992011924063381 0.232509816036955
0.0993190377598734 0.234047960669937
0.0994368831134088 0.235585251066778
0.0995547284669441 0.237121687227478
0.0996725738204795 0.238657269152036
0.0997904191740148 0.240191996840453
0.0999082645275502 0.241725870292728
0.100026109881086 0.243258889508862
0.100143955234621 0.244791054488854
0.100261800588156 0.246322365232705
0.100379645941692 0.247852821740415
0.100497491295227 0.249382424011983
0.100615336648762 0.25091117204741
0.100733182002298 0.252439065846695
0.100851027355833 0.253966105409839
0.100968872709368 0.255492290736841
0.101086718062904 0.257017621827703
0.101204563416439 0.258542098682422
0.101322408769974 0.260065721301
0.10144025412351 0.261588489683437
0.101558099477045 0.263110403829732
0.10167594483058 0.264631463739886
0.101793790184116 0.266151669413899
0.101911635537651 0.26767102085177
0.102029480891187 0.2691895180535
0.102147326244722 0.270707161019088
0.102265171598257 0.272223949748534
0.102383016951793 0.27373988424184
0.102500862305328 0.275254964499004
0.102618707658863 0.276769190520026
0.102736553012399 0.278282562304907
0.102854398365934 0.279795079853647
0.102972243719469 0.281306743166245
0.103090089073005 0.282817552242702
0.10320793442654 0.284327507083017
0.103325779780075 0.285836607687191
0.103443625133611 0.287344854055224
0.103561470487146 0.288852246187115
0.103679315840681 0.290358784082864
0.103797161194217 0.291864467742472
0.103915006547752 0.293369297165939
0.104032851901288 0.294873272353265
0.104150697254823 0.296376393304449
0.104268542608358 0.297878660019491
0.104386387961894 0.299380072498392
0.104504233315429 0.300880630741152
0.104622078668964 0.30238033474777
0.1047399240225 0.303879184518247
0.104857769376035 0.305377180052582
0.10497561472957 0.306874321350776
0.105093460083106 0.308370608412829
0.105211305436641 0.30986604123874
0.105329150790176 0.311360619828509
0.105446996143712 0.312854344182138
0.105564841497247 0.314347214299624
0.105682686850782 0.31583923018097
0.105800532204318 0.317330391826173
0.105918377557853 0.318820699235236
0.106036222911389 0.320310152408157
0.106154068264924 0.321798751344937
0.106271913618459 0.323286496045575
0.106389758971995 0.324773386510072
0.10650760432553 0.326259422738427
0.106625449679065 0.327744604730641
0.106743295032601 0.329228932486714
0.106861140386136 0.330712406006645
0.106978985739671 0.332195025290434
0.107096831093207 0.333676790338083
0.107214676446742 0.335157701149589
0.107332521800277 0.336637757724955
0.107450367153813 0.338116960064179
0.107568212507348 0.339595308167261
0.107686057860884 0.341072802034202
0.107803903214419 0.342549441665002
0.107921748567954 0.34402522705966
0.10803959392149 0.345500158218177
0.108157439275025 0.346974235140552
0.10827528462856 0.348447457826786
0.108393129982096 0.349919826276879
0.108510975335631 0.35139134049083
0.108628820689166 0.35286200046864
0.108746666042702 0.354331806210308
0.108864511396237 0.355800757715835
0.108982356749772 0.35726885498522
0.109100202103308 0.358736098018464
0.109218047456843 0.360202486815566
0.109335892810378 0.361668021376528
};
\addplot [line width=0.7000000000000001pt, color5]
table {%
0.0976692028103785 0.202920457456069
0.0977870481639138 0.20468495494879
0.0979048935174492 0.206448351717329
0.0980227388709845 0.208210647761687
0.0981405842245199 0.209971843081863
0.0982584295780552 0.211731937677857
0.0983762749315906 0.21349093154967
0.0984941202851259 0.215248824697301
0.0986119656386613 0.21700561712075
0.0987298109921966 0.218761308820018
0.098847656345732 0.220515899795104
0.0989655016992673 0.222269390046008
0.0990833470528027 0.224021779572731
0.0992011924063381 0.225773068375273
0.0993190377598734 0.227523256453632
0.0994368831134088 0.22927234380781
0.0995547284669441 0.231020330437806
0.0996725738204795 0.232767216343621
0.0997904191740148 0.234513001525254
0.0999082645275502 0.236257685982705
0.100026109881086 0.238001269715975
0.100143955234621 0.239743752725063
0.100261800588156 0.241485135009969
0.100379645941692 0.243225416570694
0.100497491295227 0.244964597407237
0.100615336648762 0.246702677519599
0.100733182002298 0.248439656907778
0.100851027355833 0.250175535571777
0.100968872709368 0.251910313511593
0.101086718062904 0.253643990727228
0.101204563416439 0.255376567218682
0.101322408769974 0.257108042985953
0.10144025412351 0.258838418029043
0.101558099477045 0.260567692347951
0.10167594483058 0.262295865942678
0.101793790184116 0.264022938813223
0.101911635537651 0.265748910959587
0.102029480891187 0.267473782381769
0.102147326244722 0.269197553079769
0.102265171598257 0.270920223053587
0.102383016951793 0.272641792303224
0.102500862305328 0.274362260828679
0.102618707658863 0.276081628629953
0.102736553012399 0.277799895707045
0.102854398365934 0.279517062059955
0.102972243719469 0.281233127688684
0.103090089073005 0.282948092593231
0.10320793442654 0.284661956773596
0.103325779780075 0.28637472022978
0.103443625133611 0.288086382961782
0.103561470487146 0.289796944969602
0.103679315840681 0.291506406253241
0.103797161194217 0.293214766812698
0.103915006547752 0.294922026647974
0.104032851901288 0.296628185759068
0.104150697254823 0.29833324414598
0.104268542608358 0.300037201808711
0.104386387961894 0.301740058747259
0.104504233315429 0.303441814961627
0.104622078668964 0.305142470451813
0.1047399240225 0.306842025217817
0.104857769376035 0.308540479259639
0.10497561472957 0.31023783257728
0.105093460083106 0.311934085170739
0.105211305436641 0.313629237040016
0.105329150790176 0.315323288185112
0.105446996143712 0.317016238606026
0.105564841497247 0.318708088302759
0.105682686850782 0.32039883727531
0.105800532204318 0.322088485523679
0.105918377557853 0.323777033047867
0.106036222911389 0.325464479847873
0.106154068264924 0.327150825923697
0.106271913618459 0.32883607127534
0.106389758971995 0.3305202159028
0.10650760432553 0.33220325980608
0.106625449679065 0.333885202985178
0.106743295032601 0.335566045440094
0.106861140386136 0.337245787170828
0.106978985739671 0.338924428177381
0.107096831093207 0.340601968459752
0.107214676446742 0.342278408017942
0.107332521800277 0.34395374685195
0.107450367153813 0.345627984961776
0.107568212507348 0.347301122347421
0.107686057860884 0.348973159008884
0.107803903214419 0.350644094946165
0.107921748567954 0.352313930159265
0.10803959392149 0.353982664648183
0.108157439275025 0.355650298412919
0.10827528462856 0.357316831453474
0.108393129982096 0.358982263769847
0.108510975335631 0.360646595362039
0.108628820689166 0.362309826230048
0.108746666042702 0.363971956373877
0.108864511396237 0.365632985793523
0.108982356749772 0.367292914488988
0.109100202103308 0.368951742460272
0.109218047456843 0.370609469707373
0.109335892810378 0.372266096230293
};
\path [draw=white, fill opacity=0] (axis cs:0,0.016519)
--(axis cs:0,0.551361);

\path [draw=white, fill opacity=0] (axis cs:1,0.016519)
--(axis cs:1,0.551361);

\path [draw=white, fill opacity=0] (axis cs:0.078,0)
--(axis cs:0.122,0);

\path [draw=white, fill opacity=0] (axis cs:0.078,1)
--(axis cs:0.122,1);

\end{axis}

\end{tikzpicture}

%% file: paper_planar_iidxz_mp.tex
\begin{tikzpicture}

\definecolor{color4}{rgb}{0.219799566082832,0.662515787685034,0.773209315931721}
\definecolor{color2}{rgb}{0.680418912779335,0.615149751467757,0.194054521114453}
\definecolor{color5}{rgb}{0.800493618642396,0.477033635337372,0.957954719600752}
\definecolor{color3}{rgb}{0.201253172212011,0.690792081537903,0.479667611892753}
\definecolor{color0}{rgb}{0.917647058823529,0.917647058823529,0.949019607843137}
\definecolor{color1}{rgb}{0.967797559291991,0.441274560091574,0.53581031550587}

\begin{axis}[
title={IID $X$/$Z$ Error Model, with multi-path summation},
xlabel={physical error probability},
ylabel={logical error probability},
xmin=0.078, xmax=0.122,
ymin=0.00931664332399626, ymax=0.509952731092437,
width=\figurewidth,
height=\figureheight,
xtick={0.075,0.08,0.085,0.09,0.095,0.1,0.105,0.11,0.115,0.12,0.125},
xticklabels={,0.080,0.085,0.090,0.095,0.100,0.105,0.110,0.115,0.120,},
ytick={0,0.1,0.2,0.3,0.4,0.5,0.6},
yticklabels={,0.1,0.2,0.3,0.4,0.5,},
tick align=outside,
tick pos=left,
xmajorgrids,
x grid style={white},
ymajorgrids,
y grid style={white},
axis line style={white},
axis background/.style={fill=color0},
legend pos=north west,
legend entries={$d = 7$, $d = 11$, $d = 15$, $d = 19$, $d = 23$}
]
\path [draw=color1, line width=0.7000000000000001pt] (axis cs:0.08,0.127789449112979)
--(axis cs:0.08,0.133718487394958);

\path [draw=color1, line width=0.7000000000000001pt] (axis cs:0.08166667,0.133403361344538)
--(axis cs:0.08166667,0.139437441643324);

\path [draw=color1, line width=0.7000000000000001pt] (axis cs:0.08333333,0.143872549019608)
--(axis cs:0.08333333,0.150105042016807);

\path [draw=color1, line width=0.7000000000000001pt] (axis cs:0.085,0.150816993464052)
--(axis cs:0.085,0.157166199813259);

\path [draw=color1, line width=0.7000000000000001pt] (axis cs:0.08666667,0.159663865546218)
--(axis cs:0.08666667,0.166164799253035);

\path [draw=color1, line width=0.7000000000000001pt] (axis cs:0.08833333,0.169094304388422)
--(axis cs:0.08833333,0.175735294117647);

\path [draw=color1, line width=0.7000000000000001pt] (axis cs:0.09,0.179446778711485)
--(axis cs:0.09,0.18625116713352);

\path [draw=color1, line width=0.7000000000000001pt] (axis cs:0.09166667,0.189075630252101)
--(axis cs:0.09166667,0.196020074696545);

\path [draw=color1, line width=0.7000000000000001pt] (axis cs:0.09333333,0.198506069094304)
--(axis cs:0.09333333,0.205567226890756);

\path [draw=color1, line width=0.7000000000000001pt] (axis cs:0.095,0.206769374416433)
--(axis cs:0.095,0.213947245564893);

\path [draw=color1, line width=0.7000000000000001pt] (axis cs:0.09666667,0.21921101774043)
--(axis cs:0.09666667,0.226528944911298);

\path [draw=color1, line width=0.7000000000000001pt] (axis cs:0.09833333,0.231886087768441)
--(axis cs:0.09833333,0.239344070961718);

\path [draw=color1, line width=0.7000000000000001pt] (axis cs:0.1,0.241970121381886)
--(axis cs:0.1,0.249544817927171);

\path [draw=color1, line width=0.7000000000000001pt] (axis cs:0.10166667,0.255018674136321)
--(axis cs:0.10166667,0.262733426704015);

\path [draw=color1, line width=0.7000000000000001pt] (axis cs:0.10333333,0.266176470588235)
--(axis cs:0.10333333,0.273984593837535);

\path [draw=color1, line width=0.7000000000000001pt] (axis cs:0.105,0.27483660130719)
--(axis cs:0.105,0.282726423902894);

\path [draw=color1, line width=0.7000000000000001pt] (axis cs:0.10666667,0.286834733893557)
--(axis cs:0.10666667,0.294829598506069);

\path [draw=color1, line width=0.7000000000000001pt] (axis cs:0.10833333,0.297794117647059)
--(axis cs:0.10833333,0.305870681605976);

\path [draw=color1, line width=0.7000000000000001pt] (axis cs:0.11,0.306034080298786)
--(axis cs:0.11,0.314169000933707);

\path [draw=color1, line width=0.7000000000000001pt] (axis cs:0.11166667,0.315884687208217)
--(axis cs:0.11166667,0.324089635854342);

\path [draw=color1, line width=0.7000000000000001pt] (axis cs:0.11333333,0.327859477124183)
--(axis cs:0.11333333,0.336146125116713);

\path [draw=color1, line width=0.7000000000000001pt] (axis cs:0.115,0.339484126984127)
--(axis cs:0.115,0.347840802987862);

\path [draw=color1, line width=0.7000000000000001pt] (axis cs:0.11666667,0.354108309990663)
--(axis cs:0.11666667,0.362546685340803);

\path [draw=color1, line width=0.7000000000000001pt] (axis cs:0.11833333,0.363620448179272)
--(axis cs:0.11833333,0.372105508870215);

\path [draw=color1, line width=0.7000000000000001pt] (axis cs:0.12,0.366795051353875)
--(axis cs:0.12,0.375291783380019);

\path [draw=color2, line width=0.7000000000000001pt] (axis cs:0.08,0.0897525676937442)
--(axis cs:0.08,0.0948412698412698);

\path [draw=color2, line width=0.7000000000000001pt] (axis cs:0.08166667,0.097140522875817)
--(axis cs:0.08166667,0.102415966386555);

\path [draw=color2, line width=0.7000000000000001pt] (axis cs:0.08333333,0.108461718020542)
--(axis cs:0.08333333,0.113993930905696);

\path [draw=color2, line width=0.7000000000000001pt] (axis cs:0.085,0.119152661064426)
--(axis cs:0.085,0.124906629318394);

\path [draw=color2, line width=0.7000000000000001pt] (axis cs:0.08666667,0.130345471521942)
--(axis cs:0.08666667,0.136332866479925);

\path [draw=color2, line width=0.7000000000000001pt] (axis cs:0.08833333,0.138387021475257)
--(axis cs:0.08833333,0.144514472455649);

\path [draw=color2, line width=0.7000000000000001pt] (axis cs:0.09,0.149054621848739)
--(axis cs:0.09,0.155368814192344);

\path [draw=color2, line width=0.7000000000000001pt] (axis cs:0.09166667,0.163818860877684)
--(axis cs:0.09166667,0.170389822595705);

\path [draw=color2, line width=0.7000000000000001pt] (axis cs:0.09333333,0.173330999066293)
--(axis cs:0.09333333,0.180042016806723);

\path [draw=color2, line width=0.7000000000000001pt] (axis cs:0.095,0.186122782446312)
--(axis cs:0.095,0.193020541549953);

\path [draw=color2, line width=0.7000000000000001pt] (axis cs:0.09666667,0.201272175536881)
--(axis cs:0.09666667,0.208380018674136);

\path [draw=color2, line width=0.7000000000000001pt] (axis cs:0.09833333,0.215184407096172)
--(axis cs:0.09833333,0.222455648926237);

\path [draw=color2, line width=0.7000000000000001pt] (axis cs:0.1,0.232551353874883)
--(axis cs:0.1,0.240032679738562);

\path [draw=color2, line width=0.7000000000000001pt] (axis cs:0.10166667,0.241433239962652)
--(axis cs:0.10166667,0.249007936507937);

\path [draw=color2, line width=0.7000000000000001pt] (axis cs:0.10333333,0.257422969187675)
--(axis cs:0.10333333,0.26516106442577);

\path [draw=color2, line width=0.7000000000000001pt] (axis cs:0.105,0.273517740429505)
--(axis cs:0.105,0.281395891690009);

\path [draw=color2, line width=0.7000000000000001pt] (axis cs:0.10666667,0.286858076563959)
--(axis cs:0.10666667,0.294852941176471);

\path [draw=color2, line width=0.7000000000000001pt] (axis cs:0.10833333,0.302275910364146)
--(axis cs:0.10833333,0.310387488328665);

\path [draw=color2, line width=0.7000000000000001pt] (axis cs:0.11,0.315067693744164)
--(axis cs:0.11,0.323272642390289);

\path [draw=color2, line width=0.7000000000000001pt] (axis cs:0.11166667,0.331349206349206)
--(axis cs:0.11166667,0.339659197012138);

\path [draw=color2, line width=0.7000000000000001pt] (axis cs:0.11333333,0.343265639589169)
--(axis cs:0.11333333,0.351645658263305);

\path [draw=color2, line width=0.7000000000000001pt] (axis cs:0.115,0.358508403361345)
--(axis cs:0.115,0.366970121381886);

\path [draw=color2, line width=0.7000000000000001pt] (axis cs:0.11666667,0.378653127917834)
--(axis cs:0.11666667,0.387208216619981);

\path [draw=color2, line width=0.7000000000000001pt] (axis cs:0.11833333,0.388725490196078)
--(axis cs:0.11833333,0.397315592903828);

\path [draw=color2, line width=0.7000000000000001pt] (axis cs:0.12,0.401505602240896)
--(axis cs:0.12,0.410142390289449);

\path [draw=color3, line width=0.7000000000000001pt] (axis cs:0.08,0.0640756302521008)
--(axis cs:0.08,0.068452380952381);

\path [draw=color3, line width=0.7000000000000001pt] (axis cs:0.08166667,0.072000466853408)
--(axis cs:0.08166667,0.0766223155929038);

\path [draw=color3, line width=0.7000000000000001pt] (axis cs:0.08333333,0.0813725490196078)
--(axis cs:0.08333333,0.0862511671335201);

\path [draw=color3, line width=0.7000000000000001pt] (axis cs:0.085,0.0922852474323063)
--(axis cs:0.085,0.0974439775910364);

\path [draw=color3, line width=0.7000000000000001pt] (axis cs:0.08666667,0.102310924369748)
--(axis cs:0.08666667,0.107703081232493);

\path [draw=color3, line width=0.7000000000000001pt] (axis cs:0.08833333,0.115172735760971)
--(axis cs:0.08833333,0.120856676003735);

\path [draw=color3, line width=0.7000000000000001pt] (axis cs:0.09,0.127369281045752)
--(axis cs:0.09,0.133298319327731);

\path [draw=color3, line width=0.7000000000000001pt] (axis cs:0.09166667,0.140686274509804)
--(axis cs:0.09166667,0.146860410830999);

\path [draw=color3, line width=0.7000000000000001pt] (axis cs:0.09333333,0.154400093370682)
--(axis cs:0.09333333,0.160807656395892);

\path [draw=color3, line width=0.7000000000000001pt] (axis cs:0.095,0.169876283846872)
--(axis cs:0.095,0.176540616246499);

\path [draw=color3, line width=0.7000000000000001pt] (axis cs:0.09666667,0.185866013071895)
--(axis cs:0.09666667,0.192763772175537);

\path [draw=color3, line width=0.7000000000000001pt] (axis cs:0.09833333,0.202941176470588)
--(axis cs:0.09833333,0.210072362278245);

\path [draw=color3, line width=0.7000000000000001pt] (axis cs:0.1,0.221043417366947)
--(axis cs:0.1,0.228384687208217);

\path [draw=color3, line width=0.7000000000000001pt] (axis cs:0.10166667,0.236169467787115)
--(axis cs:0.10166667,0.243674136321195);

\path [draw=color3, line width=0.7000000000000001pt] (axis cs:0.10333333,0.254855275443511)
--(axis cs:0.10333333,0.262570028011204);

\path [draw=color3, line width=0.7000000000000001pt] (axis cs:0.105,0.27406629318394)
--(axis cs:0.105,0.281944444444444);

\path [draw=color3, line width=0.7000000000000001pt] (axis cs:0.10666667,0.286963118580766)
--(axis cs:0.10666667,0.294957983193277);

\path [draw=color3, line width=0.7000000000000001pt] (axis cs:0.10833333,0.304260037348273)
--(axis cs:0.10833333,0.312394957983193);

\path [draw=color3, line width=0.7000000000000001pt] (axis cs:0.11,0.324474789915966)
--(axis cs:0.11,0.332738095238095);

\path [draw=color3, line width=0.7000000000000001pt] (axis cs:0.11166667,0.342588702147526)
--(axis cs:0.11166667,0.350968720821662);

\path [draw=color3, line width=0.7000000000000001pt] (axis cs:0.11333333,0.360294117647059)
--(axis cs:0.11333333,0.368767507002801);

\path [draw=color3, line width=0.7000000000000001pt] (axis cs:0.115,0.377065826330532)
--(axis cs:0.115,0.385609243697479);

\path [draw=color3, line width=0.7000000000000001pt] (axis cs:0.11666667,0.397233893557423)
--(axis cs:0.11666667,0.405859010270775);

\path [draw=color3, line width=0.7000000000000001pt] (axis cs:0.11833333,0.412546685340803)
--(axis cs:0.11833333,0.421230158730159);

\path [draw=color3, line width=0.7000000000000001pt] (axis cs:0.12,0.43109243697479)
--(axis cs:0.12,0.439822595704949);

\path [draw=color4, line width=0.7000000000000001pt] (axis cs:0.08,0.0442110177404295)
--(axis cs:0.08,0.0478991596638655);

\path [draw=color4, line width=0.7000000000000001pt] (axis cs:0.08166667,0.0534080298786181)
--(axis cs:0.08166667,0.0574346405228758);

\path [draw=color4, line width=0.7000000000000001pt] (axis cs:0.08333333,0.0631535947712418)
--(axis cs:0.08333333,0.0674953314659197);

\path [draw=color4, line width=0.7000000000000001pt] (axis cs:0.085,0.0737161531279178)
--(axis cs:0.085,0.0783846872082166);

\path [draw=color4, line width=0.7000000000000001pt] (axis cs:0.08666667,0.082843137254902)
--(axis cs:0.08666667,0.0877567693744164);

\path [draw=color4, line width=0.7000000000000001pt] (axis cs:0.08833333,0.0957049486461251)
--(axis cs:0.08833333,0.100945378151261);

\path [draw=color4, line width=0.7000000000000001pt] (axis cs:0.09,0.108543417366947)
--(axis cs:0.09,0.114075630252101);

\path [draw=color4, line width=0.7000000000000001pt] (axis cs:0.09166667,0.123144257703081)
--(axis cs:0.09166667,0.128991596638655);

\path [draw=color4, line width=0.7000000000000001pt] (axis cs:0.09333333,0.139192343604108)
--(axis cs:0.09333333,0.145343137254902);

\path [draw=color4, line width=0.7000000000000001pt] (axis cs:0.095,0.153699813258637)
--(axis cs:0.095,0.160107376283847);

\path [draw=color4, line width=0.7000000000000001pt] (axis cs:0.09666667,0.170996732026144)
--(axis cs:0.09666667,0.177672735760971);

\path [draw=color4, line width=0.7000000000000001pt] (axis cs:0.09833333,0.190021008403361)
--(axis cs:0.09833333,0.196965452847806);

\path [draw=color4, line width=0.7000000000000001pt] (axis cs:0.1,0.209348739495798)
--(axis cs:0.1,0.216549953314659);

\path [draw=color4, line width=0.7000000000000001pt] (axis cs:0.10166667,0.225093370681606)
--(axis cs:0.10166667,0.232481325863679);

\path [draw=color4, line width=0.7000000000000001pt] (axis cs:0.10333333,0.245483193277311)
--(axis cs:0.10333333,0.253104575163399);

\path [draw=color4, line width=0.7000000000000001pt] (axis cs:0.105,0.269176003734827)
--(axis cs:0.105,0.277019140989729);

\path [draw=color4, line width=0.7000000000000001pt] (axis cs:0.10666667,0.290207749766573)
--(axis cs:0.10666667,0.298225957049486);

\path [draw=color4, line width=0.7000000000000001pt] (axis cs:0.10833333,0.310259103641457)
--(axis cs:0.10833333,0.318429038281979);

\path [draw=color4, line width=0.7000000000000001pt] (axis cs:0.11,0.332586367880486)
--(axis cs:0.11,0.340908029878618);

\path [draw=color4, line width=0.7000000000000001pt] (axis cs:0.11166667,0.353408029878618)
--(axis cs:0.11166667,0.361846405228758);

\path [draw=color4, line width=0.7000000000000001pt] (axis cs:0.11333333,0.375198412698413)
--(axis cs:0.11333333,0.38374183006536);

\path [draw=color4, line width=0.7000000000000001pt] (axis cs:0.115,0.391771708683473)
--(axis cs:0.115,0.400385154061625);

\path [draw=color4, line width=0.7000000000000001pt] (axis cs:0.11666667,0.413795518207283)
--(axis cs:0.11666667,0.422478991596639);

\path [draw=color4, line width=0.7000000000000001pt] (axis cs:0.11833333,0.436788048552754)
--(axis cs:0.11833333,0.445518207282913);

\path [draw=color4, line width=0.7000000000000001pt] (axis cs:0.12,0.455999066293184)
--(axis cs:0.12,0.464775910364146);

\path [draw=color5, line width=0.7000000000000001pt] (axis cs:0.08,0.0320728291316527)
--(axis cs:0.08,0.0352474323062558);

\path [draw=color5, line width=0.7000000000000001pt] (axis cs:0.08166667,0.0382586367880486)
--(axis cs:0.08166667,0.0417133520074697);

\path [draw=color5, line width=0.7000000000000001pt] (axis cs:0.08333333,0.0464635854341737)
--(axis cs:0.08333333,0.0502450980392157);

\path [draw=color5, line width=0.7000000000000001pt] (axis cs:0.085,0.0574229691876751)
--(axis cs:0.085,0.0615896358543417);

\path [draw=color5, line width=0.7000000000000001pt] (axis cs:0.08666667,0.0658380018674136)
--(axis cs:0.08666667,0.0702731092436975);

\path [draw=color5, line width=0.7000000000000001pt] (axis cs:0.08833333,0.0803454715219421)
--(axis cs:0.08833333,0.0852007469654529);

\path [draw=color5, line width=0.7000000000000001pt] (axis cs:0.09,0.0921101774042951)
--(axis cs:0.09,0.0972572362278245);

\path [draw=color5, line width=0.7000000000000001pt] (axis cs:0.09166667,0.105322128851541)
--(axis cs:0.09166667,0.11078431372549);

\path [draw=color5, line width=0.7000000000000001pt] (axis cs:0.09333333,0.122408963585434)
--(axis cs:0.09333333,0.128232959850607);

\path [draw=color5, line width=0.7000000000000001pt] (axis cs:0.095,0.140289449112979)
--(axis cs:0.095,0.146463585434174);

\path [draw=color5, line width=0.7000000000000001pt] (axis cs:0.09666667,0.158461718020542)
--(axis cs:0.09666667,0.164939309056956);

\path [draw=color5, line width=0.7000000000000001pt] (axis cs:0.09833333,0.180333800186741)
--(axis cs:0.09833333,0.187149859943978);

\path [draw=color5, line width=0.7000000000000001pt] (axis cs:0.1,0.199964985994398)
--(axis cs:0.1,0.207049486461251);

\path [draw=color5, line width=0.7000000000000001pt] (axis cs:0.10166667,0.222175536881419)
--(axis cs:0.10166667,0.229540149393091);

\path [draw=color5, line width=0.7000000000000001pt] (axis cs:0.10333333,0.247677404295051)
--(axis cs:0.10333333,0.255322128851541);

\path [draw=color5, line width=0.7000000000000001pt] (axis cs:0.105,0.26844070961718)
--(axis cs:0.105,0.276272175536881);

\path [draw=color5, line width=0.7000000000000001pt] (axis cs:0.10666667,0.290896358543417)
--(axis cs:0.10666667,0.298914565826331);

\path [draw=color5, line width=0.7000000000000001pt] (axis cs:0.10833333,0.315557889822596)
--(axis cs:0.10833333,0.323762838468721);

\path [draw=color5, line width=0.7000000000000001pt] (axis cs:0.11,0.338492063492064)
--(axis cs:0.11,0.346848739495798);

\path [draw=color5, line width=0.7000000000000001pt] (axis cs:0.11166667,0.366514939309057)
--(axis cs:0.11166667,0.375011671335201);

\path [draw=color5, line width=0.7000000000000001pt] (axis cs:0.11333333,0.384967320261438)
--(axis cs:0.11333333,0.393545751633987);

\path [draw=color5, line width=0.7000000000000001pt] (axis cs:0.115,0.408099906629318)
--(axis cs:0.115,0.416760037348273);

\path [draw=color5, line width=0.7000000000000001pt] (axis cs:0.11666667,0.433730158730159)
--(axis cs:0.11666667,0.442460317460317);

\path [draw=color5, line width=0.7000000000000001pt] (axis cs:0.11833333,0.459920634920635)
--(axis cs:0.11833333,0.468697478991597);

\path [draw=color5, line width=0.7000000000000001pt] (axis cs:0.12,0.478396358543417)
--(axis cs:0.12,0.487196545284781);

\addplot [line width=0.7000000000000001pt, color1]
table {%
0.100601371216409 0.251336599799289
0.100719216569944 0.252070583755788
0.100837061923479 0.252804803771252
0.100954907277015 0.25353925984568
0.10107275263055 0.254273951979073
0.101190597984085 0.25500888017143
0.101308443337621 0.255744044422752
0.101426288691156 0.256479444733038
0.101544134044692 0.257215081102289
0.101661979398227 0.257950953530505
0.101779824751762 0.258687062017685
0.101897670105298 0.25942340656383
0.102015515458833 0.260159987168939
0.102133360812368 0.260896803833013
0.102251206165904 0.261633856556051
0.102369051519439 0.262371145338054
0.102486896872974 0.263108670179022
0.10260474222651 0.263846431078954
0.102722587580045 0.264584428037851
0.10284043293358 0.265322661055712
0.102958278287116 0.266061130132538
0.103076123640651 0.266799835268329
0.103193968994186 0.267538776463084
0.103311814347722 0.268277953716803
0.103429659701257 0.269017367029487
0.103547505054793 0.269757016401136
0.103665350408328 0.270496901831749
0.103783195761863 0.271237023321327
0.103901041115399 0.27197738086987
0.104018886468934 0.272717974477377
0.104136731822469 0.273458804143848
0.104254577176005 0.274199869869284
0.10437242252954 0.274941171653685
0.104490267883075 0.27568270949705
0.104608113236611 0.27642448339938
0.104725958590146 0.277166493360675
0.104843803943681 0.277908739380934
0.104961649297217 0.278651221460157
0.105079494650752 0.279393939598346
0.105197340004287 0.280136893795498
0.105315185357823 0.280880084051616
0.105433030711358 0.281623510366698
0.105550876064894 0.282367172740744
0.105668721418429 0.283111071173755
0.105786566771964 0.283855205665731
0.1059044121255 0.284599576216671
0.106022257479035 0.285344182826576
0.10614010283257 0.286089025495445
0.106257948186106 0.286834104223279
0.106375793539641 0.287579419010077
0.106493638893176 0.28832496985584
0.106611484246712 0.289070756760568
0.106729329600247 0.28981677972426
0.106847174953782 0.290563038746917
0.106965020307318 0.291309533828538
0.107082865660853 0.292056264969124
0.107200711014388 0.292803232168675
0.107318556367924 0.29355043542719
0.107436401721459 0.294297874744669
0.107554247074995 0.295045550121113
0.10767209242853 0.295793461556522
0.107789937782065 0.296541609050895
0.107907783135601 0.297289992604233
0.108025628489136 0.298038612216536
0.108143473842671 0.298787467887803
0.108261319196207 0.299536559618035
0.108379164549742 0.300285887407231
0.108497009903277 0.301035451255391
0.108614855256813 0.301785251162517
0.108732700610348 0.302535287128607
0.108850545963883 0.303285559153661
0.108968391317419 0.30403606723768
0.109086236670954 0.304786811380664
0.109204082024489 0.305537791582612
0.109321927378025 0.306289007843525
0.10943977273156 0.307040460163402
0.109557618085096 0.307792148542244
0.109675463438631 0.308544072980051
0.109793308792166 0.309296233476822
0.109911154145702 0.310048630032557
0.110028999499237 0.310801262647257
0.110146844852772 0.311554131320922
0.110264690206308 0.312307236053552
0.110382535559843 0.313060576845145
0.110500380913378 0.313814153695704
0.110618226266914 0.314567966605227
0.110736071620449 0.315322015573715
0.110853916973984 0.316076300601167
0.11097176232752 0.316830821687584
0.111089607681055 0.317585578832965
0.11120745303459 0.318340572037311
0.111325298388126 0.319095801300621
0.111443143741661 0.319851266622896
0.111560989095197 0.320606968004136
0.111678834448732 0.32136290544434
0.111796679802267 0.322119078943509
0.111914525155803 0.322875488501642
0.112032370509338 0.32363213411874
0.112150215862873 0.324389015794803
0.112268061216409 0.32514613352983
};
\addplot [line width=0.7000000000000001pt, color2]
table {%
0.100601371216409 0.23926111607219
0.100719216569944 0.240258132524374
0.100837061923479 0.241255589147918
0.100954907277015 0.24225348594282
0.10107275263055 0.243251822909082
0.101190597984085 0.244250600046702
0.101308443337621 0.245249817355682
0.101426288691156 0.246249474836021
0.101544134044692 0.247249572487718
0.101661979398227 0.248250110310775
0.101779824751762 0.24925108830519
0.101897670105298 0.250252506470965
0.102015515458833 0.251254364808099
0.102133360812368 0.252256663316592
0.102251206165904 0.253259401996443
0.102369051519439 0.254262580847654
0.102486896872974 0.255266199870224
0.10260474222651 0.256270259064153
0.102722587580045 0.257274758429441
0.10284043293358 0.258279697966087
0.102958278287116 0.259285077674093
0.103076123640651 0.260290897553458
0.103193968994186 0.261297157604182
0.103311814347722 0.262303857826265
0.103429659701257 0.263310998219707
0.103547505054793 0.264318578784508
0.103665350408328 0.265326599520668
0.103783195761863 0.266335060428187
0.103901041115399 0.267343961507065
0.104018886468934 0.268353302757303
0.104136731822469 0.269363084178899
0.104254577176005 0.270373305771854
0.10437242252954 0.271383967536168
0.104490267883075 0.272395069471841
0.104608113236611 0.273406611578874
0.104725958590146 0.274418593857265
0.104843803943681 0.275431016307015
0.104961649297217 0.276443878928124
0.105079494650752 0.277457181720593
0.105197340004287 0.27847092468442
0.105315185357823 0.279485107819606
0.105433030711358 0.280499731126152
0.105550876064894 0.281514794604056
0.105668721418429 0.28253029825332
0.105786566771964 0.283546242073942
0.1059044121255 0.284562626065924
0.106022257479035 0.285579450229264
0.10614010283257 0.286596714563964
0.106257948186106 0.287614419070022
0.106375793539641 0.28863256374744
0.106493638893176 0.289651148596217
0.106611484246712 0.290670173616352
0.106729329600247 0.291689638807847
0.106847174953782 0.292709544170701
0.106965020307318 0.293729889704914
0.107082865660853 0.294750675410485
0.107200711014388 0.295771901287416
0.107318556367924 0.296793567335706
0.107436401721459 0.297815673555355
0.107554247074995 0.298838219946363
0.10767209242853 0.299861206508729
0.107789937782065 0.300884633242455
0.107907783135601 0.30190850014754
0.108025628489136 0.302932807223984
0.108143473842671 0.303957554471787
0.108261319196207 0.304982741890949
0.108379164549742 0.30600836948147
0.108497009903277 0.30703443724335
0.108614855256813 0.308060945176589
0.108732700610348 0.309087893281187
0.108850545963883 0.310115281557145
0.108968391317419 0.311143110004461
0.109086236670954 0.312171378623136
0.109204082024489 0.31320008741317
0.109321927378025 0.314229236374563
0.10943977273156 0.315258825507315
0.109557618085096 0.316288854811427
0.109675463438631 0.317319324286897
0.109793308792166 0.318350233933726
0.109911154145702 0.319381583751915
0.110028999499237 0.320413373741462
0.110146844852772 0.321445603902369
0.110264690206308 0.322478274234634
0.110382535559843 0.323511384738258
0.110500380913378 0.324544935413242
0.110618226266914 0.325578926259584
0.110736071620449 0.326613357277286
0.110853916973984 0.327648228466346
0.11097176232752 0.328683539826766
0.111089607681055 0.329719291358544
0.11120745303459 0.330755483061682
0.111325298388126 0.331792114936179
0.111443143741661 0.332829186982034
0.111560989095197 0.333866699199249
0.111678834448732 0.334904651587823
0.111796679802267 0.335943044147756
0.111914525155803 0.336981876879047
0.112032370509338 0.338021149781698
0.112150215862873 0.339060862855708
0.112268061216409 0.340101016101077
};
\addplot [line width=0.7000000000000001pt, color3]
table {%
0.100601371216409 0.228562848898687
0.100719216569944 0.229791707045426
0.100837061923479 0.231021240201873
0.100954907277015 0.232251448368028
0.10107275263055 0.233482331543892
0.101190597984085 0.234713889729463
0.101308443337621 0.235946122924743
0.101426288691156 0.237179031129731
0.101544134044692 0.238412614344428
0.101661979398227 0.239646872568832
0.101779824751762 0.240881805802945
0.101897670105298 0.242117414046765
0.102015515458833 0.243353697300294
0.102133360812368 0.244590655563531
0.102251206165904 0.245828288836477
0.102369051519439 0.24706659711913
0.102486896872974 0.248305580411492
0.10260474222651 0.249545238713562
0.102722587580045 0.25078557202534
0.10284043293358 0.252026580346826
0.102958278287116 0.25326826367802
0.103076123640651 0.254510622018923
0.103193968994186 0.255753655369533
0.103311814347722 0.256997363729852
0.103429659701257 0.258241747099879
0.103547505054793 0.259486805479615
0.103665350408328 0.260732538869058
0.103783195761863 0.26197894726821
0.103901041115399 0.263226030677069
0.104018886468934 0.264473789095637
0.104136731822469 0.265722222523913
0.104254577176005 0.266971330961898
0.10437242252954 0.26822111440959
0.104490267883075 0.269471572866991
0.104608113236611 0.2707227063341
0.104725958590146 0.271974514810916
0.104843803943681 0.273226998297442
0.104961649297217 0.274480156793675
0.105079494650752 0.275733990299617
0.105197340004287 0.276988498815266
0.105315185357823 0.278243682340624
0.105433030711358 0.27949954087569
0.105550876064894 0.280756074420464
0.105668721418429 0.282013282974947
0.105786566771964 0.283271166539137
0.1059044121255 0.284529725113036
0.106022257479035 0.285788958696643
0.10614010283257 0.287048867289958
0.106257948186106 0.288309450892981
0.106375793539641 0.289570709505713
0.106493638893176 0.290832643128153
0.106611484246712 0.2920952517603
0.106729329600247 0.293358535402156
0.106847174953782 0.29462249405372
0.106965020307318 0.295887127714993
0.107082865660853 0.297152436385973
0.107200711014388 0.298418420066662
0.107318556367924 0.299685078757059
0.107436401721459 0.300952412457164
0.107554247074995 0.302220421166977
0.10767209242853 0.303489104886499
0.107789937782065 0.304758463615728
0.107907783135601 0.306028497354666
0.108025628489136 0.307299206103312
0.108143473842671 0.308570589861666
0.108261319196207 0.309842648629728
0.108379164549742 0.311115382407499
0.108497009903277 0.312388791194977
0.108614855256813 0.313662874992164
0.108732700610348 0.314937633799059
0.108850545963883 0.316213067615662
0.108968391317419 0.317489176441974
0.109086236670954 0.318765960277993
0.109204082024489 0.320043419123721
0.109321927378025 0.321321552979157
0.10943977273156 0.322600361844301
0.109557618085096 0.323879845719153
0.109675463438631 0.325160004603713
0.109793308792166 0.326440838497982
0.109911154145702 0.327722347401959
0.110028999499237 0.329004531315644
0.110146844852772 0.330287390239037
0.110264690206308 0.331570924172138
0.110382535559843 0.332855133114948
0.110500380913378 0.334140017067465
0.110618226266914 0.335425576029691
0.110736071620449 0.336711810001625
0.110853916973984 0.337998718983267
0.11097176232752 0.339286302974618
0.111089607681055 0.340574561975676
0.11120745303459 0.341863495986443
0.111325298388126 0.343153105006918
0.111443143741661 0.344443389037101
0.111560989095197 0.345734348076992
0.111678834448732 0.347025982126591
0.111796679802267 0.348318291185899
0.111914525155803 0.349611275254914
0.112032370509338 0.350904934333639
0.112150215862873 0.35219926842207
0.112268061216409 0.353494277520211
};
\addplot [line width=0.7000000000000001pt, color4]
table {%
0.100601371216409 0.218771801307684
0.100719216569944 0.220211845209044
0.100837061923479 0.221652824160381
0.100954907277015 0.223094738161696
0.10107275263055 0.224537587212989
0.101190597984085 0.225981371314259
0.101308443337621 0.227426090465507
0.101426288691156 0.228871744666732
0.101544134044692 0.230318333917935
0.101661979398227 0.231765858219116
0.101779824751762 0.233214317570274
0.101897670105298 0.23466371197141
0.102015515458833 0.236114041422523
0.102133360812368 0.237565305923614
0.102251206165904 0.239017505474683
0.102369051519439 0.240470640075729
0.102486896872974 0.241924709726753
0.10260474222651 0.243379714427755
0.102722587580045 0.244835654178734
0.10284043293358 0.24629252897969
0.102958278287116 0.247750338830625
0.103076123640651 0.249209083731537
0.103193968994186 0.250668763682426
0.103311814347722 0.252129378683294
0.103429659701257 0.253590928734138
0.103547505054793 0.255053413834961
0.103665350408328 0.256516833985761
0.103783195761863 0.257981189186539
0.103901041115399 0.259446479437294
0.104018886468934 0.260912704738027
0.104136731822469 0.262379865088737
0.104254577176005 0.263847960489425
0.10437242252954 0.265316990940091
0.104490267883075 0.266786956440734
0.104608113236611 0.268257856991356
0.104725958590146 0.269729692591954
0.104843803943681 0.27120246324253
0.104961649297217 0.272676168943084
0.105079494650752 0.274150809693615
0.105197340004287 0.275626385494125
0.105315185357823 0.277102896344611
0.105433030711358 0.278580342245076
0.105550876064894 0.280058723195518
0.105668721418429 0.281538039195937
0.105786566771964 0.283018290246334
0.1059044121255 0.284499476346709
0.106022257479035 0.285981597497061
0.10614010283257 0.287464653697391
0.106257948186106 0.288948644947699
0.106375793539641 0.290433571247984
0.106493638893176 0.291919432598247
0.106611484246712 0.293406228998487
0.106729329600247 0.294893960448705
0.106847174953782 0.296382626948901
0.106965020307318 0.297872228499074
0.107082865660853 0.299362765099225
0.107200711014388 0.300854236749354
0.107318556367924 0.30234664344946
0.107436401721459 0.303839985199543
0.107554247074995 0.305334261999605
0.10767209242853 0.306829473849644
0.107789937782065 0.30832562074966
0.107907783135601 0.309822702699654
0.108025628489136 0.311320719699626
0.108143473842671 0.312819671749576
0.108261319196207 0.314319558849503
0.108379164549742 0.315820380999408
0.108497009903277 0.31732213819929
0.108614855256813 0.31882483044915
0.108732700610348 0.320328457748987
0.108850545963883 0.321833020098802
0.108968391317419 0.323338517498595
0.109086236670954 0.324844949948365
0.109204082024489 0.326352317448113
0.109321927378025 0.327860619997839
0.10943977273156 0.329369857597542
0.109557618085096 0.330880030247223
0.109675463438631 0.332391137946881
0.109793308792166 0.333903180696517
0.109911154145702 0.335416158496131
0.110028999499237 0.336930071345722
0.110146844852772 0.338444919245291
0.110264690206308 0.339960702194838
0.110382535559843 0.341477420194362
0.110500380913378 0.342995073243863
0.110618226266914 0.344513661343343
0.110736071620449 0.3460331844928
0.110853916973984 0.347553642692234
0.11097176232752 0.349075035941647
0.111089607681055 0.350597364241036
0.11120745303459 0.352120627590404
0.111325298388126 0.353644825989749
0.111443143741661 0.355169959439071
0.111560989095197 0.356696027938372
0.111678834448732 0.35822303148765
0.111796679802267 0.359750970086905
0.111914525155803 0.361279843736138
0.112032370509338 0.362809652435349
0.112150215862873 0.364340396184537
0.112268061216409 0.365872074983703
};
\addplot [line width=0.7000000000000001pt, color5]
table {%
0.100601371216409 0.209644996267114
0.100719216569944 0.211281030327754
0.100837061923479 0.212918281186877
0.100954907277015 0.214556748844484
0.10107275263055 0.216196433300575
0.101190597984085 0.21783733455515
0.101308443337621 0.219479452608209
0.101426288691156 0.221122787459751
0.101544134044692 0.222767339109777
0.101661979398227 0.224413107558286
0.101779824751762 0.22606009280528
0.101897670105298 0.227708294850757
0.102015515458833 0.229357713694718
0.102133360812368 0.231008349337163
0.102251206165904 0.232660201778091
0.102369051519439 0.234313271017503
0.102486896872974 0.235967557055399
0.10260474222651 0.237623059891778
0.102722587580045 0.239279779526642
0.10284043293358 0.240937715959989
0.102958278287116 0.24259686919182
0.103076123640651 0.244257239222134
0.103193968994186 0.245918826050933
0.103311814347722 0.247581629678215
0.103429659701257 0.24924565010398
0.103547505054793 0.25091088732823
0.103665350408328 0.252577341350963
0.103783195761863 0.25424501217218
0.103901041115399 0.255913899791881
0.104018886468934 0.257584004210065
0.104136731822469 0.259255325426734
0.104254577176005 0.260927863441886
0.10437242252954 0.262601618255521
0.104490267883075 0.264276589867641
0.104608113236611 0.265952778278244
0.104725958590146 0.267630183487331
0.104843803943681 0.269308805494901
0.104961649297217 0.270988644300956
0.105079494650752 0.272669699905494
0.105197340004287 0.274351972308516
0.105315185357823 0.276035461510021
0.105433030711358 0.277720167510011
0.105550876064894 0.279406090308484
0.105668721418429 0.281093229905441
0.105786566771964 0.282781586300881
0.1059044121255 0.284471159494805
0.106022257479035 0.286161949487214
0.10614010283257 0.287853956278105
0.106257948186106 0.289547179867481
0.106375793539641 0.29124162025534
0.106493638893176 0.292937277441683
0.106611484246712 0.29463415142651
0.106729329600247 0.29633224220982
0.106847174953782 0.298031549791614
0.106965020307318 0.299732074171892
0.107082865660853 0.301433815350654
0.107200711014388 0.3031367733279
0.107318556367924 0.304840948103629
0.107436401721459 0.306546339677842
0.107554247074995 0.308252948050538
0.10767209242853 0.309960773221719
0.107789937782065 0.311669815191383
0.107907783135601 0.31338007395953
0.108025628489136 0.315091549526162
0.108143473842671 0.316804241891278
0.108261319196207 0.318518151054877
0.108379164549742 0.320233277016959
0.108497009903277 0.321949619777526
0.108614855256813 0.323667179336576
0.108732700610348 0.32538595569411
0.108850545963883 0.327105948850128
0.108968391317419 0.32882715880463
0.109086236670954 0.330549585557615
0.109204082024489 0.332273229109084
0.109321927378025 0.333998089459036
0.10943977273156 0.335724166607473
0.109557618085096 0.337451460554393
0.109675463438631 0.339179971299797
0.109793308792166 0.340909698843685
0.109911154145702 0.342640643186056
0.110028999499237 0.344372804326911
0.110146844852772 0.34610618226625
0.110264690206308 0.347840777004073
0.110382535559843 0.349576588540379
0.110500380913378 0.351313616875169
0.110618226266914 0.353051862008443
0.110736071620449 0.354791323940201
0.110853916973984 0.356532002670442
0.11097176232752 0.358273898199167
0.111089607681055 0.360017010526376
0.11120745303459 0.361761339652069
0.111325298388126 0.363506885576245
0.111443143741661 0.365253648298905
0.111560989095197 0.367001627820049
0.111678834448732 0.368750824139676
0.111796679802267 0.370501237257787
0.111914525155803 0.372252867174382
0.112032370509338 0.374005713889461
0.112150215862873 0.375759777403023
0.112268061216409 0.37751505771507
};
\path [draw=white, fill opacity=0] (axis cs:0,0.00931664332399626)
--(axis cs:0,0.509952731092437);

\path [draw=white, fill opacity=0] (axis cs:1,0.00931664332399626)
--(axis cs:1,0.509952731092437);

\path [draw=white, fill opacity=0] (axis cs:0.078,0)
--(axis cs:0.122,0);

\path [draw=white, fill opacity=0] (axis cs:0.078,1)
--(axis cs:0.122,1);

\end{axis}

\end{tikzpicture}

%% file: paper_planar_dep_no_mp.tex
\begin{tikzpicture}

\definecolor{color4}{rgb}{0.219799566082832,0.662515787685034,0.773209315931721}
\definecolor{color2}{rgb}{0.680418912779335,0.615149751467757,0.194054521114453}
\definecolor{color5}{rgb}{0.800493618642396,0.477033635337372,0.957954719600752}
\definecolor{color3}{rgb}{0.201253172212011,0.690792081537903,0.479667611892753}
\definecolor{color0}{rgb}{0.917647058823529,0.917647058823529,0.949019607843137}
\definecolor{color1}{rgb}{0.967797559291991,0.441274560091574,0.53581031550587}

\begin{axis}[
title={Depolarizing Error Model, without BP/multi-path summation},
xlabel={physical error probability},
ylabel={logical error probability},
xmin=0.133, xmax=0.177,
ymin=0.093194, ymax=0.497026,
width=\figurewidth,
height=\figureheight,
xtick={0.13,0.135,0.14,0.145,0.15,0.155,0.16,0.165,0.17,0.175,0.18},
xticklabels={,0.135,0.140,0.145,0.150,0.155,0.160,0.165,0.170,0.175,},
ytick={0.05,0.1,0.15,0.2,0.25,0.3,0.35,0.4,0.45,0.5},
yticklabels={,0.10,0.15,0.20,0.25,0.30,0.35,0.40,0.45,},
tick align=outside,
tick pos=left,
xmajorgrids,
x grid style={white},
ymajorgrids,
y grid style={white},
axis line style={white},
axis background/.style={fill=color0},
legend pos=north west,
legend entries={$d = 7$, $d = 11$, $d = 15$, $d = 19$, $d = 23$}
]
\path [draw=color1, line width=0.7000000000000001pt] (axis cs:0.135,0.19191)
--(axis cs:0.135,0.19836);

\path [draw=color1, line width=0.7000000000000001pt] (axis cs:0.13666667,0.19727)
--(axis cs:0.13666667,0.2038);

\path [draw=color1, line width=0.7000000000000001pt] (axis cs:0.13833333,0.20199)
--(axis cs:0.13833333,0.20857);

\path [draw=color1, line width=0.7000000000000001pt] (axis cs:0.14,0.20964)
--(axis cs:0.14,0.21631);

\path [draw=color1, line width=0.7000000000000001pt] (axis cs:0.14166667,0.21643)
--(axis cs:0.14166667,0.22318);

\path [draw=color1, line width=0.7000000000000001pt] (axis cs:0.14333333,0.22407)
--(axis cs:0.14333333,0.2309);

\path [draw=color1, line width=0.7000000000000001pt] (axis cs:0.145,0.23082)
--(axis cs:0.145,0.23771);

\path [draw=color1, line width=0.7000000000000001pt] (axis cs:0.14666667,0.23817)
--(axis cs:0.14666667,0.24514);

\path [draw=color1, line width=0.7000000000000001pt] (axis cs:0.14833333,0.24568)
--(axis cs:0.14833333,0.25273);

\path [draw=color1, line width=0.7000000000000001pt] (axis cs:0.15,0.25371)
--(axis cs:0.15,0.26084);

\path [draw=color1, line width=0.7000000000000001pt] (axis cs:0.15166667,0.25887)
--(axis cs:0.15166667,0.26604);

\path [draw=color1, line width=0.7000000000000001pt] (axis cs:0.15333333,0.27004)
--(axis cs:0.15333333,0.27731);

\path [draw=color1, line width=0.7000000000000001pt] (axis cs:0.155,0.27329)
--(axis cs:0.155,0.28058);

\path [draw=color1, line width=0.7000000000000001pt] (axis cs:0.15666667,0.27951)
--(axis cs:0.15666667,0.28685);

\path [draw=color1, line width=0.7000000000000001pt] (axis cs:0.15833333,0.28895)
--(axis cs:0.15833333,0.29636);

\path [draw=color1, line width=0.7000000000000001pt] (axis cs:0.16,0.29417)
--(axis cs:0.16,0.30162);

\path [draw=color1, line width=0.7000000000000001pt] (axis cs:0.16166667,0.30345)
--(axis cs:0.16166667,0.31096);

\path [draw=color1, line width=0.7000000000000001pt] (axis cs:0.16333333,0.31015)
--(axis cs:0.16333333,0.31771);

\path [draw=color1, line width=0.7000000000000001pt] (axis cs:0.165,0.31609)
--(axis cs:0.165,0.32369);

\path [draw=color1, line width=0.7000000000000001pt] (axis cs:0.16666667,0.32273)
--(axis cs:0.16666667,0.33037);

\path [draw=color1, line width=0.7000000000000001pt] (axis cs:0.16833333,0.3282)
--(axis cs:0.16833333,0.33587);

\path [draw=color1, line width=0.7000000000000001pt] (axis cs:0.17,0.33978)
--(axis cs:0.17,0.34752);

\path [draw=color1, line width=0.7000000000000001pt] (axis cs:0.17166667,0.3459)
--(axis cs:0.17166667,0.35367);

\path [draw=color1, line width=0.7000000000000001pt] (axis cs:0.17333333,0.35094)
--(axis cs:0.17333333,0.35873);

\path [draw=color1, line width=0.7000000000000001pt] (axis cs:0.175,0.35696)
--(axis cs:0.175,0.36478);

\path [draw=color2, line width=0.7000000000000001pt] (axis cs:0.135,0.16601)
--(axis cs:0.135,0.17212);

\path [draw=color2, line width=0.7000000000000001pt] (axis cs:0.13666667,0.17361)
--(axis cs:0.13666667,0.17982);

\path [draw=color2, line width=0.7000000000000001pt] (axis cs:0.13833333,0.18394)
--(axis cs:0.13833333,0.19029);

\path [draw=color2, line width=0.7000000000000001pt] (axis cs:0.14,0.19102)
--(axis cs:0.14,0.19747);

\path [draw=color2, line width=0.7000000000000001pt] (axis cs:0.14166667,0.20243)
--(axis cs:0.14166667,0.20902);

\path [draw=color2, line width=0.7000000000000001pt] (axis cs:0.14333333,0.20855)
--(axis cs:0.14333333,0.2152);

\path [draw=color2, line width=0.7000000000000001pt] (axis cs:0.145,0.21919)
--(axis cs:0.145,0.22596);

\path [draw=color2, line width=0.7000000000000001pt] (axis cs:0.14666667,0.22648)
--(axis cs:0.14666667,0.23333);

\path [draw=color2, line width=0.7000000000000001pt] (axis cs:0.14833333,0.23877)
--(axis cs:0.14833333,0.24575);

\path [draw=color2, line width=0.7000000000000001pt] (axis cs:0.15,0.24673)
--(axis cs:0.15,0.25378);

\path [draw=color2, line width=0.7000000000000001pt] (axis cs:0.15166667,0.2561)
--(axis cs:0.15166667,0.26325);

\path [draw=color2, line width=0.7000000000000001pt] (axis cs:0.15333333,0.26683)
--(axis cs:0.15333333,0.27406);

\path [draw=color2, line width=0.7000000000000001pt] (axis cs:0.155,0.27356)
--(axis cs:0.155,0.28085);

\path [draw=color2, line width=0.7000000000000001pt] (axis cs:0.15666667,0.28662)
--(axis cs:0.15666667,0.29401);

\path [draw=color2, line width=0.7000000000000001pt] (axis cs:0.15833333,0.29508)
--(axis cs:0.15833333,0.30253);

\path [draw=color2, line width=0.7000000000000001pt] (axis cs:0.16,0.30367)
--(axis cs:0.16,0.31118);

\path [draw=color2, line width=0.7000000000000001pt] (axis cs:0.16166667,0.31569)
--(axis cs:0.16166667,0.32328);

\path [draw=color2, line width=0.7000000000000001pt] (axis cs:0.16333333,0.32432)
--(axis cs:0.16333333,0.33197);

\path [draw=color2, line width=0.7000000000000001pt] (axis cs:0.165,0.33603)
--(axis cs:0.165,0.34375);

\path [draw=color2, line width=0.7000000000000001pt] (axis cs:0.16666667,0.34256)
--(axis cs:0.16666667,0.35031);

\path [draw=color2, line width=0.7000000000000001pt] (axis cs:0.16833333,0.35311)
--(axis cs:0.16833333,0.36092);

\path [draw=color2, line width=0.7000000000000001pt] (axis cs:0.17,0.36404)
--(axis cs:0.17,0.3719);

\path [draw=color2, line width=0.7000000000000001pt] (axis cs:0.17166667,0.37303)
--(axis cs:0.17166667,0.38093);

\path [draw=color2, line width=0.7000000000000001pt] (axis cs:0.17333333,0.37963)
--(axis cs:0.17333333,0.38755);

\path [draw=color2, line width=0.7000000000000001pt] (axis cs:0.175,0.39179)
--(axis cs:0.175,0.39976);

\path [draw=color3, line width=0.7000000000000001pt] (axis cs:0.135,0.14293)
--(axis cs:0.135,0.14868);

\path [draw=color3, line width=0.7000000000000001pt] (axis cs:0.13666667,0.1538)
--(axis cs:0.13666667,0.15972);

\path [draw=color3, line width=0.7000000000000001pt] (axis cs:0.13833333,0.16563)
--(axis cs:0.13833333,0.17173);

\path [draw=color3, line width=0.7000000000000001pt] (axis cs:0.14,0.17372)
--(axis cs:0.14,0.17993);

\path [draw=color3, line width=0.7000000000000001pt] (axis cs:0.14166667,0.18631)
--(axis cs:0.14166667,0.1927);

\path [draw=color3, line width=0.7000000000000001pt] (axis cs:0.14333333,0.19707)
--(axis cs:0.14333333,0.20359);

\path [draw=color3, line width=0.7000000000000001pt] (axis cs:0.145,0.20629)
--(axis cs:0.145,0.21292);

\path [draw=color3, line width=0.7000000000000001pt] (axis cs:0.14666667,0.21753)
--(axis cs:0.14666667,0.22428);

\path [draw=color3, line width=0.7000000000000001pt] (axis cs:0.14833333,0.22657)
--(axis cs:0.14833333,0.23342);

\path [draw=color3, line width=0.7000000000000001pt] (axis cs:0.15,0.24134)
--(axis cs:0.15,0.24835);

\path [draw=color3, line width=0.7000000000000001pt] (axis cs:0.15166667,0.25195)
--(axis cs:0.15166667,0.25906);

\path [draw=color3, line width=0.7000000000000001pt] (axis cs:0.15333333,0.26437)
--(axis cs:0.15333333,0.27158);

\path [draw=color3, line width=0.7000000000000001pt] (axis cs:0.155,0.27699)
--(axis cs:0.155,0.28431);

\path [draw=color3, line width=0.7000000000000001pt] (axis cs:0.15666667,0.28755)
--(axis cs:0.15666667,0.29495);

\path [draw=color3, line width=0.7000000000000001pt] (axis cs:0.15833333,0.30191)
--(axis cs:0.15833333,0.30942);

\path [draw=color3, line width=0.7000000000000001pt] (axis cs:0.16,0.3139)
--(axis cs:0.16,0.32149);

\path [draw=color3, line width=0.7000000000000001pt] (axis cs:0.16166667,0.32678)
--(axis cs:0.16166667,0.33445);

\path [draw=color3, line width=0.7000000000000001pt] (axis cs:0.16333333,0.33676)
--(axis cs:0.16333333,0.34448);

\path [draw=color3, line width=0.7000000000000001pt] (axis cs:0.165,0.34846)
--(axis cs:0.165,0.35624);

\path [draw=color3, line width=0.7000000000000001pt] (axis cs:0.16666667,0.36245)
--(axis cs:0.16666667,0.3703);

\path [draw=color3, line width=0.7000000000000001pt] (axis cs:0.16833333,0.37069)
--(axis cs:0.16833333,0.37857);

\path [draw=color3, line width=0.7000000000000001pt] (axis cs:0.17,0.38482)
--(axis cs:0.17,0.39276);

\path [draw=color3, line width=0.7000000000000001pt] (axis cs:0.17166667,0.39391)
--(axis cs:0.17166667,0.40188);

\path [draw=color3, line width=0.7000000000000001pt] (axis cs:0.17333333,0.40568)
--(axis cs:0.17333333,0.41369);

\path [draw=color3, line width=0.7000000000000001pt] (axis cs:0.175,0.41978)
--(axis cs:0.175,0.42783);

\path [draw=color4, line width=0.7000000000000001pt] (axis cs:0.135,0.12682)
--(axis cs:0.135,0.13229);

\path [draw=color4, line width=0.7000000000000001pt] (axis cs:0.13666667,0.138)
--(axis cs:0.13666667,0.14367);

\path [draw=color4, line width=0.7000000000000001pt] (axis cs:0.13833333,0.14933)
--(axis cs:0.13833333,0.15518);

\path [draw=color4, line width=0.7000000000000001pt] (axis cs:0.14,0.16087)
--(axis cs:0.14,0.1669);

\path [draw=color4, line width=0.7000000000000001pt] (axis cs:0.14166667,0.17314)
--(axis cs:0.14166667,0.17935);

\path [draw=color4, line width=0.7000000000000001pt] (axis cs:0.14333333,0.1878)
--(axis cs:0.14333333,0.19421);

\path [draw=color4, line width=0.7000000000000001pt] (axis cs:0.145,0.19843)
--(axis cs:0.145,0.20496);

\path [draw=color4, line width=0.7000000000000001pt] (axis cs:0.14666667,0.20845)
--(axis cs:0.14666667,0.2151);

\path [draw=color4, line width=0.7000000000000001pt] (axis cs:0.14833333,0.22601)
--(axis cs:0.14833333,0.23286);

\path [draw=color4, line width=0.7000000000000001pt] (axis cs:0.15,0.23601)
--(axis cs:0.15,0.24296);

\path [draw=color4, line width=0.7000000000000001pt] (axis cs:0.15166667,0.25162)
--(axis cs:0.15166667,0.25873);

\path [draw=color4, line width=0.7000000000000001pt] (axis cs:0.15333333,0.26257)
--(axis cs:0.15333333,0.26977);

\path [draw=color4, line width=0.7000000000000001pt] (axis cs:0.155,0.27669)
--(axis cs:0.155,0.284);

\path [draw=color4, line width=0.7000000000000001pt] (axis cs:0.15666667,0.29179)
--(axis cs:0.15666667,0.29922);

\path [draw=color4, line width=0.7000000000000001pt] (axis cs:0.15833333,0.3076)
--(axis cs:0.15833333,0.31515);

\path [draw=color4, line width=0.7000000000000001pt] (axis cs:0.16,0.31942)
--(axis cs:0.16,0.32704);

\path [draw=color4, line width=0.7000000000000001pt] (axis cs:0.16166667,0.33546)
--(axis cs:0.16166667,0.34317);

\path [draw=color4, line width=0.7000000000000001pt] (axis cs:0.16333333,0.34955)
--(axis cs:0.16333333,0.35734);

\path [draw=color4, line width=0.7000000000000001pt] (axis cs:0.165,0.36405)
--(axis cs:0.165,0.37191);

\path [draw=color4, line width=0.7000000000000001pt] (axis cs:0.16666667,0.37509)
--(axis cs:0.16666667,0.38299);

\path [draw=color4, line width=0.7000000000000001pt] (axis cs:0.16833333,0.39126)
--(axis cs:0.16833333,0.39922);

\path [draw=color4, line width=0.7000000000000001pt] (axis cs:0.17,0.40262)
--(axis cs:0.17,0.41062);

\path [draw=color4, line width=0.7000000000000001pt] (axis cs:0.17166667,0.41964)
--(axis cs:0.17166667,0.42769);

\path [draw=color4, line width=0.7000000000000001pt] (axis cs:0.17333333,0.43404)
--(axis cs:0.17333333,0.44212);

\path [draw=color4, line width=0.7000000000000001pt] (axis cs:0.175,0.44411)
--(axis cs:0.175,0.45221);

\path [draw=color5, line width=0.7000000000000001pt] (axis cs:0.135,0.11155)
--(axis cs:0.135,0.11673);

\path [draw=color5, line width=0.7000000000000001pt] (axis cs:0.13666667,0.12306)
--(axis cs:0.13666667,0.12846);

\path [draw=color5, line width=0.7000000000000001pt] (axis cs:0.13833333,0.13731)
--(axis cs:0.13833333,0.14296);

\path [draw=color5, line width=0.7000000000000001pt] (axis cs:0.14,0.14962)
--(axis cs:0.14,0.15548);

\path [draw=color5, line width=0.7000000000000001pt] (axis cs:0.14166667,0.16082)
--(axis cs:0.14166667,0.16685);

\path [draw=color5, line width=0.7000000000000001pt] (axis cs:0.14333333,0.17531)
--(axis cs:0.14333333,0.18154);

\path [draw=color5, line width=0.7000000000000001pt] (axis cs:0.145,0.18953)
--(axis cs:0.145,0.19596);

\path [draw=color5, line width=0.7000000000000001pt] (axis cs:0.14666667,0.20302)
--(axis cs:0.14666667,0.20961);

\path [draw=color5, line width=0.7000000000000001pt] (axis cs:0.14833333,0.21575)
--(axis cs:0.14833333,0.22248);

\path [draw=color5, line width=0.7000000000000001pt] (axis cs:0.15,0.22993)
--(axis cs:0.15,0.23682);

\path [draw=color5, line width=0.7000000000000001pt] (axis cs:0.15166667,0.24721)
--(axis cs:0.15166667,0.25428);

\path [draw=color5, line width=0.7000000000000001pt] (axis cs:0.15333333,0.26271)
--(axis cs:0.15333333,0.26991);

\path [draw=color5, line width=0.7000000000000001pt] (axis cs:0.155,0.27873)
--(axis cs:0.155,0.28606);

\path [draw=color5, line width=0.7000000000000001pt] (axis cs:0.15666667,0.29319)
--(axis cs:0.15666667,0.30064);

\path [draw=color5, line width=0.7000000000000001pt] (axis cs:0.15833333,0.31323)
--(axis cs:0.15833333,0.32081);

\path [draw=color5, line width=0.7000000000000001pt] (axis cs:0.16,0.3284)
--(axis cs:0.16,0.33607);

\path [draw=color5, line width=0.7000000000000001pt] (axis cs:0.16166667,0.34477)
--(axis cs:0.16166667,0.35253);

\path [draw=color5, line width=0.7000000000000001pt] (axis cs:0.16333333,0.36222)
--(axis cs:0.16333333,0.37007);

\path [draw=color5, line width=0.7000000000000001pt] (axis cs:0.165,0.37715)
--(axis cs:0.165,0.38506);

\path [draw=color5, line width=0.7000000000000001pt] (axis cs:0.16666667,0.39226)
--(axis cs:0.16666667,0.40023);

\path [draw=color5, line width=0.7000000000000001pt] (axis cs:0.16833333,0.41308)
--(axis cs:0.16833333,0.42111);

\path [draw=color5, line width=0.7000000000000001pt] (axis cs:0.17,0.42345)
--(axis cs:0.17,0.43151);

\path [draw=color5, line width=0.7000000000000001pt] (axis cs:0.17166667,0.43984)
--(axis cs:0.17166667,0.44794);

\path [draw=color5, line width=0.7000000000000001pt] (axis cs:0.17333333,0.45453)
--(axis cs:0.17333333,0.46265);

\path [draw=color5, line width=0.7000000000000001pt] (axis cs:0.175,0.47053)
--(axis cs:0.175,0.47867);

\addplot [line width=0.7000000000000001pt, color1]
table {%
0.1478507070006 0.247955655268017
0.147968552354135 0.248432326772465
0.148086397707671 0.248909469315316
0.148204243061206 0.24938708289657
0.148322088414741 0.249865167516226
0.148439933768277 0.250343723174284
0.148557779121812 0.250822749870745
0.148675624475347 0.251302247605608
0.148793469828883 0.251782216378873
0.148911315182418 0.252262656190541
0.149029160535953 0.252743567040611
0.149147005889489 0.253224948929084
0.149264851243024 0.253706801855959
0.149382696596559 0.254189125821237
0.149500541950095 0.254671920824917
0.14961838730363 0.255155186866999
0.149736232657166 0.255638923947484
0.149854078010701 0.256123132066371
0.149971923364236 0.25660781122366
0.150089768717772 0.257092961419352
0.150207614071307 0.257578582653447
0.150325459424842 0.258064674925944
0.150443304778378 0.258551238236843
0.150561150131913 0.259038272586144
0.150678995485448 0.259525777973849
0.150796840838984 0.260013754399955
0.150914686192519 0.260502201864464
0.151032531546054 0.260991120367375
0.15115037689959 0.261480509908689
0.151268222253125 0.261970370488405
0.15138606760666 0.262460702106523
0.151503912960196 0.262951504763044
0.151621758313731 0.263442778457968
0.151739603667266 0.263934523191293
0.151857449020802 0.264426738963021
0.151975294374337 0.264919425773152
0.152093139727873 0.265412583621685
0.152210985081408 0.26590621250862
0.152328830434943 0.266400312433958
0.152446675788479 0.266894883397698
0.152564521142014 0.267389925399841
0.152682366495549 0.267885438440386
0.152800211849085 0.268381422519333
0.15291805720262 0.268877877636683
0.153035902556155 0.269374803792435
0.153153747909691 0.26987220098659
0.153271593263226 0.270370069219147
0.153389438616761 0.270868408490107
0.153507283970297 0.271367218799468
0.153625129323832 0.271866500147233
0.153742974677367 0.2723662525334
0.153860820030903 0.272866475957969
0.153978665384438 0.27336717042094
0.154096510737974 0.273868335922314
0.154214356091509 0.27436997246209
0.154332201445044 0.274872080040269
0.15445004679858 0.27537465865685
0.154567892152115 0.275877708311834
0.15468573750565 0.27638122900522
0.154803582859186 0.276885220737008
0.154921428212721 0.277389683507199
0.155039273566256 0.277894617315792
0.155157118919792 0.278400022162788
0.155274964273327 0.278905898048186
0.155392809626862 0.279412244971986
0.155510654980398 0.279919062934189
0.155628500333933 0.280426351934794
0.155746345687468 0.280934111973802
0.155864191041004 0.281442343051212
0.155982036394539 0.281951045167024
0.156099881748075 0.282460218321239
0.15621772710161 0.282969862513857
0.156335572455145 0.283479977744876
0.156453417808681 0.283990564014299
0.156571263162216 0.284501621322123
0.156689108515751 0.28501314966835
0.156806953869287 0.285525149052979
0.156924799222822 0.286037619476011
0.157042644576357 0.286550560937445
0.157160489929893 0.287063973437282
0.157278335283428 0.287577856975521
0.157396180636963 0.288092211552162
0.157514025990499 0.288607037167206
0.157631871344034 0.289122333820652
0.157749716697569 0.289638101512501
0.157867562051105 0.290154340242752
0.15798540740464 0.290671050011405
0.158103252758176 0.291188230818461
0.158221098111711 0.29170588266392
0.158338943465246 0.29222400554778
0.158456788818782 0.292742599470043
0.158574634172317 0.293261664430709
0.158692479525852 0.293781200429777
0.158810324879388 0.294301207467247
0.158928170232923 0.29482168554312
0.159046015586458 0.295342634657395
0.159163860939994 0.295864054810072
0.159281706293529 0.296385946001152
0.159399551647064 0.296908308230635
0.1595173970006 0.29743114149852
};
\addplot [line width=0.7000000000000001pt, color2]
table {%
0.1478507070006 0.238483114264032
0.147968552354135 0.239124006038314
0.148086397707671 0.239765783813966
0.148204243061206 0.240408447590986
0.148322088414741 0.241051997369374
0.148439933768277 0.241696433149131
0.148557779121812 0.242341754930256
0.148675624475347 0.242987962712749
0.148793469828883 0.243635056496611
0.148911315182418 0.244283036281842
0.149029160535953 0.244931902068441
0.149147005889489 0.245581653856408
0.149264851243024 0.246232291645743
0.149382696596559 0.246883815436447
0.149500541950095 0.24753622522852
0.14961838730363 0.248189521021961
0.149736232657166 0.24884370281677
0.149854078010701 0.249498770612948
0.149971923364236 0.250154724410494
0.150089768717772 0.250811564209408
0.150207614071307 0.251469290009691
0.150325459424842 0.252127901811343
0.150443304778378 0.252787399614363
0.150561150131913 0.253447783418751
0.150678995485448 0.254109053224508
0.150796840838984 0.254771209031633
0.150914686192519 0.255434250840126
0.151032531546054 0.256098178649988
0.15115037689959 0.256762992461218
0.151268222253125 0.257428692273817
0.15138606760666 0.258095278087784
0.151503912960196 0.25876274990312
0.151621758313731 0.259431107719824
0.151739603667266 0.260100351537896
0.151857449020802 0.260770481357337
0.151975294374337 0.261441497178147
0.152093139727873 0.262113399000324
0.152210985081408 0.26278618682387
0.152328830434943 0.263459860648785
0.152446675788479 0.264134420475068
0.152564521142014 0.264809866302719
0.152682366495549 0.265486198131739
0.152800211849085 0.266163415962127
0.15291805720262 0.266841519793884
0.153035902556155 0.267520509627009
0.153153747909691 0.268200385461502
0.153271593263226 0.268881147297364
0.153389438616761 0.269562795134594
0.153507283970297 0.270245328973193
0.153625129323832 0.27092874881316
0.153742974677367 0.271613054654496
0.153860820030903 0.2722982464972
0.153978665384438 0.272984324341272
0.154096510737974 0.273671288186713
0.154214356091509 0.274359138033522
0.154332201445044 0.2750478738817
0.15445004679858 0.275737495731246
0.154567892152115 0.27642800358216
0.15468573750565 0.277119397434443
0.154803582859186 0.277811677288095
0.154921428212721 0.278504843143115
0.155039273566256 0.279198894999503
0.155157118919792 0.279893832857259
0.155274964273327 0.280589656716384
0.155392809626862 0.281286366576878
0.155510654980398 0.28198396243874
0.155628500333933 0.28268244430197
0.155746345687468 0.283381812166569
0.155864191041004 0.284082066032536
0.155982036394539 0.284783205899871
0.156099881748075 0.285485231768575
0.15621772710161 0.286188143638648
0.156335572455145 0.286891941510088
0.156453417808681 0.287596625382897
0.156571263162216 0.288302195257075
0.156689108515751 0.289008651132621
0.156806953869287 0.289715993009536
0.156924799222822 0.290424220887818
0.157042644576357 0.29113333476747
0.157160489929893 0.29184333464849
0.157278335283428 0.292554220530878
0.157396180636963 0.293265992414634
0.157514025990499 0.293978650299759
0.157631871344034 0.294692194186253
0.157749716697569 0.295406624074114
0.157867562051105 0.296121939963345
0.15798540740464 0.296838141853943
0.158103252758176 0.29755522974591
0.158221098111711 0.298273203639246
0.158338943465246 0.29899206353395
0.158456788818782 0.299711809430022
0.158574634172317 0.300432441327463
0.158692479525852 0.301153959226272
0.158810324879388 0.30187636312645
0.158928170232923 0.302599653027996
0.159046015586458 0.30332382893091
0.159163860939994 0.304048890835193
0.159281706293529 0.304774838740844
0.159399551647064 0.305501672647864
0.1595173970006 0.306229392556252
};
\addplot [line width=0.7000000000000001pt, color3]
table {%
0.1478507070006 0.230178753637261
0.147968552354135 0.230960508631324
0.148086397707671 0.231743630454719
0.148204243061206 0.232528119107447
0.148322088414741 0.233313974589508
0.148439933768277 0.234101196900902
0.148557779121812 0.234889786041629
0.148675624475347 0.235679742011689
0.148793469828883 0.236471064811081
0.148911315182418 0.237263754439807
0.149029160535953 0.238057810897865
0.149147005889489 0.238853234185257
0.149264851243024 0.239650024301981
0.149382696596559 0.240448181248038
0.149500541950095 0.241247705023428
0.14961838730363 0.242048595628151
0.149736232657166 0.242850853062207
0.149854078010701 0.243654477325595
0.149971923364236 0.244459468418317
0.150089768717772 0.245265826340372
0.150207614071307 0.246073551091759
0.150325459424842 0.246882642672479
0.150443304778378 0.247693101082532
0.150561150131913 0.248504926321918
0.150678995485448 0.249318118390637
0.150796840838984 0.250132677288689
0.150914686192519 0.250948603016074
0.151032531546054 0.251765895572791
0.15115037689959 0.252584554958842
0.151268222253125 0.253404581174225
0.15138606760666 0.254225974218942
0.151503912960196 0.255048734092991
0.151621758313731 0.255872860796373
0.151739603667266 0.256698354329088
0.151857449020802 0.257525214691136
0.151975294374337 0.258353441882516
0.152093139727873 0.25918303590323
0.152210985081408 0.260013996753277
0.152328830434943 0.260846324432656
0.152446675788479 0.261680018941368
0.152564521142014 0.262515080279414
0.152682366495549 0.263351508446792
0.152800211849085 0.264189303443503
0.15291805720262 0.265028465269547
0.153035902556155 0.265868993924923
0.153153747909691 0.266710889409633
0.153271593263226 0.267554151723676
0.153389438616761 0.268398780867051
0.153507283970297 0.269244776839759
0.153625129323832 0.270092139641801
0.153742974677367 0.270940869273175
0.153860820030903 0.271790965733882
0.153978665384438 0.272642429023922
0.154096510737974 0.273495259143295
0.154214356091509 0.274349456092
0.154332201445044 0.275205019870039
0.15445004679858 0.27606195047741
0.154567892152115 0.276920247914115
0.15468573750565 0.277779912180152
0.154803582859186 0.278640943275522
0.154921428212721 0.279503341200225
0.155039273566256 0.280367105954261
0.155157118919792 0.28123223753763
0.155274964273327 0.282098735950332
0.155392809626862 0.282966601192367
0.155510654980398 0.283835833263734
0.155628500333933 0.284706432164435
0.155746345687468 0.285578397894468
0.155864191041004 0.286451730453834
0.155982036394539 0.287326429842533
0.156099881748075 0.288202496060565
0.15621772710161 0.28907992910793
0.156335572455145 0.289958728984628
0.156453417808681 0.290838895690658
0.156571263162216 0.291720429226022
0.156689108515751 0.292603329590718
0.156806953869287 0.293487596784748
0.156924799222822 0.29437323080811
0.157042644576357 0.295260231660805
0.157160489929893 0.296148599342833
0.157278335283428 0.297038333854194
0.157396180636963 0.297929435194888
0.157514025990499 0.298821903364915
0.157631871344034 0.299715738364274
0.157749716697569 0.300610940192967
0.157867562051105 0.301507508850992
0.15798540740464 0.302405444338351
0.158103252758176 0.303304746655042
0.158221098111711 0.304205415801066
0.158338943465246 0.305107451776423
0.158456788818782 0.306010854581113
0.158574634172317 0.306915624215135
0.158692479525852 0.307821760678491
0.158810324879388 0.308729263971179
0.158928170232923 0.309638134093201
0.159046015586458 0.310548371044555
0.159163860939994 0.311459974825242
0.159281706293529 0.312372945435263
0.159399551647064 0.313287282874615
0.1595173970006 0.314202987143301
};
\addplot [line width=0.7000000000000001pt, color4]
table {%
0.1478507070006 0.22265815461537
0.147968552354135 0.223564832220474
0.148086397707671 0.224473411842445
0.148204243061206 0.225383893481282
0.148322088414741 0.226296277136985
0.148439933768277 0.227210562809555
0.148557779121812 0.228126750498992
0.148675624475347 0.229044840205294
0.148793469828883 0.229964831928463
0.148911315182418 0.230886725668499
0.149029160535953 0.231810521425401
0.149147005889489 0.232736219199169
0.149264851243024 0.233663818989804
0.149382696596559 0.234593320797305
0.149500541950095 0.235524724621673
0.14961838730363 0.236458030462907
0.149736232657166 0.237393238321007
0.149854078010701 0.238330348195974
0.149971923364236 0.239269360087807
0.150089768717772 0.240210273996507
0.150207614071307 0.241153089922073
0.150325459424842 0.242097807864506
0.150443304778378 0.243044427823805
0.150561150131913 0.24399294979997
0.150678995485448 0.244943373793002
0.150796840838984 0.2458956998029
0.150914686192519 0.246849927829665
0.151032531546054 0.247806057873296
0.15115037689959 0.248764089933793
0.151268222253125 0.249724024011157
0.15138606760666 0.250685860105388
0.151503912960196 0.251649598216485
0.151621758313731 0.252615238344448
0.151739603667266 0.253582780489277
0.151857449020802 0.254552224650973
0.151975294374337 0.255523570829536
0.152093139727873 0.256496819024965
0.152210985081408 0.25747196923726
0.152328830434943 0.258449021466422
0.152446675788479 0.25942797571245
0.152564521142014 0.260408831975344
0.152682366495549 0.261391590255105
0.152800211849085 0.262376250551733
0.15291805720262 0.263362812865227
0.153035902556155 0.264351277195587
0.153153747909691 0.265341643542813
0.153271593263226 0.266333911906907
0.153389438616761 0.267328082287866
0.153507283970297 0.268324154685692
0.153625129323832 0.269322129100384
0.153742974677367 0.270322005531943
0.153860820030903 0.271323783980368
0.153978665384438 0.27232746444566
0.154096510737974 0.273333046927818
0.154214356091509 0.274340531426843
0.154332201445044 0.275349917942733
0.15445004679858 0.276361206475491
0.154567892152115 0.277374397025114
0.15468573750565 0.278389489591604
0.154803582859186 0.279406484174961
0.154921428212721 0.280425380775184
0.155039273566256 0.281446179392273
0.155157118919792 0.282468880026229
0.155274964273327 0.283493482677051
0.155392809626862 0.28451998734474
0.155510654980398 0.285548394029295
0.155628500333933 0.286578702730717
0.155746345687468 0.287610913449005
0.155864191041004 0.288645026184159
0.155982036394539 0.28968104093618
0.156099881748075 0.290718957705067
0.15621772710161 0.29175877649082
0.156335572455145 0.29280049729344
0.156453417808681 0.293844120112927
0.156571263162216 0.29488964494928
0.156689108515751 0.295937071802499
0.156806953869287 0.296986400672585
0.156924799222822 0.298037631559537
0.157042644576357 0.299090764463355
0.157160489929893 0.30014579938404
0.157278335283428 0.301202736321592
0.157396180636963 0.302261575276009
0.157514025990499 0.303322316247294
0.157631871344034 0.304384959235444
0.157749716697569 0.305449504240461
0.157867562051105 0.306515951262345
0.15798540740464 0.307584300301095
0.158103252758176 0.308654551356711
0.158221098111711 0.309726704429194
0.158338943465246 0.310800759518543
0.158456788818782 0.311876716624759
0.158574634172317 0.312954575747841
0.158692479525852 0.314034336887789
0.158810324879388 0.315116000044604
0.158928170232923 0.316199565218285
0.159046015586458 0.317285032408833
0.159163860939994 0.318372401616247
0.159281706293529 0.319461672840528
0.159399551647064 0.320552846081675
0.1595173970006 0.321645921339688
};
\addplot [line width=0.7000000000000001pt, color5]
table {%
0.1478507070006 0.215720546081308
0.147968552354135 0.216740113805001
0.148086397707671 0.217762165775042
0.148204243061206 0.218786701991429
0.148322088414741 0.219813722454165
0.148439933768277 0.220843227163247
0.148557779121812 0.221875216118678
0.148675624475347 0.222909689320456
0.148793469828883 0.223946646768581
0.148911315182418 0.224986088463054
0.149029160535953 0.226028014403874
0.149147005889489 0.227072424591042
0.149264851243024 0.228119319024558
0.149382696596559 0.229168697704421
0.149500541950095 0.230220560630631
0.14961838730363 0.231274907803189
0.149736232657166 0.232331739222095
0.149854078010701 0.233391054887348
0.149971923364236 0.234452854798949
0.150089768717772 0.235517138956897
0.150207614071307 0.236583907361193
0.150325459424842 0.237653160011836
0.150443304778378 0.238724896908827
0.150561150131913 0.239799118052165
0.150678995485448 0.24087582344185
0.150796840838984 0.241955013077884
0.150914686192519 0.243036686960265
0.151032531546054 0.244120845088993
0.15115037689959 0.245207487464069
0.151268222253125 0.246296614085492
0.15138606760666 0.247388224953263
0.151503912960196 0.248482320067382
0.151621758313731 0.249578899427848
0.151739603667266 0.250677963034661
0.151857449020802 0.251779510887822
0.151975294374337 0.252883542987331
0.152093139727873 0.253990059333187
0.152210985081408 0.25509905992539
0.152328830434943 0.256210544763941
0.152446675788479 0.25732451384884
0.152564521142014 0.258440967180086
0.152682366495549 0.25955990475768
0.152800211849085 0.260681326581621
0.15291805720262 0.26180523265191
0.153035902556155 0.262931622968546
0.153153747909691 0.26406049753153
0.153271593263226 0.265191856340861
0.153389438616761 0.26632569939654
0.153507283970297 0.267462026698566
0.153625129323832 0.26860083824694
0.153742974677367 0.269742134041661
0.153860820030903 0.27088591408273
0.153978665384438 0.272032178370147
0.154096510737974 0.273180926903911
0.154214356091509 0.274332159684022
0.154332201445044 0.275485876710481
0.15445004679858 0.276642077983288
0.154567892152115 0.277800763502442
0.15468573750565 0.278961933267943
0.154803582859186 0.280125587279792
0.154921428212721 0.281291725537989
0.155039273566256 0.282460348042533
0.155157118919792 0.283631454793425
0.155274964273327 0.284805045790664
0.155392809626862 0.285981121034251
0.155510654980398 0.287159680524185
0.155628500333933 0.288340724260467
0.155746345687468 0.289524252243096
0.155864191041004 0.290710264472073
0.155982036394539 0.291898760947397
0.156099881748075 0.293089741669069
0.15621772710161 0.294283206637088
0.156335572455145 0.295479155851455
0.156453417808681 0.29667758931217
0.156571263162216 0.297878507019232
0.156689108515751 0.299081908972641
0.156806953869287 0.300287795172398
0.156924799222822 0.301496165618503
0.157042644576357 0.302707020310955
0.157160489929893 0.303920359249754
0.157278335283428 0.305136182434901
0.157396180636963 0.306354489866396
0.157514025990499 0.307575281544238
0.157631871344034 0.308798557468428
0.157749716697569 0.310024317638965
0.157867562051105 0.31125256205585
0.15798540740464 0.312483290719082
0.158103252758176 0.313716503628662
0.158221098111711 0.314952200784589
0.158338943465246 0.316190382186864
0.158456788818782 0.317431047835486
0.158574634172317 0.318674197730456
0.158692479525852 0.319919831871773
0.158810324879388 0.321167950259438
0.158928170232923 0.322418552893451
0.159046015586458 0.323671639773811
0.159163860939994 0.324927210900518
0.159281706293529 0.326185266273573
0.159399551647064 0.327445805892976
0.1595173970006 0.328708829758726
};
\path [draw=white, fill opacity=0] (axis cs:0,0.093194)
--(axis cs:0,0.497026);

\path [draw=white, fill opacity=0] (axis cs:1,0.093194)
--(axis cs:1,0.497026);

\path [draw=white, fill opacity=0] (axis cs:0.133,0)
--(axis cs:0.177,0);

\path [draw=white, fill opacity=0] (axis cs:0.133,1)
--(axis cs:0.177,1);

\end{axis}

\end{tikzpicture}

%% file: paper_planar_dep_bp.tex
\begin{tikzpicture}

\definecolor{color4}{rgb}{0.219799566082832,0.662515787685034,0.773209315931721}
\definecolor{color2}{rgb}{0.680418912779335,0.615149751467757,0.194054521114453}
\definecolor{color5}{rgb}{0.800493618642396,0.477033635337372,0.957954719600752}
\definecolor{color3}{rgb}{0.201253172212011,0.690792081537903,0.479667611892753}
\definecolor{color0}{rgb}{0.917647058823529,0.917647058823529,0.949019607843137}
\definecolor{color1}{rgb}{0.967797559291991,0.441274560091574,0.53581031550587}

\begin{axis}[
title={Depolarizing Error Model, with BP/multi-path summation},
xlabel={physical error probability},
ylabel={logical error probability},
xmin=0.1625, xmax=0.2175,
ymin=0.1341280425162, ymax=0.6230929943101,
width=\figurewidth,
height=\figureheight,
xtick={0.16,0.17,0.18,0.19,0.2,0.21,0.22},
xticklabels={,0.17,0.18,0.19,0.20,0.21,},
ytick={0.1,0.2,0.3,0.4,0.5,0.6,0.7},
yticklabels={,0.2,0.3,0.4,0.5,0.6,},
tick align=outside,
tick pos=left,
xmajorgrids,
x grid style={white},
ymajorgrids,
y grid style={white},
axis line style={white},
axis background/.style={fill=color0},
legend pos=north west,
legend entries={$d = 7$, $d = 11$, $d = 15$, $d = 19$, $d = 23$}
]
\path [draw=color1, line width=0.7000000000000001pt] (axis cs:0.165,0.21648293029872)
--(axis cs:0.165,0.223190295558717);

\path [draw=color1, line width=0.7000000000000001pt] (axis cs:0.16708333,0.22619329856172)
--(axis cs:0.16708333,0.232999446815236);

\path [draw=color1, line width=0.7000000000000001pt] (axis cs:0.16916667,0.23296981191718)
--(axis cs:0.16916667,0.239854986565513);

\path [draw=color1, line width=0.7000000000000001pt] (axis cs:0.17125,0.242788841473052)
--(axis cs:0.17125,0.249772799114904);

\path [draw=color1, line width=0.7000000000000001pt] (axis cs:0.17333333,0.250493914967599)
--(axis cs:0.17333333,0.257537142405563);

\path [draw=color1, line width=0.7000000000000001pt] (axis cs:0.17541667,0.261725541330805)
--(axis cs:0.17541667,0.268867551762289);

\path [draw=color1, line width=0.7000000000000001pt] (axis cs:0.1775,0.270043069385175)
--(axis cs:0.1775,0.277264106211475);

\path [draw=color1, line width=0.7000000000000001pt] (axis cs:0.17958333,0.276078710289237)
--(axis cs:0.17958333,0.283339260312945);

\path [draw=color1, line width=0.7000000000000001pt] (axis cs:0.18166667,0.287330093251146)
--(axis cs:0.18166667,0.294689426268374);

\path [draw=color1, line width=0.7000000000000001pt] (axis cs:0.18375,0.295104314841157)
--(axis cs:0.18375,0.302522917654497);

\path [draw=color1, line width=0.7000000000000001pt] (axis cs:0.18583333,0.30274024024024)
--(axis cs:0.18583333,0.310198356250988);

\path [draw=color1, line width=0.7000000000000001pt] (axis cs:0.18791667,0.313191480954639)
--(axis cs:0.18791667,0.320728623360202);

\path [draw=color1, line width=0.7000000000000001pt] (axis cs:0.19,0.323948948948949)
--(axis cs:0.19,0.331545361150624);

\path [draw=color1, line width=0.7000000000000001pt] (axis cs:0.19208333,0.335378141299194)
--(axis cs:0.19208333,0.343043701596333);

\path [draw=color1, line width=0.7000000000000001pt] (axis cs:0.19416667,0.339507270428323)
--(axis cs:0.19416667,0.347192587324166);

\path [draw=color1, line width=0.7000000000000001pt] (axis cs:0.19625,0.349573257467994)
--(axis cs:0.19625,0.357317844159949);

\path [draw=color1, line width=0.7000000000000001pt] (axis cs:0.19833333,0.358157499604868)
--(axis cs:0.19833333,0.365941599494231);

\path [draw=color1, line width=0.7000000000000001pt] (axis cs:0.20041667,0.367166508613877)
--(axis cs:0.20041667,0.374990121700648);

\path [draw=color1, line width=0.7000000000000001pt] (axis cs:0.2025,0.378793266951162)
--(axis cs:0.2025,0.386666271534693);

\path [draw=color1, line width=0.7000000000000001pt] (axis cs:0.20458333,0.388325825825826)
--(axis cs:0.20458333,0.396228465307413);

\path [draw=color1, line width=0.7000000000000001pt] (axis cs:0.20666667,0.393956456456456)
--(axis cs:0.20666667,0.401878852536747);

\path [draw=color1, line width=0.7000000000000001pt] (axis cs:0.20875,0.404476845266319)
--(axis cs:0.20875,0.412438754544018);

\path [draw=color1, line width=0.7000000000000001pt] (axis cs:0.21083333,0.410640904061957)
--(axis cs:0.21083333,0.418622569938359);

\path [draw=color1, line width=0.7000000000000001pt] (axis cs:0.21291667,0.417901454085665)
--(axis cs:0.21291667,0.425902876560771);

\path [draw=color1, line width=0.7000000000000001pt] (axis cs:0.215,0.426752410305042)
--(axis cs:0.215,0.434773589378853);

\path [draw=color2, line width=0.7000000000000001pt] (axis cs:0.165,0.199274932827564)
--(axis cs:0.165,0.205784732100522);

\path [draw=color2, line width=0.7000000000000001pt] (axis cs:0.16708333,0.213153943417101)
--(axis cs:0.16708333,0.21982179547969);

\path [draw=color2, line width=0.7000000000000001pt] (axis cs:0.16916667,0.223368104947052)
--(axis cs:0.16916667,0.230154496601865);

\path [draw=color2, line width=0.7000000000000001pt] (axis cs:0.17125,0.237681760708077)
--(axis cs:0.17125,0.244606448553817);

\path [draw=color2, line width=0.7000000000000001pt] (axis cs:0.17333333,0.24870594278489)
--(axis cs:0.17333333,0.25572941362415);

\path [draw=color2, line width=0.7000000000000001pt] (axis cs:0.17541667,0.258732416627153)
--(axis cs:0.17541667,0.265854670459934);

\path [draw=color2, line width=0.7000000000000001pt] (axis cs:0.1775,0.272196538643907)
--(axis cs:0.1775,0.279437332068911);

\path [draw=color2, line width=0.7000000000000001pt] (axis cs:0.17958333,0.284840761814446)
--(axis cs:0.17958333,0.29218033823297);

\path [draw=color2, line width=0.7000000000000001pt] (axis cs:0.18166667,0.295351272324957)
--(axis cs:0.18166667,0.302769875138296);

\path [draw=color2, line width=0.7000000000000001pt] (axis cs:0.18375,0.308924055634582)
--(axis cs:0.18375,0.316431563142089);

\path [draw=color2, line width=0.7000000000000001pt] (axis cs:0.18583333,0.322289394657816)
--(axis cs:0.18583333,0.329885806859491);

\path [draw=color2, line width=0.7000000000000001pt] (axis cs:0.18791667,0.331446578157104)
--(axis cs:0.18791667,0.33909238185554);

\path [draw=color2, line width=0.7000000000000001pt] (axis cs:0.19,0.346076339497392)
--(axis cs:0.19,0.353801169590643);

\path [draw=color2, line width=0.7000000000000001pt] (axis cs:0.19208333,0.360627074442864)
--(axis cs:0.19208333,0.368421052631579);

\path [draw=color2, line width=0.7000000000000001pt] (axis cs:0.19416667,0.368253121542595)
--(axis cs:0.19416667,0.376076734629366);

\path [draw=color2, line width=0.7000000000000001pt] (axis cs:0.19625,0.382882882882883)
--(axis cs:0.19625,0.390765765765766);

\path [draw=color2, line width=0.7000000000000001pt] (axis cs:0.19833333,0.393047652916074)
--(axis cs:0.19833333,0.400970048996365);

\path [draw=color2, line width=0.7000000000000001pt] (axis cs:0.20041667,0.405346135609294)
--(axis cs:0.20041667,0.413308044886992);

\path [draw=color2, line width=0.7000000000000001pt] (axis cs:0.2025,0.417921210684369)
--(axis cs:0.2025,0.425922633159475);

\path [draw=color2, line width=0.7000000000000001pt] (axis cs:0.20458333,0.42908368895211)
--(axis cs:0.20458333,0.437104868025921);

\path [draw=color2, line width=0.7000000000000001pt] (axis cs:0.20666667,0.438458195037142)
--(axis cs:0.20666667,0.446499130709657);

\path [draw=color2, line width=0.7000000000000001pt] (axis cs:0.20875,0.44810929350403)
--(axis cs:0.20875,0.456169985775249);

\path [draw=color2, line width=0.7000000000000001pt] (axis cs:0.21083333,0.462373557768295)
--(axis cs:0.21083333,0.470454006638217);

\path [draw=color2, line width=0.7000000000000001pt] (axis cs:0.21291667,0.47465228386281)
--(axis cs:0.21291667,0.482732732732733);

\path [draw=color2, line width=0.7000000000000001pt] (axis cs:0.215,0.483009325114588)
--(axis cs:0.215,0.491109530583215);

\path [draw=color3, line width=0.7000000000000001pt] (axis cs:0.165,0.184131499920974)
--(axis cs:0.165,0.190443733206891);

\path [draw=color3, line width=0.7000000000000001pt] (axis cs:0.16708333,0.197042437174016)
--(axis cs:0.16708333,0.203522601548917);

\path [draw=color3, line width=0.7000000000000001pt] (axis cs:0.16916667,0.209854591433539)
--(axis cs:0.16916667,0.21648293029872);

\path [draw=color3, line width=0.7000000000000001pt] (axis cs:0.17125,0.224099099099099)
--(axis cs:0.17125,0.230885490753912);

\path [draw=color3, line width=0.7000000000000001pt] (axis cs:0.17333333,0.24133673146831)
--(axis cs:0.17333333,0.248300932511459);

\path [draw=color3, line width=0.7000000000000001pt] (axis cs:0.17541667,0.254030346135609)
--(axis cs:0.17541667,0.261113086770981);

\path [draw=color3, line width=0.7000000000000001pt] (axis cs:0.1775,0.267494468152363)
--(axis cs:0.1775,0.274695748379959);

\path [draw=color3, line width=0.7000000000000001pt] (axis cs:0.17958333,0.279387940572151)
--(axis cs:0.17958333,0.286688003793267);

\path [draw=color3, line width=0.7000000000000001pt] (axis cs:0.18166667,0.299223565670934)
--(axis cs:0.18166667,0.306661925082978);

\path [draw=color3, line width=0.7000000000000001pt] (axis cs:0.18375,0.31493993993994)
--(axis cs:0.18375,0.322486960644855);

\path [draw=color3, line width=0.7000000000000001pt] (axis cs:0.18583333,0.329757388967915)
--(axis cs:0.18583333,0.337393314366999);

\path [draw=color3, line width=0.7000000000000001pt] (axis cs:0.18791667,0.344238975817923)
--(axis cs:0.18791667,0.351953927611822);

\path [draw=color3, line width=0.7000000000000001pt] (axis cs:0.19,0.361891496759918)
--(axis cs:0.19,0.369685474948633);

\path [draw=color3, line width=0.7000000000000001pt] (axis cs:0.19208333,0.376797850482061)
--(axis cs:0.19208333,0.38466097676624);

\path [draw=color3, line width=0.7000000000000001pt] (axis cs:0.19416667,0.390034771613719)
--(axis cs:0.19416667,0.397947289394658);

\path [draw=color3, line width=0.7000000000000001pt] (axis cs:0.19625,0.406630314525051)
--(axis cs:0.19625,0.41459222380275);

\path [draw=color3, line width=0.7000000000000001pt] (axis cs:0.19833333,0.41713094673621)
--(axis cs:0.19833333,0.425122490911965);

\path [draw=color3, line width=0.7000000000000001pt] (axis cs:0.20041667,0.435445313734787)
--(axis cs:0.20041667,0.443486249407302);

\path [draw=color3, line width=0.7000000000000001pt] (axis cs:0.2025,0.452465623518255)
--(axis cs:0.2025,0.460526315789474);

\path [draw=color3, line width=0.7000000000000001pt] (axis cs:0.20458333,0.464289947842579)
--(axis cs:0.20458333,0.472370396712502);

\path [draw=color3, line width=0.7000000000000001pt] (axis cs:0.20666667,0.476420499446815)
--(axis cs:0.20666667,0.484500948316738);

\path [draw=color3, line width=0.7000000000000001pt] (axis cs:0.20875,0.490082187450608)
--(axis cs:0.20875,0.498182392919235);

\path [draw=color3, line width=0.7000000000000001pt] (axis cs:0.21083333,0.503141299193931)
--(axis cs:0.21083333,0.511241504662557);

\path [draw=color3, line width=0.7000000000000001pt] (axis cs:0.21291667,0.515173067804647)
--(axis cs:0.21291667,0.523263394973921);

\path [draw=color3, line width=0.7000000000000001pt] (axis cs:0.215,0.527135688319899)
--(axis cs:0.215,0.535216137189821);

\path [draw=color4, line width=0.7000000000000001pt] (axis cs:0.165,0.169393077287814)
--(axis cs:0.165,0.175507744586692);

\path [draw=color4, line width=0.7000000000000001pt] (axis cs:0.16708333,0.185613244823771)
--(axis cs:0.16708333,0.191945234708393);

\path [draw=color4, line width=0.7000000000000001pt] (axis cs:0.16916667,0.20099375691481)
--(axis cs:0.16916667,0.207523312786471);

\path [draw=color4, line width=0.7000000000000001pt] (axis cs:0.17125,0.216552078394184)
--(axis cs:0.17125,0.22325944365418);

\path [draw=color4, line width=0.7000000000000001pt] (axis cs:0.17333333,0.231004030346136)
--(axis cs:0.17333333,0.237869448395764);

\path [draw=color4, line width=0.7000000000000001pt] (axis cs:0.17541667,0.249920973605184)
--(axis cs:0.17541667,0.256964201043148);

\path [draw=color4, line width=0.7000000000000001pt] (axis cs:0.1775,0.266042358147621)
--(axis cs:0.1775,0.273223881776513);

\path [draw=color4, line width=0.7000000000000001pt] (axis cs:0.17958333,0.286233602023076)
--(axis cs:0.17958333,0.293573178441599);

\path [draw=color4, line width=0.7000000000000001pt] (axis cs:0.18166667,0.300863363363363)
--(axis cs:0.18166667,0.308321479374111);

\path [draw=color4, line width=0.7000000000000001pt] (axis cs:0.18375,0.320313734787419)
--(axis cs:0.18375,0.32789039039039);

\path [draw=color4, line width=0.7000000000000001pt] (axis cs:0.18583333,0.341789157578631)
--(axis cs:0.18583333,0.349484352773826);

\path [draw=color4, line width=0.7000000000000001pt] (axis cs:0.18791667,0.355648411569464)
--(axis cs:0.18791667,0.363422633159475);

\path [draw=color4, line width=0.7000000000000001pt] (axis cs:0.19,0.375088904694168)
--(axis cs:0.19,0.382942152678995);

\path [draw=color4, line width=0.7000000000000001pt] (axis cs:0.19208333,0.391664691006796)
--(axis cs:0.19208333,0.399587087087087);

\path [draw=color4, line width=0.7000000000000001pt] (axis cs:0.19416667,0.409080132764343)
--(axis cs:0.19416667,0.417051920341394);

\path [draw=color4, line width=0.7000000000000001pt] (axis cs:0.19625,0.426831436699858)
--(axis cs:0.19625,0.434852615773668);

\path [draw=color4, line width=0.7000000000000001pt] (axis cs:0.19833333,0.441925478109689)
--(axis cs:0.19833333,0.449966413782203);

\path [draw=color4, line width=0.7000000000000001pt] (axis cs:0.20041667,0.458817370001581)
--(axis cs:0.20041667,0.466897818871503);

\path [draw=color4, line width=0.7000000000000001pt] (axis cs:0.2025,0.47689465781571)
--(axis cs:0.2025,0.484984984984985);

\path [draw=color4, line width=0.7000000000000001pt] (axis cs:0.20458333,0.492344318002213)
--(axis cs:0.20458333,0.500444523470839);

\path [draw=color4, line width=0.7000000000000001pt] (axis cs:0.20666667,0.509176940097993)
--(axis cs:0.20666667,0.517277145566619);

\path [draw=color4, line width=0.7000000000000001pt] (axis cs:0.20875,0.521742136873716)
--(axis cs:0.20875,0.529822585743638);

\path [draw=color4, line width=0.7000000000000001pt] (axis cs:0.21083333,0.53988857278331)
--(axis cs:0.21083333,0.547949265054528);

\path [draw=color4, line width=0.7000000000000001pt] (axis cs:0.21291667,0.549786628733997)
--(axis cs:0.21291667,0.557827564406512);

\path [draw=color4, line width=0.7000000000000001pt] (axis cs:0.215,0.561907302038881)
--(axis cs:0.215,0.569928481112692);

\path [draw=color5, line width=0.7000000000000001pt] (axis cs:0.165,0.156353722143196)
--(axis cs:0.165,0.162280701754386);

\path [draw=color5, line width=0.7000000000000001pt] (axis cs:0.16708333,0.173749407302039)
--(axis cs:0.16708333,0.179923344397029);

\path [draw=color5, line width=0.7000000000000001pt] (axis cs:0.16916667,0.191006796269954)
--(axis cs:0.16916667,0.197417812549392);

\path [draw=color5, line width=0.7000000000000001pt] (axis cs:0.17125,0.209844713134187)
--(axis cs:0.17125,0.216473051999368);

\path [draw=color5, line width=0.7000000000000001pt] (axis cs:0.17333333,0.227111980401454)
--(axis cs:0.17333333,0.233937885253675);

\path [draw=color5, line width=0.7000000000000001pt] (axis cs:0.17541667,0.246522838628102)
--(axis cs:0.17541667,0.253526552868658);

\path [draw=color5, line width=0.7000000000000001pt] (axis cs:0.1775,0.26578552236447)
--(axis cs:0.1775,0.272967045993362);

\path [draw=color5, line width=0.7000000000000001pt] (axis cs:0.17958333,0.285423581476213)
--(axis cs:0.17958333,0.292763157894737);

\path [draw=color5, line width=0.7000000000000001pt] (axis cs:0.18166667,0.304785048206101)
--(axis cs:0.18166667,0.312262920815552);

\path [draw=color5, line width=0.7000000000000001pt] (axis cs:0.18375,0.326408645487593)
--(axis cs:0.18375,0.334024814287972);

\path [draw=color5, line width=0.7000000000000001pt] (axis cs:0.18583333,0.345908408408408)
--(axis cs:0.18583333,0.35363323850166);

\path [draw=color5, line width=0.7000000000000001pt] (axis cs:0.18791667,0.367235656709341)
--(axis cs:0.18791667,0.375059269796112);

\path [draw=color5, line width=0.7000000000000001pt] (axis cs:0.19,0.389224751066856)
--(axis cs:0.19,0.397137268847795);

\path [draw=color5, line width=0.7000000000000001pt] (axis cs:0.19208333,0.407628022759602)
--(axis cs:0.19208333,0.4155899320373);

\path [draw=color5, line width=0.7000000000000001pt] (axis cs:0.19416667,0.427700727042832)
--(axis cs:0.19416667,0.435721906116643);

\path [draw=color5, line width=0.7000000000000001pt] (axis cs:0.19625,0.446103998735578)
--(axis cs:0.19625,0.454164691006796);

\path [draw=color5, line width=0.7000000000000001pt] (axis cs:0.19833333,0.468340050576893)
--(axis cs:0.19833333,0.476420499446815);

\path [draw=color5, line width=0.7000000000000001pt] (axis cs:0.20041667,0.484046546546547)
--(axis cs:0.20041667,0.492146752015173);

\path [draw=color5, line width=0.7000000000000001pt] (axis cs:0.2025,0.500582819661767)
--(axis cs:0.2025,0.508683025130394);

\path [draw=color5, line width=0.7000000000000001pt] (axis cs:0.20458333,0.51778093883357)
--(axis cs:0.20458333,0.525861387703493);

\path [draw=color5, line width=0.7000000000000001pt] (axis cs:0.20666667,0.534633317528054)
--(axis cs:0.20666667,0.542703888098625);

\path [draw=color5, line width=0.7000000000000001pt] (axis cs:0.20875,0.549994073020389)
--(axis cs:0.20875,0.558035008692903);

\path [draw=color5, line width=0.7000000000000001pt] (axis cs:0.21083333,0.564821400347716)
--(axis cs:0.21083333,0.572842579421527);

\path [draw=color5, line width=0.7000000000000001pt] (axis cs:0.21291667,0.581930614825352)
--(axis cs:0.21291667,0.589912280701754);

\path [draw=color5, line width=0.7000000000000001pt] (axis cs:0.215,0.592925162004109)
--(axis cs:0.215,0.600867314683104);

\addplot [line width=0.7000000000000001pt, color1]
table {%
0.170148186086966 0.243784682108514
0.170295492248582 0.244383279827113
0.170442798410198 0.244982384930644
0.170590104571814 0.245581997419107
0.170737410733431 0.246182117292502
0.170884716895047 0.246782744550829
0.171032023056663 0.247383879194087
0.171179329218279 0.247985521222278
0.171326635379895 0.2485876706354
0.171473941541511 0.249190327433454
0.171621247703128 0.24979349161644
0.171768553864744 0.250397163184358
0.17191586002636 0.251001342137208
0.172063166187976 0.25160602847499
0.172210472349592 0.252211222197704
0.172357778511208 0.252816923305349
0.172505084672825 0.253423131797927
0.172652390834441 0.254029847675436
0.172799696996057 0.254637070937877
0.172947003157673 0.25524480158525
0.173094309319289 0.255853039617555
0.173241615480905 0.256461785034792
0.173388921642522 0.25707103783696
0.173536227804138 0.257680798024061
0.173683533965754 0.258291065596093
0.17383084012737 0.258901840553058
0.173978146288986 0.259513122894954
0.174125452450602 0.260124912621782
0.174272758612218 0.260737209733542
0.174420064773835 0.261350014230234
0.174567370935451 0.261963326111857
0.174714677097067 0.262577145378413
0.174861983258683 0.2631914720299
0.175009289420299 0.26380630606632
0.175156595581915 0.264421647487671
0.175303901743532 0.265037496293954
0.175451207905148 0.265653852485169
0.175598514066764 0.266270716061316
0.17574582022838 0.266888087022394
0.175893126389996 0.267505965368405
0.176040432551612 0.268124351099347
0.176187738713229 0.268743244215222
0.176335044874845 0.269362644716028
0.176482351036461 0.269982552601766
0.176629657198077 0.270602967872436
0.176776963359693 0.271223890528038
0.176924269521309 0.271845320568572
0.177071575682926 0.272467257994037
0.177218881844542 0.273089702804435
0.177366188006158 0.273712654999764
0.177513494167774 0.274336114580025
0.17766080032939 0.274960081545219
0.177808106491006 0.275584555895343
0.177955412652623 0.2762095376304
0.178102718814239 0.276835026750389
0.178250024975855 0.27746102325531
0.178397331137471 0.278087527145162
0.178544637299087 0.278714538419947
0.178691943460703 0.279342057079663
0.178839249622319 0.279970083124311
0.178986555783936 0.280598616553891
0.179133861945552 0.281227657368403
0.179281168107168 0.281857205567847
0.179428474268784 0.282487261152223
0.1795757804304 0.28311782412153
0.179723086592016 0.283748894475769
0.179870392753633 0.284380472214941
0.180017698915249 0.285012557339044
0.180165005076865 0.285645149848079
0.180312311238481 0.286278249742046
0.180459617400097 0.286911857020945
0.180606923561713 0.287545971684775
0.18075422972333 0.288180593733538
0.180901535884946 0.288815723167232
0.181048842046562 0.289451359985859
0.181196148208178 0.290087504189417
0.181343454369794 0.290724155777907
0.18149076053141 0.291361314751329
0.181638066693027 0.291998981109683
0.181785372854643 0.292637154852968
0.181932679016259 0.293275835981186
0.182079985177875 0.293915024494335
0.182227291339491 0.294554720392417
0.182374597501107 0.29519492367543
0.182521903662724 0.295835634343375
0.18266920982434 0.296476852396252
0.182816515985956 0.297118577834061
0.182963822147572 0.297760810656802
0.183111128309188 0.298403550864474
0.183258434470804 0.299046798457079
0.18340574063242 0.299690553434615
0.183553046794037 0.300334815797083
0.183700352955653 0.300979585544483
0.183847659117269 0.301624862676815
0.183994965278885 0.302270647194079
0.184142271440501 0.302916939096275
0.184289577602117 0.303563738383403
0.184436883763734 0.304211045055462
0.18458418992535 0.304858859112454
0.184731496086966 0.305507180554377
};
\addplot [line width=0.7000000000000001pt, color2]
table {%
0.170148186086966 0.231813486901284
0.170295492248582 0.232613676067192
0.170442798410198 0.23341480345098
0.170590104571814 0.234216869052648
0.170737410733431 0.235019872872196
0.170884716895047 0.235823814909623
0.171032023056663 0.23662869516493
0.171179329218279 0.237434513638117
0.171326635379895 0.238241270329184
0.171473941541511 0.23904896523813
0.171621247703128 0.239857598364956
0.171768553864744 0.240667169709662
0.17191586002636 0.241477679272248
0.172063166187976 0.242289127052714
0.172210472349592 0.243101513051059
0.172357778511208 0.243914837267284
0.172505084672825 0.244729099701389
0.172652390834441 0.245544300353373
0.172799696996057 0.246360439223237
0.172947003157673 0.247177516310981
0.173094309319289 0.247995531616605
0.173241615480905 0.248814485140109
0.173388921642522 0.249634376881492
0.173536227804138 0.250455206840755
0.173683533965754 0.251276975017898
0.17383084012737 0.252099681412921
0.173978146288986 0.252923326025823
0.174125452450602 0.253747908856605
0.174272758612218 0.254573429905267
0.174420064773835 0.255399889171809
0.174567370935451 0.25622728665623
0.174714677097067 0.257055622358532
0.174861983258683 0.257884896278712
0.175009289420299 0.258715108416773
0.175156595581915 0.259546258772714
0.175303901743532 0.260378347346534
0.175451207905148 0.261211374138234
0.175598514066764 0.262045339147814
0.17574582022838 0.262880242375273
0.175893126389996 0.263716083820613
0.176040432551612 0.264552863483832
0.176187738713229 0.26539058136493
0.176335044874845 0.266229237463909
0.176482351036461 0.267068831780767
0.176629657198077 0.267909364315505
0.176776963359693 0.268750835068123
0.176924269521309 0.269593244038621
0.177071575682926 0.270436591226998
0.177218881844542 0.271280876633255
0.177366188006158 0.272126100257392
0.177513494167774 0.272972262099409
0.17766080032939 0.273819362159305
0.177808106491006 0.274667400437082
0.177955412652623 0.275516376932738
0.178102718814239 0.276366291646273
0.178250024975855 0.277217144577689
0.178397331137471 0.278068935726984
0.178544637299087 0.278921665094159
0.178691943460703 0.279775332679214
0.178839249622319 0.280629938482148
0.178986555783936 0.281485482502963
0.179133861945552 0.282341964741657
0.179281168107168 0.283199385198231
0.179428474268784 0.284057743872684
0.1795757804304 0.284917040765018
0.179723086592016 0.285777275875231
0.179870392753633 0.286638449203324
0.180017698915249 0.287500560749296
0.180165005076865 0.288363610513148
0.180312311238481 0.289227598494881
0.180459617400097 0.290092524694493
0.180606923561713 0.290958389111984
0.18075422972333 0.291825191747356
0.180901535884946 0.292692932600607
0.181048842046562 0.293561611671738
0.181196148208178 0.294431228960749
0.181343454369794 0.295301784467639
0.18149076053141 0.296173278192409
0.181638066693027 0.297045710135059
0.181785372854643 0.297919080295589
0.181932679016259 0.298793388673999
0.182079985177875 0.299668635270288
0.182227291339491 0.300544820084457
0.182374597501107 0.301421943116506
0.182521903662724 0.302300004366434
0.18266920982434 0.303179003834243
0.182816515985956 0.304058941519931
0.182963822147572 0.304939817423499
0.183111128309188 0.305821631544946
0.183258434470804 0.306704383884274
0.18340574063242 0.307588074441481
0.183553046794037 0.308472703216568
0.183700352955653 0.309358270209535
0.183847659117269 0.310244775420381
0.183994965278885 0.311132218849107
0.184142271440501 0.312020600495713
0.184289577602117 0.312909920360199
0.184436883763734 0.313800178442564
0.18458418992535 0.31469137474281
0.184731496086966 0.315583509260935
};
\addplot [line width=0.7000000000000001pt, color3]
table {%
0.170148186086966 0.221363957889034
0.170295492248582 0.222336912575556
0.170442798410198 0.223311297793778
0.170590104571814 0.224287113543702
0.170737410733431 0.225264359825326
0.170884716895047 0.226243036638651
0.171032023056663 0.227223143983678
0.171179329218279 0.228204681860405
0.171326635379895 0.229187650268833
0.171473941541511 0.230172049208962
0.171621247703128 0.231157878680792
0.171768553864744 0.232145138684323
0.17191586002636 0.233133829219554
0.172063166187976 0.234123950286487
0.172210472349592 0.235115501885121
0.172357778511208 0.236108484015455
0.172505084672825 0.23710289667749
0.172652390834441 0.238098739871227
0.172799696996057 0.239096013596664
0.172947003157673 0.240094717853802
0.173094309319289 0.241094852642641
0.173241615480905 0.242096417963181
0.173388921642522 0.243099413815422
0.173536227804138 0.244103840199364
0.173683533965754 0.245109697115007
0.17383084012737 0.24611698456235
0.173978146288986 0.247125702541395
0.174125452450602 0.24813585105214
0.174272758612218 0.249147430094587
0.174420064773835 0.250160439668734
0.174567370935451 0.251174879774582
0.174714677097067 0.252190750412132
0.174861983258683 0.253208051581382
0.175009289420299 0.254226783282333
0.175156595581915 0.255246945514985
0.175303901743532 0.256268538279337
0.175451207905148 0.257291561575391
0.175598514066764 0.258316015403146
0.17574582022838 0.259341899762601
0.175893126389996 0.260369214653758
0.176040432551612 0.261397960076615
0.176187738713229 0.262428136031173
0.176335044874845 0.263459742517433
0.176482351036461 0.264492779535393
0.176629657198077 0.265527247085054
0.176776963359693 0.266563145166416
0.176924269521309 0.267600473779479
0.177071575682926 0.268639232924243
0.177218881844542 0.269679422600707
0.177366188006158 0.270721042808873
0.177513494167774 0.271764093548739
0.17766080032939 0.272808574820307
0.177808106491006 0.273854486623575
0.177955412652623 0.274901828958544
0.178102718814239 0.275950601825215
0.178250024975855 0.277000805223586
0.178397331137471 0.278052439153658
0.178544637299087 0.279105503615431
0.178691943460703 0.280159998608905
0.178839249622319 0.28121592413408
0.178986555783936 0.282273280190955
0.179133861945552 0.283332066779532
0.179281168107168 0.284392283899809
0.179428474268784 0.285453931551788
0.1795757804304 0.286517009735467
0.179723086592016 0.287581518450847
0.179870392753633 0.288647457697929
0.180017698915249 0.289714827476711
0.180165005076865 0.290783627787194
0.180312311238481 0.291853858629378
0.180459617400097 0.292925520003263
0.180606923561713 0.293998611908849
0.18075422972333 0.295073134346135
0.180901535884946 0.296149087315123
0.181048842046562 0.297226470815811
0.181196148208178 0.298305284848201
0.181343454369794 0.299385529412291
0.18149076053141 0.300467204508083
0.181638066693027 0.301550310135575
0.181785372854643 0.302634846294768
0.181932679016259 0.303720812985662
0.182079985177875 0.304808210208257
0.182227291339491 0.305897037962553
0.182374597501107 0.306987296248549
0.182521903662724 0.308078985066247
0.18266920982434 0.309172104415646
0.182816515985956 0.310266654296745
0.182963822147572 0.311362634709546
0.183111128309188 0.312460045654047
0.183258434470804 0.313558887130249
0.18340574063242 0.314659159138153
0.183553046794037 0.315760861677757
0.183700352955653 0.316863994749062
0.183847659117269 0.317968558352068
0.183994965278885 0.319074552486774
0.184142271440501 0.320181977153182
0.184289577602117 0.321290832351291
0.184436883763734 0.3224011180811
0.18458418992535 0.323512834342611
0.184731496086966 0.324625981135822
};
\addplot [line width=0.7000000000000001pt, color4]
table {%
0.170148186086966 0.211923740711442
0.170295492248582 0.213050083418846
0.170442798410198 0.214178399099091
0.170590104571814 0.215308687752177
0.170737410733431 0.216440949378105
0.170884716895047 0.217575183976874
0.171032023056663 0.218711391548485
0.171179329218279 0.219849572092936
0.171326635379895 0.22098972561023
0.171473941541511 0.222131852100364
0.171621247703128 0.22327595156334
0.171768553864744 0.224422023999157
0.17191586002636 0.225570069407816
0.172063166187976 0.226720087789316
0.172210472349592 0.227872079143657
0.172357778511208 0.22902604347084
0.172505084672825 0.230181980770864
0.172652390834441 0.231339891043729
0.172799696996057 0.232499774289436
0.172947003157673 0.233661630507984
0.173094309319289 0.234825459699374
0.173241615480905 0.235991261863605
0.173388921642522 0.237159037000677
0.173536227804138 0.238328785110591
0.173683533965754 0.239500506193345
0.17383084012737 0.240674200248942
0.173978146288986 0.241849867277379
0.174125452450602 0.243027507278658
0.174272758612218 0.244207120252779
0.174420064773835 0.245388706199741
0.174567370935451 0.246572265119544
0.174714677097067 0.247757797012188
0.174861983258683 0.248945301877674
0.175009289420299 0.250134779716001
0.175156595581915 0.25132623052717
0.175303901743532 0.25251965431118
0.175451207905148 0.253715051068031
0.175598514066764 0.254912420797724
0.17574582022838 0.256111763500258
0.175893126389996 0.257313079175633
0.176040432551612 0.25851636782385
0.176187738713229 0.259721629444908
0.176335044874845 0.260928864038807
0.176482351036461 0.262138071605548
0.176629657198077 0.26334925214513
0.176776963359693 0.264562405657554
0.176924269521309 0.265777532142819
0.177071575682926 0.266994631600925
0.177218881844542 0.268213704031873
0.177366188006158 0.269434749435661
0.177513494167774 0.270657767812292
0.17766080032939 0.271882759161763
0.177808106491006 0.273109723484076
0.177955412652623 0.274338660779231
0.178102718814239 0.275569571047227
0.178250024975855 0.276802454288064
0.178397331137471 0.278037310501742
0.178544637299087 0.279274139688262
0.178691943460703 0.280512941847624
0.178839249622319 0.281753716979826
0.178986555783936 0.28299646508487
0.179133861945552 0.284241186162755
0.179281168107168 0.285487880213482
0.179428474268784 0.28673654723705
0.1795757804304 0.28798718723346
0.179723086592016 0.28923980020271
0.179870392753633 0.290494386144802
0.180017698915249 0.291750945059736
0.180165005076865 0.293009476947511
0.180312311238481 0.294269981808127
0.180459617400097 0.295532459641584
0.180606923561713 0.296796910447883
0.18075422972333 0.298063334227024
0.180901535884946 0.299331730979005
0.181048842046562 0.300602100703828
0.181196148208178 0.301874443401493
0.181343454369794 0.303148759071998
0.18149076053141 0.304425047715346
0.181638066693027 0.305703309331534
0.181785372854643 0.306983543920564
0.181932679016259 0.308265751482435
0.182079985177875 0.309549932017147
0.182227291339491 0.310836085524701
0.182374597501107 0.312124212005097
0.182521903662724 0.313414311458333
0.18266920982434 0.314706383884411
0.182816515985956 0.316000429283331
0.182963822147572 0.317296447655091
0.183111128309188 0.318594438999693
0.183258434470804 0.319894403317137
0.18340574063242 0.321196340607422
0.183553046794037 0.322500250870548
0.183700352955653 0.323806134106515
0.183847659117269 0.325113990315324
0.183994965278885 0.326423819496974
0.184142271440501 0.327735621651466
0.184289577602117 0.329049396778799
0.184436883763734 0.330365144878973
0.18458418992535 0.331682865951989
0.184731496086966 0.333002559997846
};
\addplot [line width=0.7000000000000001pt, color5]
table {%
0.170148186086966 0.203227360933094
0.170295492248582 0.204492644502507
0.170442798410198 0.205760486471404
0.170590104571814 0.207030886839783
0.170737410733431 0.208303845607647
0.170884716895047 0.209579362774994
0.171032023056663 0.210857438341824
0.171179329218279 0.212138072308138
0.171326635379895 0.213421264673935
0.171473941541511 0.214707015439216
0.171621247703128 0.21599532460398
0.171768553864744 0.217286192168228
0.17191586002636 0.218579618131959
0.172063166187976 0.219875602495174
0.172210472349592 0.221174145257873
0.172357778511208 0.222475246420055
0.172505084672825 0.22377890598172
0.172652390834441 0.225085123942868
0.172799696996057 0.226393900303501
0.172947003157673 0.227705235063617
0.173094309319289 0.229019128223216
0.173241615480905 0.230335579782299
0.173388921642522 0.231654589740865
0.173536227804138 0.232976158098915
0.173683533965754 0.234300284856448
0.17383084012737 0.235626970013465
0.173978146288986 0.236956213569965
0.174125452450602 0.238288015525949
0.174272758612218 0.239622375881416
0.174420064773835 0.240959294636367
0.174567370935451 0.242298771790801
0.174714677097067 0.243640807344719
0.174861983258683 0.24498540129812
0.175009289420299 0.246332553651005
0.175156595581915 0.247682264403373
0.175303901743532 0.249034533555225
0.175451207905148 0.25038936110656
0.175598514066764 0.251746747057379
0.17574582022838 0.253106691407681
0.175893126389996 0.254469194157467
0.176040432551612 0.255834255306736
0.176187738713229 0.257201874855489
0.176335044874845 0.258572052803725
0.176482351036461 0.259944789151445
0.176629657198077 0.261320083898648
0.176776963359693 0.262697937045335
0.176924269521309 0.264078348591505
0.177071575682926 0.265461318537159
0.177218881844542 0.266846846882296
0.177366188006158 0.268234933626917
0.177513494167774 0.269625578771021
0.17766080032939 0.271018782314609
0.177808106491006 0.27241454425768
0.177955412652623 0.273812864600235
0.178102718814239 0.275213743342273
0.178250024975855 0.276617180483795
0.178397331137471 0.2780231760248
0.178544637299087 0.279431729965288
0.178691943460703 0.280842842305261
0.178839249622319 0.282256513044716
0.178986555783936 0.283672742183655
0.179133861945552 0.285091529722078
0.179281168107168 0.286512875659984
0.179428474268784 0.287936779997374
0.1795757804304 0.289363242734247
0.179723086592016 0.290792263870604
0.179870392753633 0.292223843406444
0.180017698915249 0.293657981341768
0.180165005076865 0.295094677676575
0.180312311238481 0.296533932410866
0.180459617400097 0.29797574554464
0.180606923561713 0.299420117077898
0.18075422972333 0.300867047010639
0.180901535884946 0.302316535342863
0.181048842046562 0.303768582074571
0.181196148208178 0.305223187205763
0.181343454369794 0.306680350736438
0.18149076053141 0.308140072666597
0.181638066693027 0.309602352996239
0.181785372854643 0.311067191725365
0.181932679016259 0.312534588853974
0.182079985177875 0.314004544382066
0.182227291339491 0.315477058309642
0.182374597501107 0.316952130636702
0.182521903662724 0.318429761363245
0.18266920982434 0.319909950489272
0.182816515985956 0.321392698014782
0.182963822147572 0.322878003939775
0.183111128309188 0.324365868264252
0.183258434470804 0.325856290988213
0.18340574063242 0.327349272111657
0.183553046794037 0.328844811634585
0.183700352955653 0.330342909556996
0.183847659117269 0.33184356587889
0.183994965278885 0.333346780600268
0.184142271440501 0.33485255372113
0.184289577602117 0.336360885241475
0.184436883763734 0.337871775161304
0.18458418992535 0.339385223480616
0.184731496086966 0.340901230199411
};
\path [draw=white, fill opacity=0] (axis cs:0,0.1341280425162)
--(axis cs:0,0.6230929943101);

\path [draw=white, fill opacity=0] (axis cs:1,0.1341280425162)
--(axis cs:1,0.6230929943101);

\path [draw=white, fill opacity=0] (axis cs:0.1625,0)
--(axis cs:0.2175,0);

\path [draw=white, fill opacity=0] (axis cs:0.1625,1)
--(axis cs:0.2175,1);

\end{axis}

\end{tikzpicture}

%% file: smooth_rough_traversal.tex
\begin{tikzpicture}[rotate=90, transform shape, x = 17pt, y = 17pt, gridline/.style = {black, line width = 2pt, line join = round, line cap = round}]

\begin{scope}[shift = {(0, 0)}]
    \foreach \x in {0, 2, ..., 8}{
        \draw[gridline, black!50!white] (\x, 0) -- (\x, 10);
    }
    \foreach \y in {2, 4, ..., 8}{
        \draw[gridline, black!50!white] (0, \y) -- (8, \y);
    }
    \draw[gridline, step=2, shift = {(-1,-1)}] (2, 2) grid (8, 10);
    \foreach \y in {1, 3, ..., 9}{
        \draw[gridline] (-1, 5) -- (1, \y);
        \draw[gridline] (9, 5) -- (7, \y);
    }
    \foreach \x in {1, 3, ..., 7}{
        \foreach \y in {1, 3, ..., 9}{
            \node (check_\x_\y) at (\x, \y) [circle, line width = 2pt, draw = black, fill=white, minimum size = 6 pt, inner sep = 0 pt] {};
        }
    }
    \node (check_-1_5) at (-1, 5) [circle, line width = 2pt, draw = black, fill=quantumviolet, minimum size = 6 pt, inner sep = 0 pt] {};
    \node (check_9_5) at (9, 5) [circle, line width = 2pt, draw = black, fill=quantumviolet, minimum size = 6 pt, inner sep = 0 pt] {};
\end{scope}

\end{tikzpicture}

%% file: rotated_traversal.tex
\begin{tikzpicture}[x = 12pt, y = 12pt, gridline/.style = {black, line width = 2pt, line join = round, line cap = round}]

\pgfdeclarelayer{foreground}
\pgfsetlayers{main,foreground}

\begin{scope}[shift = {(0, 0)}, opacity=0.5]
\foreach \x/\y in {1/1, 1/5, 3/3, 3/7, 5/1, 5/5, 7/3, 7/7}{
    \fill[white] (\x, \y) rectangle +(2,2);
}
\foreach \x/\y in {1/3, 1/7, 3/1, 3/5, 5/3, 5/7, 7/1, 7/5}{
    \fill[quantumgray!50!white] (\x, \y) rectangle +(2,2);
}
\draw[gridline, step=2, shift = {(-1,-1)}] (2, 2) grid (10, 10);
\foreach \x/\y in {3/9, 7/9}{
    \filldraw[gridline, fill = white] (\x, \y) arc (0:180:1) -- cycle;
}
\foreach \x/\y in {9/3, 9/7}{
    \filldraw[gridline, fill = quantumgray!50!white] (\x, \y) arc (-90:90:1) -- cycle;
}
\foreach \x/\y in {1/3, 1/7}{
    \filldraw[gridline, fill = quantumgray!50!white] (\x, \y) arc (90:270:1) -- cycle;
}
\foreach \x/\y in {3/1, 7/1}{
    \filldraw[gridline, fill = white] (\x, \y) arc (180:360:1) -- cycle;
}
\end{scope}


\begin{pgfonlayer}{foreground}
\foreach \x/\y in {0/2, 0/6, 0/10, 2/0, 2/4, 2/8, 4/2, 4/6, 4/10, 6/0, 6/4, 6/8, 8/2, 8/6, 8/10, 10/0, 10/4, 10/8}{
    \node (check_\x_\y) at (\x, \y) [circle, line width = 2pt, draw = black, fill=white, minimum size = 6 pt, inner sep = 0 pt] {};
}
    \node (check_6_-2) at (6,-2) [circle, line width = 2pt, draw = black, fill=quantumviolet, minimum size = 6 pt, inner sep = 0 pt] {};
    \node (check_4_12) at (4,12) [circle, line width = 2pt, draw = black, fill=quantumviolet, minimum size = 6 pt, inner sep = 0 pt] {};
\end{pgfonlayer}
\foreach \x/\y in {2/4, 2/8, 4/2, 4/6, 6/4, 6/8, 8/2, 8/6}{
    \foreach \sx/\sy in {2/2, -2/2, 2/-2, -2/-2} {
        \draw[gridline] (check_\x_\y) --++ (\sx, \sy); 
    }
}

\draw[gridline] (2, 0) -- (0, 2);
\draw[gridline] (10, 8) -- (8, 10);

\foreach \x/\y in {2/0, 6/0, 10/0}{
    \draw[gridline, quantumgray] (\x, \y) -- (6, -2);
}

\foreach \x/\y in {0/10, 4/10, 8/10}{
    \draw[gridline, quantumgray] (\x, \y) -- (4, 12);
}


\end{tikzpicture}